\documentclass[fleqn,usenatbib]{mnras}

\usepackage{newtxtext,newtxmath}
\usepackage[T1]{fontenc}

\DeclareRobustCommand{\VAN}[3]{#2}
\let\VANthebibliography\thebibliography
\def\thebibliography{\DeclareRobustCommand{\VAN}[3]{##3}\VANthebibliography}

\usepackage{graphicx}	% Including figure files
\usepackage{amsmath}	% Advanced maths commands
\usepackage{threeparttable}
\usepackage{pdflscape}

\newcommand{\rev}[1]{#1}

\title[Variability survey of young objects]{A Near-infrared Variability Survey of Young Planetary-mass Objects}

\author[P. Liu et al.]{
Pengyu Liu,$^{1,2,3}$\thanks{E-mail: pengyu.liu@ed.ac.uk}
Beth A. Biller,$^{1,2}$
Johanna M. Vos,$^{4,5}$
Niall Whiteford, $^{5}$
Zhoujian Zhang,$^{6}$\thanks{NASA Sagan Fellow}
Michael C. Liu,$^{7}$
\newauthor Cl\'emence Fontanive,$^{8,9}$
Elena Manjavacas,$^{10,11}$
Thomas Henning,$^{12}$
Matthew A. Kenworthy,$^{3}$
\newauthor Mariangela Bonavita,$^{1,2}$
Mickaël Bonnefoy,$^{13}$
Emma Bubb,$^{1,2}$
Simon Petrus,$^{13, 14}$
and Joshua Schlieder$^{15}$
\\
$^{1}$Institute for Astronomy, University of Edinburgh, Royal Observatory, Edinburgh EH9 3HJ, UK\\
$^{2}$Centre for Exoplanet Science, University of Edinburgh, Edinburgh, UK\\
$^{3}$Leiden Observatory, Leiden University, PO Box 9513, 2300 RA Leiden, The Netherlands\\
$^{4}$School of Physics, Trinity College Dublin, The University of Dublin, Dublin 2, Ireland\\
$^{5}$Department of Astrophysics, American Museum of Natural History, New York, NY 10024, USA\\
$^{6}$Department of Astronomy \& Astrophysics, University of California, Santa Cruz, CA 95064, USA\\
$^{7}$Institute for Astronomy, University of Hawaii at Manoa, Honolulu, HI 96822, USA\\
$^{8}$D\'epartement de Physique and Observatoire du Mont-M\'egantic, Universit\'e de Montr\'eal, C.P. 6128, Succ. Centre-ville, Montr\'eal, H3C 3J7, Québec, Canada
\\
$^{9}$Institut Trottier de Recherche sur les exoplan\`etes, Universit\'e de Montr\'eal, Québec, Canada
\\
$^{10}$AURA for the European Space Agency (ESA), ESA Office, Space Telescope Science Institute, 3700 San Martin Drive, Baltimore, MD 21218 USA\\
$^{11}$Department of Physics and Astronomy, Johns Hopkins University, Baltimore, MD 21218, USA\\
$^{12}$Max-Planck-Institut für Astronomie, Königstuhl 17, 69117 Heidelberg, Germany\\
$^{13}$Universit\'e Grenoble Alpes, CNRS, IPAG, 38000 Grenoble, France\\
$^{14}$N\'ucleo Milenio Formaci\'on Planetaria–NPF, Universidad de Valpara\'iso, Av.GranBreta\~na1111, Valpara\'iso, Chile\\
$^{15}$Exoplanets and Stellar Astrophysics Laboratory, Code 667, NASA Goddard Space Flight Center, Greenbelt, MD, USA 
}

\date{Accepted XXX. Received YYY; in original form ZZZ}

\pubyear{2023}

\defcitealias{Vos2019}{Vos2019}
\defcitealias{Radigan2014a}{Radigan2014a}
\defcitealias{Vos2022}{Vos2022}
\begin{document}
\label{firstpage}
\pagerange{\pageref{firstpage}--\pageref{lastpage}}
\maketitle

\begin{abstract}
% This is a simple template for authors to write new MNRAS papers.
% The abstract should briefly describe the aims, methods, and main results of the paper.
% It should be a single paragraph not more than 250 words (200 words for Letters).
% No references should appear in the abstract.
We present a photometric variability survey of young planetary-mass objects using the New Technology Telescope in the $J_S$ and $K_S$ bands. 
Surface gravity plays an important role in the atmospheric structure of brown dwarfs, as young low gravity L dwarfs have a higher variability rate than field L dwarfs.
In this study, we extend variability studies to young T-type planetary-mass objects and investigate the effects of surface gravity on the variability of L and T dwarfs across a large sample. 
We conduct continuous monitoring for 18 objects with spectral types from L5 to T8 and detect four new variables and two variable candidates.
Combining with previous variability surveys of field and young L and T objects, we find that young objects tend to be more variable than field objects within peak-to-peak variability amplitude ranges of 0.5\%--10\% and period ranges of 1.5--20\,hr.
For the first time, we constrain the variability rate of young T dwarfs to be $56_{-18}^{+20}$\% compared to \rev{$25_{-7}^{+8}$\%} for field T dwarfs. 
Both field and young samples have higher variability rates at the L/T transition than outside the L/T transition. \rev{The differences in the variability rates between field and young samples are about 1$\sigma$ and therefore larger sample sizes are needed to confirm and refine the results.}
Besides the L/T transition, young L dwarfs with strong variability tend to assemble in a narrow spectral type range of L6--L7.5. This work supports the critical role of surface gravity on the atmospheric structure from L to T spectral types.
% In addition, we detect different periods in the $J_S$- and $K_S$-band light curves of a young L7 object W1147-2040, indicating it may have multi-layer heterogeneous clouds with different wind speeds. 
% Another young L7 binary 2M1119-1147 may have a second shortest rotation period of 3.02\,hr between young objects if confirmed with resolved observations of the two components.
\end{abstract}


%%%%%%%%%%%%%%%%%%%%%%%%%%%%%%%%%%%%%%%%%%%%%%%%%%

%%%%%%%%%%%%%%%%% BODY OF PAPER %%%%%%%%%%%%%%%%%%

\section{Introduction}
\label{sec:introduction}
\rev{The first direct imaging and spectroscopy observations of planetary mass objects outside of the solar system can be dated back more than 20 years ago \citep{Oasa1999, Lucas2000, ZapateroOsorio2000, Lucas2001}. These objects were discovered in nearby star-forming regions and are the first of a large population of free-floating exoplanets which are now discovered in young stellar associations such as Upper Scorpius and Ophiuchus \citep{MiretRoig2022,Bouy2022}.
In recent years, dozens of young brown dwarfs with planetary mass have been discovered by spectroscopy or kinematic characterisation in the field or nearby young moving groups \citep[e.g.][]{LiuM2013, Allers2013, Gagne2015, Gagne2017, Gagne2018, LiuM2016, Faherty2016, Schneider2016, Schneider2017, ZhangZ2021}}
These objects share similar properties with directly-imaged exoplanets, including mass, surface gravity, effective temperature, and spectral type. The study of these ultracool planetary-mass objects provides a unique opportunity to understand the atmospheres of giant exoplanets, as observations of giant exoplanets are often hindered by the brightness of the central stars. In contrast, young planetary-mass brown dwarfs are free-floating analogues to giant exoplanets. This makes them ideal targets for atmospheric characterization of planetary-mass objects, which opens a window to understand the atmospheres of giant exoplanets.

%weather on brown dwarfs
Without a sustainable heat source, brown dwarfs cool as they age. 
L dwarfs have an effective temperature from $\sim$2500 to $\sim$1300 K, while T dwarfs have an effective temperature from $\sim$1300 to $\sim$400 K \citep{Kirkpatrick2005}.
As their temperature decreases, their atmospheres undergo drastic changes. 
From early-L to late-L spectral types, they become fainter and redder in the near-infrared \cite{Kirkpatrick2005}.
However, from late-L to mid-T spectral types, their near-infrared magnitudes span in a small range, indicating an almost constant effective temperature ($\sim$1400 $\pm$ 200 K), but their $J-K$ colours dramatically turn blue by $\sim$ 2 mag \citep{Kirkpatrick2005}. This phenomenon is known as the L/T transition \citep{Golimowski2004, Stephens2009}. Beyond mid-T spectral types, they continue becoming fainter.
The prevailing explanation for the change in magnitude and colour at the L/T transition is that refractory materials, such as silicate and iron compounds, condense in the atmosphere of L dwarfs when their effective temperature falls below $\sim$2300 K \citep{Tsuji1996,Lodders1999,Burrows2006}. This leads to increased opacity due to dust and clouds. When the temperature falls below $\sim$1300 K, the cloud grains and particles dissipate, resulting in a clear atmosphere in T dwarfs and a shift towards bluer colours at the L/T transition \citep{Burrows1999,Tsuji2003,Knapp2004,Cushing2008,Marley2010}.

%variability
As brown dwarfs rotate, inhomogeneous atmospheric structures may cause variability in their light curves. Brown dwarfs are usually fast-rotators with periods varying from 1--2 h to 1--2 d \citep[e.g.][]{Zapatero-Osorio2004,Metchev2015, Scholz2018, Moore2019}.
Numerous observations have been conducted to search for atmospheric variability in field L and T brown dwarfs. 
The first continuous monitoring survey, by \cite{Koen2004}, detects low-level variability of less than 0.02\,mag in 18 L and T dwarfs observed simultaneously in the $J H K_S$ bands.
A similar result is reported in their follow-up survey of ultracool dwarfs, with the exception of one T dwarf which shows a variability of 0.03\,mag \citep{Koen2005}.
A large survey of L4--T9 dwarfs in the $J$-band by \cite{Radigan2014a} detects variability in 9 out of 57 dwarfs with over 99\% confidence. The strongest signals, with peak-to-peak amplitudes over 2\%, are all detected at the L/T transition (L9--T3.5), indicating that variability is most common among L/T transition brown dwarfs.
Combining two large surveys by \cite{Radigan2014a} and \cite{Wilson2014}, \cite{Radigan2014b} report the observed frequency of strong variability is 24\% at the L/T transition compared with 3.2\% outside the L/T transition. These results support the theory that inhomogeneous atmospheric structures such as patchy clouds at the L/T transition are the driving sources of the observed variability. 

%surface gravity
Brown dwarfs contract as they age, and thus low surface gravity is often associated with young brown dwarfs.
Surface gravity plays an important role in shaping the atmospheric structures of brown dwarfs, such as affecting dust particle size and non-equilibrium chemistry \citep{Barman2011,Barman2011b,Marley2012}. 
% Low-gravity objects may have lower effective temperature at the L/T transition than field objects \citep{Metchev2006,Saumon2008,Stephens2009,Marley2012}.
\cite{Morley2014} raise a possible link between low surface gravity and variability as planetary-mass objects tend to have thicker clouds than high-mass brown dwarfs.
\cite{Metchev2015} present a variability survey of 44 L3--L8 dwarfs with the \textit{Spitzer Space Telescope} and confirm that variability is common in L and T dwarfs with 80\% of L dwarfs varying more than 0.2\% and 36\% of T dwarfs varying more than 0.4\% in the mid-infrared. They suggest a tentative association between low surface gravity and strong variability based on six L3--L5.5 dwarfs with low gravity.
\cite{Vos2019} report the first variability survey of 30 young and low-gravity L dwarfs in the $J_S$ band and find a variability occurrence rate of 30\% for low-gravity L0--L8.5 dwarfs, significantly higher than the 11\% variability occurrence rate of the field L0--L8.5 dwarfs reported in \cite{Radigan2014a, Radigan2014b}. \cite{Vos2022} also find higher maximum variability amplitudes in young objects than field dwarfs in \textit{Spitzer} data.

% Similar to the young free-floating objects, directly-imaged planet companions are also usually young and have low surface gravity.
% A few directly-imaged planets have been monitored for variability in the near-infrared with the Hubble Space Telescope (HST), achieving sub-percent precision. 
% \cite{ZhouY2016} detect a variability level $<$ 2\% in the near-infrared for 2M1207b. 
% HST observations make tentative variability detection for AB~Pic~B, 2M0122~B and HD~106906~b \citep{ZhouY2019,ZhouY2020a}.
% \cite{Biller2021} conduct ground-based observations of the multi-planet system HR~8799 and achieve a variability sensitivity with amplitude $>$ 5\% for HR~8799~b and 25\% for HR~8799~c with no confirmed variability on them. 

%young strong variables
Several young low-surface gravity objects with strong variability have been discovered in recent years. For instance, PSO318.5-22 \citep{Biller2015,Vos2019}, VHS1256-1257b \citep{ZhouY2020b}, WISEP~J004701.06+680352.1\citep{Lew2016}, and 2MASS~J2244316+204343 \citep{Vos2019} are detected with peak-to-peak amplitudes $>$ 5\% in the near-infrared, which are young L6--L7.5 objects. Additionally, two strong variable T dwarfs, SIMP~J013656.5+093347 (T2.5) and 2MASS J21392676+0220226 (T1.5), were initially classified as field dwarfs but were later confirmed as planetary-mass objects of the 200-Myr-old Carina-Near moving group \citep{Gagne2017b,ZhangZ2021}. 
SIMP~J013656.5+093347 is detected with a peak-to-peak amplitude of 8\% in the $J$ band with a period of 2.4\,hr \citep{Artigau2009, Radigan2014a}. 2MASS~J21392676+0220226 has a 26\% peak-to-peak amplitude in the $J$ band with a 7.7-hour periodic modulation and night-to-night variations \citep{Radigan2012}. 

%this paper
Although we have a number of young L-type planetary-mass objects, there were few known young T-type objects. Only one of them has been monitored for variability in the near-infrared \citep[Ross~458c,][]{Manjavacas2019} and several have been monitored for variability in the mid-infrared \citep{Vos2022}.
Identifying young T-type objects is more challenging, as they do not have prominent spectral features associated with surface gravity.
\cite{ZhangZ2021} identify 30 new T0--T9 planetary-mass candidates of nearby young moving groups (YMG) based on their proper motions, parallaxes and available radial velocities, providing a sizeable sample of young T-type objects suited for time-resolved photometric studies.
In this work, we present a first near-infrared variability survey of these T-type planetary-mass objects, which also includes several young mid-late L objects without existing variability monitoring, with the aim of estimating their variability rates, as it was performed earlier with young L dwarfs.
Combining the results with previous surveys of field and young low-gravity L and T dwarfs, we make a statistical analysis of the variability of field and young objects from L0--T9 and investigate how variability properties depend on the spectral type and surface gravity.

\section{Sample}
\label{sec:sample}
From the new planetary-mass candidate members detected in \cite{ZhangZ2021}, we selected 12 objects with spectral types of T2.5--T8 that are bright enough ($J$ < $\sim$17.5 mag) for variability monitoring with a ground-based 4-m class telescope in the southern hemisphere. We also included six young L5--L7 dwarf candidate members identified with spectral and kinematic information from \cite{Kellogg2016}, \cite{Schneider2016} and \cite{Schneider2017}. 
% Due to some observation constraints, we lacked objects to observe on certain nights and therefore observed an additional L5 dwarf candidate member WISEA~J043718.77-550944.0 from \cite{Schneider2017}, which is a possible member of the $\beta$~ Pictoris moving group.
In total, our sample consists of 18 L5--T8 YMG candidate members with masses $\leq$ 20 $M_\textrm{J}$ and ages $\leq$ 200 Myr without previous near-infrared variability observations.
They are members of the AB Doradus \citep[149$^{+51}_{-19}$\,Myr,][]{Bell2015}, Argus \citep[40--50\,Myr,][]{Zuckerman2019}, $\beta$~Pictoris \citep[22$\pm$6\,Myr,][]{Shkolnik2017}, Carina-Near \citep[200$\pm$50\,Myr,][]{Zuckerman2006} and TW Hydrae \citep[TWA, 10$\pm$3\,Myr,][]{Bell2015} moving groups. There are two known planetary-mass binaries in our sample, 2MASS~J11193254–1137466AB \citep{Best2017} and 2MASSI~J1553022+153236AB \citep{Dupuy2012}, but neither are resolved in our observations.
Table~\ref{tab:SOFIobjects} lists the key information of these objects.

\begin{table*}
\centering
\caption{Object information. Spectral type (SpT), magnitude, and membership with BANYAN II or BANYAN $\Sigma$ (ref 4) probability are from the literature (Ref). The last column is the variability results in this work. \rev{Four} variables are detected with two potential variable candidates.}
\begin{tabular}{l l l l l l l l l l}
\hline\hline
Name & SpT & $J$ & $H$ & $K$ & Mag & Binary & Membership & Ref & Variable\\
\hline
WISEA~J004403.39+022810.6 & L7 & 16.997$\pm$0.187 & 15.822$\pm$0.169 & 14.876$\pm$0.105 & 2MASS & N & Beta Pic (78\%) & 1 & N \\
WISEA~J020047.29-510521.4 & L6 & 16.414$\pm$0.120 & 14.941$\pm$0.069 & 13.871$\pm$0.050 & 2MASS & N & AB Dor (98\%) & 1 & \rev{N}\\
WISEA~J022609.16-161000.4 & L6 & 17.334$\pm$0.266 & 15.750$\pm$0.142 & 14.581$\pm$0.093 & 2MASS & N & AB Dor (85\%) & 1 & ?\\
WISEA~J114724.10-204021.3 & L7 & 17.637$\pm$0.058 & 15.764$\pm$0.106 & 14.872$\pm$0.106 & 2MASS & N & TWA (96\%) & 2 & Y\\
2MASS~J11193254-1137466 & L7 & 17.474$\pm$0.058 & 15.788$\pm$0.034 & 14.751$\pm$0.012 & 2MASS & Y & TWA (92\%) & 3 & Y\\
WISE~J024124.73-365328.0 & T7 & 16.59$\pm$0.04 & 17.04$\pm$0.07 & --- & MKO & N & Argus (87.7\%) & 4 & N\\
CFBDS~J232304.41-015232.3 & T6 & 17.23$\pm$0.03 & 17.46$\pm$0.04 & 17.30$\pm$0.03 & MKO & N & Beta Pic (89.1\%) & 4 & Y\\
WISEPCJ225540.74-311841.8 & T8 & 17.33$\pm$0.01 & 17.66$\pm$0.03 & 17.42$\pm$0.05 & MKO & N & Beta Pic (98.7\%) & 4 & N\\
SDSSJ020742.48+000056.2 & T4.5 & 16.73$\pm$0.01 & 16.81$\pm$0.04 & 16.72$\pm$0.05 & MKO & N & Arg (95.6\%) & 4 & N\\
WISEPAJ081958.05-033529.0 & T4 & 14.78$\pm$0.02 & 14.60$\pm$0.05 & 14.64$\pm$0.05 & MKO & N & Beta Pic (86.3\%) & 4 & Y\\
ULASJ131610.13+031205.5 & T3 & 16.75$\pm$0.02 & 16.13$\pm$0.02 & 15.43$\pm$0.02 & MKO & N & Carina-Near (91.7\%) & 4 & N\\
ULASJ081918.58+210310.4 & T6 & 16.95$\pm$0.01 & 17.28$\pm$0.04 & 17.18$\pm$0.16 & MKO & N & AB Dor (86.3\%) & 4 & N\\
ULASJ075829.83+222526.7 & T6.5 & 17.62$\pm$0.02 & 17.91$\pm$0.02 & 17.87$\pm$0.12 & MKO & N & Arg (92.7\%) & 4 & N\\
PSOJ168.1800-27.2264 & T2.5 & 17.12$\pm$0.03 & 16.75$\pm$0.03 & 16.65$\pm$0.06 & MKO & N & Arg (83.4\%) & 4 & N\\
SDSSJ152103.24+013142.7 & T3 & 16.10$\pm$0.01 & 15.68$\pm$0.01 & 15.57$\pm$0.02 & MKO & N & Arg (82.3\%) & 4 & N\\
WISE~J163645.56-074325.1 & T4.5 & 16.42$\pm$0.02 & 16.28$\pm$0.05 & 16.32$\pm$0.05 & MKO & N & AB Dor (81.6\%) & 4 & N\\
2MASSI~J1553022+153236AB & T7 & 15.34$\pm$0.03 & 15.76$\pm$0.03 & 15.94$\pm$0.03 & MKO & Y & Carina-Near (89.6\%) & 4 & \rev{?}\\
WISEA~J043718.77-550944.0 & L5 & 16.985$\pm$0.192 & 15.583$\pm$0.157 & 14.640$\pm$0.098 & 2MASS & N & Beta Pic (94\%) & 1 & N\\
\hline
\end{tabular}
\begin{tablenotes}
\item References:~(1)~\cite{Schneider2017}; (2)~\cite{Schneider2016}; (3)~\cite{Kellogg2015}; (4)~\cite{ZhangZ2021}.
\end{tablenotes}
\label{tab:SOFIobjects}
\end{table*}

\section{Observations and data reduction}
\label{sec:observations and dara reduction}
% Normally the next section describes the techniques the authors used.
% It is frequently split into subsections, such as Section~\ref{sec:maths} below.

\subsection{Observations}
\label{sec:observations} % used for referring to this section from elsewhere
% Simple mathematics can be inserted into the flow of the text e.g. $2\times3=6$
% or $v=220$\,km\,s$^{-1}$, but more complicated expressions should be entered
% as a numbered equation:

% \begin{equation}
%     x=\frac{-b\pm\sqrt{b^2-4ac}}{2a}.
% 	\label{eq:quadratic}
% \end{equation}

% Refer back to them as e.g. equation~(\ref{eq:quadratic}).
We conducted the first epoch survey of our 18 targets for 17 nights between October 2021 and June 2022 with the infrared spectrograph and imaging camera, Son of ISAAC (SOFI) on the 3.58\,m ESO New Technology Telescope (NTT) at the La Silla Observatory \citep{Moorwood1998}. Brown dwarfs tend to have strong variability in the $J$ band \citep{Radigan2014a}, so our targets were primarily monitored in the $J_S$ band. 
For targets that are extremely faint in $J$ but much brighter in $K$, we observed them in the  $K_S$ band.
For the two TWA objects with known variability at mid-infrared wavelengths, we obtained interleaved observations in the $J_S$ and $K_S$ band. 
The $J_S$ band has a center wavelength at 1.24\,$\mu$m with a width of 0.29\,$\mu$m, avoiding the water band at 1.4\,$\mu$m in the $J$ band; the $K_S$ band is at 2.16 $\mu$m with a width of 0.28\,$\mu$m, avoiding the atmospheric absorption feature at 1.9\,$\mu$m and elevated thermal background beyond 2.3\,$\mu$m in the $K$ band.
The field of view of SOFI is 4\farcm92 $\times$ 4\farcm92 with a pixel scale of 0\farcs288. Each target has continuous observations of 1.5--7.5\,hr. The observations were affected by poor seeing in the October 2022 run and by clouds in the June 2022 run. We conducted a second epoch of ten nights between October 2022 and May 2023 to re-observe objects observed in poor conditions and confirm variables detected in the first epoch. Table~\ref{tab:obs log} summarises the observing log. 

Our targets were observed at airmass $<$2. We used an ABBA nod pattern with three exposures at each position and 12 exposures in a loop. For each exposure, the detector integration time (DIT) is 20\,s and the number of DIT (NDIT) is 3 in the $J_S$ band and 10\,s and 6 in the $K_S$ band, respectively. For the interleaved observation, every 12 exposures in the $J_S$ band were followed by 12 exposures in the $K_S$ band. The peak intensity of the point spread function (PSF) of the target was kept below 10,000 ADU to prevent non-linearity effects.

\begin{table*}
\centering
\caption{Observing log. We measured the median full width at half maximum (FWHM) of all stars in each target's field of view. We took the median of all frames as a seeing representative value for each target as listed in column FWHM. Column FWHM std is the standard deviation of all frames, representing seeing variations during observations.}
\begin{tabular}{l l l l l l l}
\hline\hline
Date & Object & Band & DIT$\times$NDIT [s] & FWHM ["] & FWHM std ["] & Elapsed time [hr] \\
\hline
2021-10-20 & J0044+0228 & $J_S$ & 20$\times$3 & 1.45 & 0.18 & 3.75 \\
2021-10-20 & J0200-5105 & $J_S$ & 20$\times$3 & 1.48 & 0.32 & 4.91 \\
2021-10-21 & J0241-3653 & $J_S$ & 20$\times$3 & 1.59 & 0.11 & 3.27 \\
2021-10-21 & J2323-0152 & $J_S$ & 20$\times$3 & 1.24 & 0.18 & 5.58 \\
2021-10-22 & J0226-1610 & $J_S$ & 20$\times$3 & 1.36 & 0.15 & 4.35 \\
2021-10-22 & J2255-3118 & $J_S$ & 20$\times$3 & 1.12 & 0.14 & 4.52 \\
2021-10-23 & J0044+0228 & $J_S$ & 20$\times$3 & 1.54 & 0.20 & 2.04 \\
2021-10-23 & J0207+0000 & $J_S$ & 20$\times$3 & 1.45 & 0.17 & 3.91 \\
2021-10-23 & J0241-3653 & $J_S$ & 20$\times$3 & 1.30 & 0.16 & 3.20 \\
2021-10-24 & J2323-0152 & $J_S$ & 20$\times$3 & 0.79 & 0.22 & 4.11 \\
2021-10-24 & J0226-1610 & $J_S$ & 20$\times$3 & 1.02 & 0.13 & 4.73 \\
2022-02-10 & J0819-0335 & $J_S$ & 20$\times$3 & 0.89 & 0.09 & 4.63 \\
2022-02-10 & J1316+0312 & $J_S$ & 20$\times$3 & 0.92 & 0.11 & 4.37 \\
2022-02-11 & J0819+2103 & $J_S$ & 20$\times$3 & 1.29 & 0.14 & 3.53 \\
2022-02-11 & 2M1119-1137AB & $K_S$ & 10$\times$6 & 0.88 & 0.10 & 4.47 \\
2022-02-12 & J0819-0335 & $J_S$ & 20$\times$3 & 1.17 & 0.18 & 3.13 \\
2022-02-12 & 2M1119-1137AB & $J_S$ & 20$\times$3 & 1.26 & 0.17 & 5.28 \\
2022-02-12 & 2M1119-1137AB & $K_S$ & 10$\times$6 & 1.11 & 0.14 & 5.29 \\
2022-02-13 & J0200-5105 & $J_S$ & 20$\times$3 & 1.58 & 0.28 & 1.92 \\
2022-02-13 & W1147-2040 & $K_S$ & 10$\times$6 & 1.34 & 0.13 & 7.31 \\
2022-02-14 & J0437-5509 & $K_S$ & 10$\times$6 & 0.96 & 0.10 & 2.80 \\
2022-02-14 & W1147-2040 & $J_S$ & 20$\times$3 & 0.93 & 0.11 & 5.80 \\
2022-02-14 & W1147-2040 & $K_S$ & 10$\times$6 & 0.91 & 0.10 & 6.32 \\
2022-02-17 & J0758+2225 & $J_S$ & 20$\times$3 & 0.94 & 0.12 & 3.19 \\
% 2022-02-17 & VHS1256-1257b & $J_S$ & 10 & 0.91 & 0.16 & 5.68 \\
% 2022-02-17 & VHS1256-1257b & H & 10 & 0.87 & 0.12 & 4.90 \\
% 2022-02-17 & VHS1256-1257b & $K_S$ & 10 & 0.87 & 0.15 & 4.90 \\
2022-02-18 & PSO168-27 & $J_S$ & 20$\times$3 & 1.06 & 0.24 & 4.10 \\
2022-02-19 & PSO168-27 & $J_S$ & 20$\times$3 & 0.84 & 0.16 & 4.96 \\
2022-06-16 & SDSSJ1521+0131 & $J_S$ & 20$\times$3 & 1.67 & 0.26 & 5.14 \\
2022-06-16 & J2323-0152 & $J_S$ & 20$\times$3 & 1.41 & 0.09 & 1.62 \\
2022-06-17 & WISEJ1636-0743 & $J_S$ & 20$\times$3 & 0.91 & 0.16 & 5.50 \\
2022-06-17 & J2323-0152 & $J_S$ & 20$\times$3 & 0.90 & 0.06 & 1.81 \\
2022-06-18 & SDSSJ1521+0131 & $J_S$ & 20$\times$3 & 0.71 & 0.05 & 2.06 \\
2022-06-18 & 2M1553+1532 & $J_S$ & 20$\times$3 & 0.81 & 0.15 & 4.00 \\
% 2022-06-18 & PSO318 & $J_S$ & 10$\times$3 & 0.91 & 0.22 & 4.59 \\
% 2022-06-19 & VHS1256-1257b & $J_S$ & 10 & 1.13 & 0.08 & 1.40 \\
2022-06-19 & WISEJ1636-0743 & $J_S$ & 20$\times$3 & 0.92 & 0.12 & 1.90 \\
2022-10-09 & J2323-0152 & $J_S$ & 20$\times$3 & 1.09 & 0.17 & 5.00 \\
2022-10-09 & J0200-5105 & $J_S$ & 20$\times$3 & 1.36 & 0.17 & 3.94 \\
2022-10-10 & J2323-0152 & $J_S$ & 20$\times$3 & 1.09 & 0.16 & 4.48 \\
2022-10-10 & J0200-5105 & $J_S$ & 20$\times$3 & 0.92 & 0.09 & 2.98 \\
2022-11-02 & J0226-1610 & $J_S$ & 20$\times$3 & 0.72 & 0.10 & 3.01 \\
2022-12-03 & J0200-5105 & $J_S$ & 20$\times$3 & 0.70 & 0.07 & 3.96 \\
2022-12-03 & J0819-0335 & $J_S$ & 20$\times$3 & 0.68 & 0.09 & 3.99 \\
2022-12-04 & J0226-1610 & $J_S$ & 20$\times$3 & 0.75 & 0.12 & 3.00 \\
2022-12-04 & J0819-0335 & $J_S$ & 20$\times$3 & 0.91 & 0.11 & 3.25 \\
2023-05-06 & W1147-2040 & $K_S$ & 10$\times$6 & 1.34 & 0.23 & 3.38 \\
% 2023-05-06 & PSO318 & $J_S$ & 10 & 1.3 & 0.23 & 3.81 \\
2023-05-07 & 2M1119-1137AB & $K_S$ & 10$\times$6 & 0.81 & 0.12 & 4.18 \\
2023-05-07 & 2M1553+1532 & $J_S$ & 20$\times$3  & 1.43 & 0.11 & 1.56 \\
% 2023-05-07 & PSO318 & $J_S$ & 10 & 1.15 & 0.2 & 3.31 \\
2023-05-08 & 2M1119-1137AB & $K_S$ & 10$\times$6 & 0.98 & 0.15 & 4.88 \\
% 2023-05-08 & PSO318 & $J_S$ & 10 & 1.2 & 0.16 & 2.33 \\
2023-05-08 & SDSSJ1521+0131 & $J_S$ & 20$\times$3  & 0.98 & 0.07 & 1.34 \\
2023-05-09 & 2M1553+1532 & $J_S$ & 20$\times$3  & 1.18 & 0.23 & 3.00 \\
2023-05-09 & WISEJ1636-0743 & $J_S$ & 20$\times$3  & 0.99 & 0.14 & 1.87 \\
2023-05-09 & W1147-2040 & $K_S$ & 10$\times$6 & 0.89 & 0.20 & 5.80 \\
2023-05-09 & W1147-2040 & $J_S$ & 20$\times$3  & 0.94 & 0.24 & 5.28 \\
2023-05-10 & 2M1119-1137AB & $K_S$ & 10$\times$6 & 0.84 & 0.11 & 4.92 \\
2023-05-10 & 2M1553+1532 & $J_S$ & 20$\times$3  & 1.10 & 0.09 & 3.05 \\
2023-05-10 & WISEJ1636-0743 & $J_S$ & 20$\times$3  & 0.95 & 0.08 & 2.49 \\
\hline
\end{tabular}
\label{tab:obs log}
\end{table*}

% \begin{table*}
% \centering
% \caption{Observed objects list. FWHM: median seeing.}
% \begin{tabular}{c c c c c c c}
% \hline\hline
% Object & Spectral type & Jmag & Date & Band & FWHM & Variable\\ 
% \hline
% J0044+0228 & L7 & 17.00 & 2021-10-20 & Js & 1.43 & No\\
%  & & & 2021-10-23 & Js & 1.53 & No\\
% J0200-5105 & L6 & 16.41 & 2021-10-20 & Js & 1.48 & ?\\
%  & & & 2022-02-13 & Js & 1.57 & No\\
% J0241-3653 & T7 & 16.59 & 2021-10-21 & Js & 1.58 & No\\
%  & & & 2021-10-23 & Js & 1.31 & No\\
% J2323-0152 & T6 & 17.23 & 2021-10-21 & Js & 1.22 & No\\
%  & & & 2021-10-24 & Js & 0.78 & ?\\
% J0226-1610 & L6 & 17.33 & 2021-10-22 & Js & 1.31 & No\\
%  & & & 2021-10-24 & Js & 1.03 & ?\\
% J2255-3118 & T8 & 17.33 & 2021-10-22 & Js & 1.12 & No\\
% J0207+0000 & T4.5 & 16.73 & 2021-10-23 & Js & 1.45 & No\\
% J0819-0335 & T4 & 14.78 & 2022-02-10 & Js & 0.89 & \\
%  & & & 2022-02-12 & Js & 1.17 & ?\\
% J1316+0312 & T3 & 16.75 & 2022-02-10 & Js & 0.92 & No\\
% J0819+2103 & T6 & 16.95 & 2022-02-11 & Js & 1.29 & No\\
% 2M1119-1127AB & L7 & 17.23 & 2022-02-11 & Ks & 0.88 & Yes\\
%  & & & 2022-02-12 & Js & 1.26 & No\\
%  & & & 2022-02-12 & Ks & 1.11 & No\\
% W1147-2040 & L7 & 17.44 & 2022-02-13 & Ks & 1.32 & No\\
%  & & & 2022-02-14 & Js & 0.93 & Yes\\
%  & & & 2022-02-14 & Ks & 0.91 & Yes\\
% J0758+2225 & T6.5 & 17.62 & 2022-02-17 & Js & 0.94 & ?\\
% PSO168-27 & T2.5 & 17.12 & 2022-02-18 & Js & 1.06 & No\\
%  & & & 2022-02-19 & Js & 0.84 & No\\
% J0437-5509(backup) & L5 & 17.00 & 2022-02-14 & Ks & 0.96 & No\\
% \hline
% \end{tabular}
% \label{tab:obs log}
% \end{table*}

\subsection{Raw image reduction}
We followed the data reduction steps in the SOFI manual to reduce raw images, which is also presented in \cite{Vos2019}. 
1) Cross talk removal: a bright source can cause a ghost affecting the row where the source is and also the row in the other half of the detector, referred to as inter-quadrant row cross talk. These can be removed by subtracting 1.4$\times 10^{-5}$ times the integrated flux of the row. 
2) Flat fielding and shade pattern removal: when taking a flat, there is a difference between the shade pattern in the image with the lamp on and lamp off. Therefore, eight special dome flats were taken to do the flat-fielding including removing the residual shade pattern with the lamp on and off. 
3) Illumination correction: the illumination of the dome panel is different from that of the sky, so a grid of 16 observations of a standard star was taken and we fitted a 2D surface to the photometry of the 16 positions to correct the difference in illumination between the dome panel and sky.
4) Sky subtraction and dark current correction: we subtracted frames by frames of different nods and closest in time to remove the fast-varying thermal background and also the dark current. We scaled the frame by the median flux ratio between it and the subtracted frame before the subtraction.
5) Bad pixel flagging: bad pixels were identified in the flat frame. The median of the flat frame was obtained after 3 sigma clipping. Pixels with flux deviation larger than 10 sigma from the median value were identified as bad pixels. The bad pixel map provided by ESO, which was created in 2012, was also combined into the final bad pixel map. 

\subsection{Aperture photometry}
We used \textsc{DAOStarFinder} from the \textsc{photutils} python package to detect sources in an image and fitted a 2D Gaussian model to the detected sources to accurately measure their positions. Then we performed aperture photometry on the detected stars with a series of aperture sizes fixed to all frames. We also took the median after 3-$\sigma$ clip of a concentric annulus as the local background of the star and subtracted it from the aperture measurement. The inner radius of the annulus is 18 pixels and the outer radius of the annulus is 24 pixels.
The final aperture size we used is determined in the light curve analysis as described below.

\subsection{Light curve analysis}
The raw light curves contain conspicuous systematics, including the effects of seeing, airmass, atmosphere, and instrument. 
We selected reference stars in the field of view of the target to calibrate and remove these systematics. 
At first, we excluded extremely faint stars and bright stars with flux in the non-linear regime ($>$ 10,000 ADU). Second, for each star, the raw light curves of different nods were normalised by their own medians. Then the normalised light curves of different nods were corrected to the same baseline. This step scales the light curves of different stars to the same level. We then selected a set of well-behaved reference stars to build a calibration light curve for the target. We used the same iteration algorithm from \cite{Radigan2014a} and \cite{Vos2019}. First, stars affected by bad pixels were discarded. Then for each star, its calibration light curve was created from the median light curve of the other candidate reference stars. We divided the calibration light curve from the light curve of the star to remove the variations caused by systematics. Good reference stars should have no intrinsic variations and have flat light curves after detrending. To exclude the effects of outliers, we calculated the robust standard deviation and robust linear slope of detrended light curves. Stars with sigma and slope $< \sim$1.2--3 times that of the target were retained for the next iteration. These steps were repeated several times until a set of well-behaved reference stars was chosen. \rev{We further removed some reference stars that showed variability after the iteration algorithm by visual inspection.}
We applied this iteration algorithm to light curves of different aperture sizes. For each target, the final aperture we used was the same as or close to the median full width at half maximum (FWHM) of the PSFs of all stars in all frames of the target, which results in the smallest photometric noise. 
The photometric noise of the detrended light curve, $\sigma_\textrm{pt}$, was estimated by the robust standard deviation of the subtraction of a light curve shifted by one time bin from the original light curve, divided by $\sqrt{2}$, the same method used in \cite{Radigan2014a}. This method is sensitive to high-frequency noise but not sensitive to low-frequency noise such as intrinsic astrophysical variability trends in light curves. It performs better than the standard deviation in quantifying the photometric noise and uncertainty of detrended light curves, especially for variable targets.

\section{Variability analysis}
%periodogram: LG and BGLS
The variability of the targets was detected using a periodogram analysis of their detrended light curves.
We used the Lomb-Scargle (LS) periodogram method as the primary analysis tool \citep{Lomb1976, Scargle1982}. We also used a secondary analysis, the Bayesian Generalized Lomb-Scargle (BGLS) periodogram, to independently verify the peaks in the LS periodogram \citep{Mortier2015}.
\cite{Manjavacas2018} find that unlike LS, BGLS is insensitive to gaps in light curves. BGLS calculates the relative probability between peaks rather than the power spectrum calculated by the LS periodogram. The peaks detected in the BGLS periodogram agree with those detected in the LS periodogram of our light curves. We also calculated the LS periodogram of detrended reference light curves, the seeing curve over the observation and the window function of the observation cadence.
If the peaks in these periodograms matched those in the target light curve, they were considered false detections due to residual systematic effects.
The window function was calculated as a light curve with a flux of 1 without pre-centering or using a floating-mean model in the LS calculation \citep{VanderPlas2018}.

\begin{figure}
	\includegraphics[width=1\columnwidth]{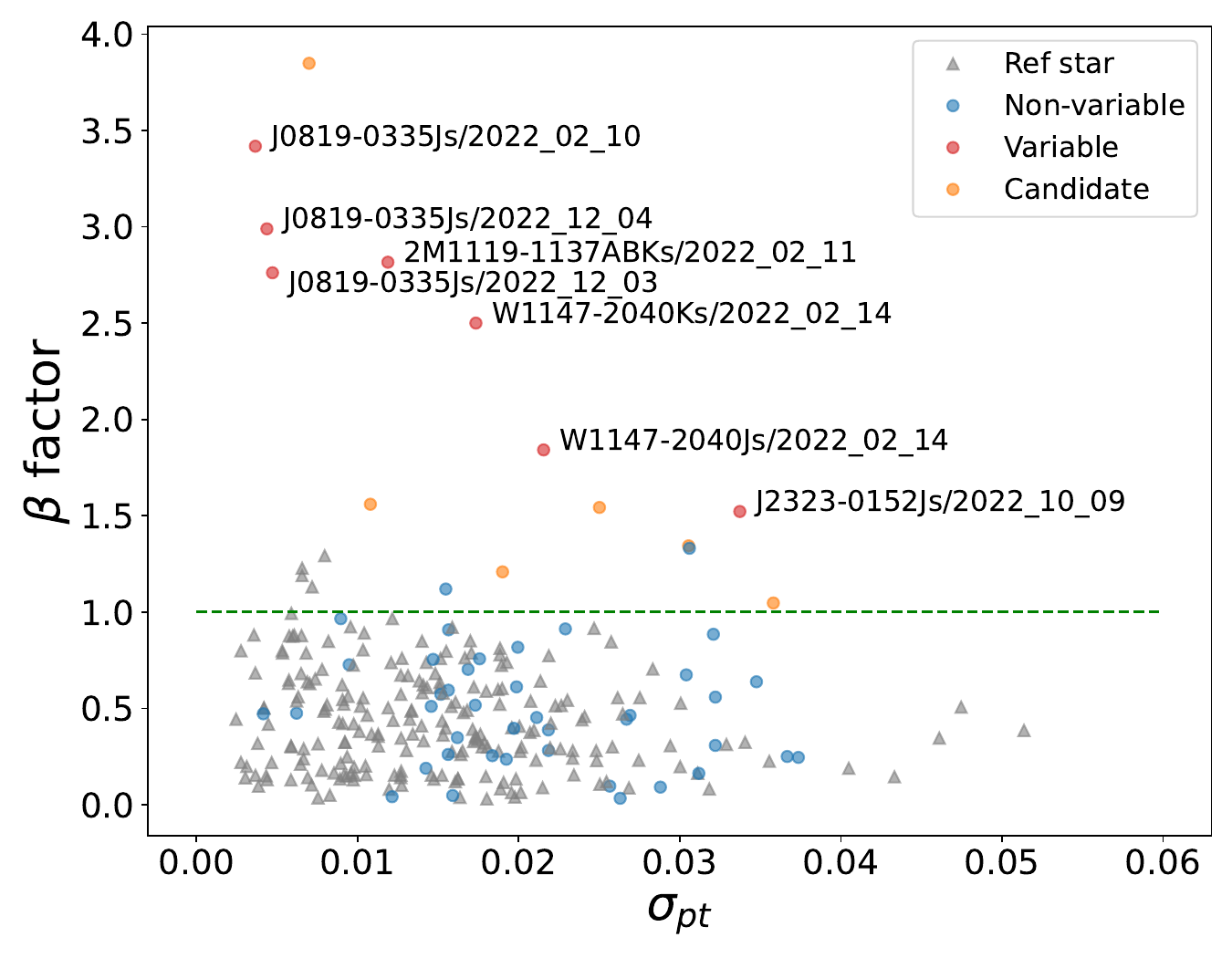}
    \caption{$\beta$ factor against the photometric error of a detrended light curve, $\sigma_\textrm{pt}$. The $\beta$ factor is calculated by the periodogram power peak of every light curve divided by its own 1\% FAP level. The four confirmed variable targets (red) fall above $\beta$ = 1. 1.8\% of reference stars fall above $\beta$ = 1, which is very close to the expected value, 1\%. Potential variable candidates are orange circles and non-variable targets are blue circles. The reference stars are shown by grey triangles.}
    \label{fig:betaplot}
\end{figure}

To assess the significance of the peaks in the periodogram, we calculated the 1\% false-alarm probability (FAP) level using the Astropy.timeseries Python package. We used the bootstrap option in that routine to calculate the peak level of 1\% FAP, which is equivalent to simulating the light curve and calculating the periodogram over 10$^3$ times. \cite{Radigan2014a} and \cite{Vos2019} calculated the 1\% FAP level also by randomly permuting reference star light curves 1000 times and the $\beta$ factor of every light curve which is the peak value in its periodogram divided by the 1\% FAP level. They expected to have 1\% reference stars with a $\beta$ factor above 1 but found that more than 1\% reference stars peaked above this level. Therefore, they scaled the 1\% FAP level by a factor between 1.4 and 3.4.
Because we included the light curve uncertainty in the LS periodogram calculation and applied the `standard' normalisation in that routine, the power spectra value should not be compared directly between the target and reference star light curves. We also randomly permuted the target light curve 1000 times and found that the 1\% FAP level calculated automatically by the bootstrap routine in Astropy.timeseries is always higher than the 1\% FAP level calculated by our version of 1000 random permutations of the target light curve. We adopted the 1\% FAP level calculated with Astropy.timeseries. \rev{We also calculated the $\beta$ factor by dividing the periodogram power peak of every light curve by its own 1\% FAP level. Four out of 216 reference stars fall above $\beta$ = 1 in Fig.~\ref{fig:betaplot}, which is close to 1\%.} \cite{Eriksson2019} used the 0.1\% FAP level with the \textsc{Astropy} routine for the variability detection in their light curves. We also calculated the 0.1\% FAP level and found our confirmed detections remain significant when using this level.

%beta factor

If the peak in the target periodogram is above the 1\% FAP level and the detrended light curve has a well-behaved appearance upon visual inspection, the target is identified as a variable object; if the peak in the target periodogram is above the 1\% FAP level but the detrended light curve is not well-behaved, it is identified as a potential variable candidate; if no peak in the periodogram exceeds the 1\% FAP level, the object is identified as non-variable in this survey. While the period of the detected variability can also be estimated from the periodogram, many targets did not have a clearly defined peak in their periodogram, indicating the presence of long-term variability that exceeds the duration of our observations. In these cases, we can only place lower limits on the period.

\begin{figure*}
      \includegraphics[width=1\columnwidth]{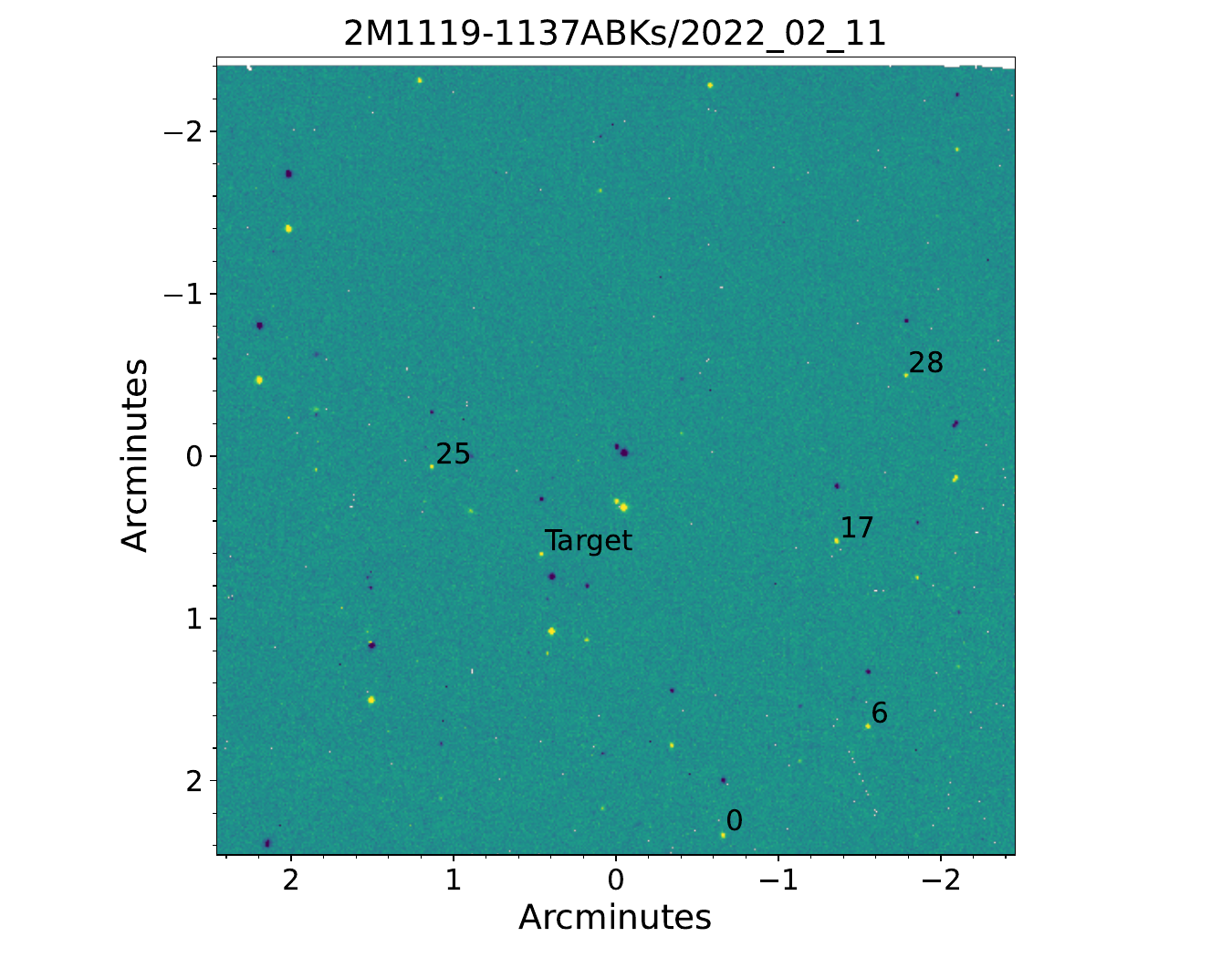}
      \includegraphics[width=1\columnwidth]{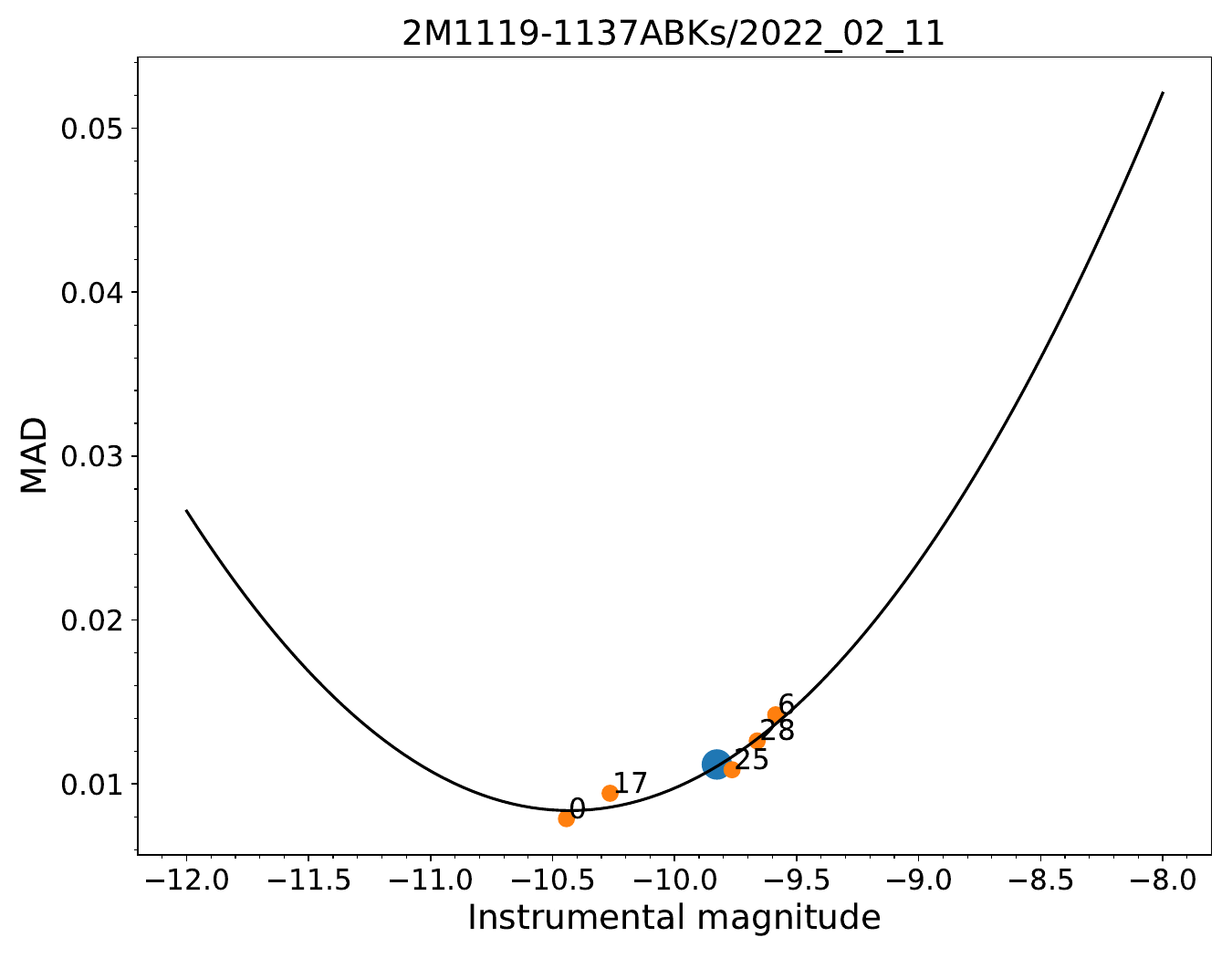}
    \caption{Target 2M1119-1137AB in the $K_S$ band and the selected reference stars of the observation on 11 Feb 2022. Left: the sky subtracted image. It is one nod subtracted from the other nod, which results in these dark sources. The textual label of the reference stars is their original number before the reference star selection. North is left and east is down. Right: the MAD and instrumental magnitude of the reference stars (orange) and target (blue). The black line is a second-order polynomial fitting to the reference stars. This target meets the MAD-magnitude relationship of the reference stars but it is variable.}
    \label{fig:2M1119-1137ABKs_star_mag}
\end{figure*}

%add discussion on the reference star selections
\rev{We also calculated the instrumental magnitude and median absolute deviation (MAD) of the detrended light curve which is similar to the standard deviation but not sensitive to outliers. Although MAD increases generally with fainter stars, we do not find a consistent relationship between the MAD and the instrumental magnitude between different observations. For some observations, their relationship can be fitted with a second-order polynomial as the relationship presented \cite{Martin2001a}. For some observations, there is not a monotonous relationship between MAD and the instrumental magnitude. Figure~\ref{fig:2M1119-1137ABKs_star_mag} shows one object 2M1119-1137AB and the selected reference stars and their MAD-magnitude relationship. The same figures for all other observations can be found in Appendix~\ref{apd:stars_mag}. Therefore, using the relationship between the standard deviation and the magnitude to do variability analysis as the method used in \cite{Martin2001a} is not applicable to our datasets. 
}

\section{Sensitivity calculation}
\label{sec:senscal}
To estimate our detection sensitivity for each observation, we injected artificial sinusoidal curves into randomly permuted detrended target light curves. For variable light curves, we fitted a low-order polynomial fitting or a sinusoidal curve before the injection to remove its variability. The injected peak-to-peak amplitude (2$\times$amplitude of the sinusoidal curve) varies from 0.5\% to 10\% and the period varies from 1.5 to 20 hours. In each grid, we injected the sinusoidal curves 1000 times and calculated the detection rate. The successful retrieval criterion is that the peak of the injected curves in the periodogram is above the 1\% FAP level.

% Figures and tables should be placed at logical positions in the text. Don't
% worry about the exact layout, which will be handled by the publishers.

% Figures are referred to as e.g. Fig.~\ref{fig:example_figure}, and tables as
% e.g. Table~\ref{tab:example_table}.

% % Example figure
% \begin{figure}
% 	% To include a figure from a file named example.*
% 	% Allowable file formats are eps or ps if compiling using latex
% 	% or pdf, png, jpg if compiling using pdflatex
% 	\includegraphics[width=\columnwidth]{example}
%     \caption{This is an example figure. Captions appear below each figure.
% 	Give enough detail for the reader to understand what they're looking at,
% 	but leave detailed discussion to the main body of the text.}
%     \label{fig:example_figure}
% \end{figure}

% % Example table
% \begin{table}
% 	\centering
% 	\caption{This is an example table. Captions appear above each table.
% 	Remember to define the quantities, symbols and units used.}
% 	\label{tab:example_table}
% 	\begin{tabular}{lccr} % four columns, alignment for each
% 		\hline
% 		A & B & C & D\\
% 		\hline
% 		1 & 2 & 3 & 4\\
% 		2 & 4 & 6 & 8\\
% 		3 & 5 & 7 & 9\\
% 		\hline
% 	\end{tabular}
% \end{table}

\section{Results}
\label{sec:results}
\begin{figure}
	\includegraphics[width=1\columnwidth]{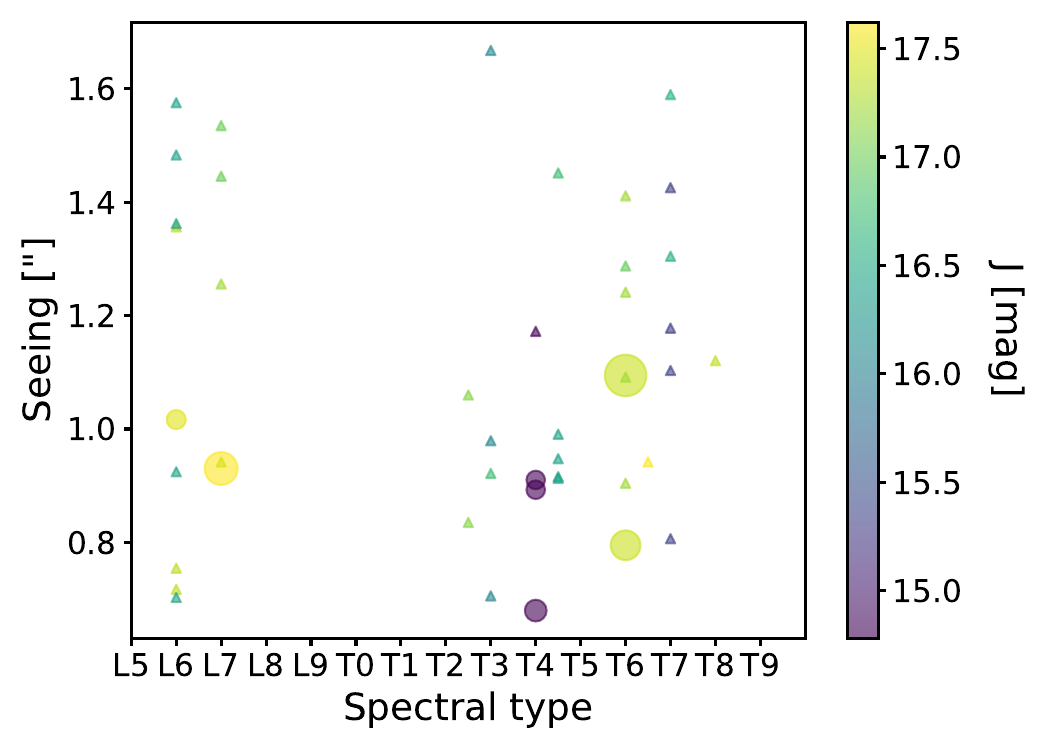}
    \caption{Observation and detection diagram in the $J_S$ band. Circles represent positive detections with the radius proportional to the variability amplitude. Triangles represent non-detection observations. The colour represents the apparent magnitude of the target in the $J$ band. All variable detections were observed under a seeing $<$1\farcs1, while they do not have a strong dependence on the magnitude.}
    \label{fig:seeing_spt}
\end{figure}

\rev{We detect four new variables, two variable candidates and twelve non-variables.} Table~\ref{tab:SOFIobjects} lists the variability detection results of all objects in this work.
Table~\ref{tab:Varinfo} summarises the amplitude and period information of variable objects. The detections are mainly limited by seeing. As shown in Fig.~\ref{fig:seeing_spt}, all the positive detections were observed under a seeing $<$1\farcs1. They do not have a strong correlation on the apparent magnitude as we have positive detections from faint to bright targets.

\begin{table}
\centering
\caption{Detected variables in this work. The amplitude is the peak-to-peak amplitude. Binaries are not resolvable with NTT.}
\label{tab:Varinfo}
\begin{tabular}{l l l l l l} % four columns, alignment for each
\hline\hline
Target & SpT & Band & Amplitude & Period [hr] & Binary\\
\hline
2M1119-1137AB & L7 & $K_S$ & 3.2$\pm$0.8\% & 6.9$\pm$1.6 & Y\\
W1147-2040 & L7 & $J_S$ & \rev{4.6$\pm$1.0\%} & \rev{11.2$\pm$3.8} & N\\
           &    & $K_S$ & 4.8$\pm$0.4\% & 5.5$\pm$0.2  & \\
J0819-0335 & T4 & $J_S$ & \rev{1.8$\pm$0.8\%} & Long & N\\
J2323-0152 & T6 & $J_S$ & $\sim7.6\%$ & Long & N\\
\hline
\end{tabular}
\end{table}

\subsection{Significant detections}
WISEPAJ081958.05-033529.0: this T4 dwarf is a high-probability member of the $\beta$ Pictoris young moving group identified by \cite{ZhangZ2021} with trigonometric parallax, though it needs an RV measurement for confirmation. Assigning an age of $\sim$20 Myr to it, it has an estimated mass of $\sim 5.7\,M_\textrm{J}$ \citep{ZhangZ2021}. We observed it on four nights in the $J_S$ band: 10 Feb 2022 and 12 Feb 2022 in the first epoch and 03 Dec 2022 and 04 Dec 2022 in the second epoch. We detected variability on three nights but not on 12 Feb 2022. The variable light curves are shown in Fig.~\ref{fig:J0819-0335_curves} and the non-variable light curve is in Appendix~\ref{apd:non-variable}.
The light curve of 10 Feb 2022 presents an obvious downward slope. We fit a line to it using the least-squares algorithm and find a peak-to-peak variability amplitude (max-min) of $1.2\pm0.1\%$ in the 4.63-hour observation. This variability is far above the 1\% FAP level in its periodogram. If it is a periodic signal, the period is longer than the observation length. As our observation does not cover a full rotation, the fitted variability amplitude is a lower limit.
We detected no variability on 12 Feb 2022.
The light curve of 12 Feb 2022 has a marginal downward slope by visual inspection but it is not detected in its periodogram.
The relatively poorer seeing of the second night degrades the detection sensitivity, which can be seen from its sensitivity plot. In fact, the variability of the first night would not be detectable on the second night.
It is one of the reasons why there is no variability detection on the second night. 
Another reason could be that the object has a long rotation period and reached the peak of its light curve on 12 Feb 2022 where relative variability would be lower as compared to the slope between extrema and thus it does not present an apparent relative variability during the three-hour observation.

We fit a second-order polynomial curve to the light curve of 03 Dec 2022 and measure a peak-to-peak amplitude of 1.8$\pm$0.8\%. If it is a periodic signal, its period is longer than the observation length, 4\,hr. Fitting a sinusoidal curve to the light curve of 04 Dec 2022, we measure a peak-to-peak amplitude of 1.2$\pm$0.1\% with a period of 2.6$\pm$0.1\,hr. We suspect that J0819-0335 has a period $>$5\,hr and the light curve was transitioning from a downward to an upward trend during the observation on 04 Dec 2022. This is why a short-period sinusoidal light curve was observed. Further longer continuous observations are necessary to determine the true period of this young object.

\begin{figure*}
	\includegraphics[width=2\columnwidth]{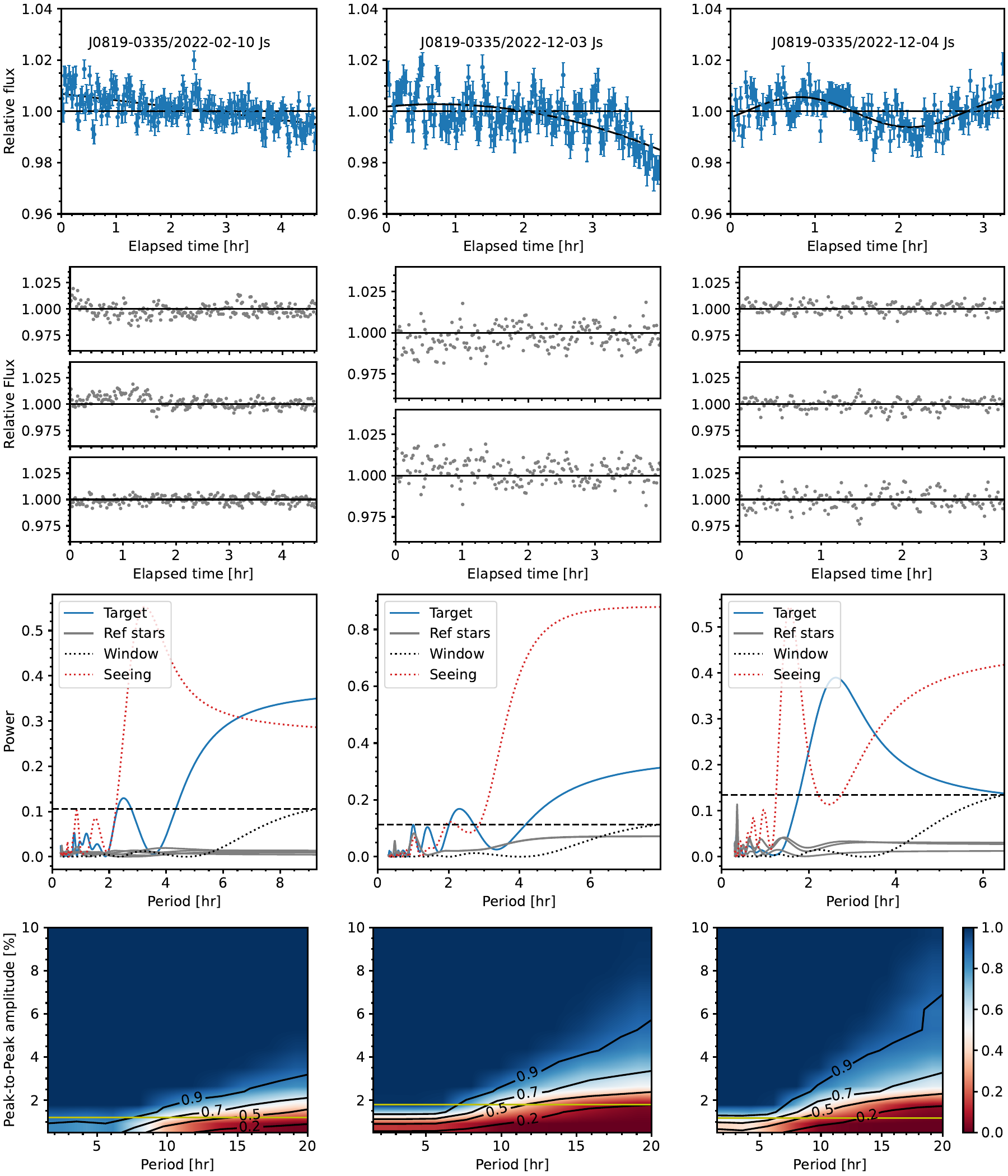}
    \caption{Results of a variable object, J0819-0335. Top row: detrended light curve. Second row: detrended light curve of its reference stars (show up to three stars). Third row: \rev{Lomb-Scargle periodogram of the detrended light curve of the target. We also include the periodograms of the detrended light curves of reference stars, seeing curve, and window function. They help to identify unremoved systematic variability in the detrended light curve of the target.}
    Dashed line is the 1\% FAP level of the target. Bottom row: sensitivity plot calculated from injected signals. The yellow line shows the measured amplitude. The X-axis and Y-axis are the period and peak-to-peak amplitude of the injected sinusoidal signal, respectively. The colour bar is the retrieval rate by our method, ranging from 0 to 1. The signal is injected into the variability-removed light curve of the target.}
    \label{fig:J0819-0335_curves}
\end{figure*}

\begin{figure*}
	\includegraphics[width=2\columnwidth]{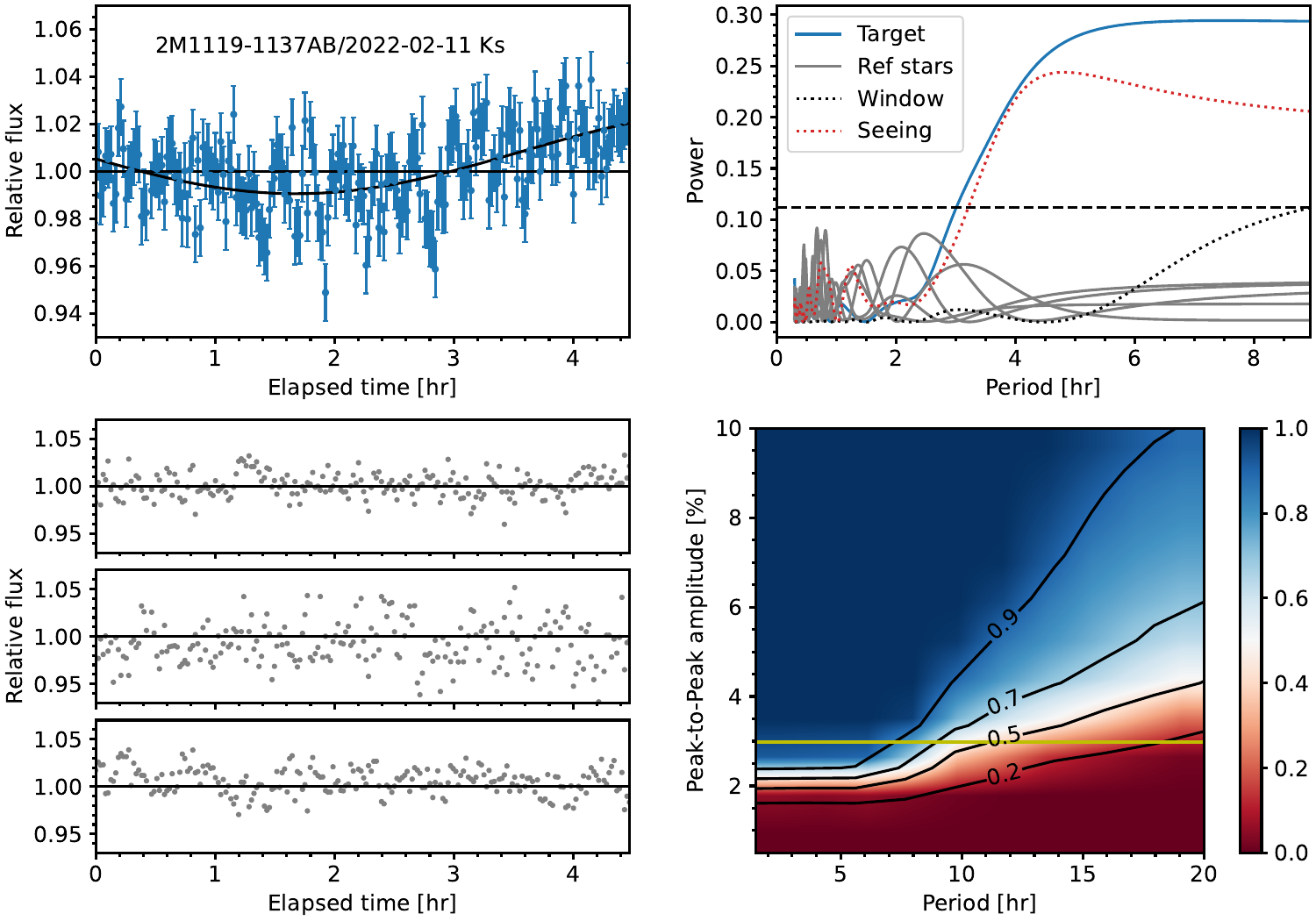}
    \caption{Variable light curve, periodogram and sensitivity plot of 2M1119-1137AB, including detrended light curves and periodograms of its reference stars. The variability is detected in the $K_S$ band.}
    \label{fig:2M1119-1137ABKs_curves}
\end{figure*}

2MASS~J11193254-1137466AB: 2M1119-1137AB has extremely red optical and near-infrared colours \citep{Kellogg2015}. It was first characterised as a low-mass L7 dwarf and a high probability candidate of the TW Hydrae Association (TWA) by \cite{Kellogg2015,Kellogg2016}. \cite{Best2017} resolve it to be a binary system of two similar $\sim3.7\,M_\textrm{J}$ L7 brown dwarfs with a separation of 0\farcs14, adopting the 10 Myr age of TWA. The orbital period of this system is about 90 years. It is a flux reversal binary as one component is slightly brighter in the $J$ band but fainter in the $K$ band. 
\cite{Schneider2018} report mid-infrared variability with a period of $3.02^{+0.04}_{-0.03}$\,hr and semi-amplitudes of $0.230^{+0.036}_{-0.035}\%$ at 3.6\,$\mu$m and $0.453 \pm 0.037\%$ at 4.5\,$\mu$m for this system in \textit{Spitzer} observations.
These light curves have also been suggested to show evidence of an exomoon \citep{Limbach2021}.
SOFI was unable to resolve this system. We observed it on two continuous nights in the first epoch: 11 Feb 2022 in the $K_S$ band and 12 Feb 2022 interleaved in the $J_S$ and $K_S$ bands. We detected significant variability on the first night as shown in Fig.~\ref{fig:2M1119-1137ABKs_curves}. We fit a sinusoidal curve to the light curve in the $K_S$ band using the Levenberg-Marquardt least-squares (LM) method and find a peak-to-peak amplitude of 3.2$\pm$0.8\% and a period of 6.9$\pm$1.6\,hr. Our observation does not cover a complete cycle and thus this period needs further confirmation. 

\begin{figure}
	\includegraphics[width=\columnwidth]{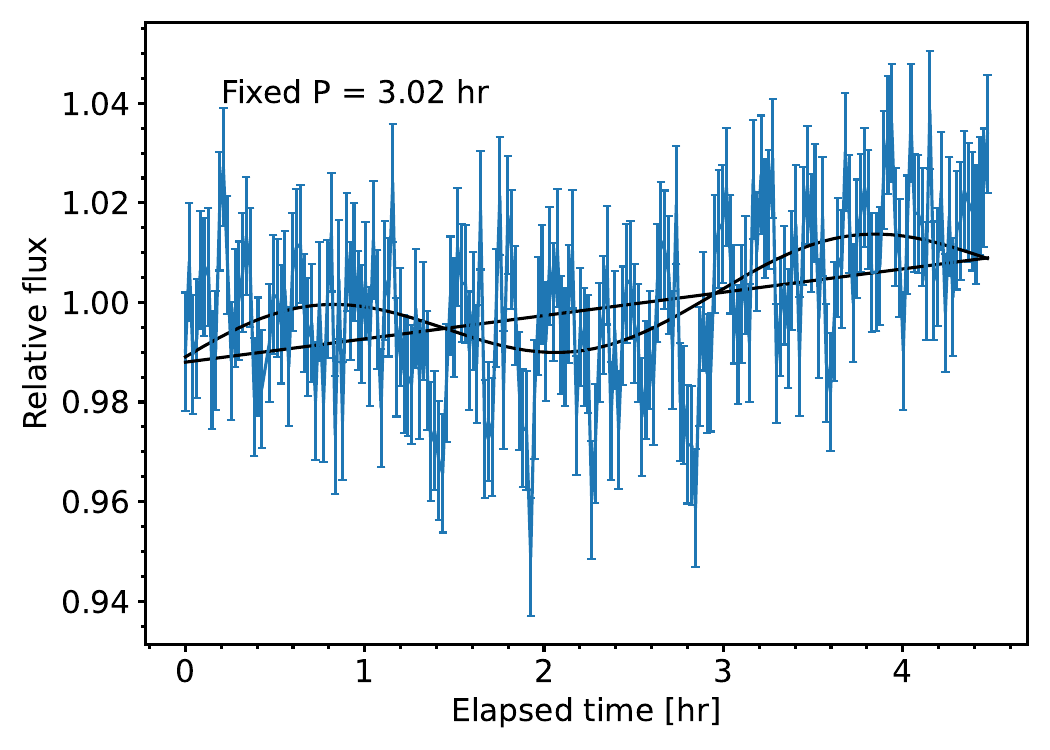}
    \caption{Variable light curve of 2M1119-1147AB, fitted with a linear term and a sinusoidal curve with a fixed period of 3.02\,hr, which is its variability period at 3.6 and 4.5\,$\mu$m. Its $K_S$-band light curve can be depicted by two variability terms which may come from the two components}
    \label{fig:2M1119-1137ABfixed}
\end{figure}

This period is longer than the variability period of 3.02\,hr at 3.6 and 4.5\,$\mu$m reported by \citep{Schneider2018}. To fit a sinusoidal curve with a fixed period of 3.02\,hr, we need to add a linear term of 0.0047t to the sinusoidal curve as shown in Fig.~\ref{fig:2M1119-1137ABfixed}, yielding a semi-amplitude of 0.86\% for the sinusoidal curve. As 2M1119-1137AB is a binary system, the linear term and sinusoidal term could be attributed to the two components respectively. \cite{Schneider2018} suspect that their measured variability with the period of 3.02\,hr possibly comes from one component of the binary, similar to the L7.5+T0.5 binary WISE~J104915.57-531906.1AB \citep{Burgasser2014}.
Further observations with larger telescopes such as the very large telescope (VLT) are necessary to resolve the variability of the two components. 
If confirmed, 2M1119-1137AB would be one of the few young L and T dwarfs with periods $<$ $\sim$3 hours. 
As brown dwarfs are expected to rotate faster as they age and contract because of angular momentum conservation \citep{Schneider2018}, young objects with short periods are rare.
% Many field brown dwarfs have periods  $<$2\,hr.
% The confirmed member of Carina-Near, SIMP0136+0933, has a period of 2.4\,hr while other three young objects candidates from \cite{Vos2022} have periods shorter than two hours.

The second night was observed under poorer seeing conditions than the previous night and we do not detect any variability. The $\sim$3\% variability of the first night could not be reliably detected under the conditions of the second night. We observed it another three times in May 2023 in the $K_S$ band but detected no variability due to poor seeing or clouds. The non-detection results can be found in Appendix~\ref{apd:non-variable}.

\begin{figure*}
	\includegraphics[width=2\columnwidth]{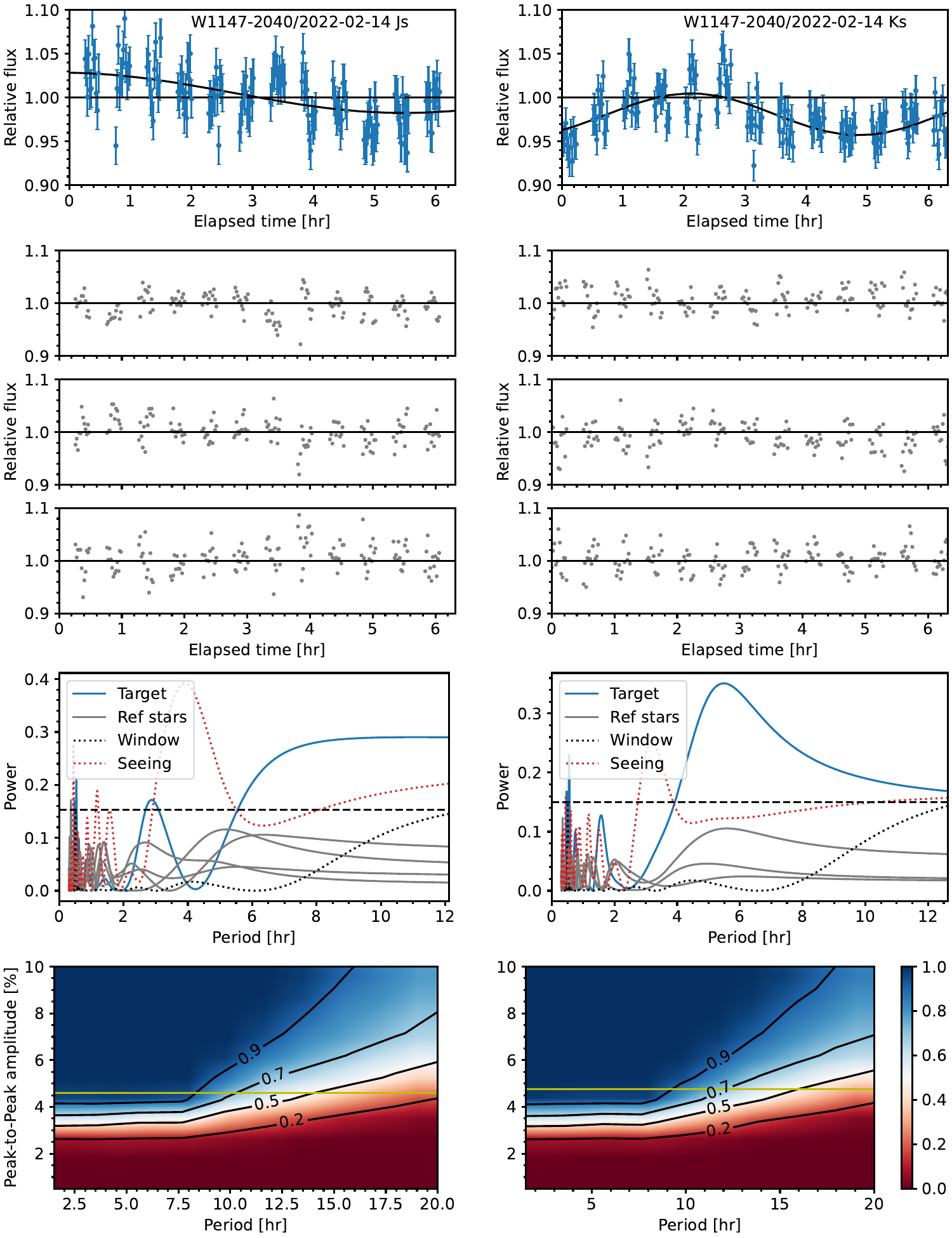}
    \caption{Variable light curves, periodograms, and sensitivity plots of W1147-2040, including detrended light curves and periodograms of its reference stars. The observations were interleaved in the $J_S$ and $K_S$ bands.}
    \label{fig:W1147-2040KsJs_curves}
\end{figure*}

WISEA~J114724.10-204021.3: W1147-2040 is an L7 dwarf with an extremely red 2MASS $J-K_S$ colour and a mass of 5-13$\,M_\textrm{J}$ \citep{Schneider2016}. Its spectrum has obvious evidence of youth. \cite{Schneider2016} identify it as a high probability member of the TWA using its sky position and proper motion. \cite{Schneider2018} find variability with a period of 19.39$^{+0.33}_{-0.28}$\,hr by combining its \textit{Spitzer} light curves at 3.6\,$\mu$m and 4.5\,$\mu$m. The light curves have a semi-amplitude of 0.798$^{+0.081}_{-0.083}$\% and 1.108$^{+0.093}_{-0.094}$\%, respectively. We observed it on two consecutive nights: 13 Feb 2022 and 14 Feb 2022. It was monitored in the $K_S$ band on the first night under relatively poor seeing conditions and we did not detect significant variability. These plots are in Appendix~\ref{apd:non-variable}. The observations on the second night were interleaved in the $J_S$ and $K_S$ bands with seeing about 0\farcs9. We detect variability in both bands with significance higher than 99\% as shown in Fig.~\ref{fig:W1147-2040KsJs_curves}. We fit a sinusoidal curve to each light curve using the LM method. We find a peak-to-peak amplitude of 4.6$\pm$1.0\% with a period of 11.2$\pm$3.8\,hr in the $J_S$ band and a peak-to-peak amplitude of 4.8$\pm$0.4\% with a period of 5.5$\pm$0.2\,hr in the $K_S$ band. The $K_S$ band light curve just covers a full period while the $J_S$ band light curve does not. The periods in the $J_S$ band, $K_S$ band, and mid-infrared of this target are quite different. We are cautious about whether these differences are astrophysical since these observations are relatively short compared to their measured periods. We also fit the curves with a sinusoidal curve with the period in the mid-infrared. The fitting in the $J_S$ band is acceptable while the $K_S$ band is poorly-fitted.
We did follow-up observations of it on 06 May 2023 in the $K_S$ band and on 09 May 2023 interleaved in the $K_S$ and $J_S$ band. Only the $J_S$ band on the second night presents variability. Its periodogram shows a peak above the 1\%FAP level around 6 hr but it coincides with the seeing curve. We consider this night as a possible detection and the light curves are presented in Appendix~\ref{apd:potential}. 
The light curves of 06 May can be found in Appendix~\ref{apd:non-variable}.

\begin{figure*}
    \centering
    \includegraphics[width=2\columnwidth]{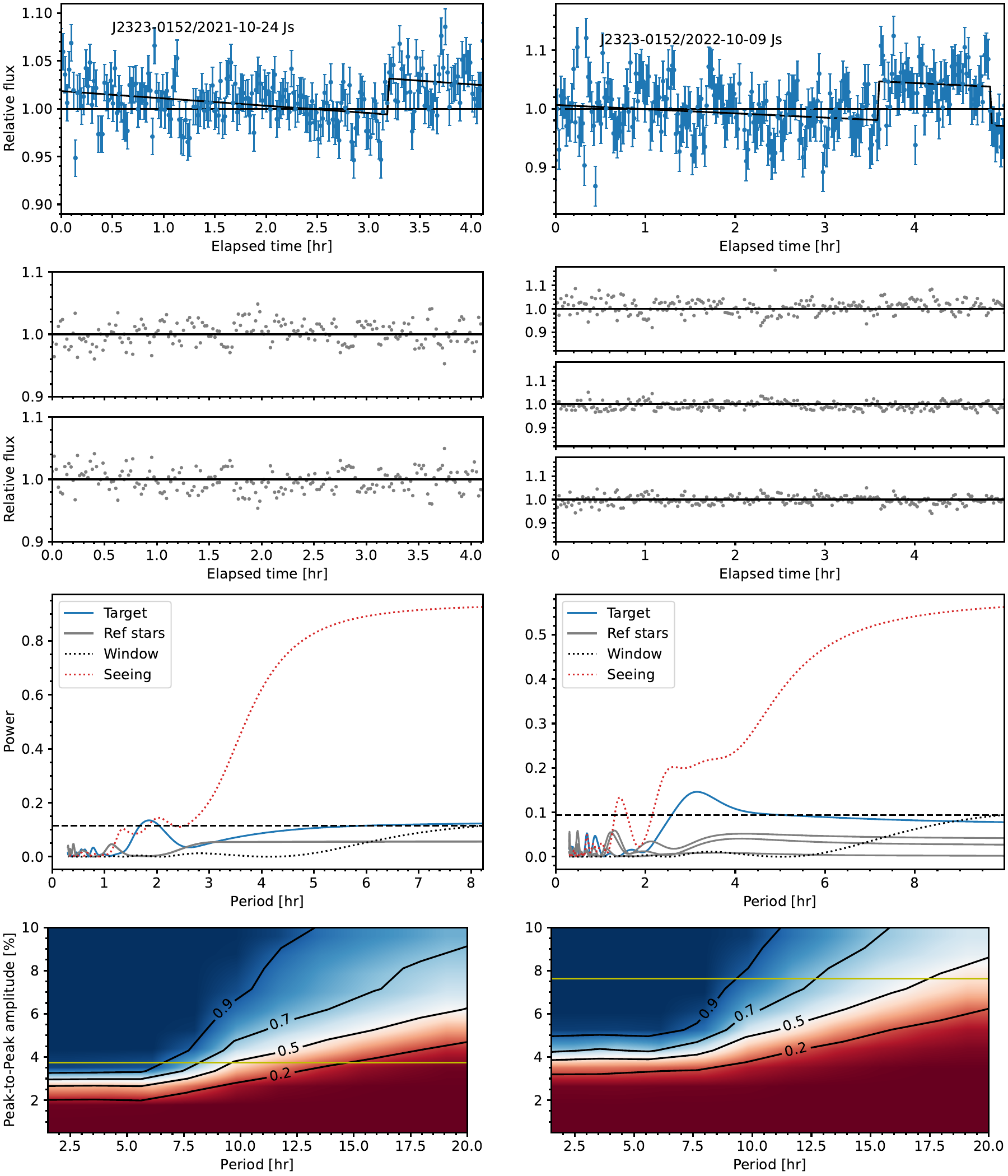}
    \caption{Variable light curves, periodograms, and sensitivity plots of J2323-0152 on 24 Oct 2021 and 09 Oct 2022, including detrended light curves and periodograms of its reference stars.}
    \label{fig:J2323-0152_var}
\end{figure*}

CFBDS~J232304.41-015232.3: discovered by \cite{Albert2011}, this T6 dwarf is identified as a high-probability member of $\beta$ Pictoris with a mass of $\sim$4.8\,$M_\textrm{J}$ \citep{ZhangZ2021}. We observed J2323-0152 on 21 Oct 2021 and 24 Oct 2021 in the first epoch and 09 Oct 2023 and 10 Oct 2022 in the second epoch.
We also observed J2323-0152 on 16 Jun 2022 and 17 Jun 2022 less than two hours per night. But these observations were taken under high humidity or cloudy conditions without any variability detection.
We detected no variability on 21 Oct 2021 and 10 Oct 2022. These non-variable light curves are shown in Appendix~\ref{apd:non-variable}.
We have positive detections on 24 Oct 2021 and 09 Oct 2022. The results are shown in Fig.~\ref{fig:J2323-0152_var}.
We find marginal variability just above the 1\% FAP level on 24 Oct 2021. The peak around $\sim$2\,hr in the periodogram of 24 Oct 2021 is close to the peak of the seeing curve. This peak is likely related to the residual seeing effect since we were only able to pick two good reference stars.
There is another long-term variability according to the periodogram. 
The light curve shows a variable pattern with a decreasing trend and a plateau-shaped enhancement on 09 Oct 2022, which is confirmed to be significant in the periodogram with a reported period of 3.15\,hr, but this period is not evident from visual inspection of the light curve. 
If we assume the variability is caused by this decreasing trend with a plateau-shaped enhancement, the period is longer than the observing length of 5\,hr. The max-min amplitude of the light curve of 09 Oct 2022 is $\sim7.6\%$.
We also notice that the light curve of 24 Oct 2021 begins to rise at the end, which is likely another plateau-shaped enhancement. We also fit a linear trend with a plateau-shaped enhancement to this light curve and measure a max-min amplitude of $\sim3.7\%$.
Therefore, J2323-0152 is classified as a variable with a long period.
% What causes the enhancement of this target is interesting to investigate deeply. 

\subsection{Variable candidates}
We identify two potentially variable candidates. Although they present variability above the 1\% FAP level, we are cautious in our interpretation of these results due to the low quality of the data. Their light curves, periodograms, and sensitivity plots can be found in Appendix~\ref{apd:potential}.

\rev{2MASSI~J1553022+153236AB: 2M1553+1532 is a T7 candidate of the Carina-Near YMG with a probability of 89.6\% and an estimated mass of 12\,$M_\textrm{J}$ according to \cite{ZhangZ2021} using the trigonometric parallax. It was initially resolved to be a field binary system of two T6.5+T7.5 dwarfs \citep{Burgasser2002,Burgasser2006,Dupuy2012}. 
The separation is about 0\farcs6 in the resolved image of \cite{Dupuy2012}. It was not resolvable with SOFI. 
We detected long-term variability on 18 Jun 2022 and 09 May 2023. However, its periodogram shows a similar trend as the periodogram of the seeing curve on both nights and we were only able to select two good reference stars on both nights. The variability is suspicious. It might be a coincidence because the impact of the seeing variations should be removed after detrending with the reference stars as the detrended light curves of the reference stars are flat. 
We also observed it on 07 May 2023 for 1.56\,hr and 10 May 2023 for 3.02\,hr but did not detect any variability. Therefore, we consider it a potential variable.}

WISEA~J022609.16-161000.4: this L6 dwarf is a high-probability member of AB Doradus and has a mass range of 16--28\,$M_\textrm{J}$ \citep{Schneider2017}. We observed it on two nights: 22 Oct 2021 and 24 Oct 2021.
On the first night, poor weather conditions resulted in the loss of two hours of data and no variability was detected.
The weather conditions were much better on the second night.
The periodogram detects two peaks above the 1\% FAP level. The first peak around 2.3\,hr coincides with the peaks of reference stars, raising suspicions of systematic variability. The second peak is actually a plateau towards long periods, showing long-term variability. 
However, selecting suitable reference stars for this target is challenging as there are few point sources in the field. Only three faint reference stars were chosen and their own detrended light curves are quite noisy, reducing the quality of the calibration curve created with them. 
We re-observed it on 02 Nov 2022 and 04 Dec 2022. There is a peak at 2.19\,hr in the periodogram of 02 Nov 2022 but this variability is not evident in the visual inspection of the light curve. The light curve on 04 Dec 2022 did not show any variability.
We consider J0226-1610 a potential variable.

\subsection{Non-detections}
We do not detect significant variability in eleven targets, including two L dwarfs and nine T dwarfs. Their light curves, periodograms and sensitivity plots are presented in Appendix~\ref{apd:non-variable}. Several noteworthy targets are discussed below.

WISEA~J004403.39+022810.6: discovered by \cite{Skrzypek2016}, this L7 dwarf is a high-probability member of $\beta$ Pictoris with a mass of 7-11\,$M_\textrm{J}$ \citep{Schneider2017}. We observed it on two nights: 3.75\,hr on 20 Oct 2021 and 2.04\,hr on 23 Oct 2021. We detect no variability in its light curves.

\rev{WISEA~J020047.29-510521.4: J0200-5105 is a high-probability member of AB Doradus and is identified as an L6--L9 dwarf with a mass range of 16--28\,$M_\textrm{J}$ \citep{Schneider2017}. We observed it on 20 Oct 2021 in the first epoch and did not detect variability above the 1\% FAP level. The seeing changed from 1\farcs0 to 2\farcs3 during the observation. Due to the poor seeing conditions, we re-observed it on four nights: 13 Feb 2022, 09 Oct 2022, 10 Oct 2022, and 03 Dec 2022.
No variability is detected in the light curves of these observations.}

ULASJ075829.83+222526.7: discovered by \cite{Burningham2013}, this T6.5 dwarf is a high-probability member of Argus with a mass of approximately 4.8\,$M_\textrm{J}$ \citep{ZhangZ2021}. With a $J$ magnitude of 17.62, it is the faintest target in our survey. It was observed under favourable conditions on 17 Feb 2022 for 3.2\,hr. 
While its light curve appears to exhibit a decreasing trend, this variability was below the 1\% FAP level according to the periodogram. Further observations may provide evidence to confirm this variability.

PSOJ168.1800-27.2264: the T2.5 dwarf, discovered by \cite{Best2015}, is a likely member of the Argus group identified with photometric parallax \citep{ZhangZ2021}. It has a mass of $\sim$8\,$M_\textrm{J}$. We observed it on two consecutive nights: 4.1\,hr on 18 Feb 2022 and 5.0\,hr on 19 Feb 2022. No significant variability is detected in its light curves.

WISEA~J043718.77-550944.0: this L5 dwarf was identified as a high-probability member of $\beta$ Pictoris by \cite{Schneider2017}. However, its predicted distance and surface gravity have conflicting results from different methods, making its youth and membership status uncertain. Due to a shortage of suitable observation targets in Feb 2022, we observed J0437-5509 on 14 Feb 2022 for 2.8 hours. The periodogram analysis of its light curve does not detect any variability.

\section{Statistical analysis}
\label{sec:statistical analysis}
Our sample provides a first investigation of the variability of young planetary-mass T dwarfs. 
To gain a comprehensive understanding of the variability of both young and field L and T dwarfs at near-infrared wavelengths, we combine our survey with previous studies. These include the variability survey of young L dwarfs from \cite{Vos2019}, the variability survey of field L and T dwarfs from \cite{Radigan2014a}, and the smaller variability survey of field objects at the L/T transition from \cite{Eriksson2019}. All of these studies were conducted using ground-based photometric monitoring campaigns in the $J$ band, similar to our own. They also employed similar variability identification criteria, with a variability significance level higher than the 1\% FAP level in the LS periodogram (except for 0.1\% FAP in \cite{Eriksson2019}). 
The sensitivity plots of samples from our survey, \cite{Vos2019} and \cite{Radigan2014a} are also calculated in a similar way by injecting and detecting the simulated sinusoidal signals in the light curves. We extract the light curves from \cite{Eriksson2019} and calculate their sensitivity plots using the same method. Therefore, it is reasonable to compare them statistically.

\begin{figure}
	\includegraphics[width=\columnwidth]{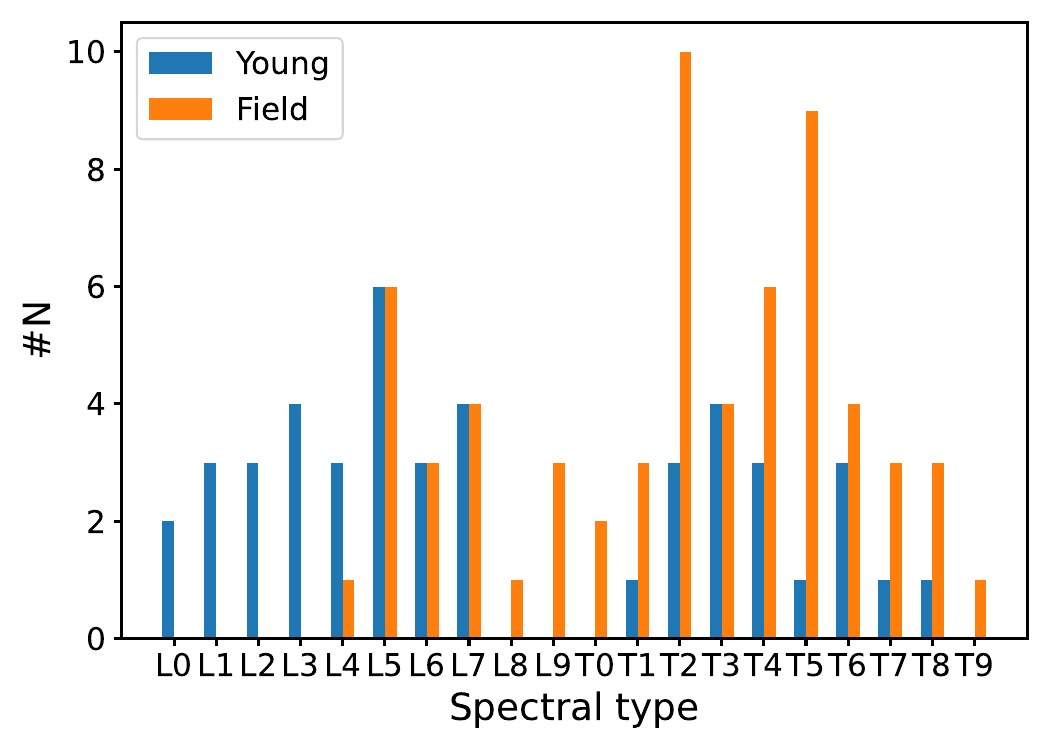}
    \caption{Spectral distribution of brown dwarfs included in the statistical analysis.}
    \label{fig:spt_dis}
\end{figure}

\begin{figure*}
	\includegraphics[width=2\columnwidth]{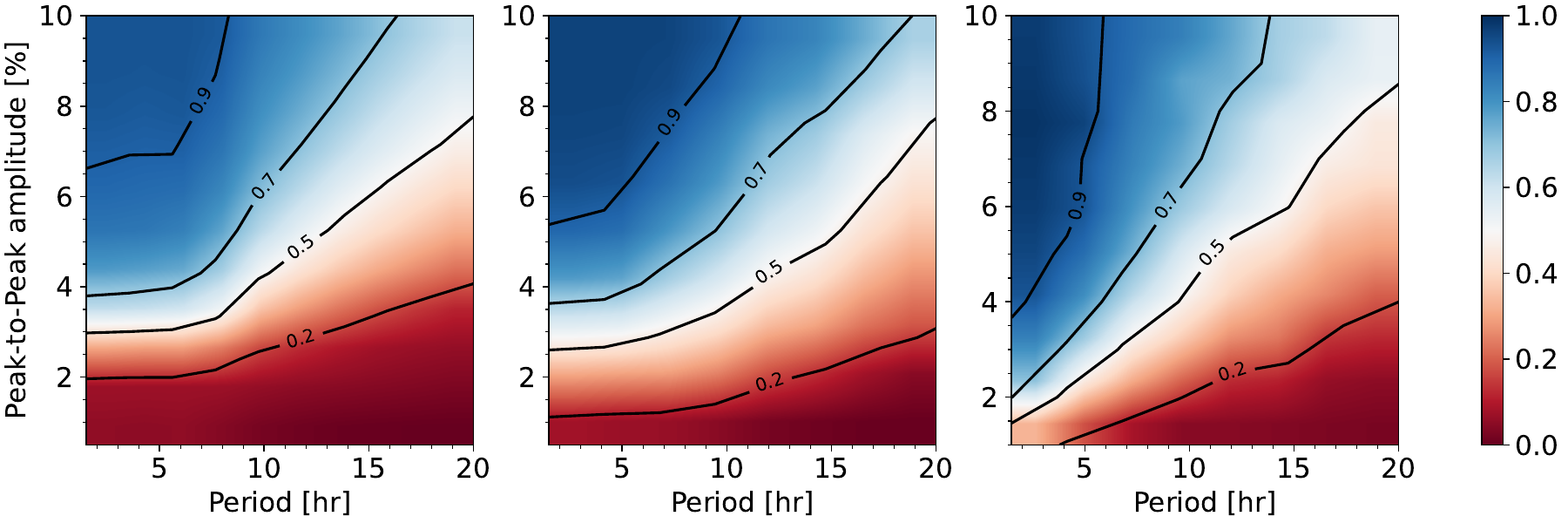}
    \caption{Average sensitivity maps of this survey, \citetalias{Vos2019}, and \citetalias{Radigan2014a} for objects included in the statistical analysis. The colour represents the retrieval rate. Our objects are fainter than objects in \citetalias{Vos2019} and we also have poorer seeing. Field dwarfs in \citetalias{Radigan2014a} are bright and thus they are more sensitive to weak variability.
    }
    \label{fig:average_sensi3}
\end{figure*}

We exclude known binary objects in the statistical analysis, as their variability may be due to one or both components or eclipsing binaries. 
We also exclude objects with uncertain youth, as our goal is to compare variability between young and field objects. 
The two variables in \cite{Radigan2014a}, SIMP0136+0933 and 2MASS2139+0220, were originally classified as field T dwarfs but were later found to be members of the Carina-Near YMG \citep{Gagne2017b, ZhangZ2021}. We include the two variables in the sample of young objects instead of the field sample. Another variable T dwarf in \cite{Eriksson2019}, 2MASS0013-1143, also turned out to be a candidate member of the Argus YMG \citep{ZhangZ2021} and is added to the young sample.
One variable T2 dwarf in \cite{Vos2019}, PSO071, was initially identified as a likely member of $\beta$ Pictoris by \cite{Best2015}, but was later classified as a field dwarf by \cite{Best2020, Marocco2021}. Thus we include it in the field T sample.
We consider only significant variability detections as variables in the statistical analysis.
Marginal detections or potential variable candidates are considered non-variables. We also exclude two objects that were observed for less than 2\,hr in \cite{Eriksson2019}.
Additionally, one variable from their sample, 2M2239+1617, has variability significance below the 1\% FAP level in our periodogram analysis and thus we consider it non-variable.
In total, we have 45 (10 variables) young objects consisting of 15 (3) from our survey, 26 (4) from \cite{Vos2019}, 2 (2) from \cite{Radigan2014a} and 1 (1) from \cite{Eriksson2019}, and 63 (10 variables) field objects including 55 (7) from \cite{Radigan2014a}, 7 (2) from \cite{Eriksson2019} and 1 (1) from \cite{Vos2019}. Table~\ref{tab:YFsurvey} lists the numbers of L and T brown dwarfs included in the statistical analysis. Figure.~\ref{fig:spt_dis} shows the spectral distribution. Young objects span from L0 to T8 and field objects span from L4 to T9. The averaged sensitivity maps of this survey, \cite{Vos2019}, and \cite{Radigan2014a} are shown in Fig.~\ref{fig:average_sensi3}.
Our objects are fainter than objects in \cite{Vos2019} and were observed under poorer conditions, while field dwarfs in \cite{Radigan2014a} are much brighter than young objects. Therefore, it is easier for \cite{Radigan2014a} to detect weaker variability.

% Our young sample lack objects from L8 to T0.

\begin{table}
	\centering
	\caption{Brown dwarfs included in the statistical analysis.}
	\label{tab:YFsurvey}
	\begin{tabular}{l l l l} % four columns, alignment for each
		\hline\hline
		Type & Variable & Non-Variable & Total\\
		\hline
		Young L & 5 & 23 & 28\\
		Young T & 5 & 12 & 17\\
		Young L and T & 10 & 35 & 45\\
		Field L & 1 & 17 & 18\\
		Field T & 9 & 36 & 45\\
		Field L and T & 10 & 53 & 63\\
		\hline
	\end{tabular}
\end{table}

\begin{figure}
	\includegraphics[width=\columnwidth]{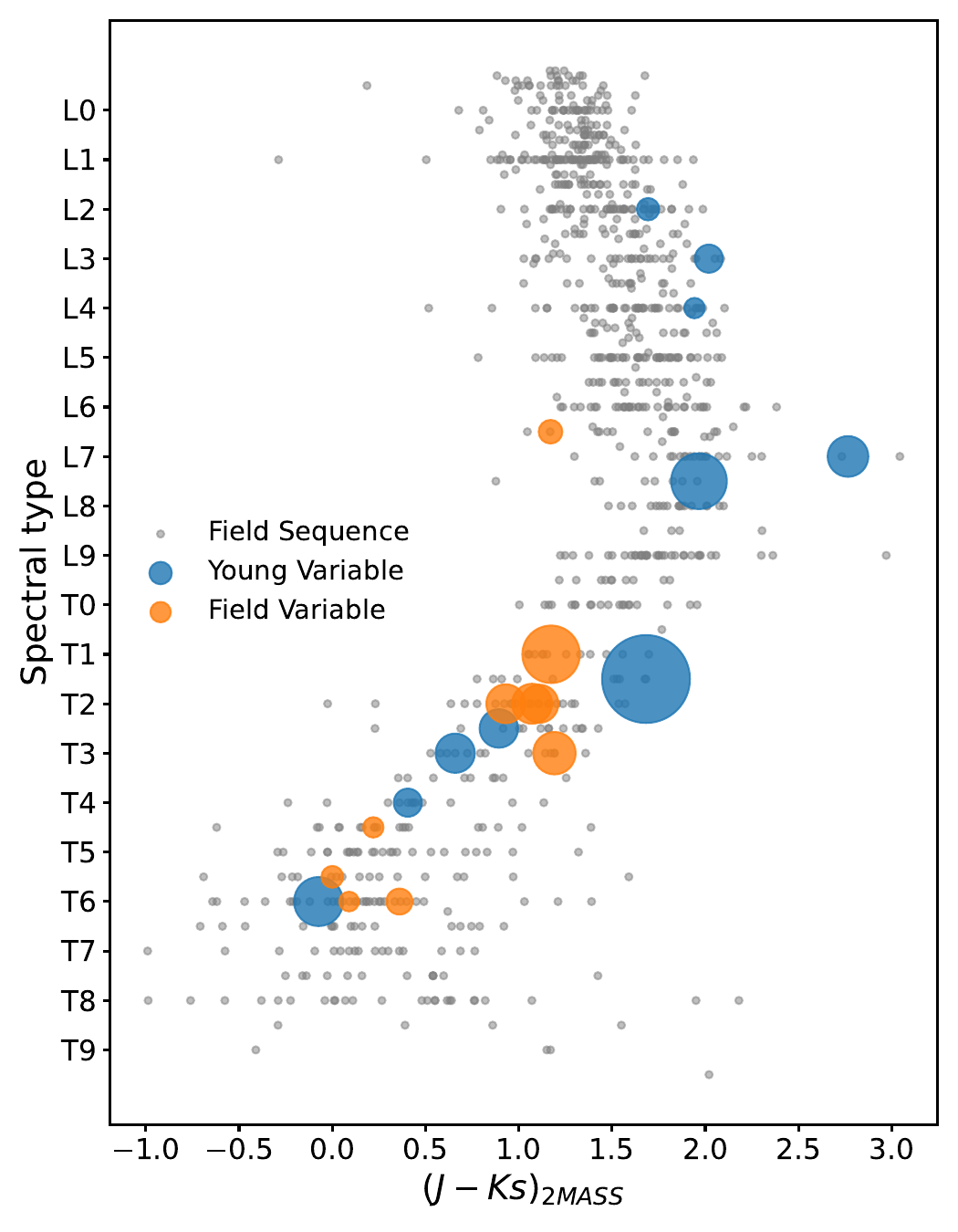}
    \caption{Spectral type against $(J-K_S)_{2MASS}$ colour of variable objects from the four surveys. The grey dots are the field sequence from The UltracoolSheet. 
    The dots in blue and orange represent the variables from the four surveys, with their size proportional to the variability amplitudes.}
    % While variables span in a broad spectral type range, strong variables with peak-to-peak amplitude $>$ 2\% tend to locate in a narrower range. Field variables with strong variability assemble from T1 to T3, while young variables with strong variability assemble in a broader range from L7 to T6.
    \label{fig:spt_color_var}
\end{figure}

Figure.~\ref{fig:spt_color_var} illustrates the relationship between spectral type and 2MASS $J-K_S$ colour of variable objects from the four surveys.
The field sequence objects are taken from the UltracoolSheet\footnote{\url{http://bit.ly/UltracoolSheet}} \citep{Best2020}, a catalogue of over 3,000 ultracool dwarfs and directly-imaged exoplanets. Some young T dwarfs in our sample have only MKO magnitudes and we convert their MKO $J-K$ colour to the 2MASS photometric system using the transformation equation\footnote{\url{https://www.ipac.caltech.edu/2mass/releases/allsky/doc/sec6_4b.html}}. Strong variables with peak-to-peak amplitude $>$ 2\% are concentrated within a narrow range, with strong young variables assembling in L7--T6 and strong field variables gathering from T1 to T3, a narrower distribution compared to strong young variables.

\subsection{Statistical formalism}
Although we can estimate the variability rate of brown dwarfs by the ratio of the number of variables to the total number, this is biased by the detection sensitivity which can vary significantly between different observations. To include the effect of detection sensitivity, we adopted the statistical formalism used in \cite{Vos2019} and \cite{Vos2022}, which is a Bayesian method based on \cite{Lafreniere2007b} and \cite{Bonavita2013}. $f$ is the variability frequency with amplitude and rotation period in the interval $[a_{min},a_{max}] \cap [r_{min},r_{max}]$. If we number the observations of N objects by $j = 1...N$, $p_j$ is the probability that such variability would be detected in observation $j$. In the sensitivity map calculated for each observation, $g(r,a)$ is the detection rate of the injected sinusoidal signals in each grid. $p_j$ is the integral of $g(r,a)$ over the considered amplitude and period ranges of the injected signals normalised by the area of the map:
\begin{equation}
   p_j = \frac{\int_{a_{min}}^{a_{max}} \int_{r_{min}}^{r_{max}} g(a,r) \,da\,dr}{\int_{a_{min}}^{a_{max}} \int_{r_{min}}^{r_{max}} 1 \,da\,dr}
\end{equation}

In our case, $0.005 \le a \le 0.1$ and $1.5 \le r \le 20\,hr$ as we inject these signals to calculate the sensitivity map. We choose the amplitude and period boundaries to keep the sensitivity map consistent with the maps of \cite{Vos2019} and \cite{Radigan2014a}. Though our average sensitivity rate is smaller than 0.2 when the amplitude is smaller than 2\% in Fig.~\ref{fig:average_sensi3}, we are able to detect amplitude as low as 0.5\% when the observation conditions are good and the targets are bright such as the observations of J0819-0335 in Fig.~\ref{fig:J0819-0335_curves}.
The lower boundary of the period is the shortest observation length and the upper boundary is an arbitrarily long period.

The probability of detecting one object to be variable is therefore $fp_j$ and non-variable is $1-fp_j$. The detection made in observation $j$ is $d_j$: $d_j = 1$ for positive detection and $d_j = 0$ for non-detection. The probability of observing detections in N observations for a given $f$ is:
\begin{equation}
   L(d_j|f) = \prod_{j=1}^{N} (1-fp_j)^{(1-d_j)} (fp_j)^{d_j}
\end{equation}
According to Bayes's theorem, the posterior distribution (the probability density of $f$ for a given $d_j$) is:
\begin{equation}
   p(f|d_j) = \frac{L(d_j|f)p(f)}{\int_{0}^{1} L(d_j|f)p(f)\,df}
\end{equation}
The likelihood function is the previously calculated $L(d_j|f)$. Since we know little about the prior distribution of the variability occurrence rate $f$, we used the non-informative Jeffreys prior \citep{Vos2022}:
\begin{equation}
   J(f) = \sqrt{\sum_{j} \frac{p_j}{f(1-fp_j)}}
\end{equation}

We calculate $p(f|d_j)$, the probability density function (PDF) of the variability occurrence rate of brown dwarfs using the above equations. We also calculate the 68\% and 95\% confidence intervals of $f$ with the maximum likelihood following the method used in \cite{Kraft1991} and \cite{Vos2022}. If both upper and lower boundaries exist, the confidence interval of credibility $\alpha$ in $[f_{min}, f_{max}]$ is given by:
\begin{equation}
  \alpha = \int_{f_{min}}^{f_{max}} p(f|d_j)\,df; \\
   p(f_{min}|d_j) = p(f_{max}|d_j)
\end{equation}
If only one side boundary can be calculated, the upper or lower boundary is given by:
\begin{equation}
  \alpha = \int_{f_{min}}^{1} p(f|d_j)\,df; \\
  or\, \alpha = \int_{0}^{f_{max}} p(f|d_j)\,df
\end{equation}

\subsection{Variability occurrence rates of field and young L and T dwarfs}
We calculate the sensitivity map for each observation in our survey. For objects observed multiple times, we use the most sensitive sensitivity map or the one with a positive detection. 
We obtained sensitivity maps of the objects from \cite{Vos2019}. For field objects from \cite{Radigan2014a}, we are only able to obtain sensitivity maps for part of the sample via private communication. For the rest of the objects that have light curves presented in \cite{Radigan2014a}, we extract these curves and calculate their sensitivity maps using our routine. These give us sensitivity maps of 23 field objects with a good representation of L and T spectral types.
We use the average sensitivity map for the corresponding spectral type interval when calculating $f$ for field objects as a function of spectral type.
This is a reasonable approach since \cite{Radigan2014a} demonstrate that their survey sensitivity does not vary significantly with spectral type. For objects from \cite{Eriksson2019}, we extract the light curves and calculate sensitivity maps for each object.

\begin{figure}
	\includegraphics[width=\columnwidth]{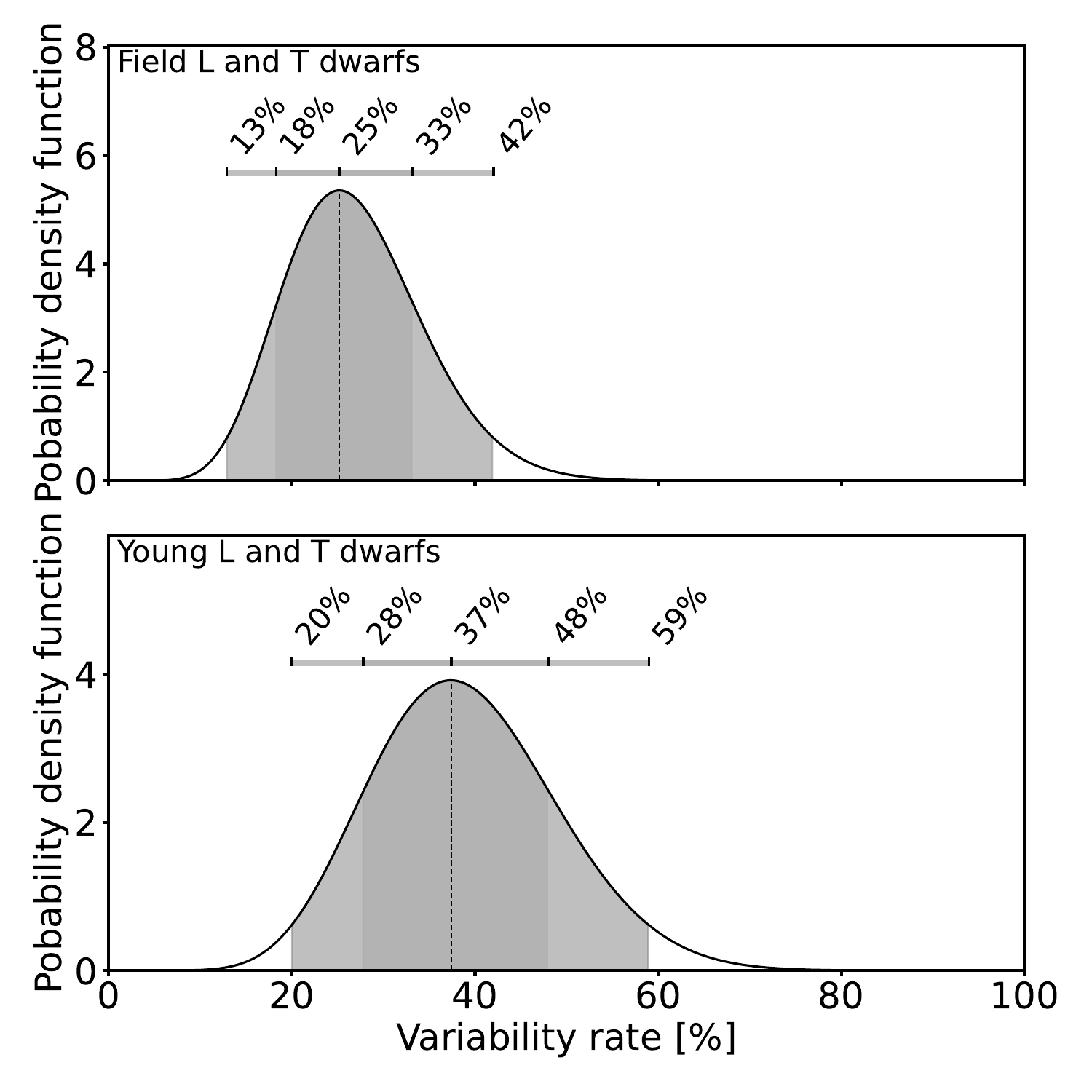}
    \caption{Probability density function of variability rate of the field and young samples. The dark grey area shows the 68\% confidence interval and the light grey area shows the 95\% confidence interval.}
    \label{fig:compare_LT}
\end{figure}

\begin{figure*}
	\includegraphics[width=2\columnwidth]{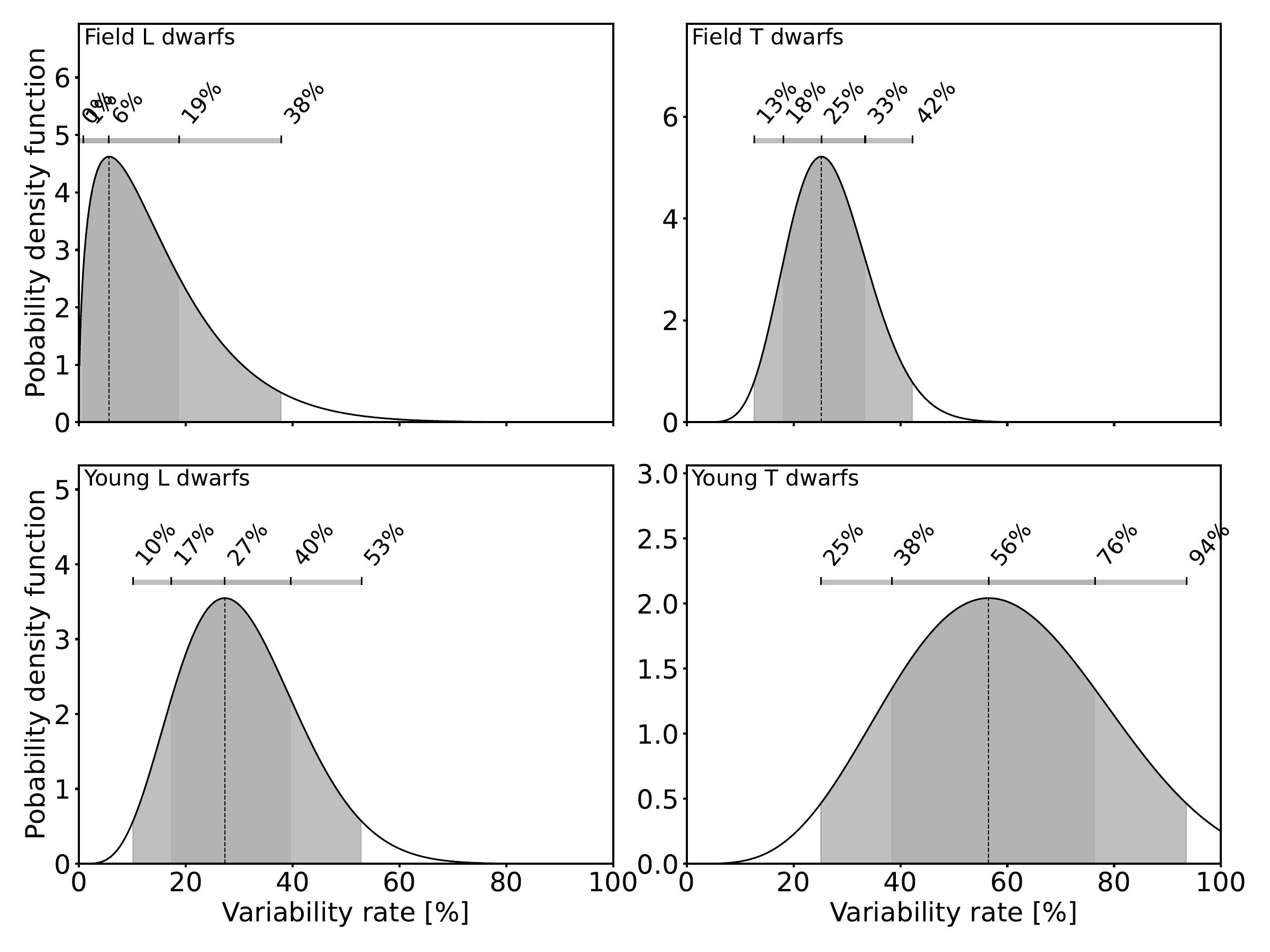}
    \caption{Probability density function of variability rate in the field and young samples for L and T spectral types, respectively. The dark grey area shows the 68\% confidence interval and the light grey area shows the 95\% confidence interval.}
    \label{fig:compare_LandT}
\end{figure*}

First, we calculate the total variability rates in the field and young samples with a variability amplitude between 0.5\% and 10\% and a period between 1.5 and 20 hours.
The field sample has a variability rate of $25_{-7}^{+8}$\% and the young sample has a variability rate of $37_{-9}^{+11}$\% as shown in Fig.~\ref{fig:compare_LT}. Yong L and T objects tend to be more variable than field objects in ground-based near-infrared observations but the difference is not significant as the rates overlap within 1$\sigma$.
We also calculate the variability rate for L (L0--L9.5) and T (T0--T9.5) spectral types separately as compared in Fig.~\ref{fig:compare_LandT}. The field L dwarfs have a variability rate of $6_{-5}^{+13}$\% while the young L dwarfs have a variability rate of $27_{-10}^{+13}$\%, which is consistent with the previous result reported by \cite{Vos2019}.
But since the difference in rate is within 1$\sigma$, this trend is not significant.
We find that young T dwarfs are also more variable than field T dwarfs with a variability rate of $56_{-18}^{+20}$\% compared with $25_{-7}^{+8}$\%. Though the difference is larger than 1$\sigma$, we are cautious about it as the young T sample is small. 

In both field and young samples, T dwarfs have a tendency to be more likely to be variable than L dwarfs. We suspect that this may be biased by the L/T transition, which is from L9 to T3.5 and covers more T spectral types than L types. Hence we also calculate the variability rate of field and young dwarfs with spectral types later than L9 and earlier than T3.5. 
Field T4--T9.5 dwarfs have a variability rate of $17_{-7}^{+9}$\% which is higher than the rate of $7_{-6}^{+15}$\% for field L0--L8.5 dwarfs. Young T4--T9.5 dwarfs are also still more likely to be variable than young L0--L8.5 dwarfs, with a variability rate of $44_{-22}^{+18}$\% and $27_{-10}^{+13}$\% separately. But these rates overlap significantly within 1$\sigma$, therefore whether T dwarfs are more likely to be variable than L dwarfs needs larger samples to refine. Nevertheless, the comparable variability rates of T dwarfs to L dwarfs suggest that clouds are also common in mid-late T spectral types. After the condensation of silicate clouds at the L/T transition, clouds composed of other species can form in the atmospheres of T dwarfs, such as sulfide clouds \citep{Morley2012}. The relatively higher rates in young samples suggest that low surface gravity is more favourable to cloud formation. Our work is from a statistical view. Characterisation of individual objects, such as time-resolved spectroscopy observations and atmospheric simulations can help study the impact of surface gravity in detail \citep[e.g.][]{Marley2012, Manjavacas2014, Filippazzo2015, Vos2023}.

\subsection{Variability occurrence rates at and outside the L/T transition}
\begin{figure*}
	\includegraphics[width=2\columnwidth]{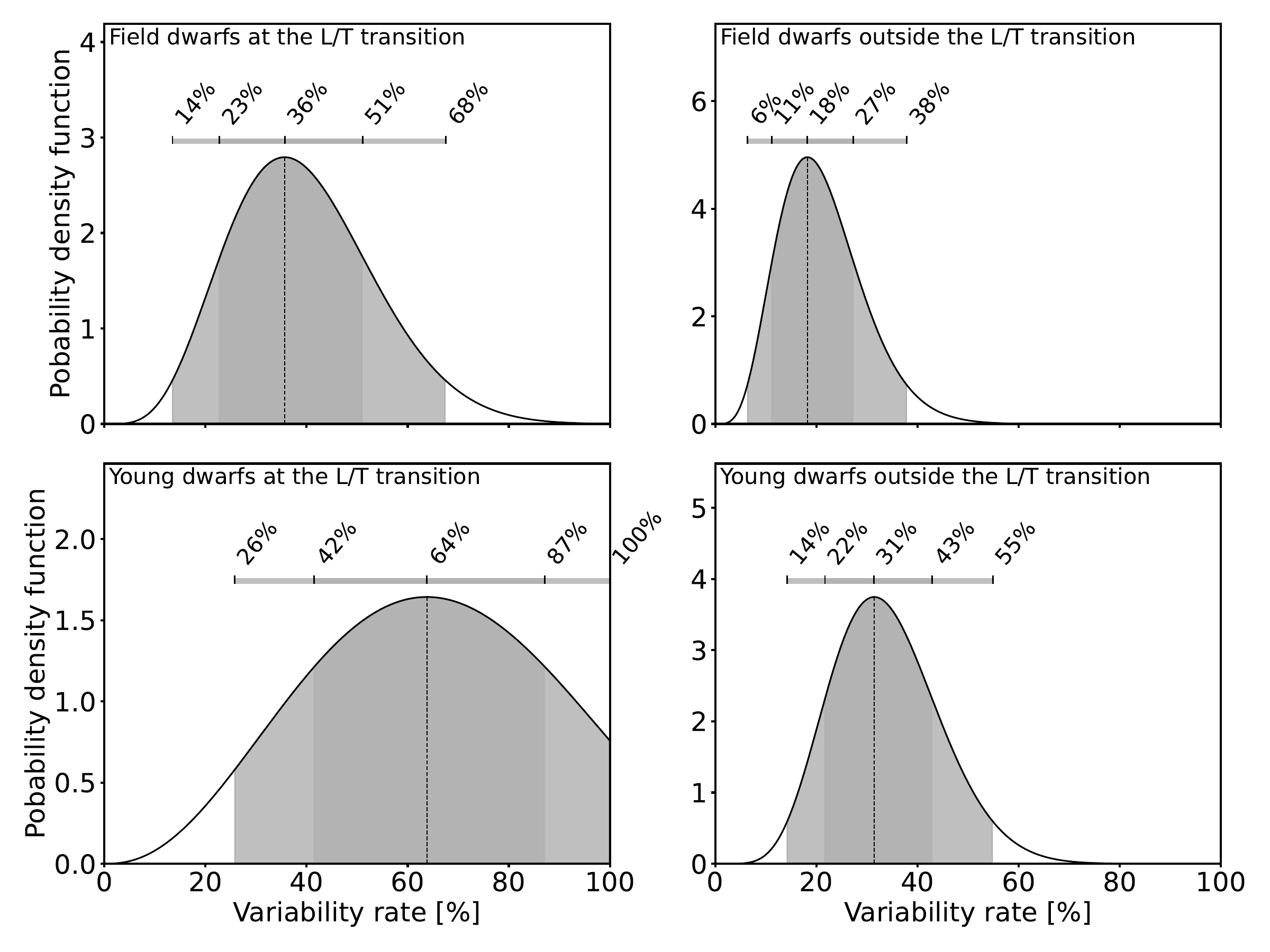}
    \caption{Probability density function of variability rate at the L/T transition and outside the L/T transition for field and young samples. The dark grey area shows the 68\% confidence interval and the light grey area shows the 95\% confidence interval.}
    \label{fig:compare_transition}
\end{figure*}

We also investigate how the variability occurrence rate varies at and outside the L/T transition for field and young samples with a variability amplitude between 0.5\% and 10\% and a period between 1.5 and 20 hours.
While there is no clear definition of the spectral type range of the L/T transition, we adopted the spectral type range of L9--T3.5 used in \cite{Radigan2014a} for a fair comparison with their results. This definition is also used in \cite{Vos2022} for their analysis of young object variability at mid-infrared wavelengths. As presented in Fig.~\ref{fig:compare_transition}, for field objects, the variability rate at the L/T transition is higher than that outside the L/T transition, with a value of $36_{-13}^{+15}$\% and $18_{-7}^{+9}$\%, respectively. 
\cite{Radigan2014a} also report that strong variables (peak-to-peak amplitudes $>$ 2\%) are more frequent at the L/T transition than outside of the L/T transition. 
This contradicts the variability observations at mid-infrared wavelengths by \textit{Spitzer}. \cite{Metchev2015} report that they did not observe a variability enhancement or stronger amplitudes at the L/T transition, though their sample consists of field and young objects. \cite{Vos2022} separate field and young objects and they find a higher variability rate for field L0--L8.5 dwarfs than field dwarfs at the L/T transition. This indicates that the variability of field dwarfs in the near-infrared and mid-infrared may originate from different sources since different wavelengths probe different heights in atmospheres. 
$J$ band detects a deeper and higher-pressure region, $\sim$10 bar, while mid-infrared wavelengths probe lower-pressure layers, $\sim$1 bar \citep{Vos2022}. \rev{It could also be due to the sensitivity difference between the ground-based observations which detect variability over 1\% and the space observations which achieve subpercent sensitivity.}

For young objects, we find that variables are also more likely to be located at the L/T transition than outside it, with a variability rate of $64_{-22}^{+23}$\% and $31_{-9}^{+12}$\% separately. The large uncertainty of the variability rate at the L/T transition is due to the small sample size. There are only three variables out of eight objects at the L/T transition compared with seven variables out of 37 outside the L/T transition in the young sample. Again, the variability rates of objects at the L/T transition and outside the L/T transitions overlap slightly within 1$\sigma$ for both field and young samples. This tentative enhancement indicates that the L/T transition has an impact on the near-infrared variability properties of field brown dwarfs and planetary-mass objects.

If we compare variability rates of the three spectral ranges, L0--L8.5, L9--T3.5, and T4--T9.5, we find that the rate is the highest at the L/T transition, then T4--T9.5 and the lowest at L0--L8.5. This trend is the same for field and young objects in the near-infrared. \cite{Vos2022} find that the variability rate is high for young objects from L to T but it drops from L to T for field objects in the mid-infrared for amplitudes of 0.05\%--3\% and periods of 0.5--40\,hr.
Table~\ref{tab:variabilitycom} provides the statistical variability rate of field and young objects from this work and \cite{Vos2022}. 
The variability rates of objects with spectral types later than L8.5 are consistent with each other in the near-infrared and mid-infrared for both field and young samples. For L dwarfs (L0--L8.5), there is a big discrepancy in the variability rates between the near-infrared and mid-infrared for both field and young samples. The L dwarfs have a higher variability rate in the mid-infrared but we are cautious about this discrepancy because the mid-infrared observations are from \textit{Spitzer}. These space observations have a much higher sensitivity and much longer observation time than ground-based observations. They could detect weak variables with long periods, thus resulting in higher detection rates. 

% Wavelength could also contribute to the discrepancy of the variability rates between young and field objects in the near-infrared and mid-infrared.
% Longer wavelengths probe lower pressure layers. Young L0--L8.5 dwarfs have a higher variability rate than field L dwarfs at near-infrared wavelengths, but they have similar variability rates at mid-infrared wavelengths. For T4--T9.5 spectral types, the discrepancy of variability rates between young and field objects is similar at near-infrared wavelengths and mid-infrared wavelengths. 
% These differences between L0--L8.5 and T4--T9.5 may reveal some atmospheric differences between L and T dwarfs. L dwarfs may have more cloud layers and complex atmospheric structures, while T dwarfs may have fewer and thinner cloud layers. As such, the variability detected at near-infrared and mid-infrared wavelengths may be more correlated in T4--T9.5 than L0--L8.5, causing the variability rates in the near-infrared and mid-infrared to be similar in T4--T9.5.
% Mid-infrared wavelengths probe high-altitude, low-pressure regions of the atmosphere that are less affected by surface gravity so that the variability rates are similar between field and young L objects. 
% Near-infrared wavelengths probe deeper atmospheric layers where the surface gravity plays a more important role, especially in L dwarfs, and thus result in different variability rates between young and field objects.

\begin{table}
	\centering
	\caption{Variability rates of field and young objects in the near-infrared (NIR) and mid-infrared (MIR). The NIR results are from this work for amplitudes of 0.5\%--10\% and periods of 1.5--20\,hr. The MIR results are from \citetalias{Vos2022} for amplitudes of 0.05\%--3\% and periods of 0.5--40\,hr.}
	\label{tab:variabilitycom}
	\begin{tabular}{l l l l}
		\hline\hline
		Type & L0--L8.5 & L9--T3.5 & T4--T9.5\\
		\hline
        Field NIR & $7_{-6}^{+15}$\% & $36_{-13}^{+15}$\% & $17_{-7}^{+9}$\% \\
        Young NIR & $27_{-10}^{+13}$\% & \rev{$64_{-22}^{+23}$\%} & \rev{$44_{-22}^{+28}$\%} \\
        Field MIR & 83--100\% & $41_{-19}^{+23}$\% & $18_{-15}^{+32}$\% \\
        Young MIR & 81--100\% & 85--100\% & 44--100\% \\
		\hline
	\end{tabular}
\end{table}

%effective digital number
% \begin{table}
% 	\centering
% 	\caption{Variability rates of field and young objects in the near-infrared (NIR) and mid-infrared (MIR). The NIR results are from this work and the MIR results are from \cite{Vos2022}.}
% 	\label{tab:variabilitycom}
% 	\begin{tabular}{l l l l}
% 		\hline\hline
% 		Type & L0--L8.5 & L9--T3.5 & T4--T9.5\\
% 		\hline
%         Field NIR & $8.5_{-7.3}^{+19.2}$\% & $41.9_{-15.0}^{+18.1}$\% & $17.5_{-7.4}^{+9.7}$\% \\
%         Young NIR & $31.0_{-11.4}^{+13.9}$\% & $72.1_{-21.9}^{+22.9}$\% & $49.6_{-23.1}^{+39.0}$\% \\
%         Field MIR & 83--100\% & $41_{-19}^{+23}$\% & $18_{-15}^{+32}$\% \\
%         Young MIR & 81--100\% & 85-100\% & 44--100\% \\
% 		\hline
% 	\end{tabular}
% \end{table}

\section{A census of variable objects}
Besides the four surveys discussed in the previous section, there are a few variable brown dwarfs monitored in other surveys or small sample observations \citep[e.g.][]{Buenzli2014, Metchev2015, Yang2015, Yang2016, Manjavacas2019b}. To gain a comprehensive understanding of the variability in L and T objects, it is necessary to have a census of known variables. \cite{Vos2020} summarise a list of known variables from the literature and we update that list with newly discovered variables and new results from existing variables. \rev{They are presented in Table~\ref{tab:allvariables}.}
Variability is common in L and T dwarfs from near-infrared to mid-infrared. 
Since our survey is conducted in the near-infrared, we focus our variability census in the $J$ band including ground-based $J$ band observations and HST/WFC3/NIR observations using filters in 800--1700\,nm.
% While we find that variability rate is higher at the L/T transition than outside the L/T transition for both field and young objects in the $J$ band, \cite{Vos2022} find that variability rate of field objects decreases after L9 but young objects maintain the variability rate from spectral type L to the L/T transition in the mid-infrared. They find a variability rate of 83--100\% for field L objects and 22--64\% at the L/T transition with 68\% confidence interval compared to 81--100\% and 85--100\% for young objects, separately. Their rates are significantly higher than our analysis, since their samples are from space observations of \textit{Spitzer} which are more sensitive than ground-based observations.
% \cite{Metchev2015} did not find an enhancement of the variability rate at the L/T transition than outside the L/T transition in the mid-infrared for a sample including both field and young objects. The discrepancy between these surveys indicates that L and T objects have different variability behaviour between the near-infrared and mid-infrared. It is reasonable since different wavelengths detect different layers in the atmospheres of brown dwarfs. $J$ band detects a deeper region than the mid-infrared. 

\begin{landscape}
\begin{table}
\caption{Variable brown dwarf detections in the literature. $A_{Jground}$: $J$ band peak-to-peak amplitude measured from ground observations; $A_{WFC3 NIR}$: peak-to-peak amplitude measured in the IR channel of HST/Wide Field Camera 3 (WFC3); $A_{3.6}$ and $A_{4.5}$: peak-to-peak amplitude measured in the 3.6\,$\mu$m and 4.5\,$\mu$m channels of \textit{Spitzer}.}
\label{tab:allvariables}
\begin{tabular}{llllllllllll}
\hline\hline
Target & SpT & $A_{Jground}$ [\%] & $A_{WFC3 NIR}$ [\%] & $A_{3.6}$ [\%] & $A_{4.5}$ [\%] & Period [hr] & Youth & Companion & Age [Myr] & $vsini$ [$km\:s^{-1}$]& Variability ref \\
\hline
2MASS J00011217+1535355 & L4 &- &-& 0.69 $\pm$ 0.04 &- & 15.75 $\pm$ 0.37 & 1 & 0 & 120 $\pm$ 30 & - & 1 \\
2MASS J00132229-1143006 & T3 (T3.5+T4.5) & 4.6 $\pm$ 0.2 & -  &  -  & -  &  -  & 1 & 0 & 45 $\pm$ 5 & - & 2 \\
2MASSW J0030300-145033 & L7 &  -  &  -  & 1.52 $\pm$ 0.06 &  -  & 4.22 $\pm$ 0.02 & 1 & 0 & 45 $\pm$ 5 & - & 1 \\
2MASS J00310928+5749364 & L9 &  -  &  -  & 0.35 $\pm$ 0.03 &  -  & 1.64 $\pm$ 0.01 & 1 & 0 & 200 $\pm$ 50 & - & 1 \\
LSPM J0036+1821 & L3.5 & 1.22 $\pm$ 0.04 &  -  & 0.47 $\pm$ 0.05 & 0.19 $\pm$ 0.04 & 2.7 $\pm$ 0.3 & 0 & 0 & 1000 & 36.0 $\pm$ 0.2 & 3, 4 \\
2MASS J00452143+1634446 & L2 & 1.0 $\pm$ 0.1 &  -  & 0.18 $\pm$ 0.04 & 0.16 $\pm$ 0.04 & 2.4 $\pm$ 0.1 & 1 & 0 & 45 $\pm$ 5 & $31.76_{-0.25}^{+0.22}$ & 5, 25 \\
2MASS J00470038+6803543 & L6 &  -  & 8.0  & 1.07 $\pm$ 0.04 &  -  & 16.4 $\pm$ 0.2 & 1 & 0 & 130 $\pm$ 20 & 9.8 $\pm$ 0.3 & 6, 7 \\
2MASS J00501994-3322402 & T7 &  -  &  -  &  <0.59 & 1.07 $\pm$ 0.11 & 1.55 $\pm$ 0.02 & 0 & 0 & 1000 & - & 3 \\
2MASSI J0103320+193536 & L6 &  -  &  - & 0.56 $\pm$ 0.03 & 0.87 $\pm$ 0.09 & 2.7 $\pm$ 0.1 & 1 & 0 & 300 $\pm$ 200 & $40.0_{-4.7}^{+3.7}$ & 3 \\
2MASS J01075242+0041563 & L8 &  -  &  -  & 1.27 $\pm$ 0.13 & 1.0 $\pm$ 0.2 & 5.0 $\pm$ 2 & 0 & 0 & 1000 & 19.4 $\pm$ 0.8 & 3 \\
GU PSC B & T3.5 & 4.0 &  -  &  -  &  - &  -  & 1 & 0 & 130 $\pm$ 20 & - & 8 \\
SIMP J013656.57+093347.3 & T2.5 & 5.0 & 4.5 & 1.5 $\pm$ 0.2 &  -  & 2.414 $\pm$ 0.078 & 1 & 0 & 200 $\pm$ 50 & 50.9 $\pm$ 0.8 & 9, 10 \\
2MASS J01383648-0322181 & T3 & 5.5 $\pm$ 1.2 &  - &  -  &  -  &  -  & 0 & 0 & 1000 & - & 2\\
2MASS J01531463-6744181 & L2 &  -  &  -  & 0.48 $\pm$ 0.07 &  -  & 17.63 $\pm$ 1.13 & 1 & 0 & 45 $\pm$ 4 & - & 1 \\
2MASSW J0310599+164816 & L8 &  -  &  -  &  -  &  -  &  -  & 0 & 0 & 1000 & - & 11\\
2MASSI J0342162-681732 & L4 &  -  &  -  & 0.73 $\pm$ 0.07 &  -  & 14.73 $\pm$ 0.51 & 1 & 0 & 45 $\pm$ 4 & - & 1 \\
2MASS J03480772-6022270 & T7 &  -  &  -  &  -  & 1.5 $\pm$ 0.1 & 1.08 $\pm$ 0.004 & 0 & 0 & 1000 & - & 12 \\
2MASS J03492367+0635078 & L5 &  -  &  -  & 0.53 $\pm$ 0.09 &  -  & 14.62 $\pm$ 1.08 & 1 & 0 & 45 $\pm$ 4 & - & 1 \\
2MASS J03552337+1133437 & L5 &  -  &  - & 0.26 $\pm$ 0.02 &  -  & 9.53 $\pm$ 0.19 & 1 & 0 & 130 $\pm$ 20 & - & 1 \\
2MASS J04070752+1546457 & L3.5 &  -  &  -  & 0.36 $\pm$ 0.1 &  -  & 1.23 $\pm$ 0.01 & 0 & 0 & 1000 & - &  12 \\
SDSS J042348.57-041403.5 & T0 & 0.8 $\pm$ 0.08 &  -  &  -  &  -  & 2.0 $\pm$ 0.4 & 0 & 0 & 1000 & 68.0 $\pm$ 0.9 & 13 \\
PSO J071.8769-12.2713 & T2 & 4.5 $\pm$ 0.6 &  -  & 0.93 &  -  &  -  & 0 &  0 & 1000  & - & 1, 5 \\
2MASS J05012406-0010452 & L3 & 2.0 $\pm$ 1.0 &  -  & 0.36 $\pm$ 0.04 & 0.24 $\pm$ 0.04 & 15.7 $\pm$ 0.2 & 1 & 0 & 300 $\pm$ 200 & $9.57_{-0.58}^{+0.67}$ & 5, 25\\
2MASS J05065012+5236338 & T4.5 &  -  &  -  & 0.58 $\pm$ 0.1 &  -  & 13.77 $\pm$ 0.82 & 0 &  0 & 2000 & - & 1 \\
2MASS J05591914-1404488 & T4.5 & 0.7 $\pm$ 0.5 &  -  &  -  &  -  & 10.0 $\pm$ 3 & 0 & 0  & 1000 & - & 10 \\
2MASS J06244595-4521548 & L6.5 &  -  & 1.5 &  -  &  -  &  -  & 0 & 0 & 1000 & -  & 11 \\
2MASS J06420559+4101599 & L9 &  -  &  -  & 2.16 $\pm$ 0.16 &  -  & 10.11 $\pm$ 0.07 & 1 & 0 & 42 $\pm$ 6 & - & 1 \\
2M0718-6415 & T5 &  -  &  - & 2.14 $\pm$ 0.21 &  -  & 1.08 $\pm$ 0.01 & 1 & 0 & 23 $\pm$ 3 & - & 1 \\
SDSS J075840.33+324723.4 & T2(T0+T3.5) & 4.8 $\pm$ 0.2 &  -  &  -  &  - & 4.9 $\pm$ 0.2 & 0 & 0 & 1000 & - & 10 \\
2MASS J08095903+4434216 & L6 &  - &  -  & 0.77 $\pm$ 0.06 &  -  & 1.365 $\pm$ 0.004 & 1 & 0 & 300 $\pm$ 200 & - & 1 \\
DENIS J081730.0-615520 & T6 & 0.6 $\pm$ 0.1 &  -  &  -  &  -  & 2.8 $\pm$ 0.2 & 0 & 0 & 1000 & - & 10 \\
2MASSI J0825196+211552 & L7.5 &  >1  &  -  & 0.81 $\pm$ 0.08 & 1.4 $\pm$ 0.3 & 7.6 $\pm$ 5 & 0 & 0 & 1000 & - & 3 \\
WISE J085510.83-071442.5 & Y0+ &  -  &  -  & 4.0 $\pm$ 1.0 & 4.0 $\pm$ 1.0 & 12.5 $\pm$ 3.5 & 0 &  0 & 1000 & - & 14\\
LP261-75B & L6 &  -  & 2.4 $\pm$ 0.14 &  -  &  - & 4.78 $\pm$ 0.98 & 0 & 0 & 300 $\pm$ 50 & - & 15 \\
SDSS J104335.08+121314.1 & L9 &  -  &  -  & 1.54 $\pm$ 0.15 & 1.2 $\pm$ 0.2 & 3.8 $\pm$ 0.2 & 0 & 0 & 1000 & - & 3 \\
2MASS J10475385+2124234 & T6.5 &  - &  -  &  -  & 1.04 $\pm$ 0.07 & 1.741 $\pm$ 0.007 & 0 &  0 & 1000 & - & 16 \\
WISE 1049B & L7.5+T0.5 &  -  & 11.0  &  -  &  -  & 4.87 $\pm$ 0.01 & 0 & 1 & 1000 & 21.6 $\pm$ 0.2 & 17, 18 \\
SDSS J105213.51+442255.7 & T0.5 & 2.2 $\pm$ 0.5 &  -  &  -  &  -  & 3.0 $\pm$ 0.5 & 0 & 0 & 1000 & - & 17 \\
DENIS-P J1058.7-1548 & L3 & 0.843 $\pm$ 0.098 &  -  & 0.39 $\pm$ 0.04 & <0.3  & 4.1 $\pm$ 0.2 & 0 & 0 & 1000 & 37.5 $\pm$ 2.5 & 3 \\
2MASS J11193254-1137466AB & L7 &  -  &  -  & 0.46 $\pm$ 0.036 & 0.96 $\pm$ 0.037 & 3.02 $\pm$ 0.04 & 1 & 1 & 10 $\pm$ 3 & - & 19, 20 \\
2MASS J11263991-5003550 & L4.5 & 1.2 $\pm$ 0.1 &  -  & 0.21 $\pm$ 0.04 & 0.29 $\pm$ 0.15 & 3.2 $\pm$ 0.3 & 0 & 0 & 1000 & $22.8_{-2.4}^{+1.6}$ & 3, 10 \\
WISEA J114724.10-204021.3 & L7 & \rev{4.6$\pm$1.0} &  -  & 1.596 $\pm$ 0.08 & 2.216 $\pm$ 0.09 & 19.39 $\pm$ 0.3 & 1 & 0 & 10 $\pm$ 3 & - & 19, 20 \\
2M1207b & L6 & - & 1.36 &  -  &  -  & 10.7 $\pm$1 & 1 & 0 & 10 $\pm$ 3 & - & 21 \\
2MASS J12195156+3128497 & L8 &  -  & 4.8 & 0.55 $\pm$ 0.13 &  -  & 1.14 $\pm$ 0.03 & 0 & 0 & 1000 & - & 11, 12 \\
VHS1256-1257AB b & L7 &  >10 & 38  & -  & 5.76 $\pm$ 0.04 & 22.04 $\pm$ 0.05 & 1 & 0 & 225 $\pm$ 75 & $13.5_{-4.1}^{+3.6}$ & 22, 23, 24 \\
\hline
\end{tabular}
\end{table}
\end{landscape}

\begin{landscape}
\begin{table}
\contcaption{Variable brown dwarfs detections in the literature. $A_{Jground}$: $J$ band peak-to-peak amplitude measured from ground observations; $A_{WFC3 NIR}$: peak-to-peak amplitude measured in the IR channel of HST/Wide Field Camera 3 (WFC3); $A_{3.6}$ and $A_{4.5}$: peak-to-peak amplitude measured in the 3.6\,$\mu$m and 4.5\,$\mu$m channels of \textit{Spitzer}.}
\label{tab:variables}
\begin{tabular}{llllllllllll}
\hline\hline
Target & SpT & $A_{Jground}$ [\%] & $A_{WFC3 NIR}$ [\%] & $A_{3.6}$ [\%] & $A_{4.5}$ [\%] & Period [hr] & Youth & Companion & Age [Myr] & $vsini$ [$km\:s^{-1}$]& Variability ref \\
\hline
Ross 458C & T8 &  -  & 2.62 $\pm$ 0.02 & <1.37  & <0.72 & 6.75 $\pm$ 1.58 & 1 & 0 & 125 $\pm$ 75 & - & 3 \\
2MASS J13243559+6358284 & T2.5 &  -  &  - & 3.05 $\pm$ 0.15 & 3.0 $\pm$ 0.3 & 13.0 $\pm$ 1 & 1 & 0 & 130 $\pm$ 20 & - & 3 \\
WISE J140518.39+553421.3 & Y0.5 &  -  &  -  & 3.6 $\pm$ 0.4 & 3.54 $\pm$ 0.09 & 8.2 $\pm$ 0.3 & 0 & 0 & 1000 & - & 26 \\
2MASS J14252798-3650229 & L4 & 0.7 $\pm$ 0.3 &  -  &  <0.16  &  <0.18  &  - & 1 & 0 & 130 $\pm$ 20 & $33.08_{-0.49}^{+0.53}$ & 25, 25 \\
2MASSW J1507476-162738 & L5 &  -  & 0.7  & 0.53 $\pm$ 0.11 & 0.45 $\pm$ 0.09 & 2.5 $\pm$ 0.1 & 0 & 0 & 1000 & 19.1$\pm$ 0.5 & 3, 9 \\
SDSS J151114.66+060742.9 & L5.5+T5.0 &  -  &  -  & 0.67 $\pm$ 0.07 & <0.49  & 11.0 $\pm$ 2 & 0 & 1 & 1000 & - &3 \\
SDSS J151643.01+305344.4 & T0.5 &  -  &  -  & 2.4 $\pm$ 0.2 & 1.3 $\pm$ 0.7 & 6.7 $\pm$ 5 & 0 & 0 & 1000 & - & 3 \\
2MASS J16154255+4953211 & L4 &  -  &  -  & 0.9 $\pm$ 0.2 & <0.39 & 24.0 $\pm$ 5 & 1 & 0 & 300 $\pm$ 200 & $9.5_{-1.2}^{+1.3}$ & 3 \\
2MASS J16291840+0335371 & T2 & 4.3 $\pm$ 2.4 &  -  &  - &  -  & 6.9 $\pm$ 2.4 & 0 &  0 & 1000 & $19.7_{-0.8}^{+0.7}$ & 10 \\
2MASSW J1632291+190441 & L8 &  -  &  - & 0.42 $\pm$ 0.08 & 0.5 $\pm$ 0.3 & 3.9 $\pm$ 0.2 & 0 & 0 & 1000 & - & 3 \\
2MASS J16471580+5632057 & L7 &  -  &  -  & 0.468 $\pm$ 0.06 &  -  & 9.234 $\pm$ 0.25 & 1 & 0 & 300 $\pm$ 200 & - & 1 \\
2MASSI J1721039+334415 & L3 &  -  &  -  & 0.33 $\pm$ 0.07 & <0.29 & 2.6 $\pm$ 0.1 & 0 & 0 & 1000 & 21.5 $\pm$ 0.3 & 3 \\
WISEP J173835.52+273258.9 & Y0 &  -  &  -  &  -  & 3.0 $\pm$ 0.1 & 6.0 $\pm$ 0.1 & 0 & 0 & 1000 & - & 27 \\
2MASS J17410280-4642218 & L7 &  -  &  -  & 0.35 $\pm$ 0.03 &  - & 15.0 $\pm$ 0.71 & 1 & 0 & 130 $\pm$ 20 & - & 1 \\
2MASS J17502484-0016151 & L5 &  >1  &  -  & -  & -  &  -  & 0 & 0 & 1000 & - & 11 \\
2MASS J17503293+1759042 & T3.5 &  >1  & - &  -  &  -  &  -  & 0 & 0 & 1000 & - & 11 \\
2MASS J17534518-6559559 & L4 &  -  &  -  & <0.25  & -  &  - & 0 & 0 & 1000 & - & 3\\
2MASS J18212815+1414010 & L4.5 &  -  & 2.5  & 0.54 $\pm$ 0.05 & 0.71 $\pm$ 0.14 & 4.2 $\pm$ 0.1 & 0 &  0& 1000 & 30.7 $\pm$ 0.2 & 3, 9 \\
2MASS J18283572-4849046 & T5.5 & 0.9 $\pm$ 0.1 &  -  &  -  &  -  & 5.0 $\pm$ 0.6 & 0 & 0 & 1000 & - & 10 \\
2MASS J20025073-0521524 & L5 & 1.7 $\pm$ 0.2 &  -  & 0.28 &  -  &  -  & 1 & 0 & 300 $\pm$ 200 & - & 1 \\
PSO 318.5-22 & L7.5 & 10.0 $\pm$ 1.3 & - &  -  & 3.4 $\pm$ 0.08 & 8.61 $\pm$ 0.06 & 1 & 0 & 22 $\pm$ 6 & $17.5_{-2.8}^{+2.3}$ & 5, 28, 29, \\
HD 203030B & L7.5 &  -  & 1.1 $\pm$ 0.3 &  -  &  -  & 7.5 $\pm$ 0.6 & 1 & 0 & 90 $\pm$ 60 & - & 30 \\
2MASS J21392676+0220226 & T1.5 & 26 & 27 & 11.0 $\pm$ 1.0 & 10.0 $\pm$ 1.0 & 7.618 $\pm$ 0.18 & 1 & 0 & 200 $\pm$ 50  & 18.7 $\pm$ 0.3 & 9, 11, 31 \\
HN PegB & T2.5 & - & 1.2 & 0.77 $\pm$ 0.15 & 1.1 $\pm$ 0.5 & 15.4 $\pm$ 0.5 & 1 & 0 & 237 $\pm$ 33 & - & 3 \\
2MASS J21481628+4003593 & L6 &  -  &  -  & 1.33 $\pm$ 0.07 & 1.03 $\pm$ 0.1 & 19.0 $\pm$ 4 & 0 &  1 & 300 $\pm$ 200 & $9.2_{-0.3}^{+0.4}$ & 3 \\
2MASS J22062520+3301144 & T1.5 &  -  &  -  & 1.2 $\pm$ 0.13 &  -  & 15.91 $\pm$ 0.62 & 1 & 0 & 45 $\pm$ 5 & - & 1 \\
2MASSW J2208136+292121 & L3 &  -  &  -  & 0.69 $\pm$ 0.07 & 0.54 $\pm$ 0.11 & 3.5 $\pm$ 0.2 & 1 & 0 & 22 $\pm$ 6 & $40.6_{-1.4}^{+1.3}$ & 3 \\
2MASS J22153705+2110554 & T1 (T0+T2) & 10.7 $\pm$ 0.4 &  - &  -  &  -  & 5.2 $\pm$ 0.5 & 0 & 0 & 1000 & - & 2\\
2MASS J22282889-4310262 & T6 & 1.6 $\pm$ 0.3 & 5.3 & 4.6 $\pm$ 0.2 & 1.51 $\pm$ 0.15 & 1.41 $\pm$ 0.01 & 0 & 0 & 1000 & - & 3, 10, 32 \\
2MASS J2244316+204343 & L6 & 5.5 $\pm$ 0.6 &  -  & 0.8 $\pm$ 0.2 &  -  & 11.0 $\pm$ 2 & 1 &  0 & 130 $\pm$ 20 & $14.3_{-1.5}^{+1.4}$ & 5, 7, 33\\
WISEPAJ081958.05-033529.0 & T4 & 2.0 $\pm$ 0.7 &  -  &  -  &  -  &  - & 1 & 0 & 22 $\pm$ 6 & - & 20 \\
2MASSIJ1553022+153236 & T7 & 1.1 $\pm$ 0.2 &  -  &  -  &  -  &  -  & 1 & 1 & 200 $\pm$ 50 & - & 20 \\
CFBDS J232304.41-015232.3 & T6 & 7.6 &  -  &  -  &  -  &  -  & 1 & 0 & 22 $\pm$ 6 & - & 20 \\
\hline
\end{tabular}
\begin{tablenotes}
\item Varaibility references:~(1)~\citet{Vos2022};
~(2)~\citet{Eriksson2019};
~(3)~\citet{Metchev2015};
~(4)~\citet{Croll2016};
~(5)~\citet{Vos2019};
~(6)~\citet{Lew2016};
~(7)~\citet{Vos2018};
~(8)~\citet{Naud2017}
~(9)~\citet{Yang2016};
~(10)~\citet{Radigan2014a};
~(11)~\citet{Buenzli2014};
~(12)~\citet{Tannock2021};
~(13)~\citet{Clarke2008};
~(14)~\citet{Esplin2016};
~(15)~\citet{Manjavacas2018};
~(16)~\citet{Allers2020};
~(17)~\citet{Girardin2013};
~(18)~\citet{Buenzli2015a};
~(19)~\citet{Schneider2018};
~(20)~This work;
~(21)~\citet{ZhouY2016};
~(22)~\citet{ZhouY2020b};
~(23)~\citet{ZhouY2022};
~(24)~\citet{Bowler2020a};
~(25)~\citet{Vos2020};
~(26)~\citet{Cushing2016};
~(27)~\citet{Leggett2016};
~(28)~\citet{Allers2016};
~(29)~\citet{Biller2018};
~(30)~\citet{Miles-Paez2019};
~(31)~\citet{Apai2017};
~(32)~\citet{Buenzli2012};
~(33)~\citet{Morales-Calderon2006}
\end{tablenotes}
\end{table}
\end{landscape}

\begin{figure}
	\includegraphics[width=\columnwidth]{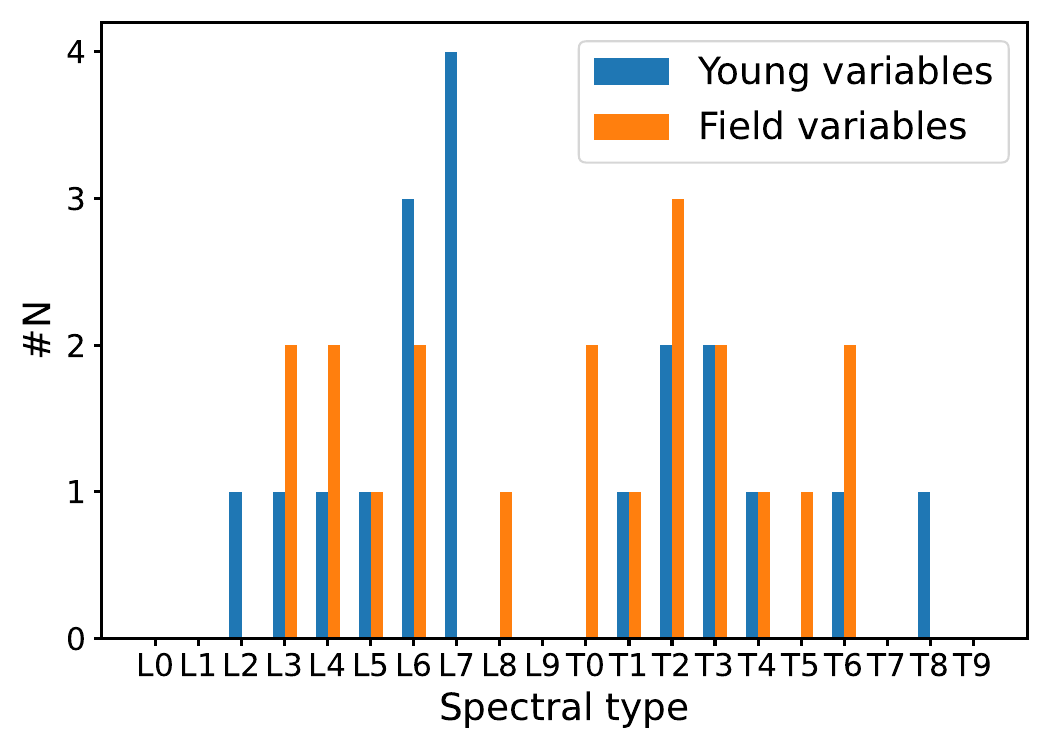}
    \caption{Spectral type distribution of known young and field variables in the $J$ band from the literature.}
    \label{fig:variable_nir_hist}
\end{figure}

\begin{figure}
	\includegraphics[width=\columnwidth]{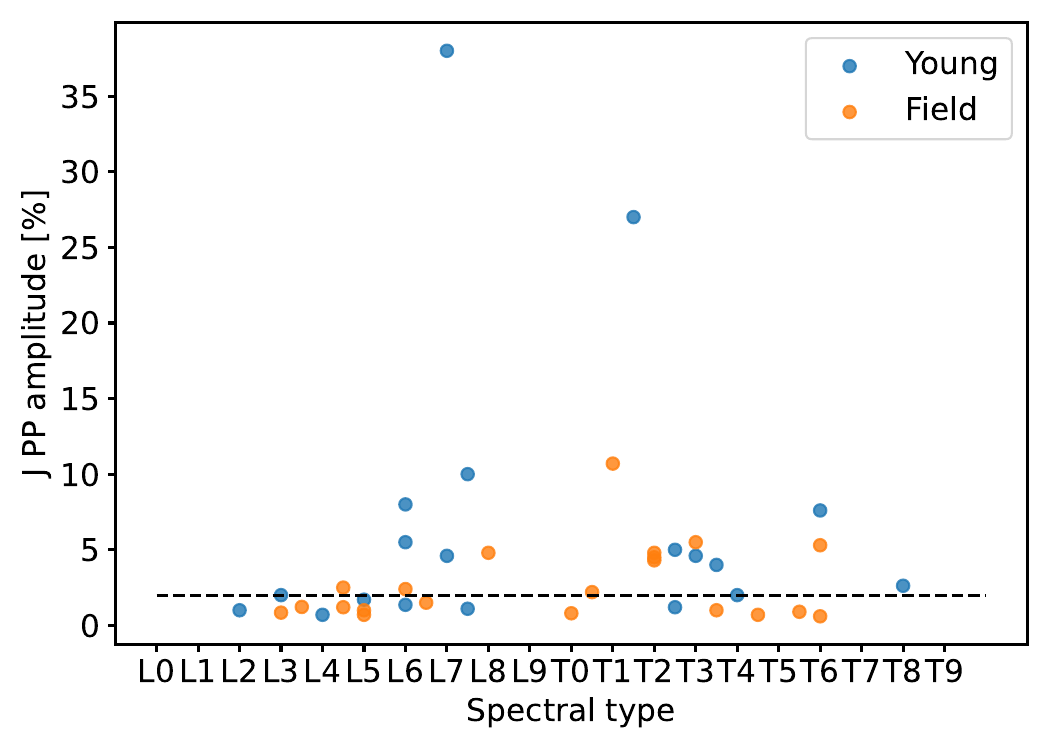}
    \caption{Peak-to-peak amplitude in the $J$ band versus spectral type of known young and field variables from the literature. Variables with amplitudes $>\sim$2\% are strong variables and are above the dashed line. Young strong variables span from L6 to T3.5, while field strong variables span in a narrower range, from L8 to T3. Furthermore, young objects have the strongest amplitudes.}
    \label{fig:variable_nir_amp}
\end{figure}

The spectral type distribution of known variables in the $J$ band is illustrated in Fig.~\ref{fig:variable_nir_hist}. Variability is observed across a wide range of spectral types in field brown dwarfs and an even wider range in young objects. Field objects exhibit variability from L3 to T6, with a weak peak around T2--T2.5. Meanwhile, young variable objects have a range of spectral types from L2 to T8, with a peak at L6--L7.5.
Figure.~\ref{fig:variable_nir_amp} shows the maximum peak-to-peak amplitudes of these objects versus their spectral types. Strong field variables with an amplitude $>\sim$2\% tend to assemble from L8 to T3, with three exceptions having spectral types L4.5, L6, and T6, respectively. The strongest field variable is a T1 object, 2MASS~J2215+2110, with an amplitude of 10.7$\pm$0.4\% measured by \cite{Eriksson2019}. Strong young variables tend to assemble in L6--T3.5, though the latest spectral type of strong young variables can be T8. 
The two strongest variables in Fig.~\ref{fig:variable_nir_amp} are young objects. The L7 young object, VHS~1256-1257b, has a recorded largest variability amplitude of 38\% in the $J$ band measured by \cite{ZhouY2022}. The second-strongest young variable is a T1.5 object, 2MASS~J2139+0220, with an amplitude of 26\% in the $J$ band measured by \cite{Radigan2012}.

While young objects also show strong variability at the L/T transition as field objects, young objects have even stronger variability in the spectral range of L6--L7.5.
In this narrow spectral range, there are at least five of the known strongest variables, including the one with the known maximum variability. They are VHS~1256-1257b (L7) with an amplitude of 38\% \citep{ZhouY2022}, PSO 318.5-22 (L7.5) with an amplitude of 10\% \citep{Vos2019}, 2MASS~J00470038+6803543 (L6) with an amplitude of 8\% \citep{Lew2016}, 2MASS~J2244316+204343 (L6) with an amplitude of 5.5\% \citep{Vos2019} and WISEA~J114724.10-204021.3 (L7) with an amplitude of 4.6\% (this work). 
These objects have extremely red $J-K$ colours, indicating that they have thicker clouds than field dwarfs. 
If we posit the patchy clouds scenario with thick and thin cloud patches \citep{Marley2010,Apai2013}, a break-up in thicker clouds could result in higher contrast and thus stronger variability in light curves.
\cite{Filippazzo2015} find that low gravity L dwarfs have a cooler $T_\mathrm{eff}$ than field L dwarfs of the same spectral type, with a difference of up to 300\,K. This implies that the start of cloud condensation in young L dwarfs may occur at an earlier spectral type than field L dwarfs if we assume that the cloud condensation occurs at a certain $T_\mathrm{eff}$.
However, \cite{Marley2012} suggest that the L/T transition of low gravity objects occurs at a lower $T_\mathrm{eff}$ range than that of field dwarfs.
Therefore, what drives the extreme variability of young L dwarfs in this spectral range is still unclear and needs more observations and atmospheric models to study.

\section{Colour of young L and T objects}
\begin{figure}
	\includegraphics[width=\columnwidth]{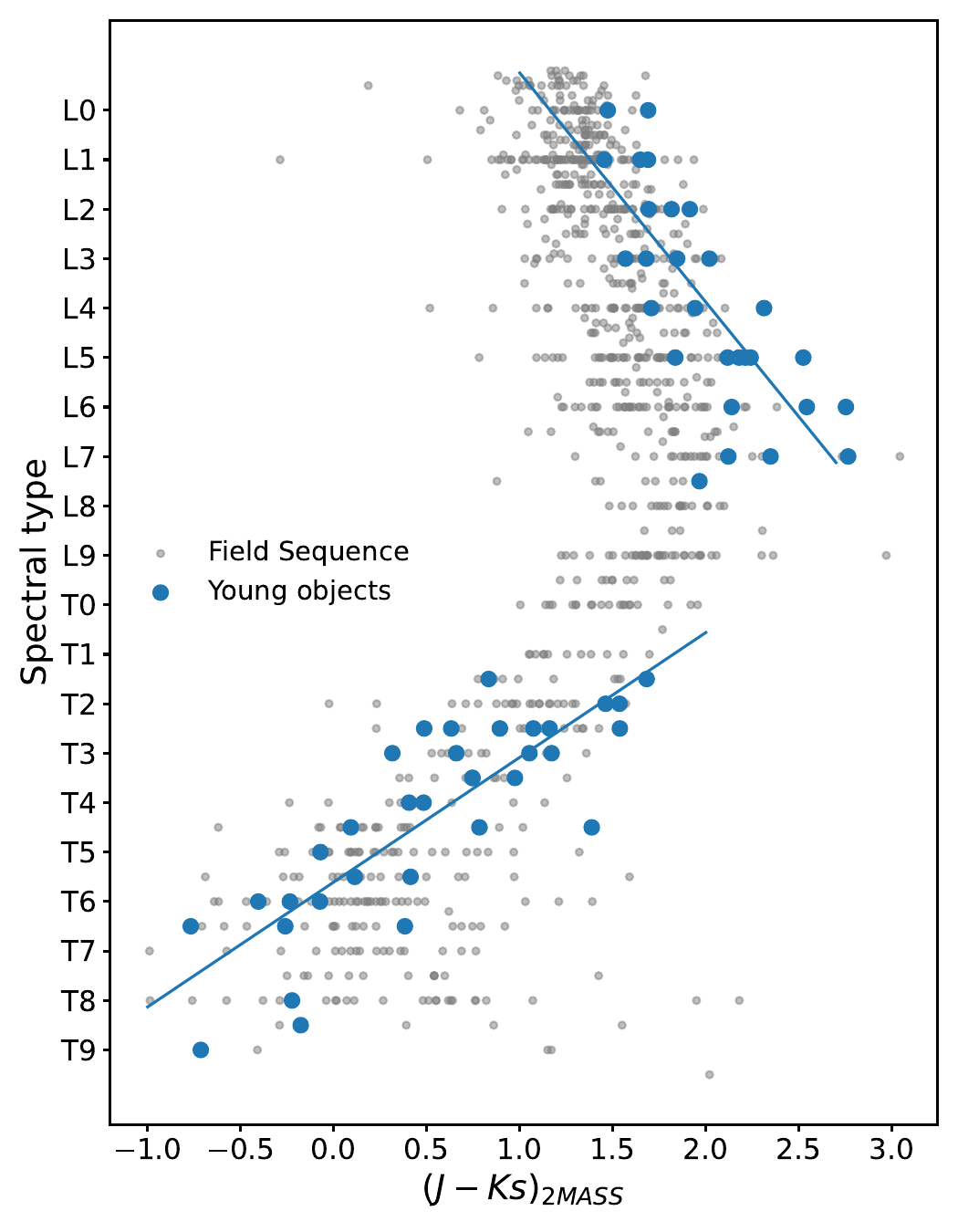}
    \caption{Spectral type and colour plot of young objects. The young L dwarfs have redder colour than the field dwarfs, while young T dwarfs do not have this reddening trend and even become slightly bluer from T4 to T7. Blue lines are a linear fit to young L and T dwarfs, respectively.}
    \label{fig:youngLT_census_sptcolor}
\end{figure}

The sample of young L dwarfs are redder in the near-infrared compared to their field counterparts, as noted in previous studies \citep[e.g.][]{LiuM2016,Faherty2016}, while the young T dwarfs do not have this reddening trend. This agrees with the prevailing scenario that young L dwarfs tend to have thicker clouds than field L dwarfs, and these silicate clouds condense below the photosphere in T dwarfs.
We compare the \textit{2MASS} $J-K_{S}$ colour of all the known young T dwarfs and suspect that later T types gradually become bluer than their field dwarf counterparts from T4 to T7 in Fig.~\ref{fig:youngLT_census_sptcolor}. 
With $J-K$ MKO colours versus spectral types, \cite{ZhangZ2021} also mention that earlier T are redder than their field dwarfs counterparts but this trend vanishes in later spectral types.
We fit a line to young L and T dwarfs respectively and calculate the root mean square distance (rms) of the young and field samples to the fitting. Young L dwarfs have a rms of 1.40 while field L dwarfs have a rms of 3.00. Young T dwarfs have a rms of 1.11 while field T dwarfs have a rms of 2.04. Since the difference in T spectral types is not as significant as it is in L spectral types for field and young dwarfs, we are cautious about this slight bluing trend in T spectral types as it could be biased by the small number of objects known to date. 
Future detections of more young objects in this range are needed to confirm it.
The colours of young objects from T1.5 to T4 fall within a similar range as their field counterparts. The absence of young objects between L8 and T1 hinders the understanding of how the colour of young objects changes at this critical spectral range, for example, where the young objects reach their reddest point.

\rev{It is worth noticing that the colour of variable brown dwarfs is also related to their inclination angle. \cite{Vos2017} find a positive correlation between the inclination angle and the $J-K_S$ colour of variable brown dwarfs, which suggests that there are thicker or larger-grained clouds at the equator than at the poles. \cite{Suarez2023} support this by finding a positive correlation between the inclination angle and silicate absorption index.
To measure the inclination angle, we need to measure the $vsini$ from the high-resolution spectra of our objects. However since our objects are mainly young T-type planetary mass objects, most of them are fainter than 16.5 magnitudes in the $J$ band, which makes it difficult to obtain high S/N spectroscopic observations under reasonable integration time with telescopes such as VLT. Future telescopes such as E-ELT can achieve $vsini$ measurements of these objects with high S/N, which can also place an upper limit on the rotation period of our objects. Combined with the period from variability observations which is the lower limit for objects detected with long-term variability, the rotation period of the variable objects and their inclination angle can be constrained.
}

\section{Conclusions}
We report a near-infrared survey for photometric variability in young planetary-mass objects, including the largest sample of young T dwarfs monitored to date.
We conduct continuous $J_S$- or $K_S$-band monitoring of 18 objects, with observation time ranging from 1.5 to 7.5 hours per object. One variable 2M1119-1137AB is an unresolved binary.
We detected variability in \rev{four} variables with significance higher than 99\%, as well as two variable candidates.
The shortest period found among the other three variables is 5.5$\pm$0.2 hours, consistent with the trend that young objects have longer rotation periods than field objects, which is due to angular momentum conservation during contraction \citep{Schneider2018}.

We combine our survey with three previous $J$-band photometric variability surveys of field L and T objects and young low gravity L objects, comprising a total number of 108 objects. 
From the statistical calculation, we find that young dwarfs have a tendency to be more variable than field dwarfs within peak-to-peak variability amplitude ranges of 0.5\%--10\% and period ranges of 1.5--20\,hr.
The variability rate of young L dwarfs is $27_{-10}^{+13}$\% compared to $6_{-5}^{+13}$\% for field L dwarfs, consistent with the previous result reported by \cite{Vos2019}.
We constrain the near-infrared variability rate of young T dwarfs for the first time.
Young T dwarfs have a variability rate of $56_{-18}^{+20}$\% compared to $25_{-7}^{+8}$\% for field T dwarfs. Both young L and T samples tend to be more variable than their field dwarf counterparts, which are of $\sim$ 1$\sigma$ difference.
Moreover, both young and field samples also tend to be more variable at the L/T transition than objects outside the L/T transition, suggesting the strong impact of the L/T transition on atmospheric structures.
The variability rate of field objects is $36_{-13}^{+15}$\% at the L/T transition compared to $18_{-7}^{+9}$\% outside the L/T transition. The variability rate of young objects is $64_{-22}^{+23}$\% at the L/T transition compared to $31_{-9}^{+12}$\% outside the L/T transition. 
Besides the L/T transition, our analysis of known variables in the $J$ band in the literature finds that young low gravity L dwarfs with high variability amplitudes tend to congregate in a narrow spectral range of L6-L7.5, while field L dwarfs do not have this trend.

This study once again demonstrates that young dwarfs and field dwarfs are likely distinct groups with differing variability properties at least in terms of strong variability (> $\sim$1\%), with surface gravity playing a crucial role in the variability of brown dwarfs for both L-type and T-type dwarfs.
Future multi-wavelength time-resolved observations with JWST, along with atmospheric modelling, will enable a deep understanding of how gravity impacts atmospheric structures of L and T dwarfs, and will also be critical for understanding the atmospheres of directly imaged exoplanets, which are similar to young low gravity dwarfs.

\section*{Acknowledgements}
Based on observations collected at the European Organisation for Astronomical Research in the Southern Hemisphere under ESO programmes 108.2256, 109.22XF, 110.23Y0, and 111.24LQ.
B.B acknowledges funding by the UK Science and Technology Facilities Council (STFC) grant no. ST/M001229/1.
J. M. V. acknowledges support from a Royal Society - Science Foundation Ireland University Research Fellowship (URF$\backslash$1$\backslash$221932).
This research made use of \textsc{Astropy}, a community-developed core \textsc{Python} package for Astronomy \citep{Astropy2013,Astropy2018}, \textsc{SciPy} \citep{Scipy2020}, \textsc{NumPy} \citep{Numpy2020} and \textsc{Matplotlib}, a Python library for publication quality graphics \citep{Matplotlib2007}.

%%%%%%%%%%%%%%%%%%%%%%%%%%%%%%%%%%%%%%%%%%%%%%%%%%
\section*{Data Availability}
% The inclusion of a Data Availability Statement is a requirement for articles published in MNRAS. Data Availability Statements provide a standardised format for readers to understand the availability of data underlying the research results described in the article. The statement may refer to original data generated in the course of the study or to third-party data analysed in the article. The statement should describe and provide means of access, where possible, by linking to the data or providing the required accession numbers for the relevant databases or DOIs.
The raw data of our observations can be accessed via ESO Science Archive Facility.



\bibliographystyle{mnras}
\bibliography{main} % if your bibtex file is called example.bib

%\begin{thebibliography}{99}
%\end{thebibliography}

%%%%%%%%%%%%%%%%%%%%%%%%%%%%%%%%%%%%%%%%%%%%%%%%%%

%%%%%%%%%%%%%%%%% APPENDICES %%%%%%%%%%%%%%%%%%%%%

\appendix
\section{Reference stars and MAD-magnitude plots}
\label{apd:stars_mag}
\rev{We show the target and selected reference stars of all observations, including the MAD and instrumental magnitude relationship plots. If not specified, observations are in the $J_S$ band.}

\begin{figure*}
    \includegraphics[width=0.5\columnwidth]{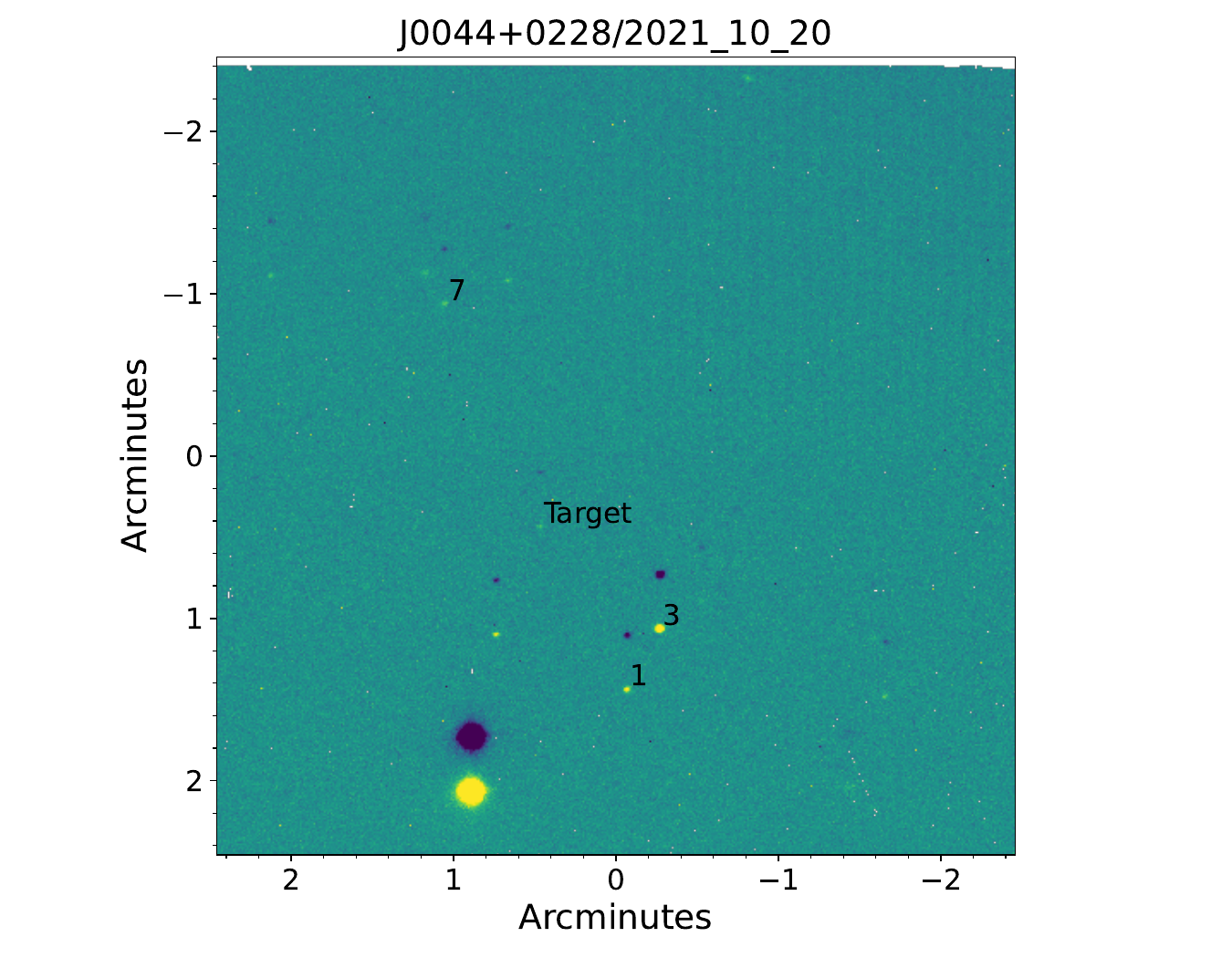}
    \includegraphics[width=0.5\columnwidth]{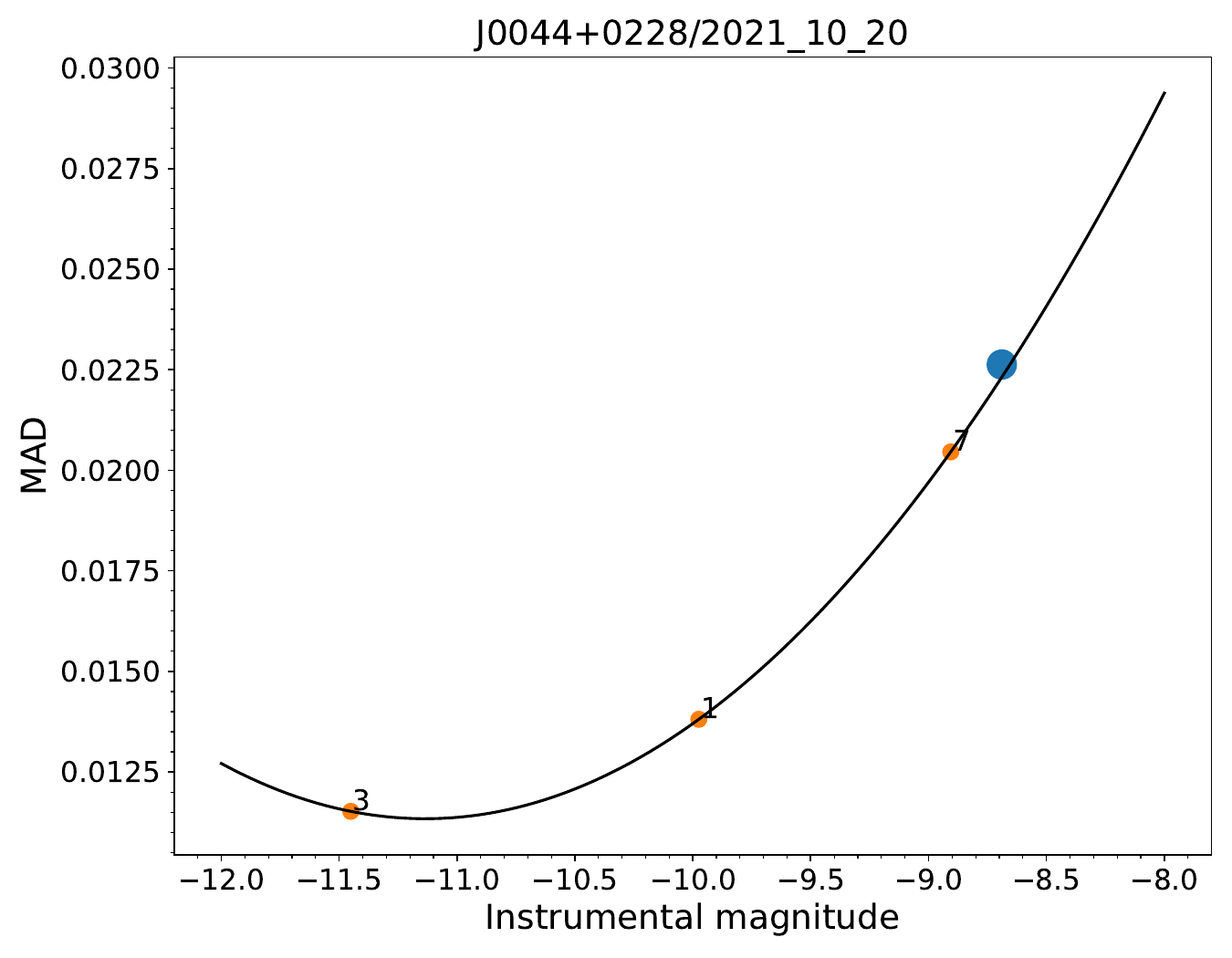}
    \includegraphics[width=0.5\columnwidth]{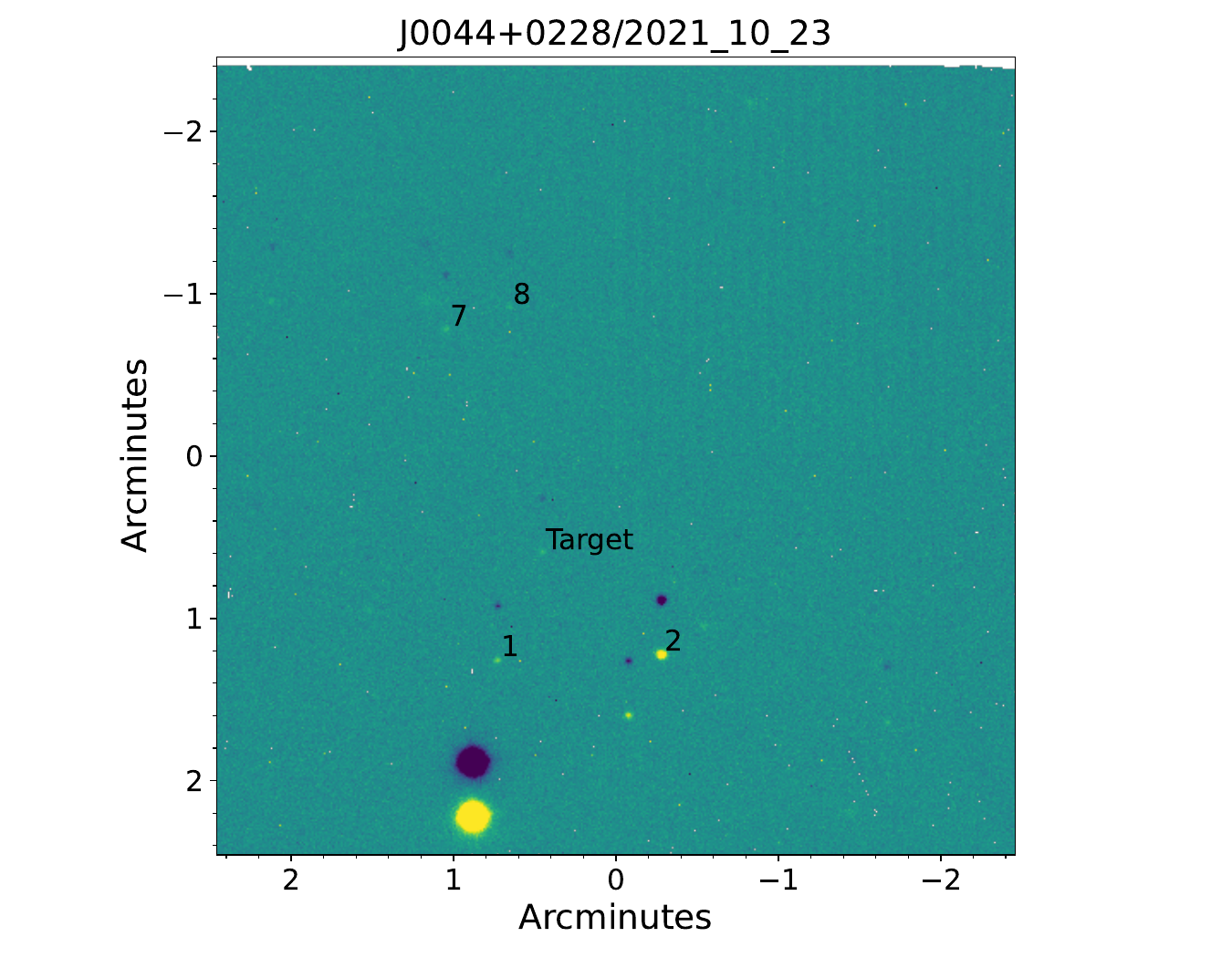}
    \includegraphics[width=0.5\columnwidth]{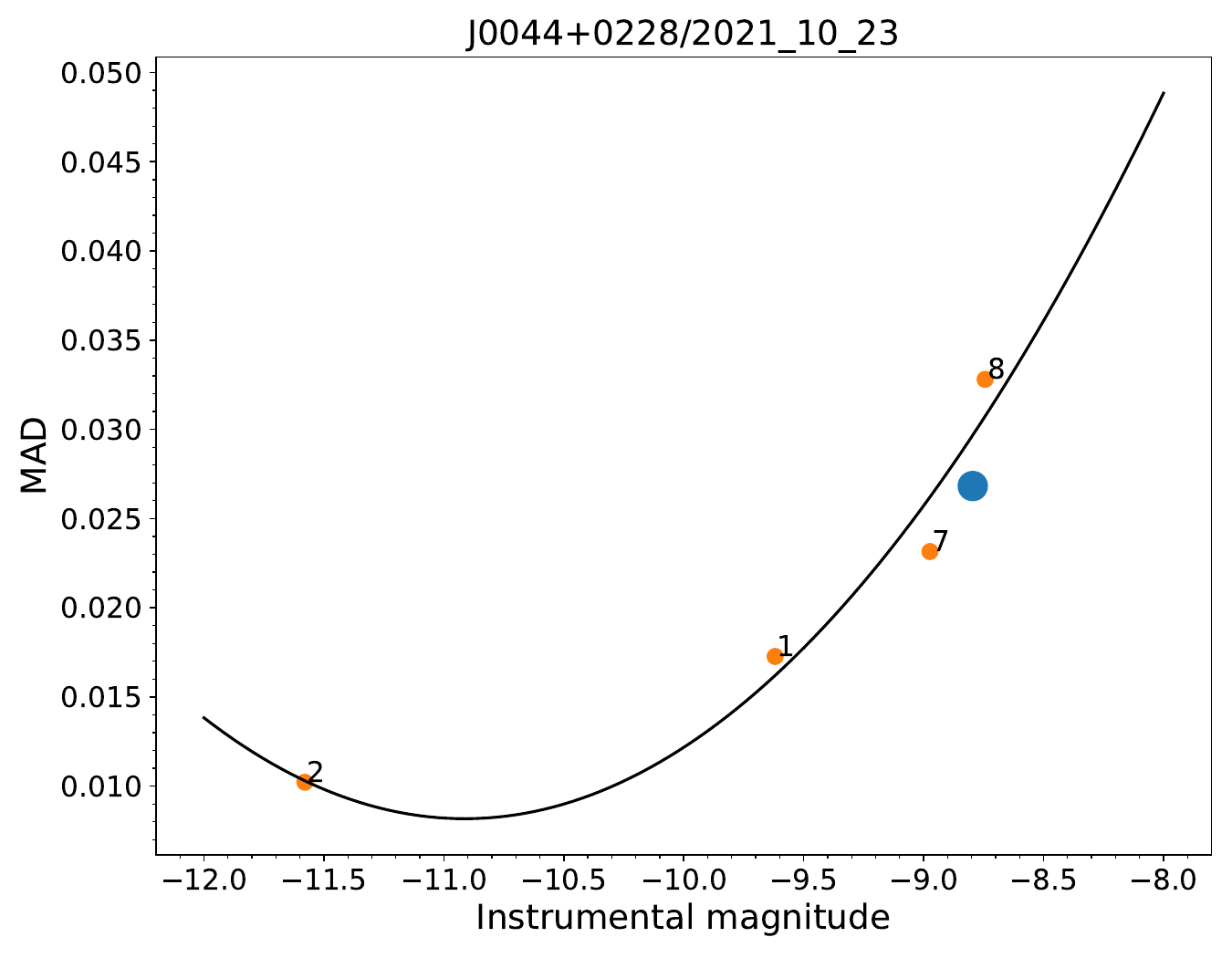}
    %J0200-5205
    \includegraphics[width=0.5\columnwidth]{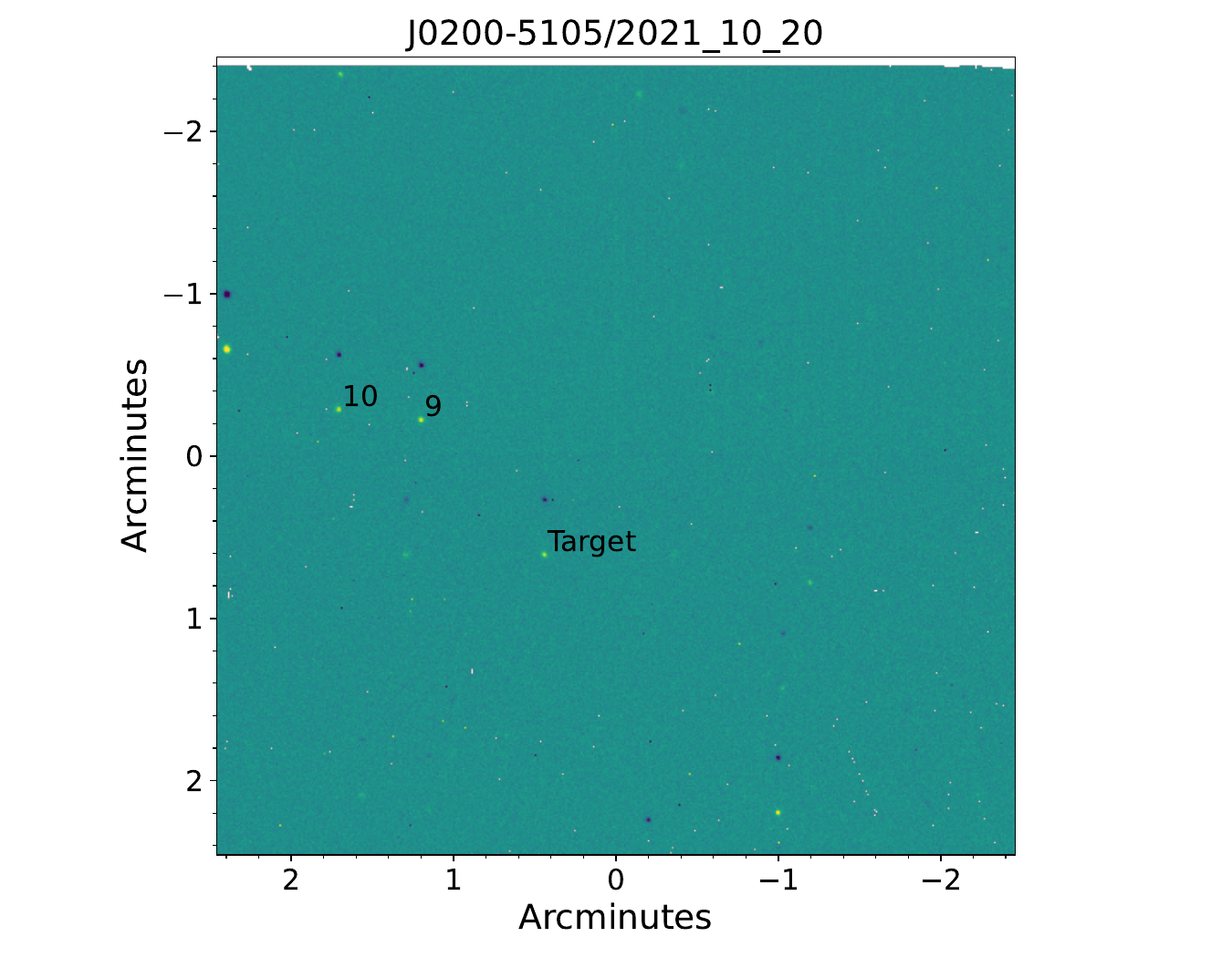}
    \includegraphics[width=0.5\columnwidth]{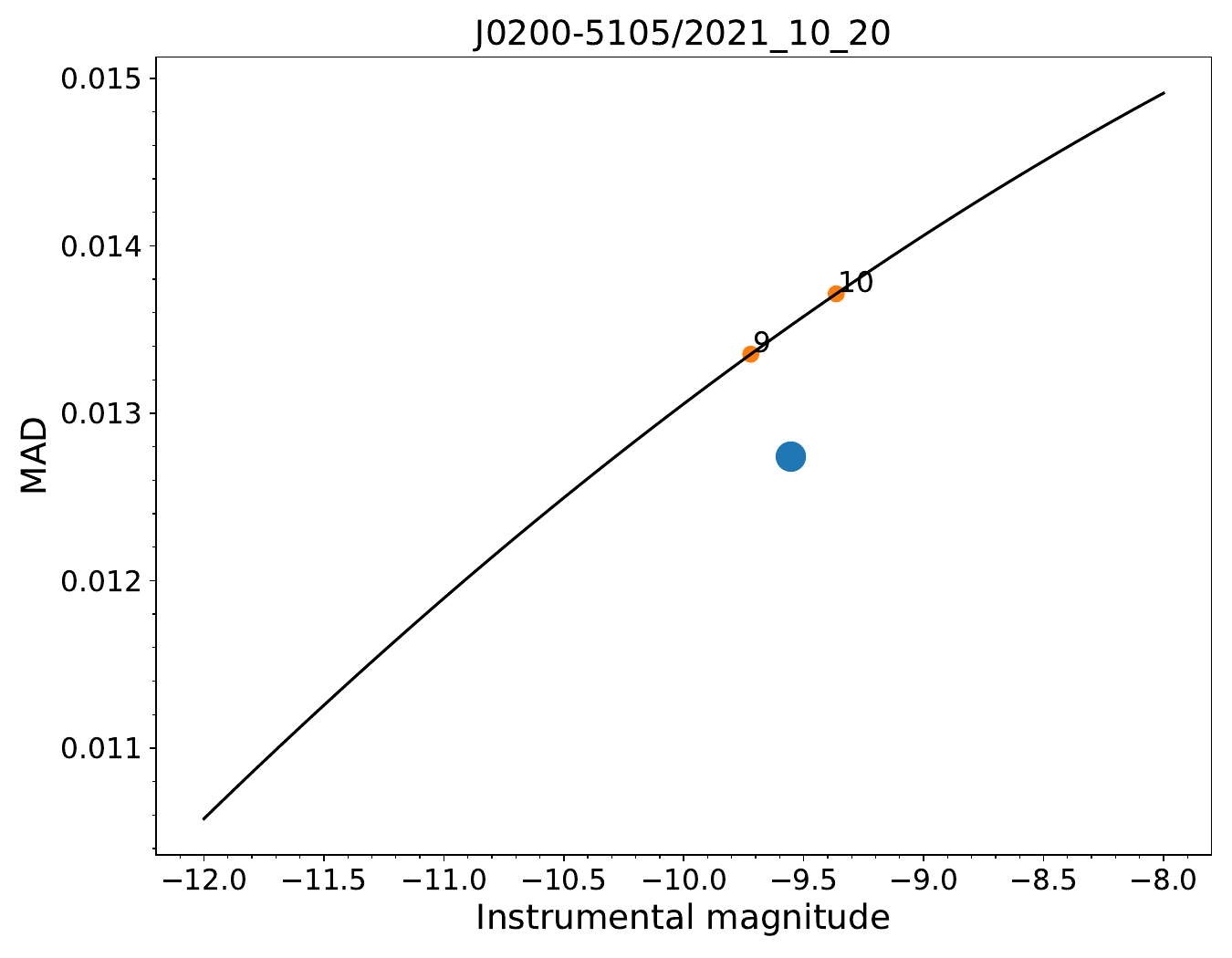}
    \includegraphics[width=0.5\columnwidth]{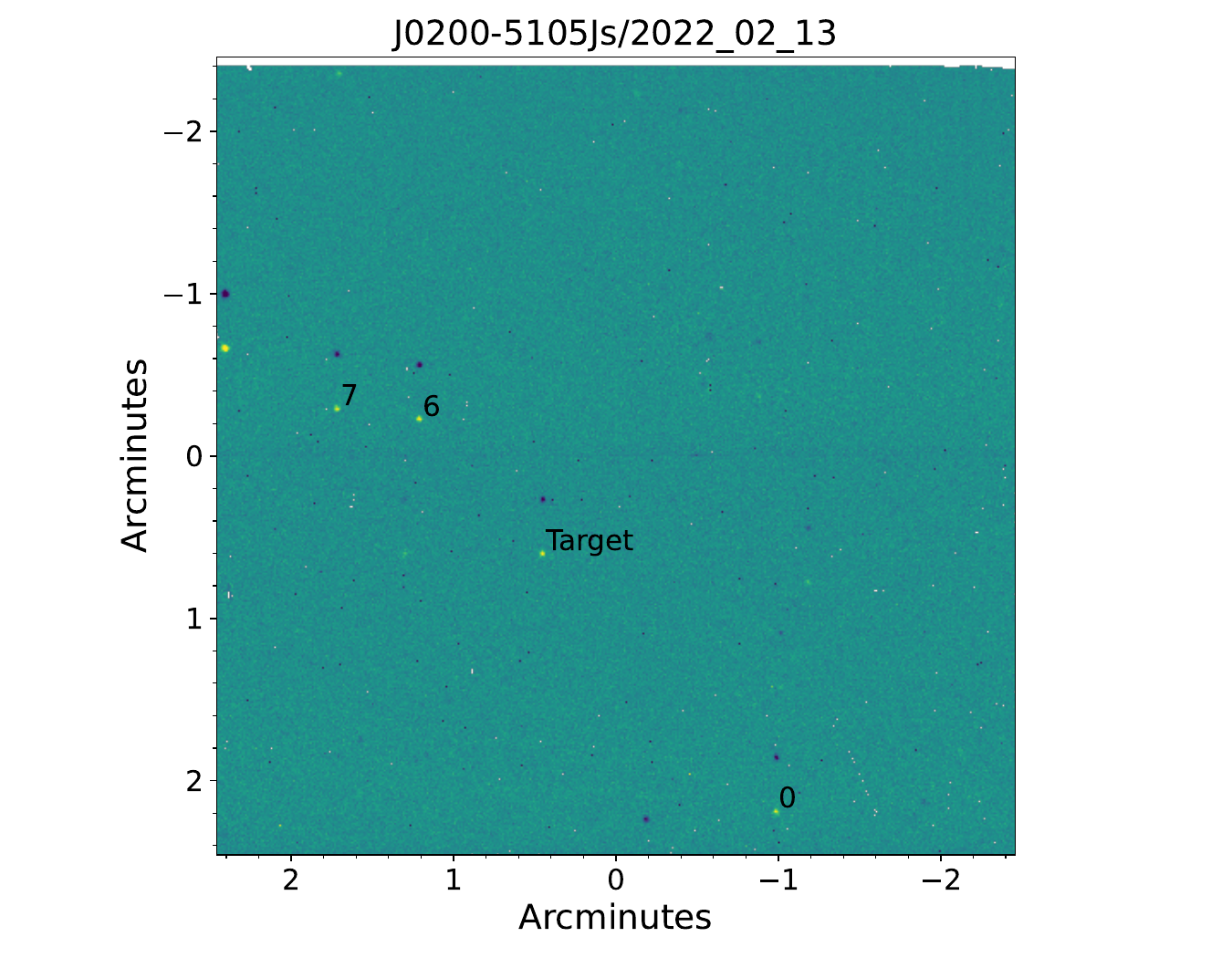}
    \includegraphics[width=0.5\columnwidth]{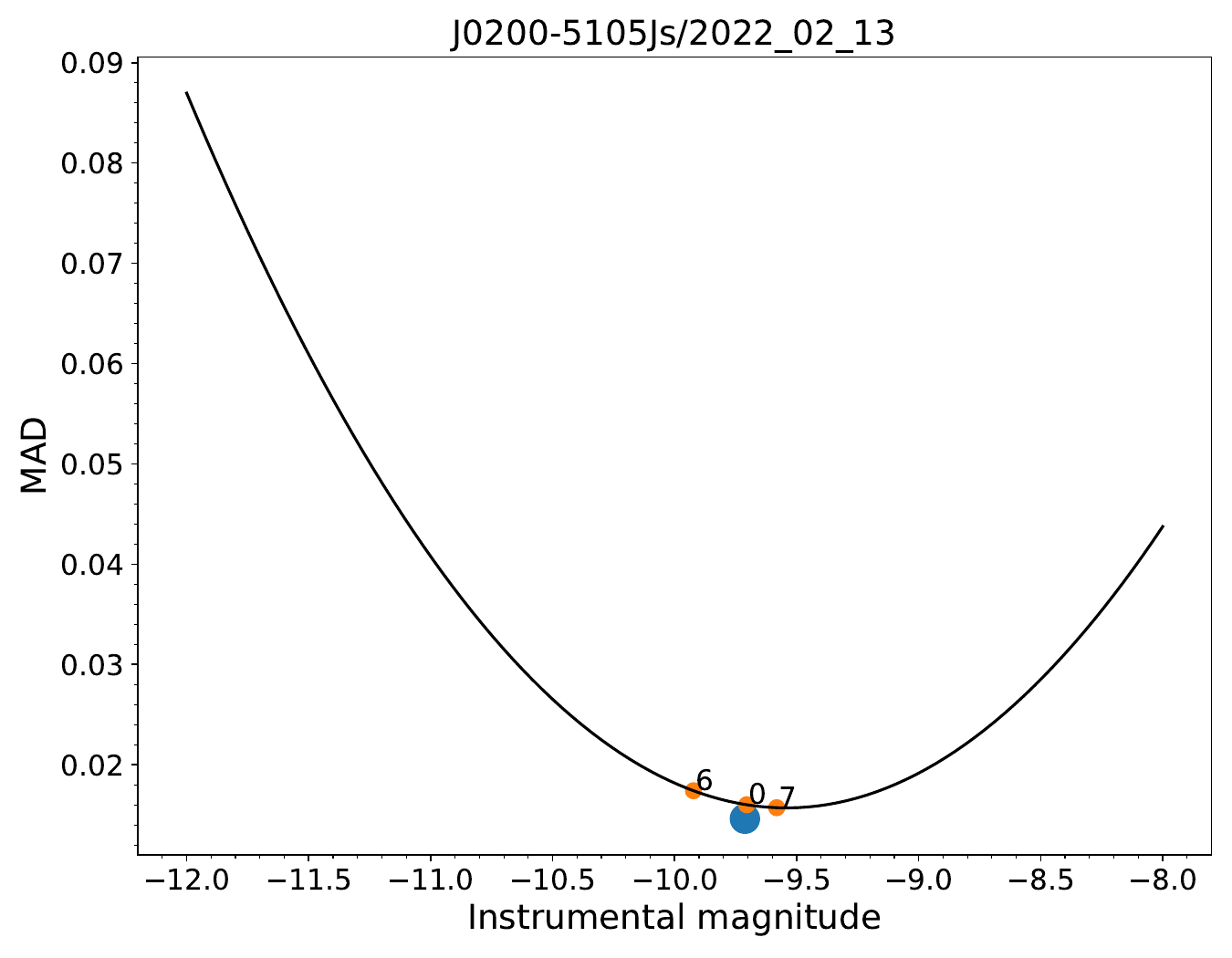}
    \includegraphics[width=0.5\columnwidth]{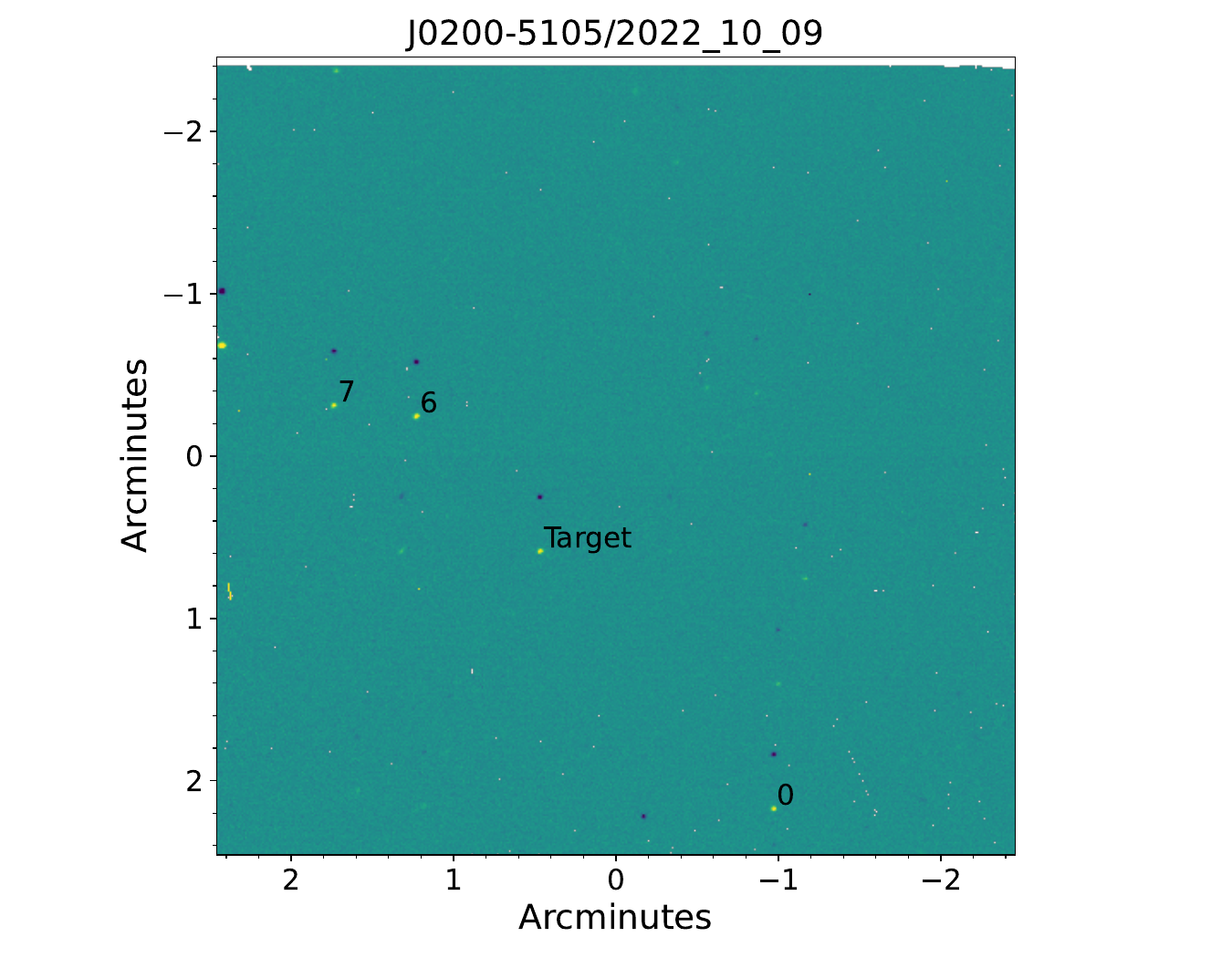}
    \includegraphics[width=0.5\columnwidth]{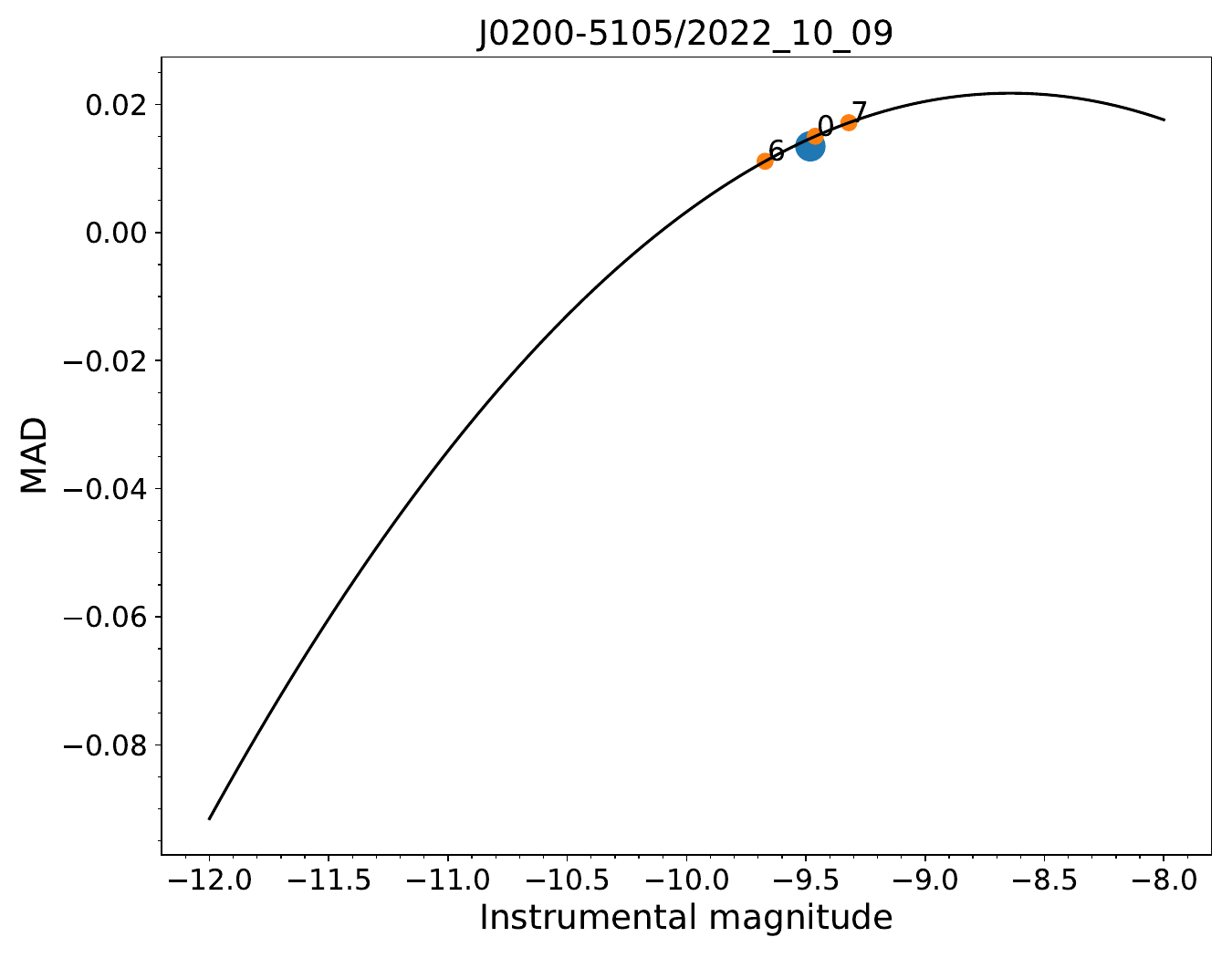}
    \includegraphics[width=0.5\columnwidth]{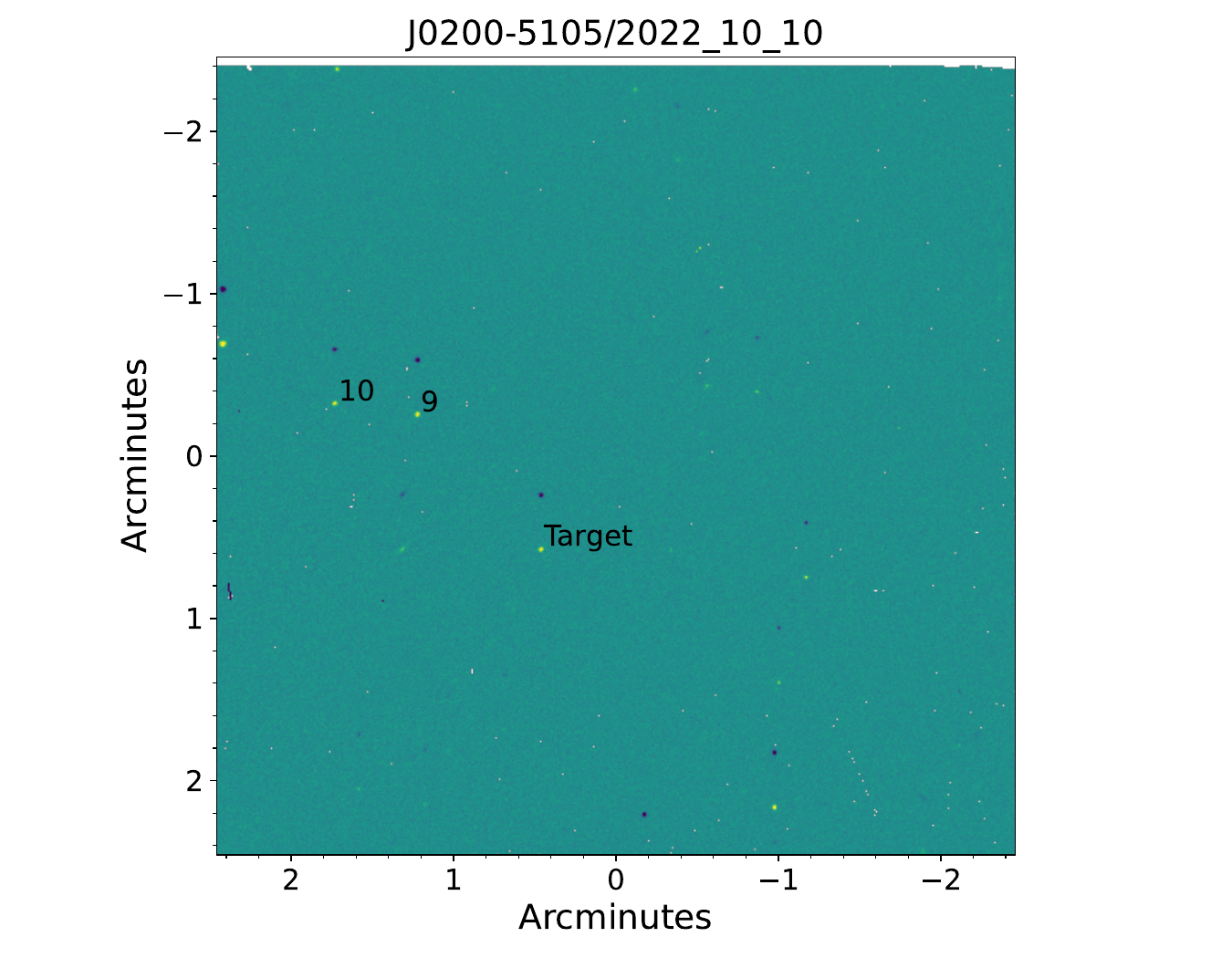}
    \includegraphics[width=0.5\columnwidth]{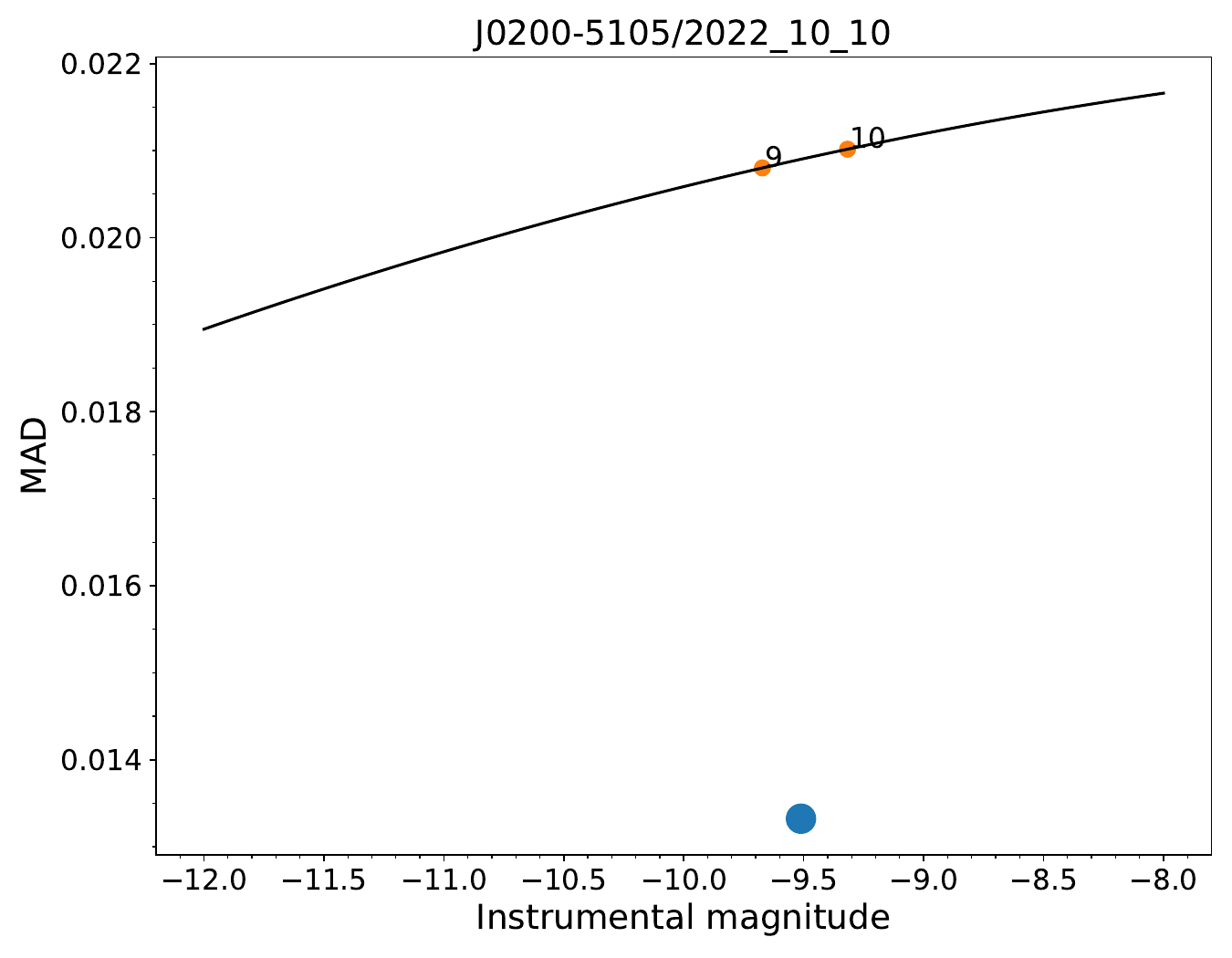}
    \includegraphics[width=0.5\columnwidth]{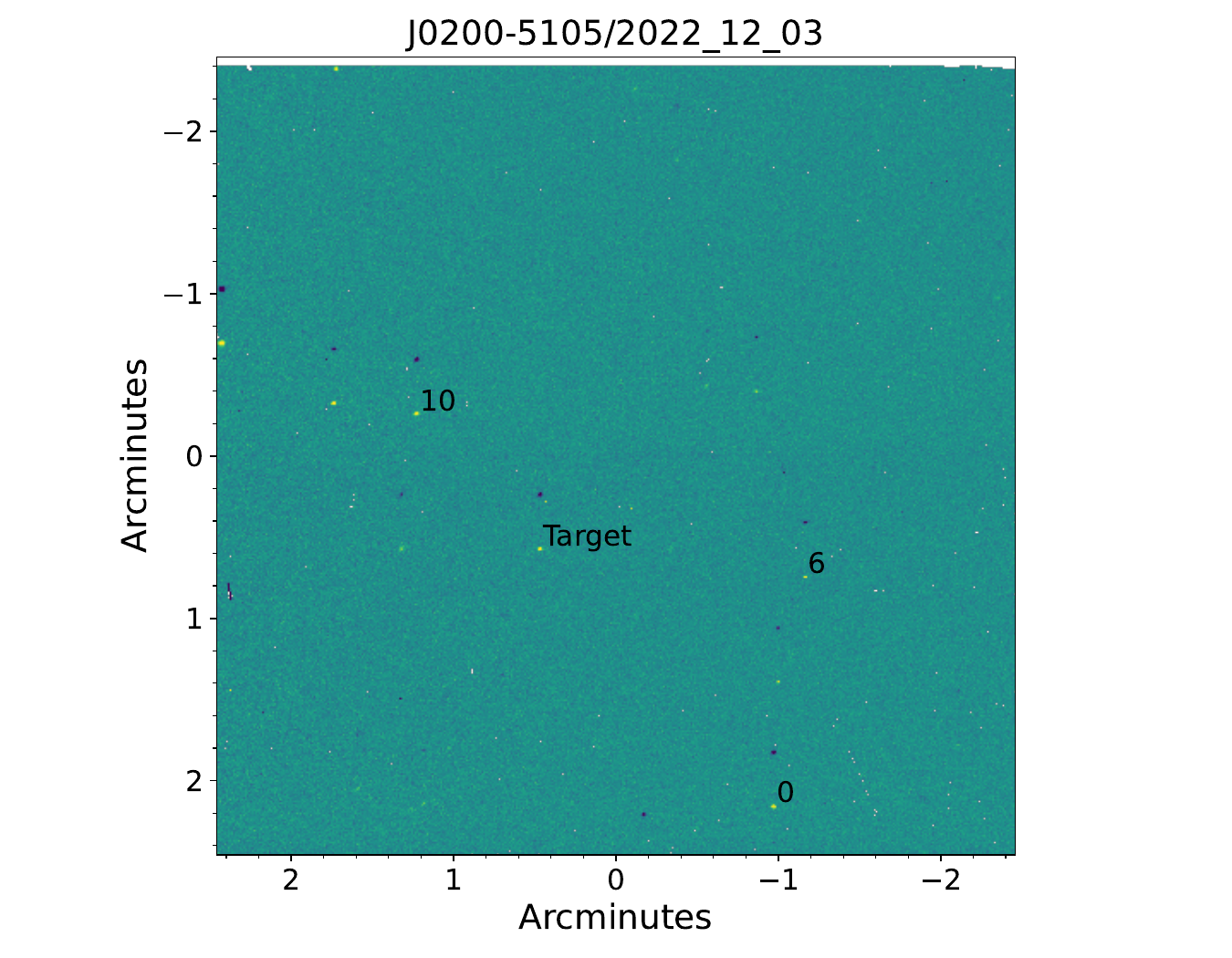}
    \includegraphics[width=0.5\columnwidth]{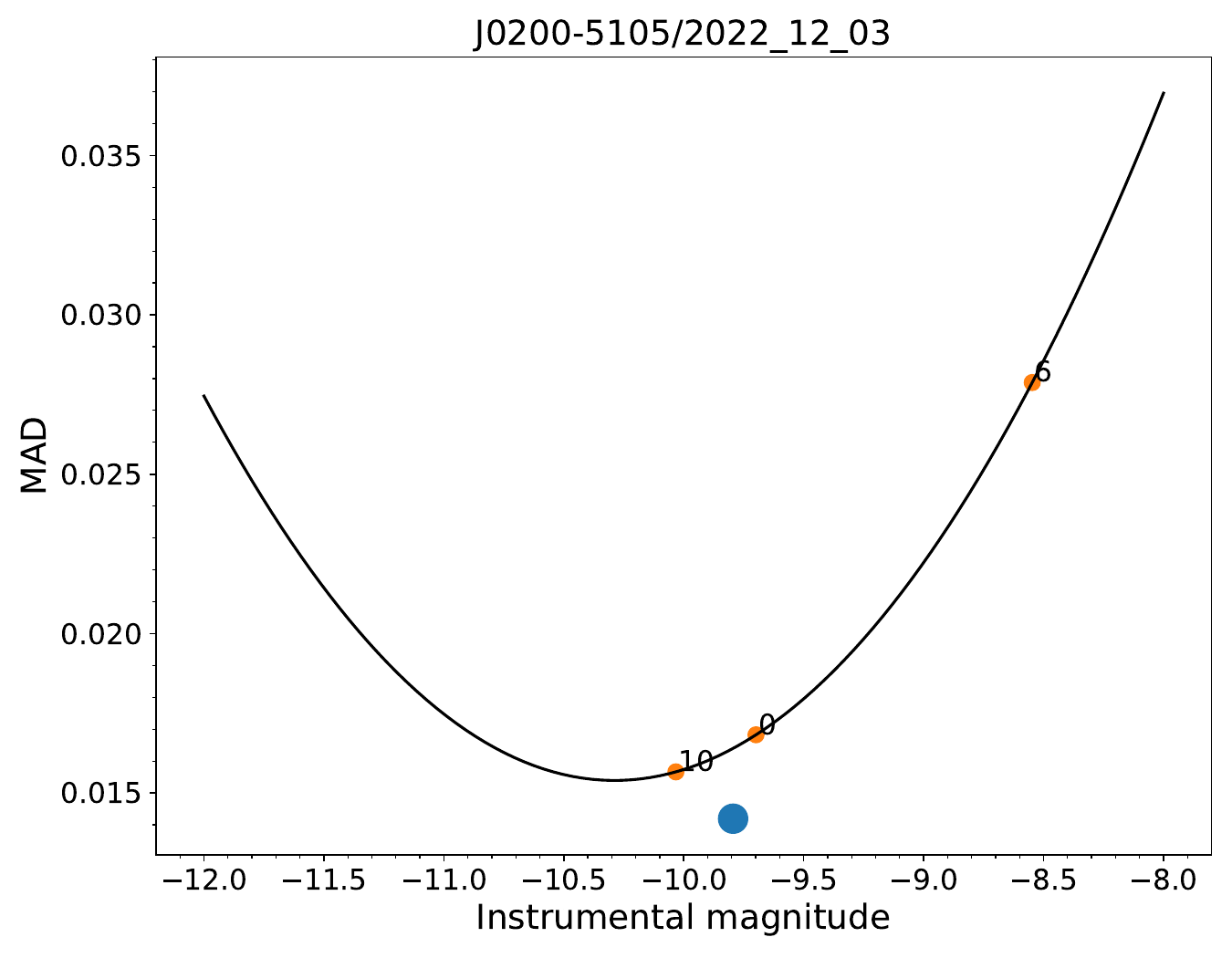}
    %J0200+0000
    \includegraphics[width=0.5\columnwidth]{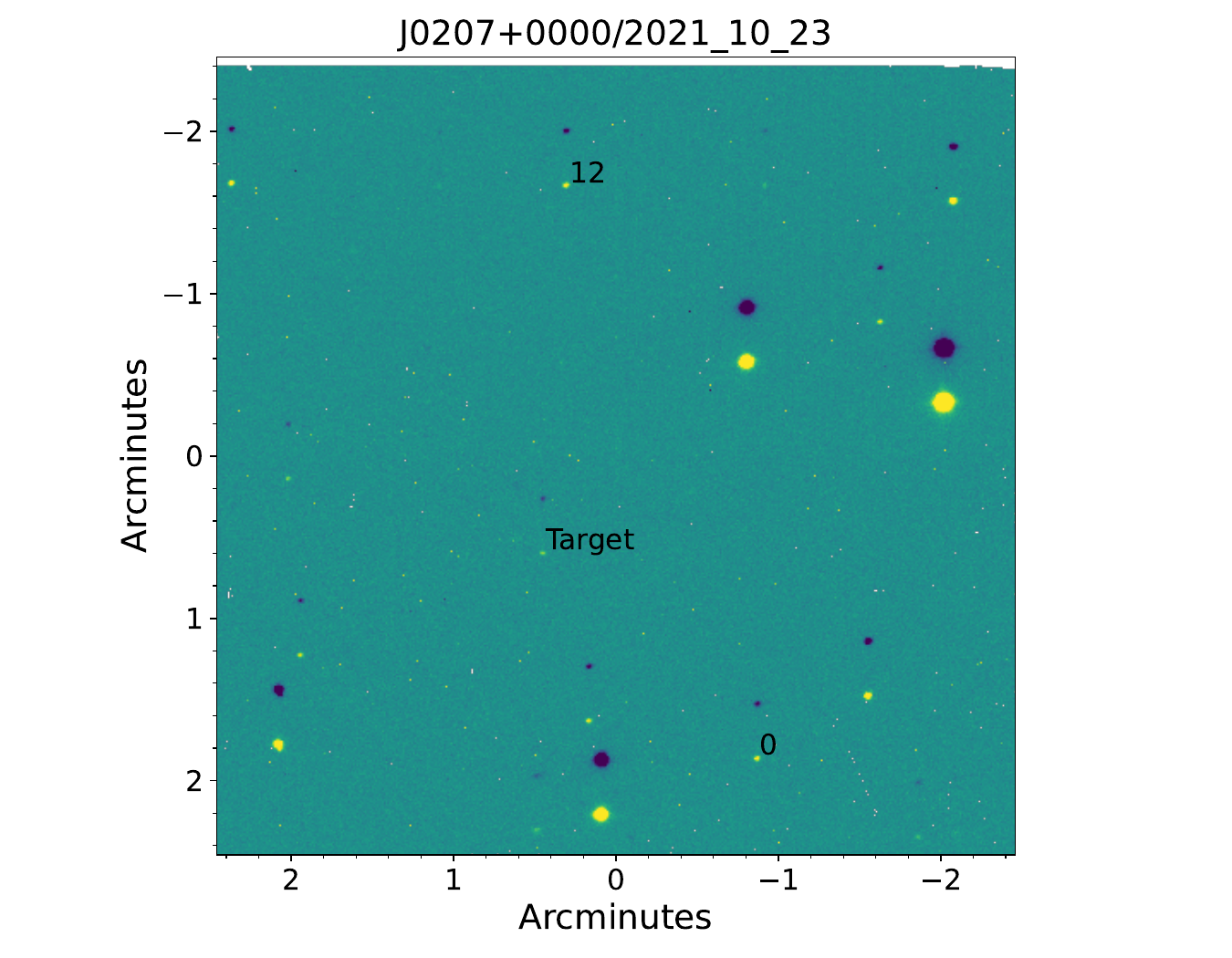}
    \includegraphics[width=0.5\columnwidth]{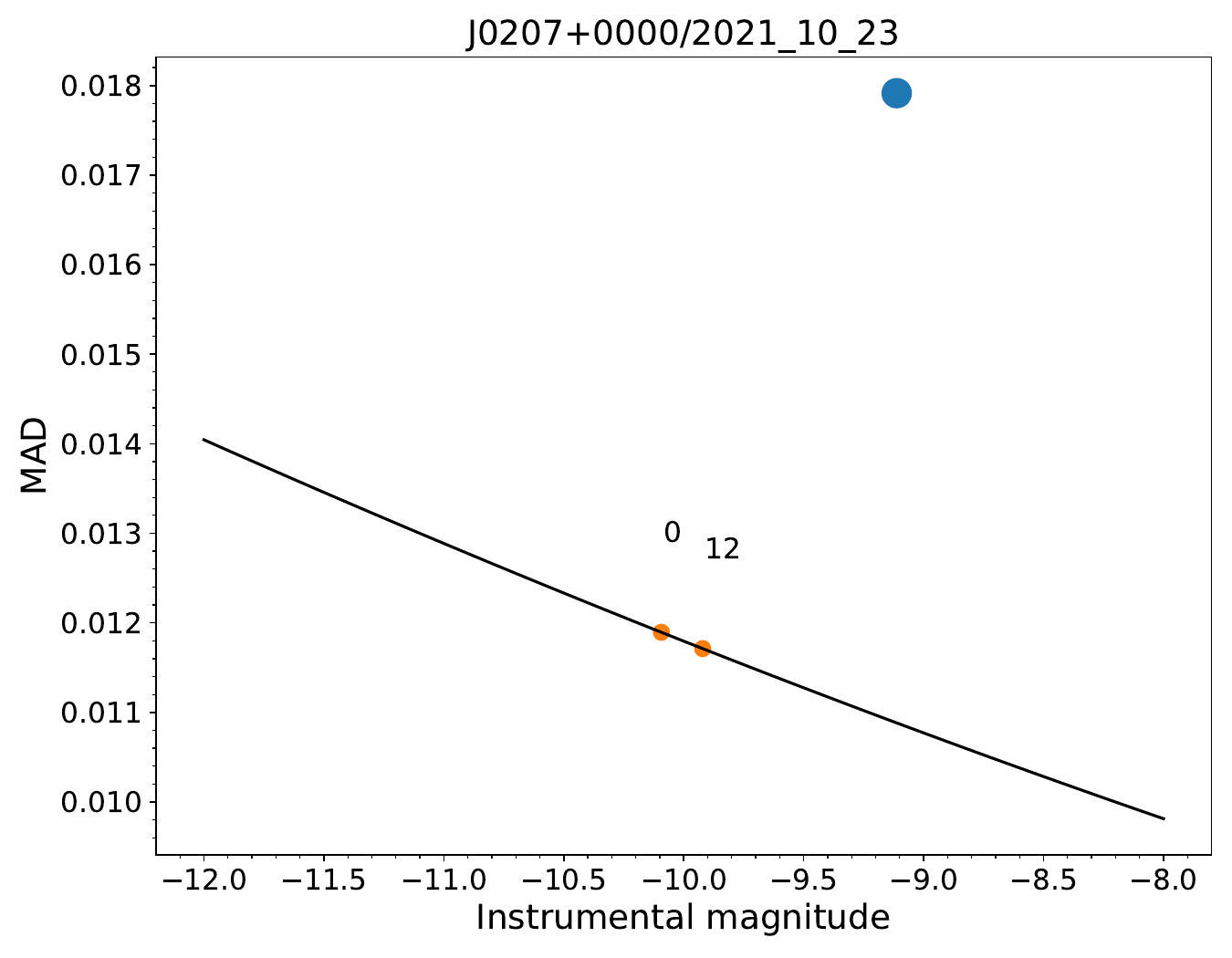}
    %J0226-1610
    \includegraphics[width=0.5\columnwidth]{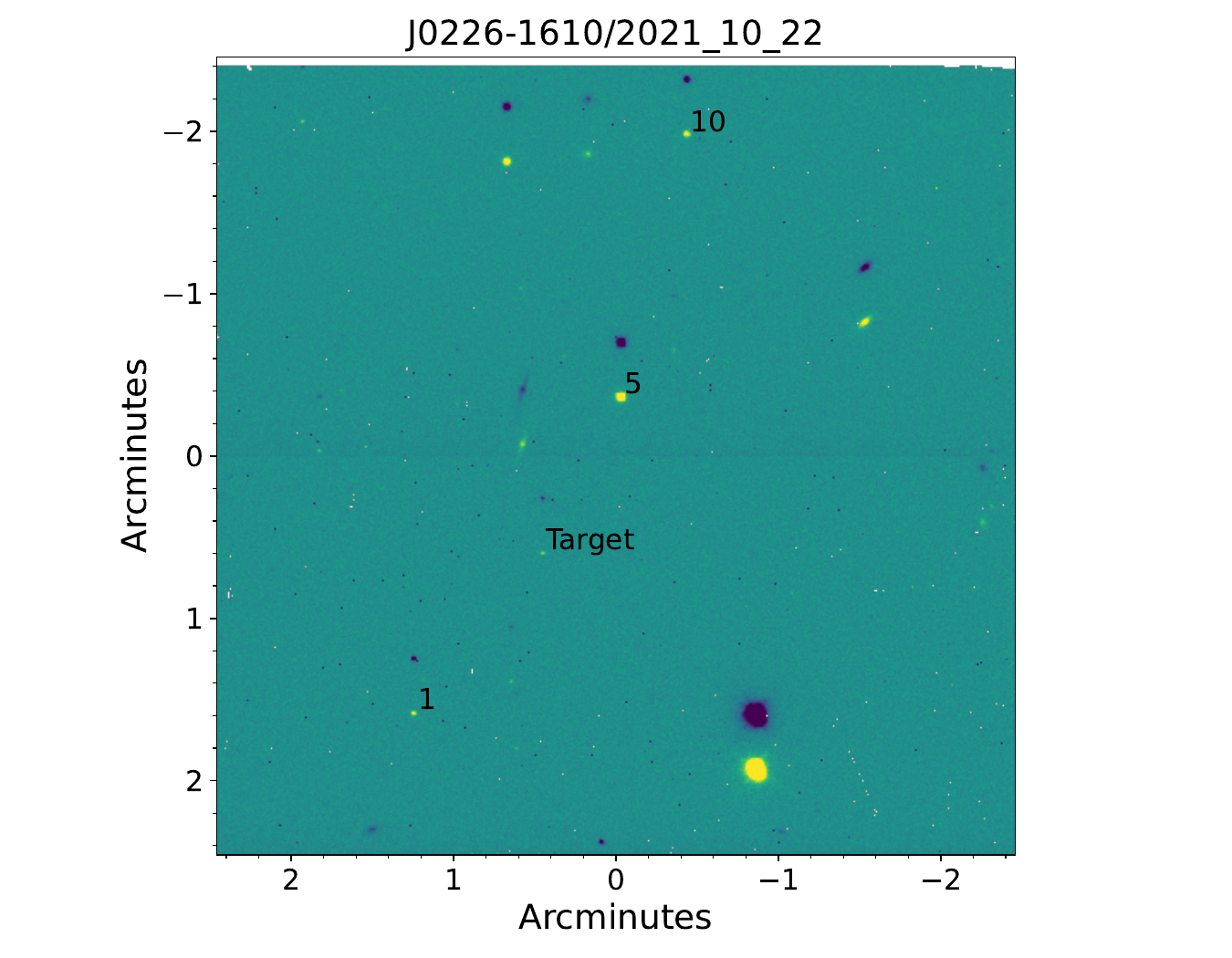}
    \includegraphics[width=0.5\columnwidth]{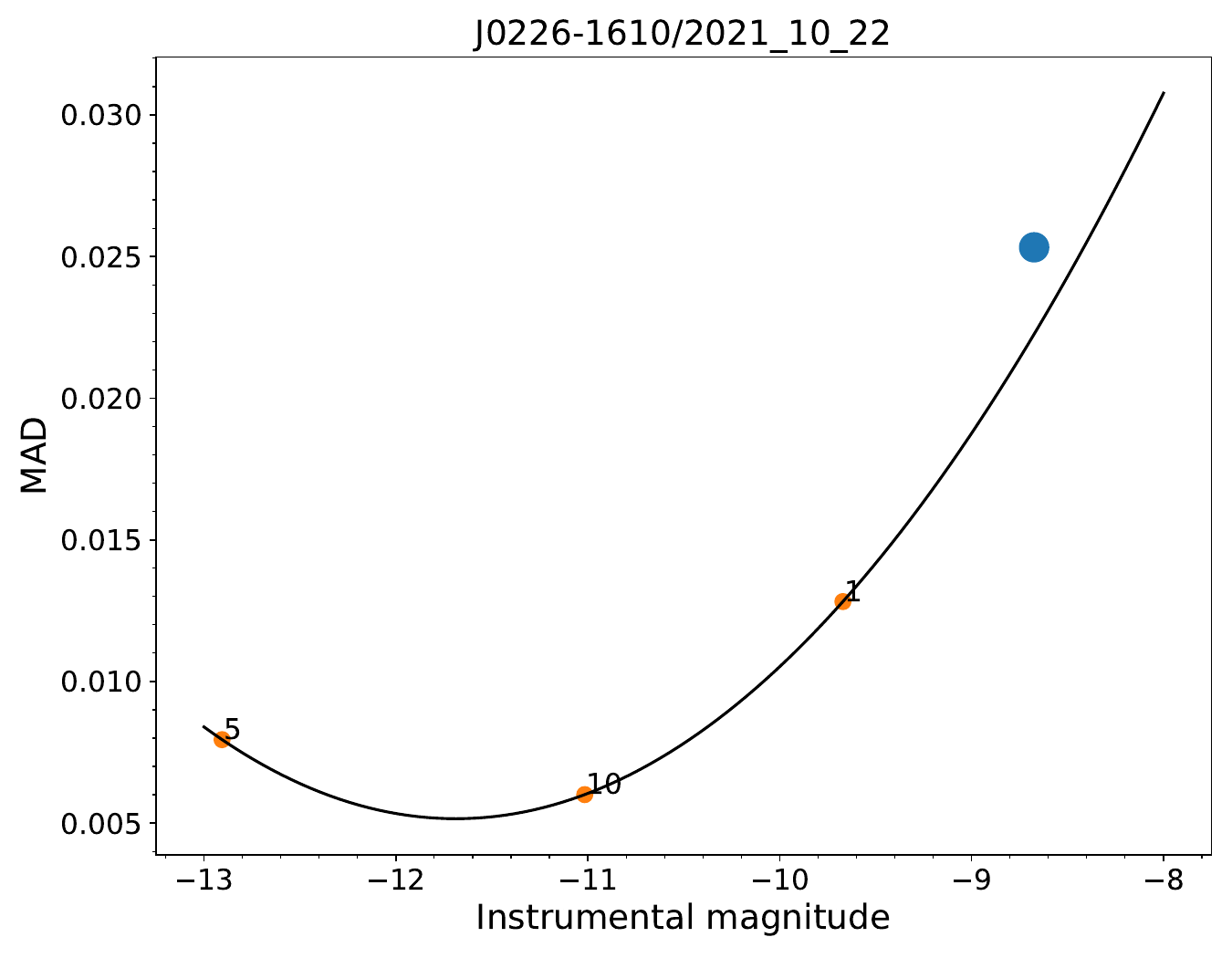}
    \includegraphics[width=0.5\columnwidth]{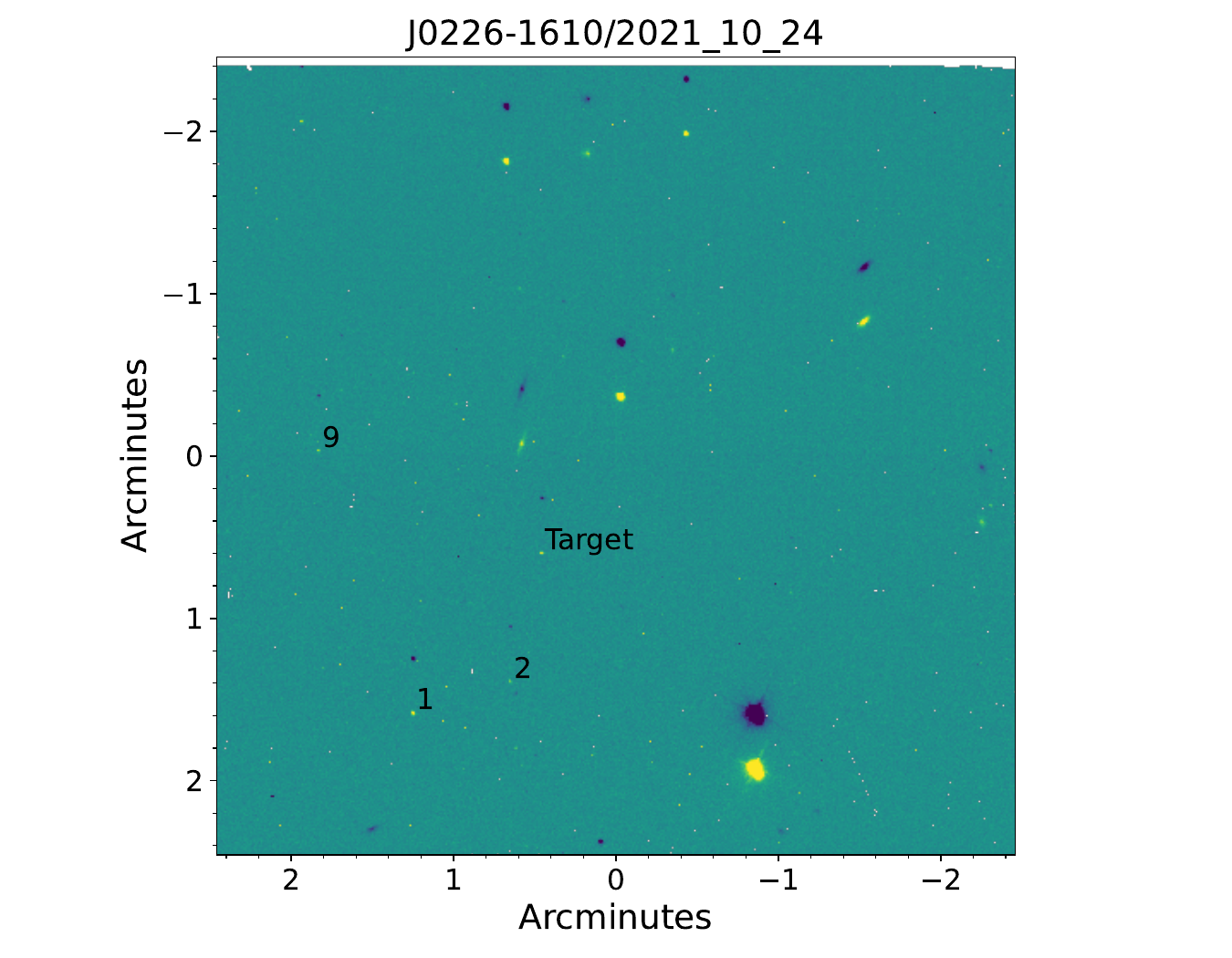}
    \includegraphics[width=0.5\columnwidth]{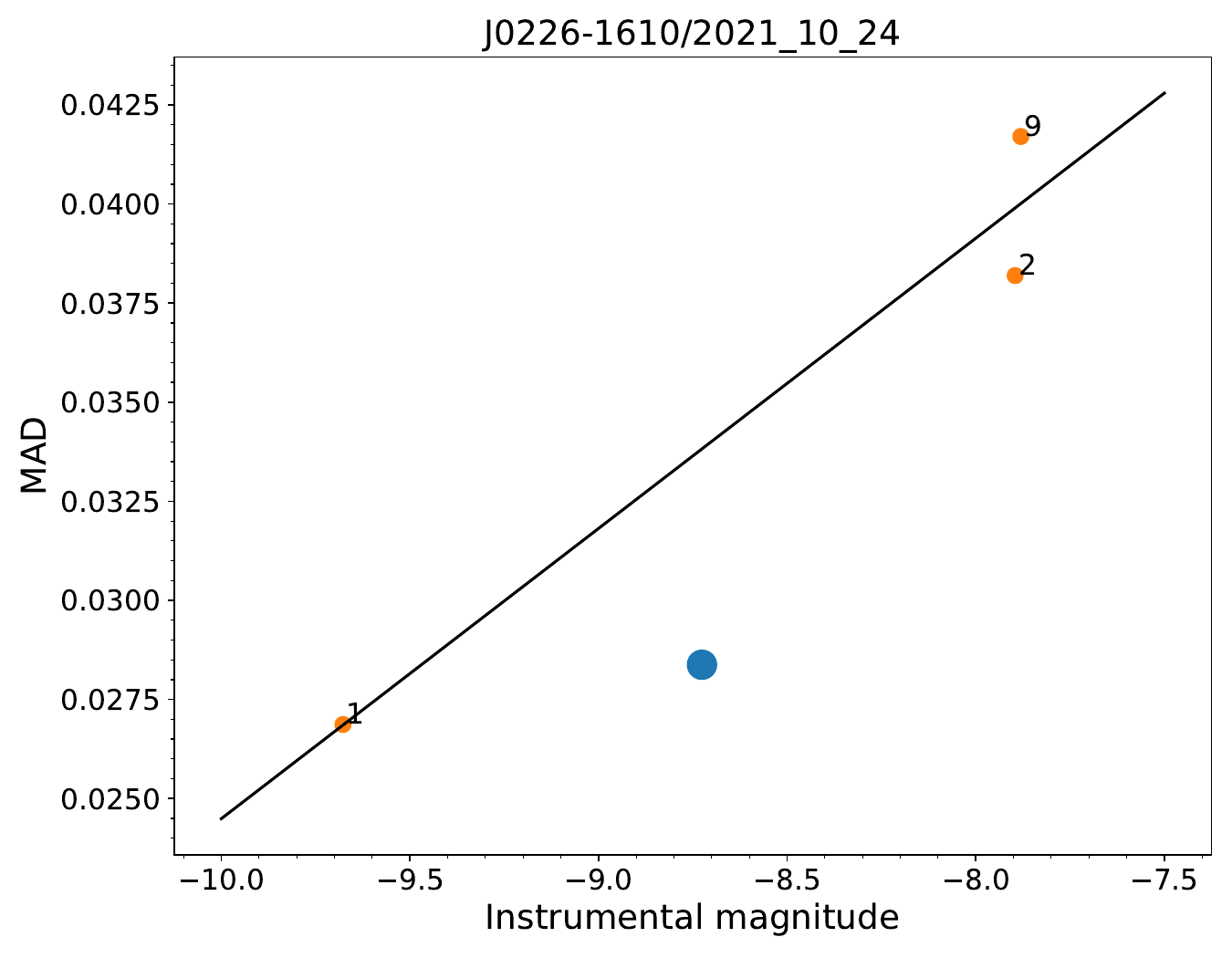}
    \includegraphics[width=0.5\columnwidth]{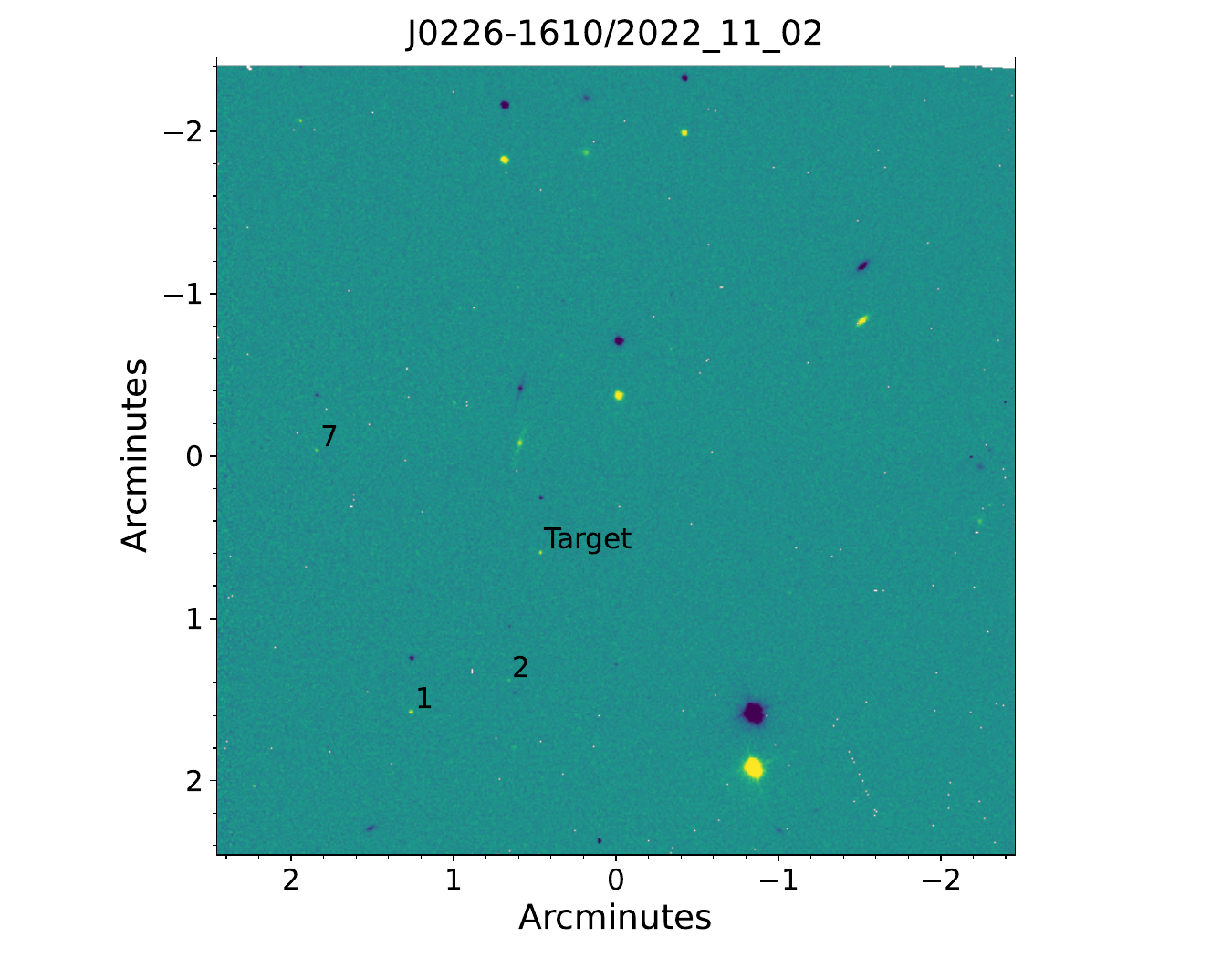}
    \includegraphics[width=0.5\columnwidth]{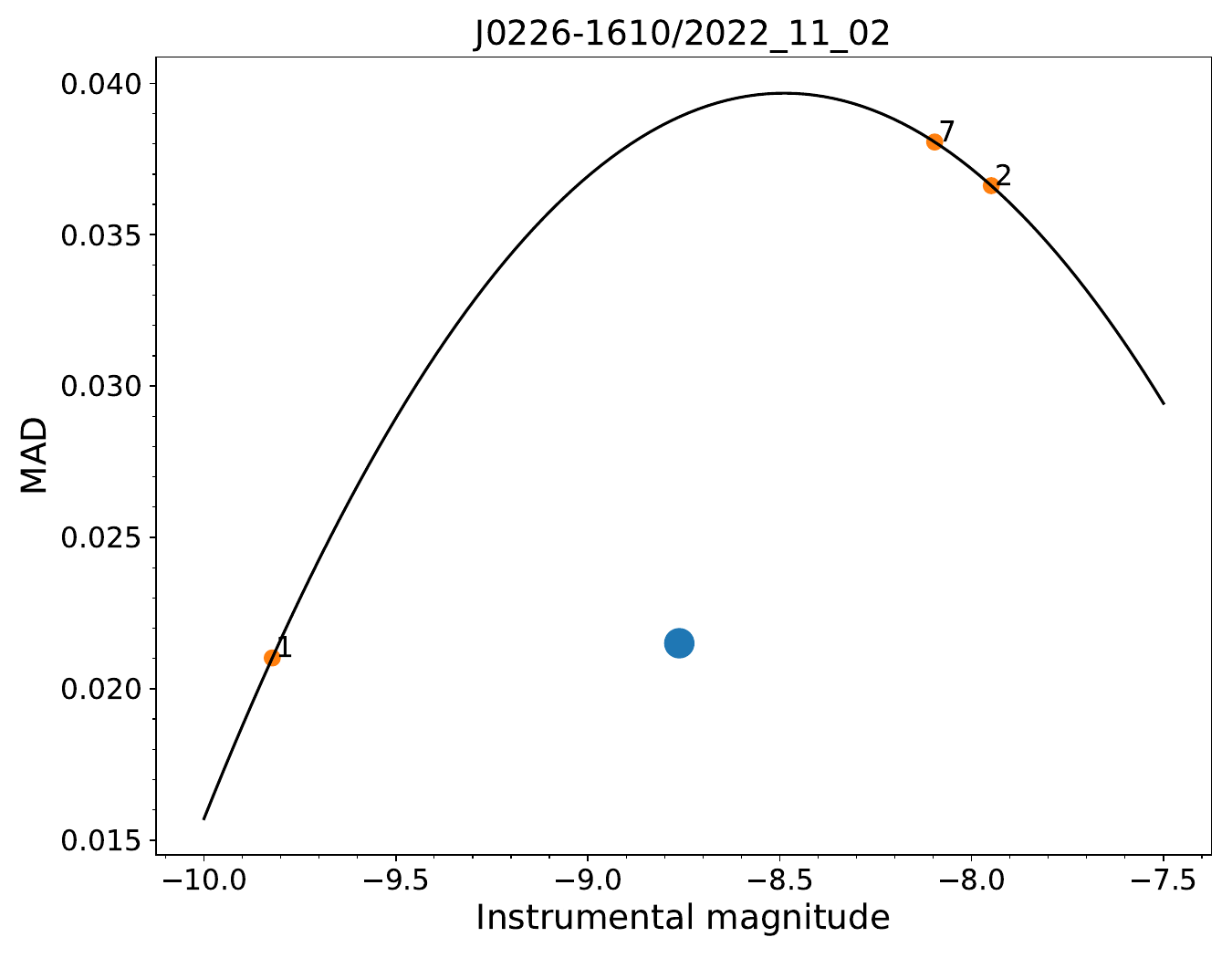}
    \includegraphics[width=0.5\columnwidth]{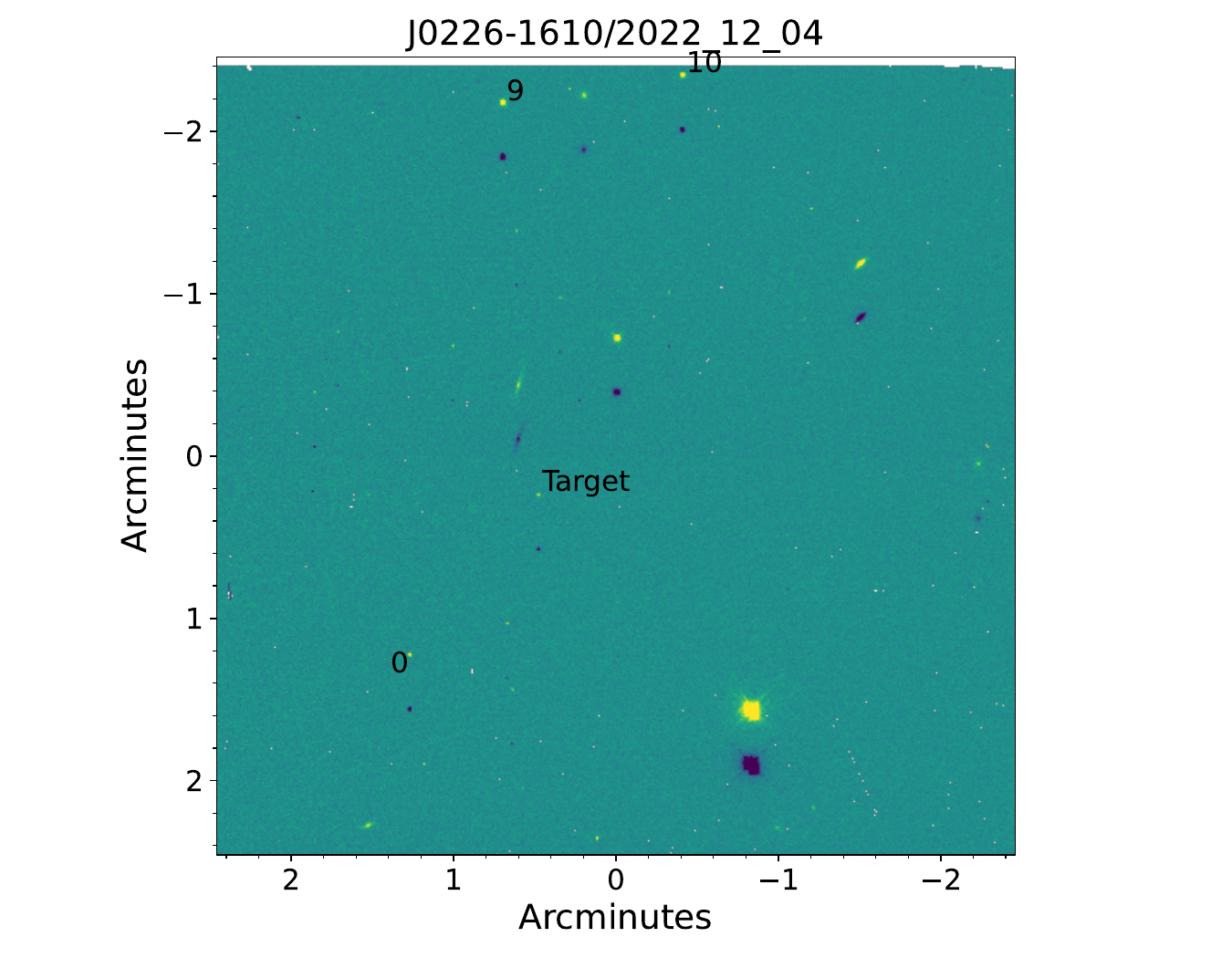}
    \includegraphics[width=0.5\columnwidth]{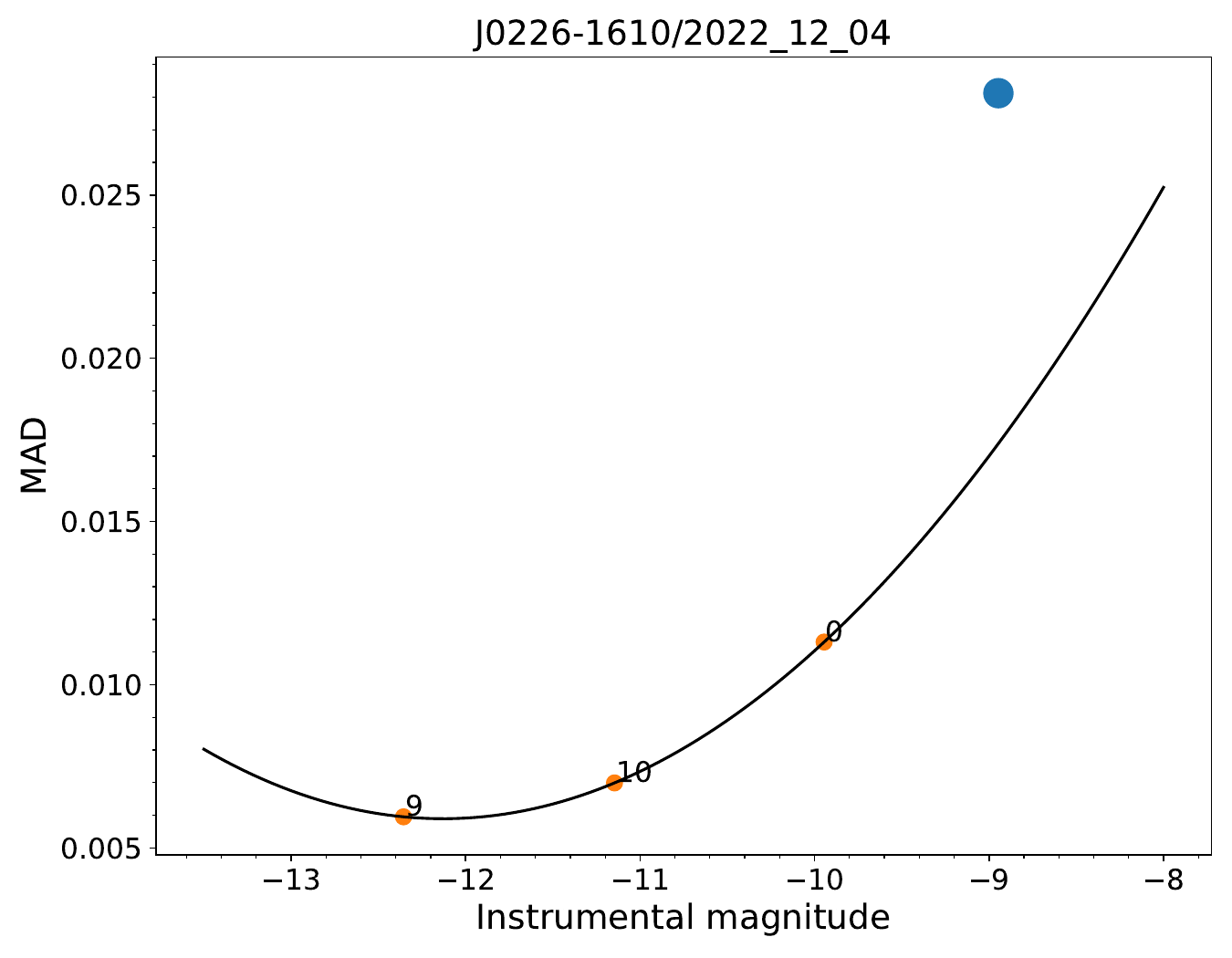}
    %J0241-3653
    \includegraphics[width=0.5\columnwidth]{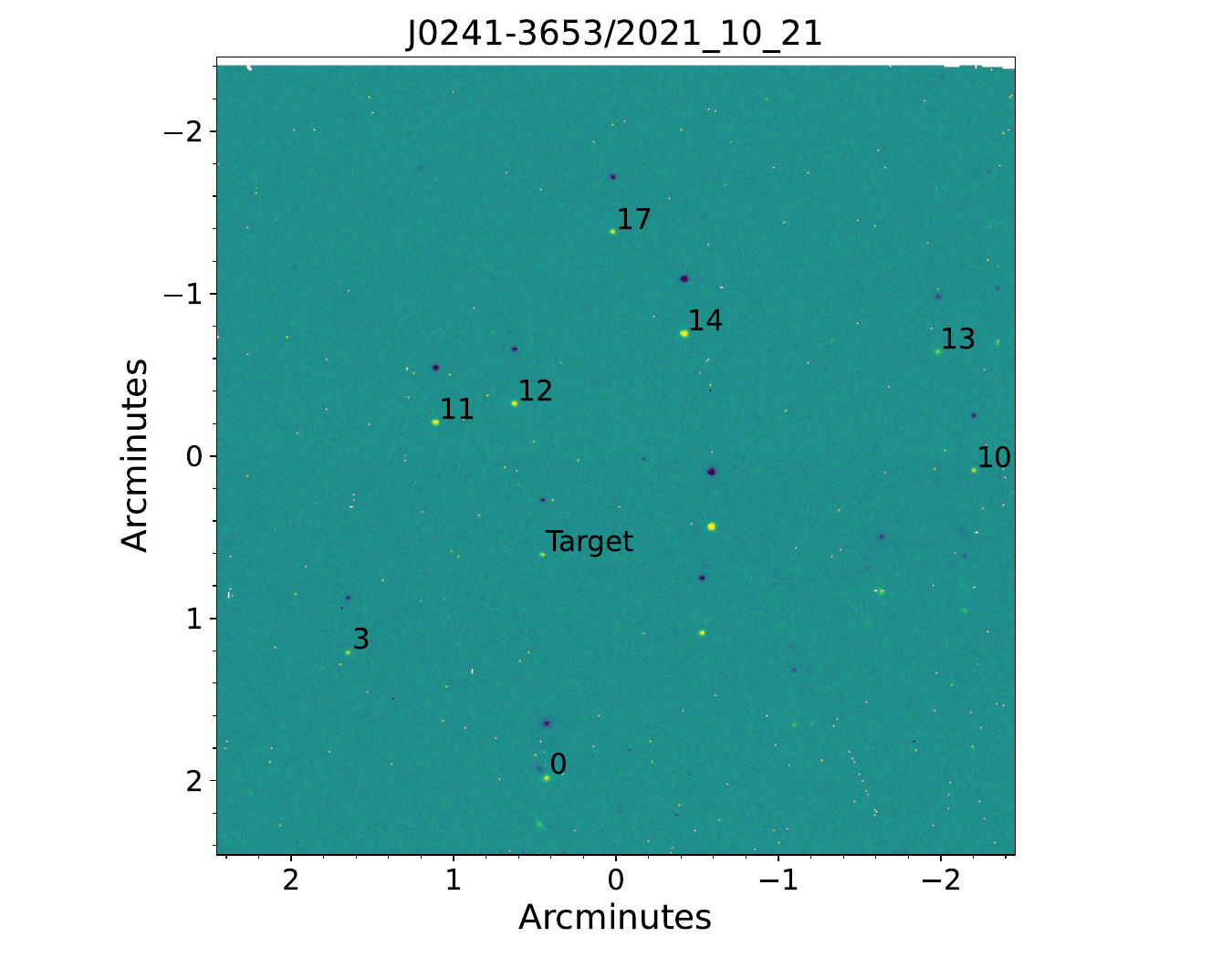}
    \includegraphics[width=0.5\columnwidth]{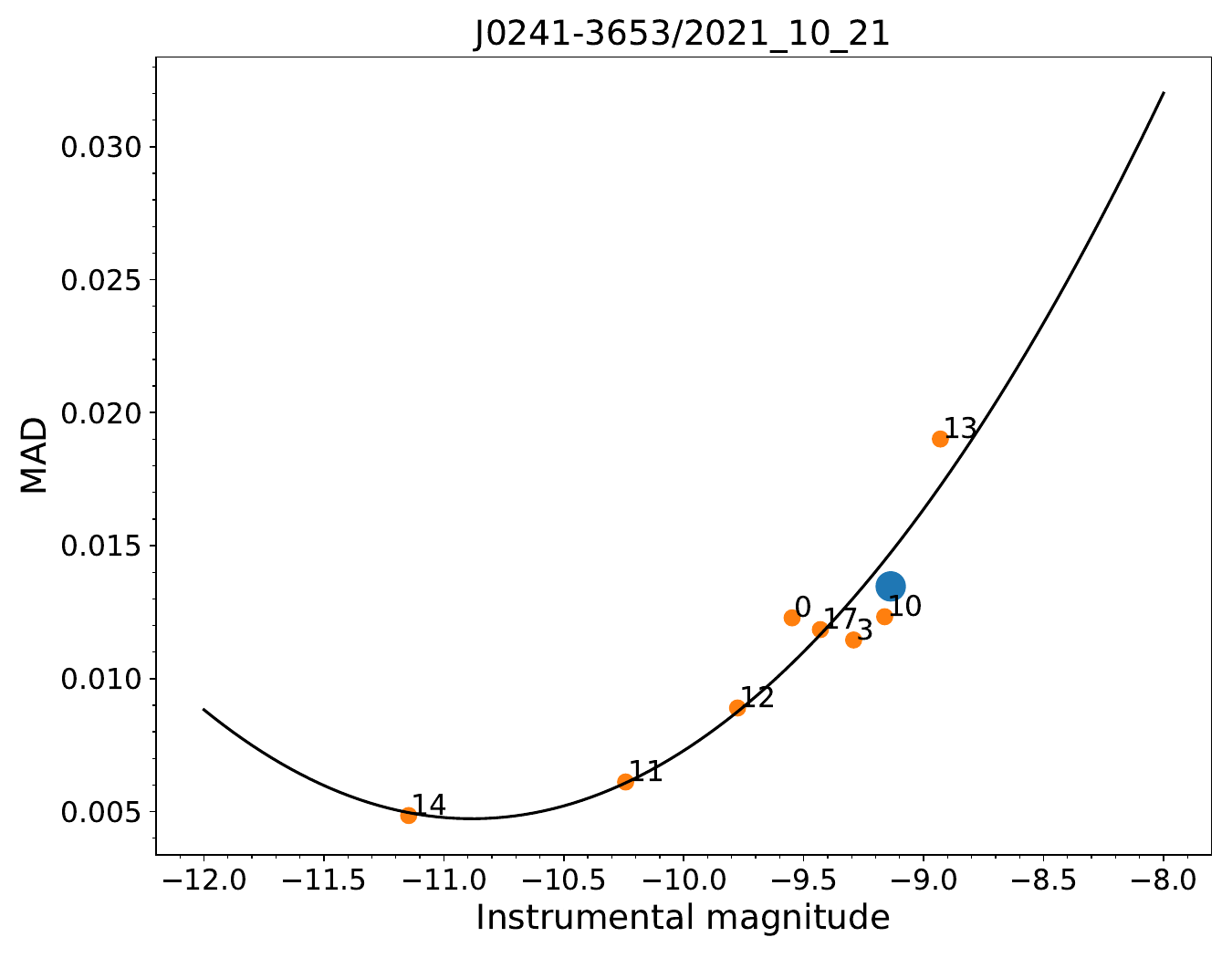}
     \includegraphics[width=0.5\columnwidth]{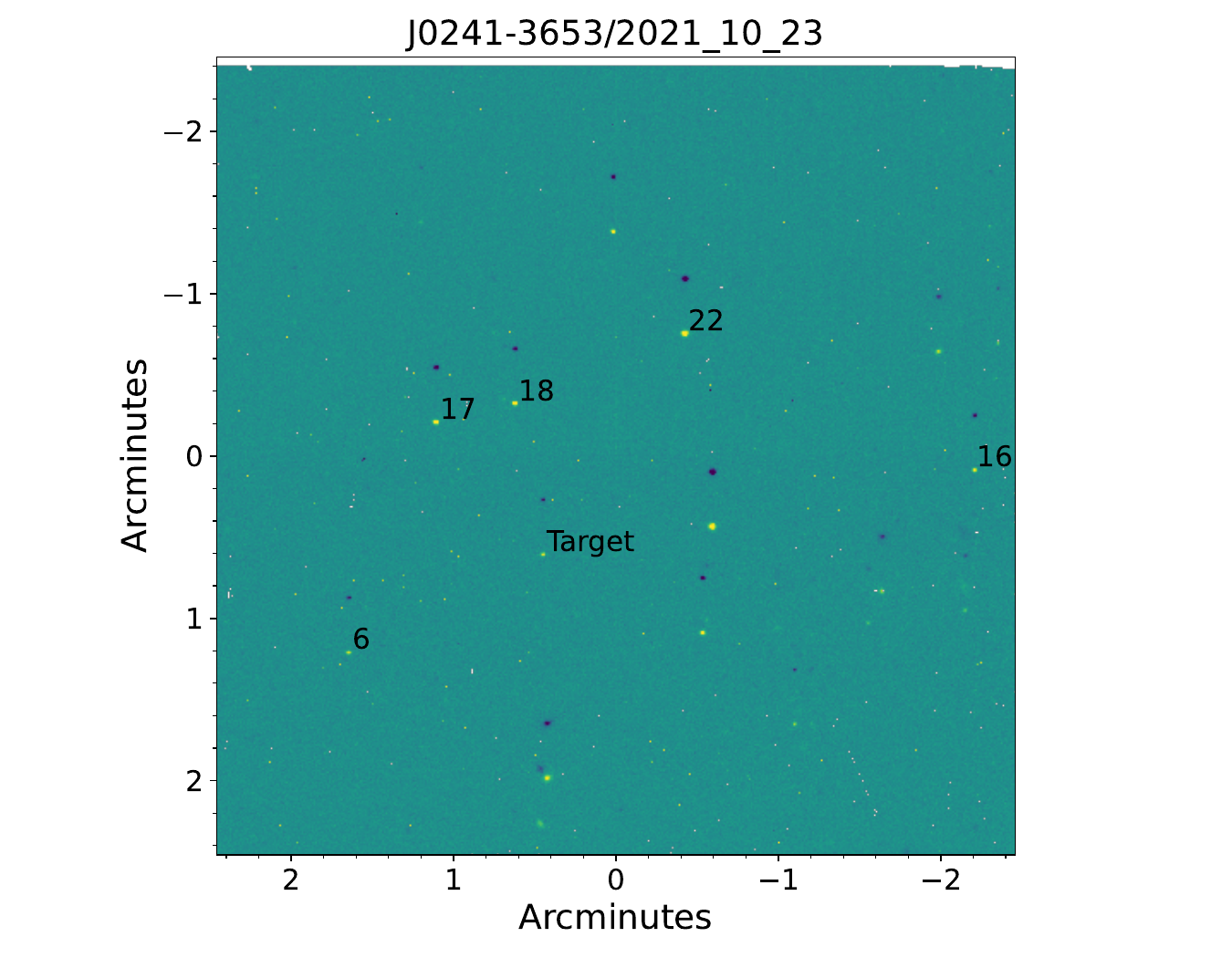}
    \includegraphics[width=0.5\columnwidth]{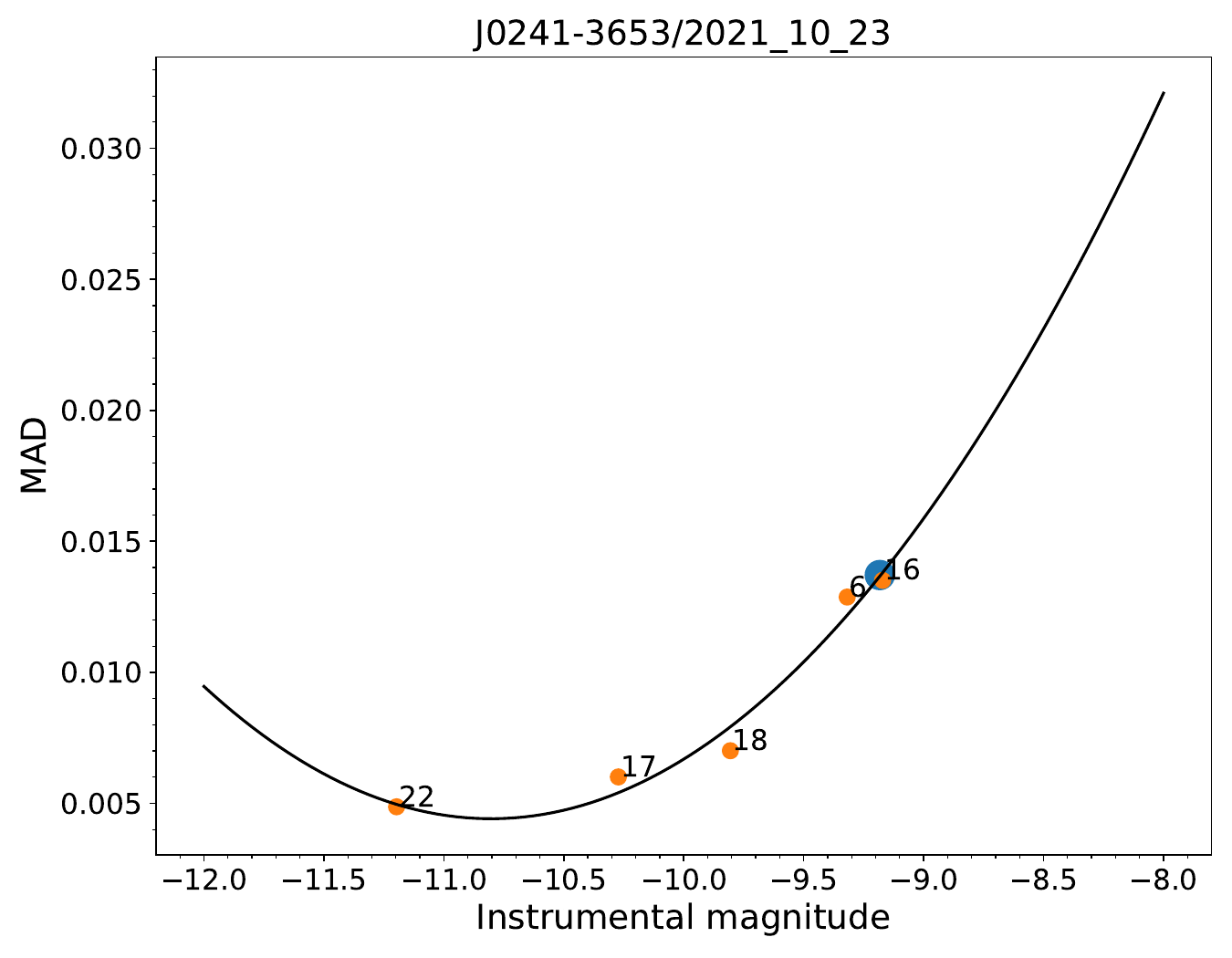}
    \caption{Targets and their selected reference stars. North is left and east is down.}
    \label{fig:stars_mag}
\end{figure*}

\begin{figure*}
    \includegraphics[width=0.5\columnwidth]{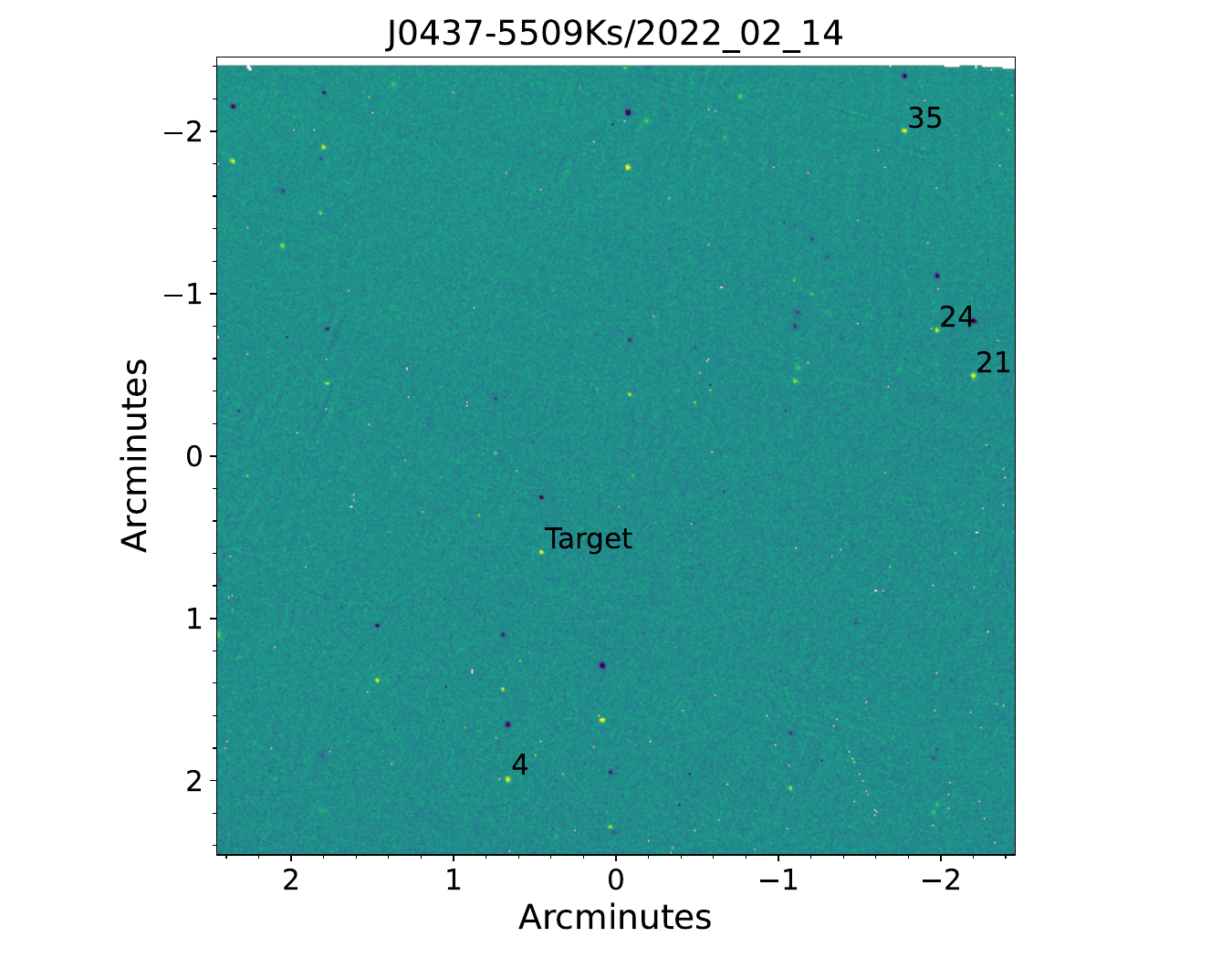}
    \includegraphics[width=0.5\columnwidth]{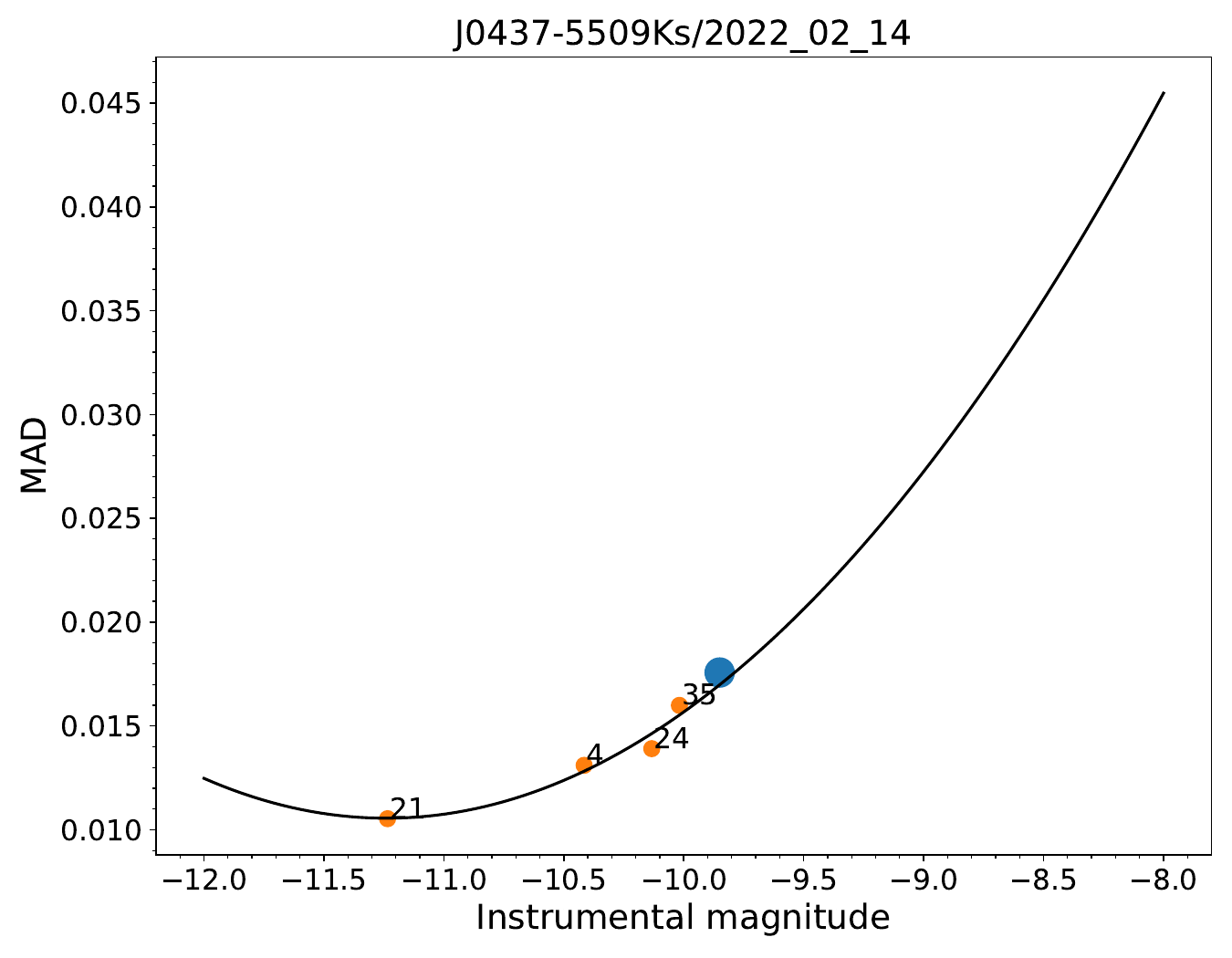}
    %J0758+2225
    \includegraphics[width=0.5\columnwidth]{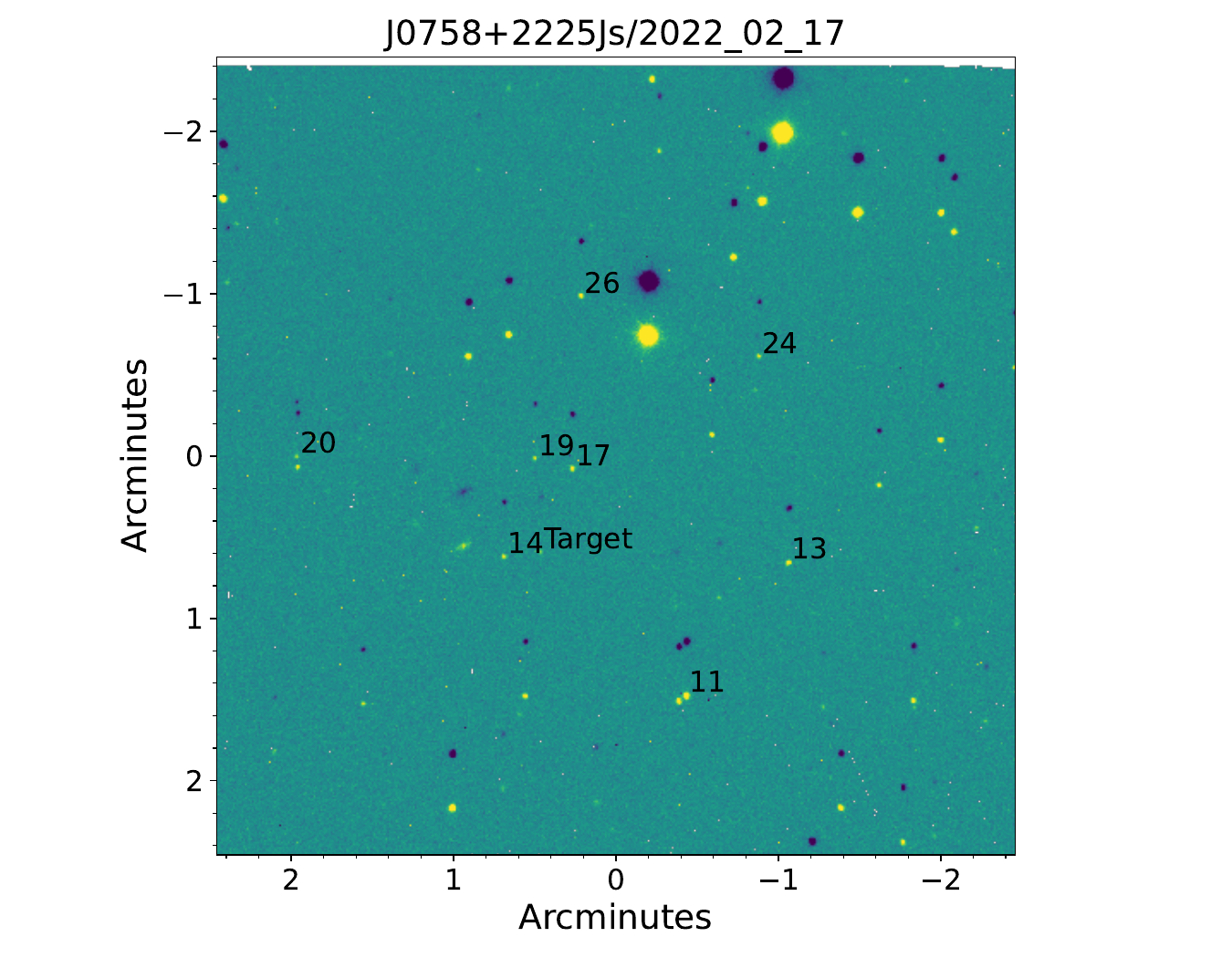}
    \includegraphics[width=0.5\columnwidth]{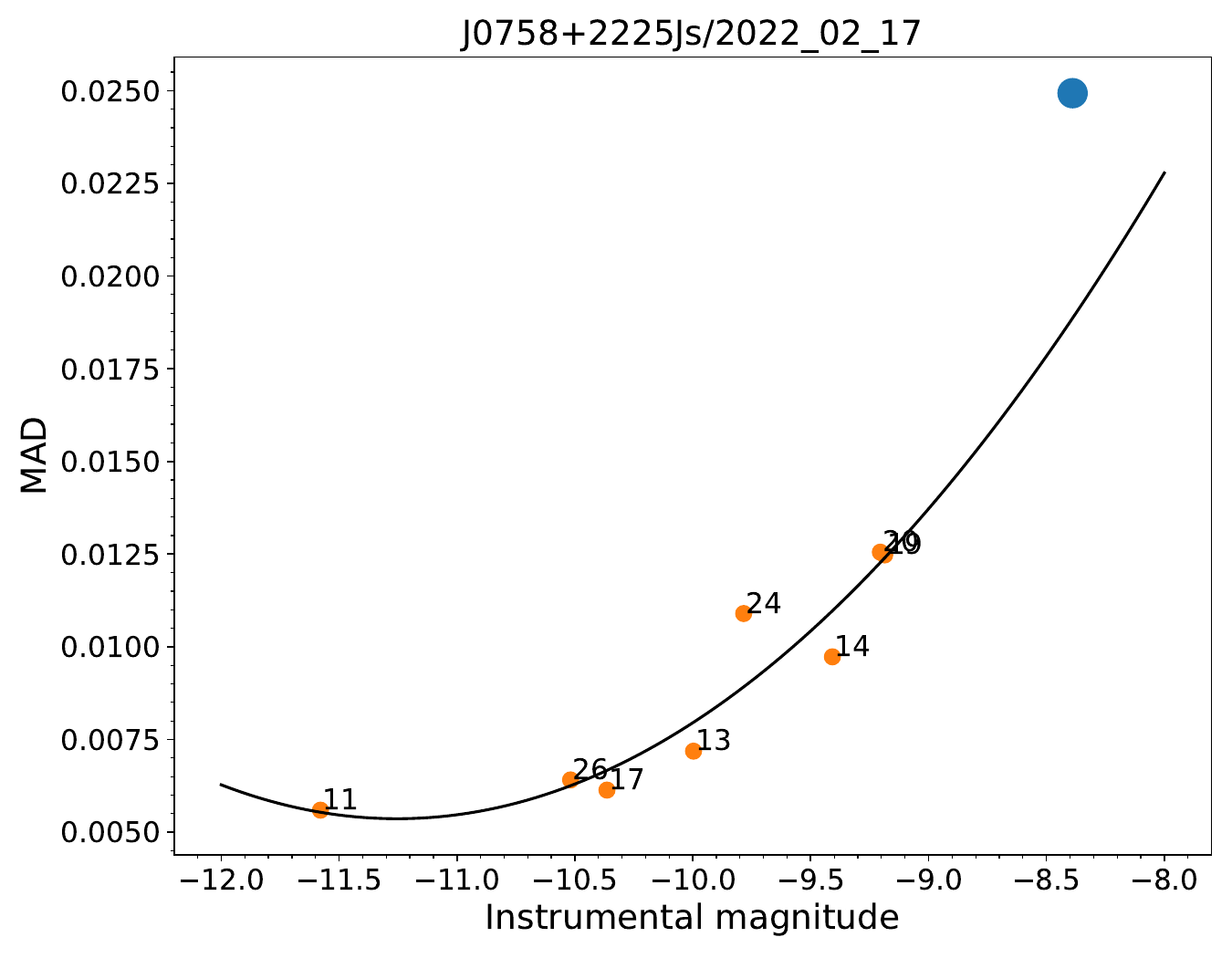}
    %J0819-0335
    \includegraphics[width=0.5\columnwidth]{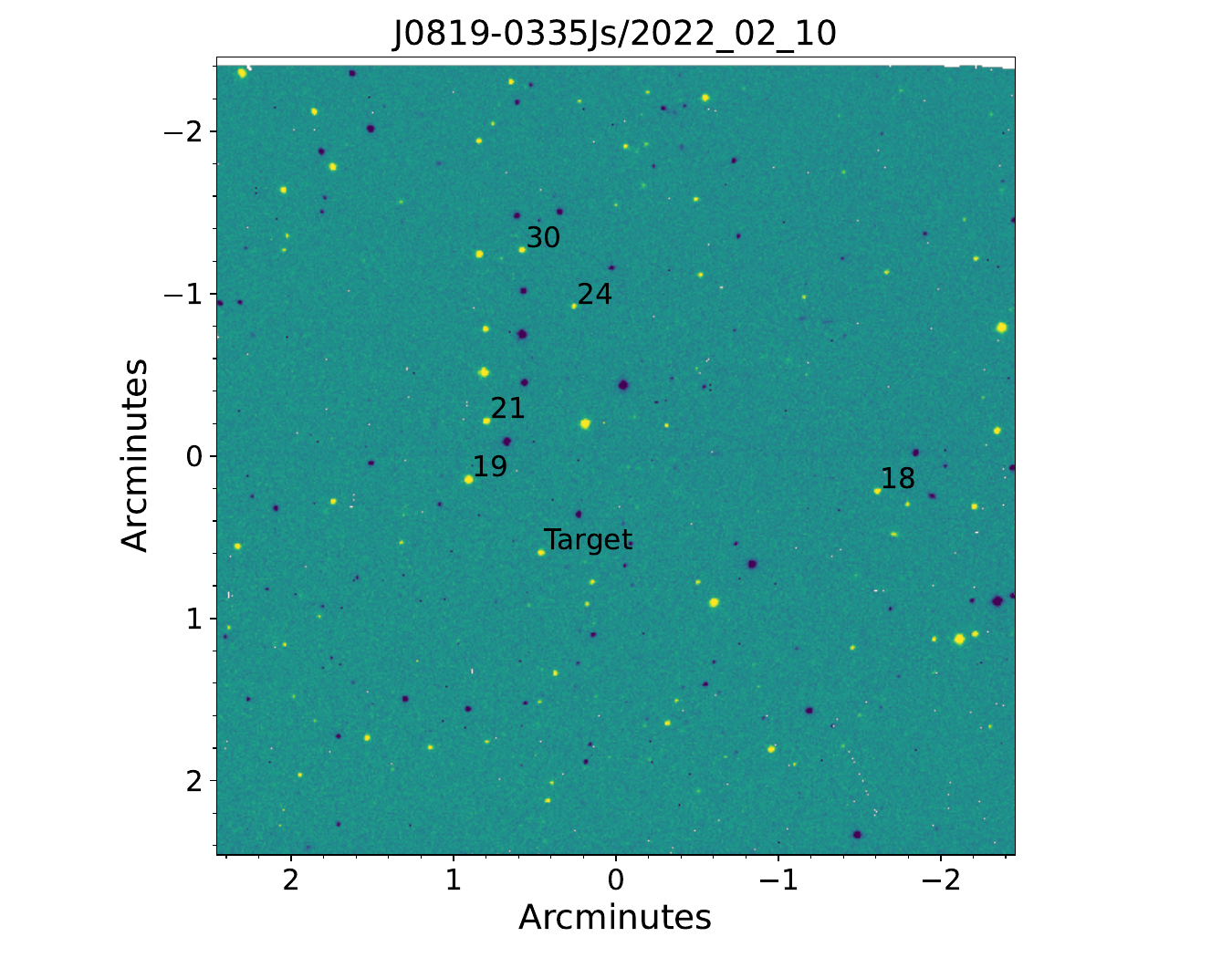}
    \includegraphics[width=0.5\columnwidth]{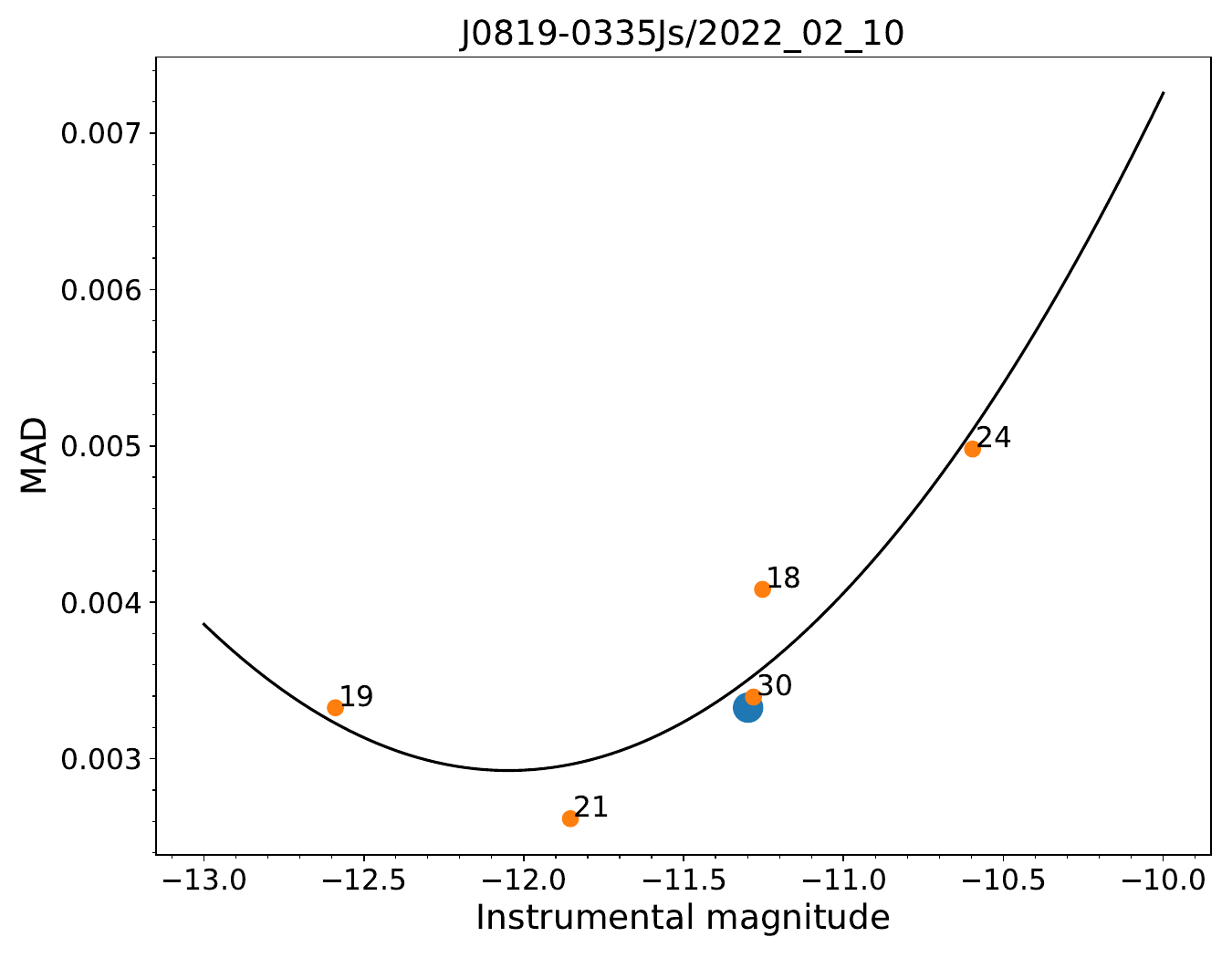}
    \includegraphics[width=0.5\columnwidth]{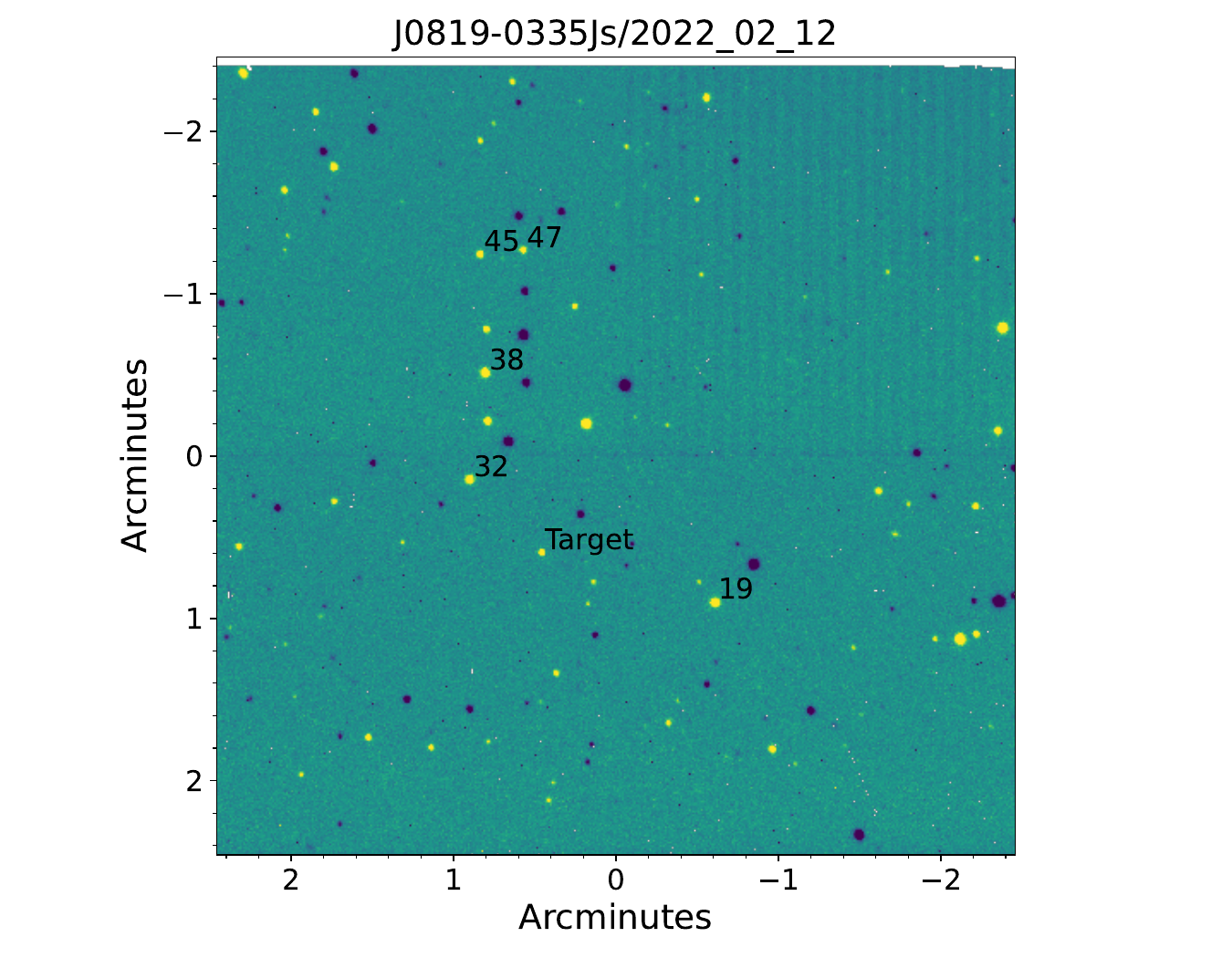}
    \includegraphics[width=0.5\columnwidth]{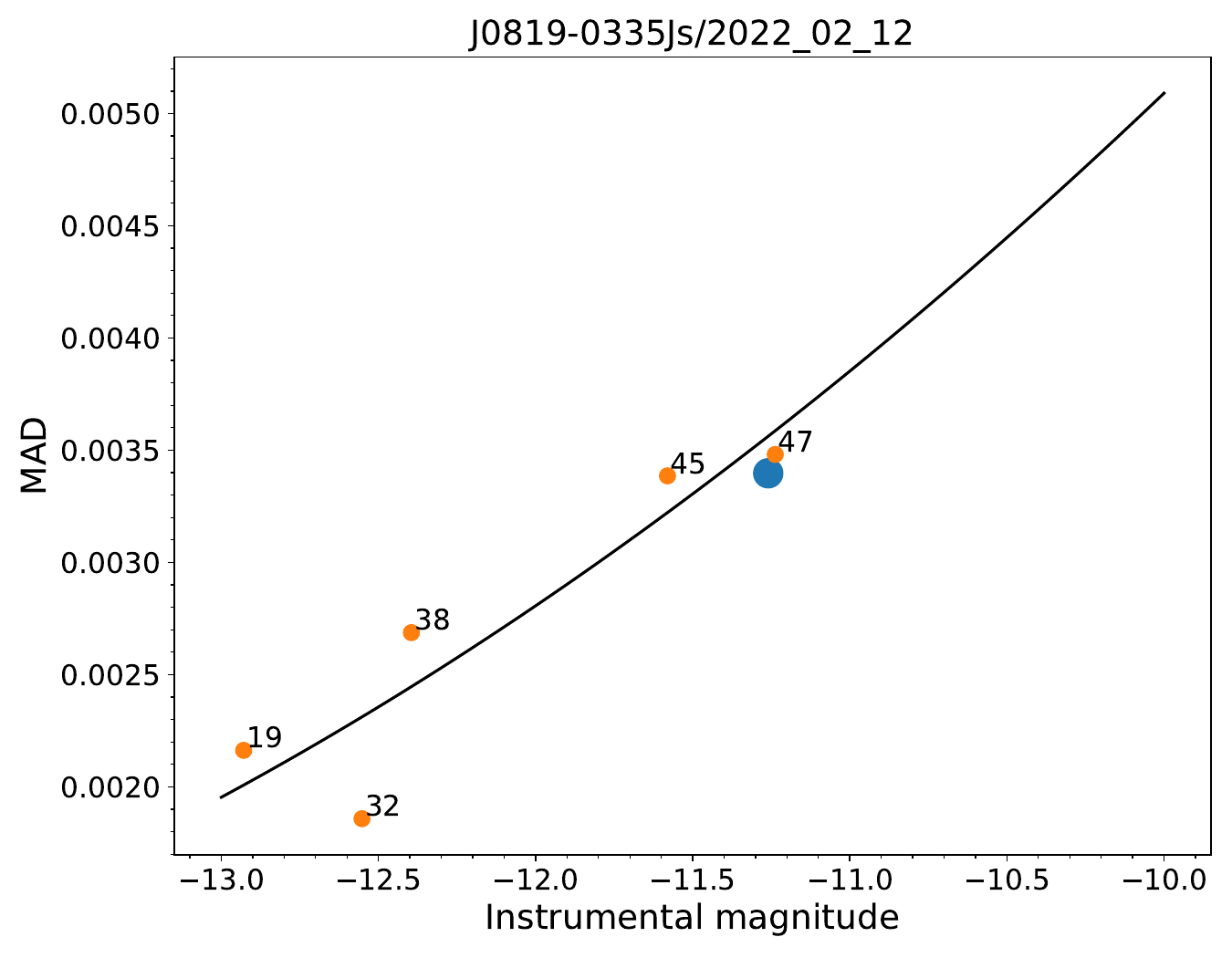}
    \includegraphics[width=0.5\columnwidth]{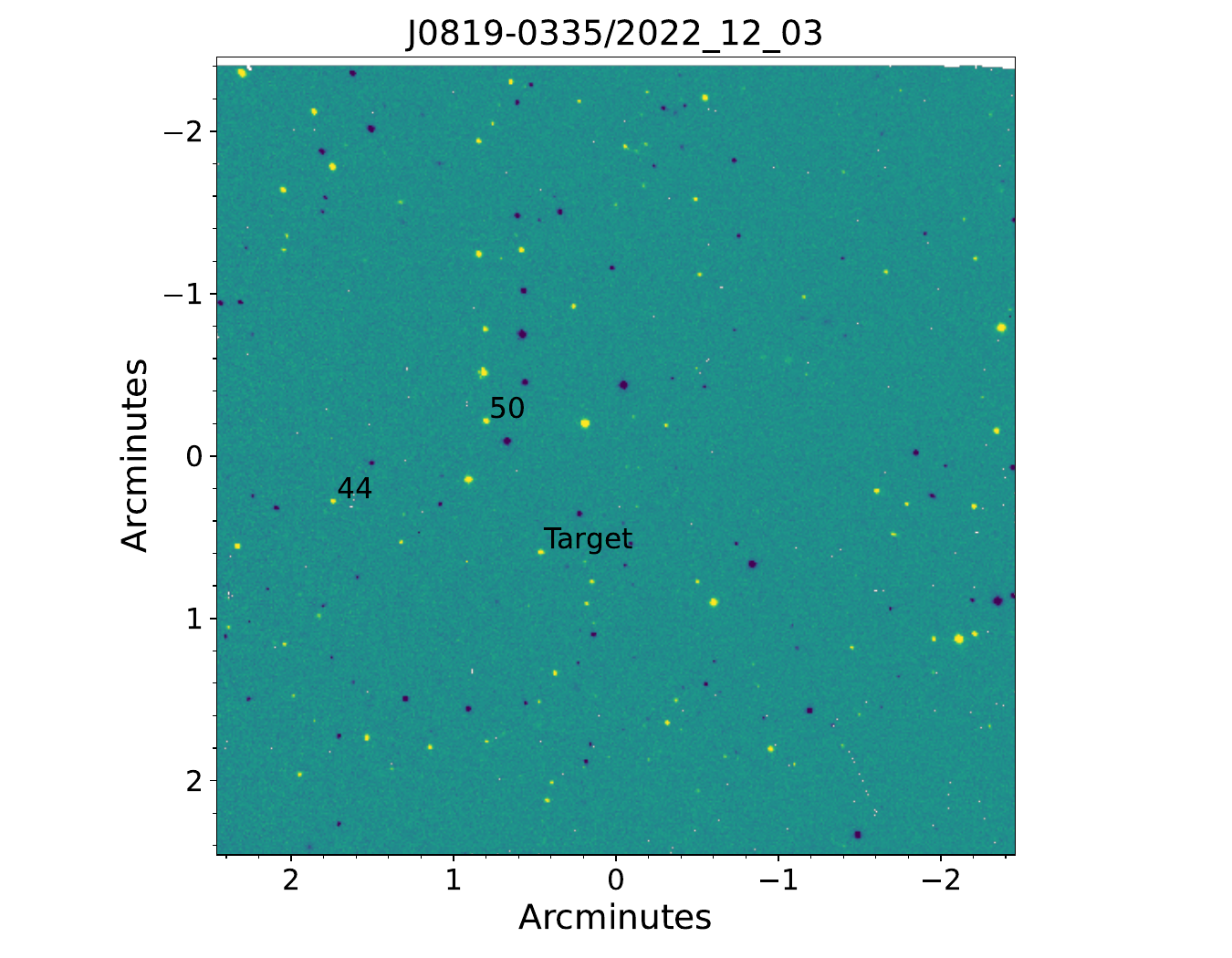}
    \includegraphics[width=0.5\columnwidth]{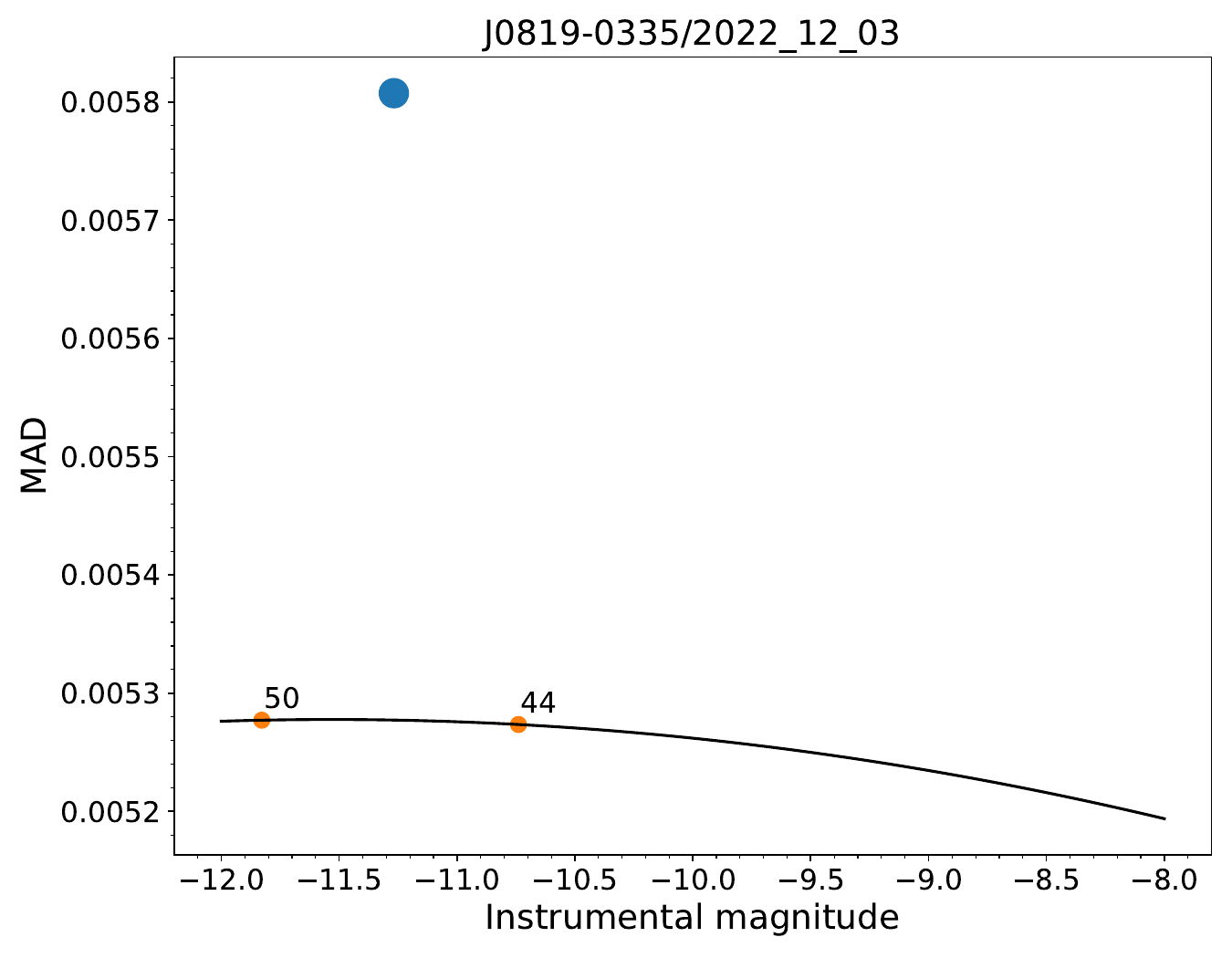}
    \includegraphics[width=0.5\columnwidth]{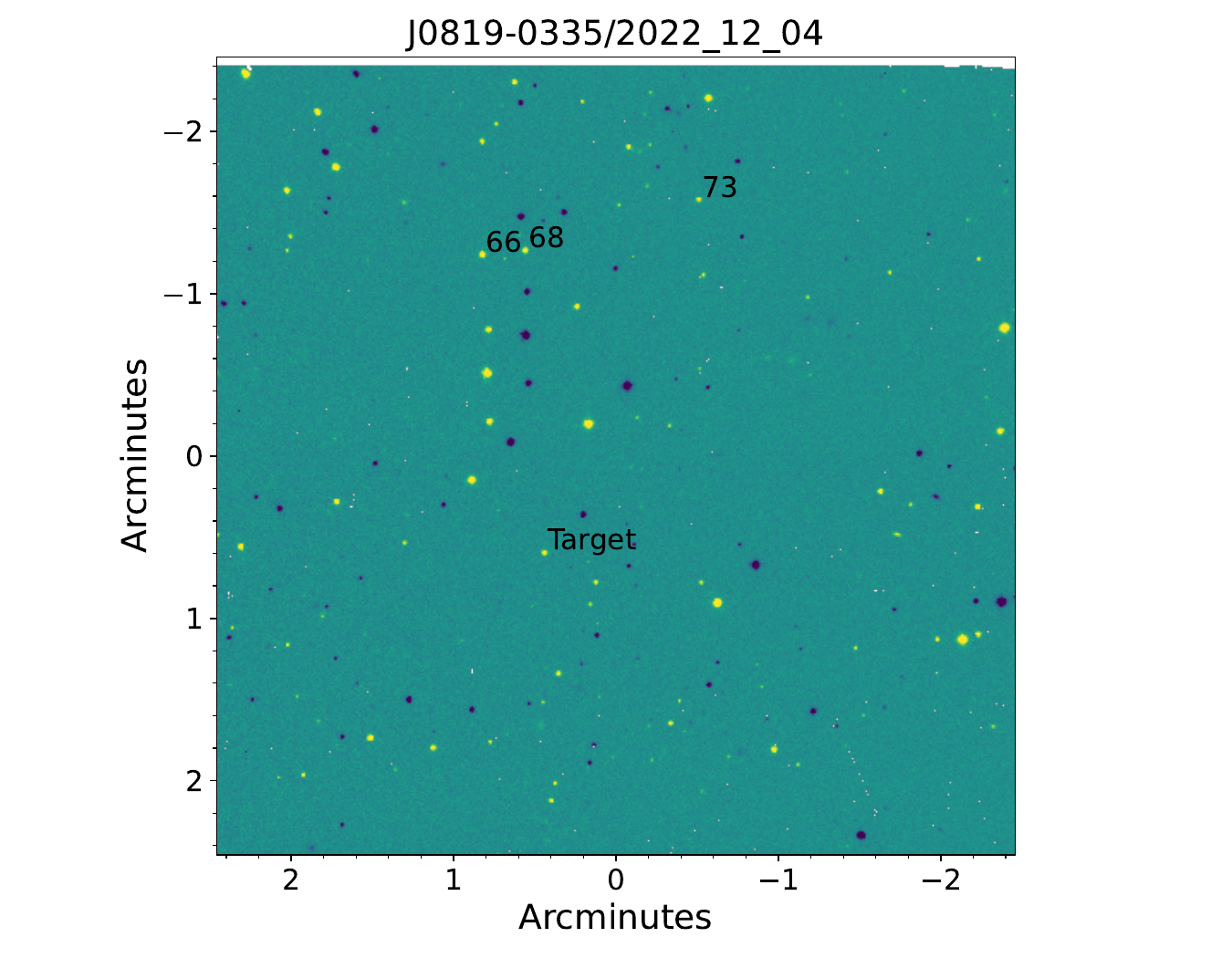}
    \includegraphics[width=0.5\columnwidth]{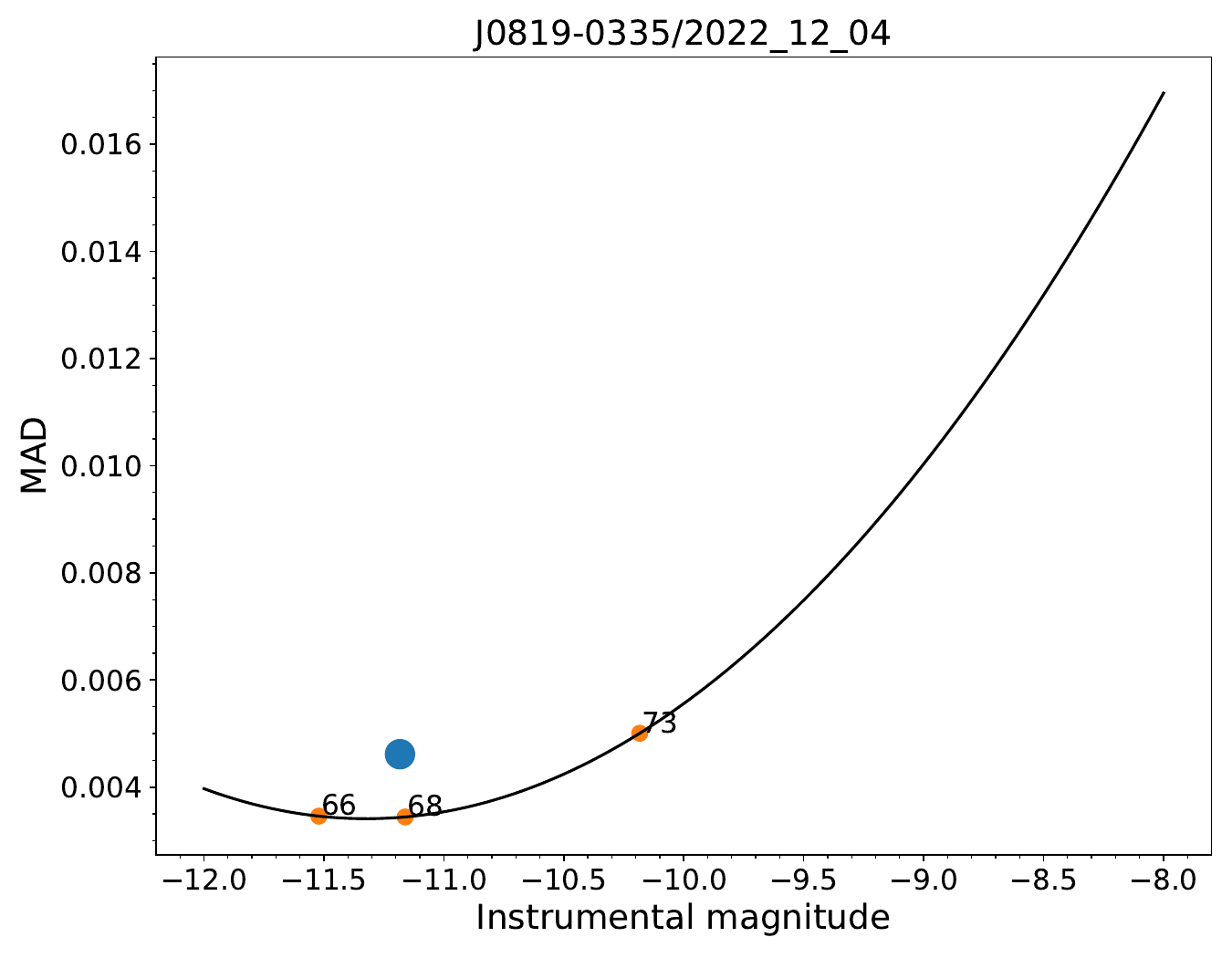}
    %J0819+2103
    \includegraphics[width=0.5\columnwidth]{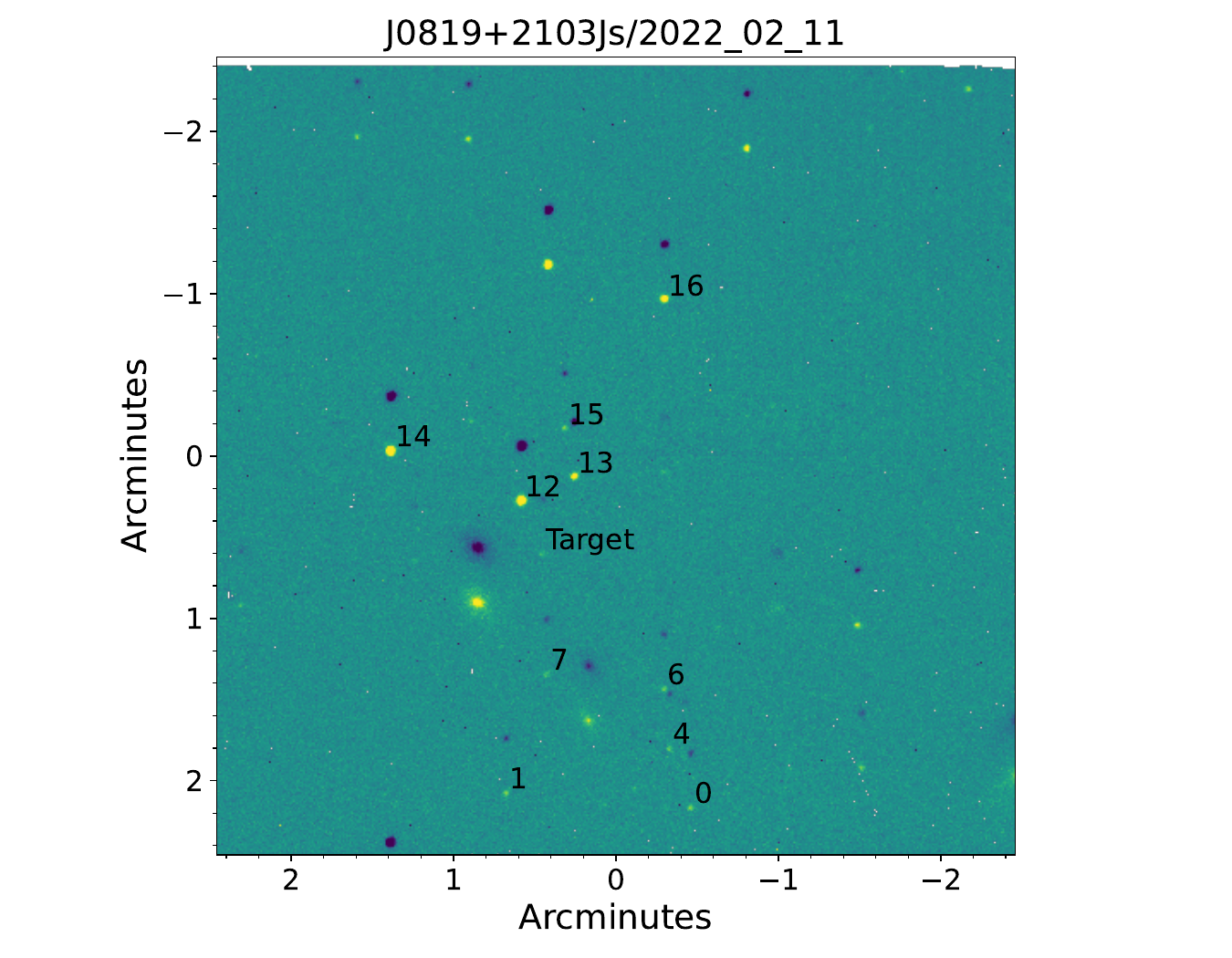}
    \includegraphics[width=0.5\columnwidth]{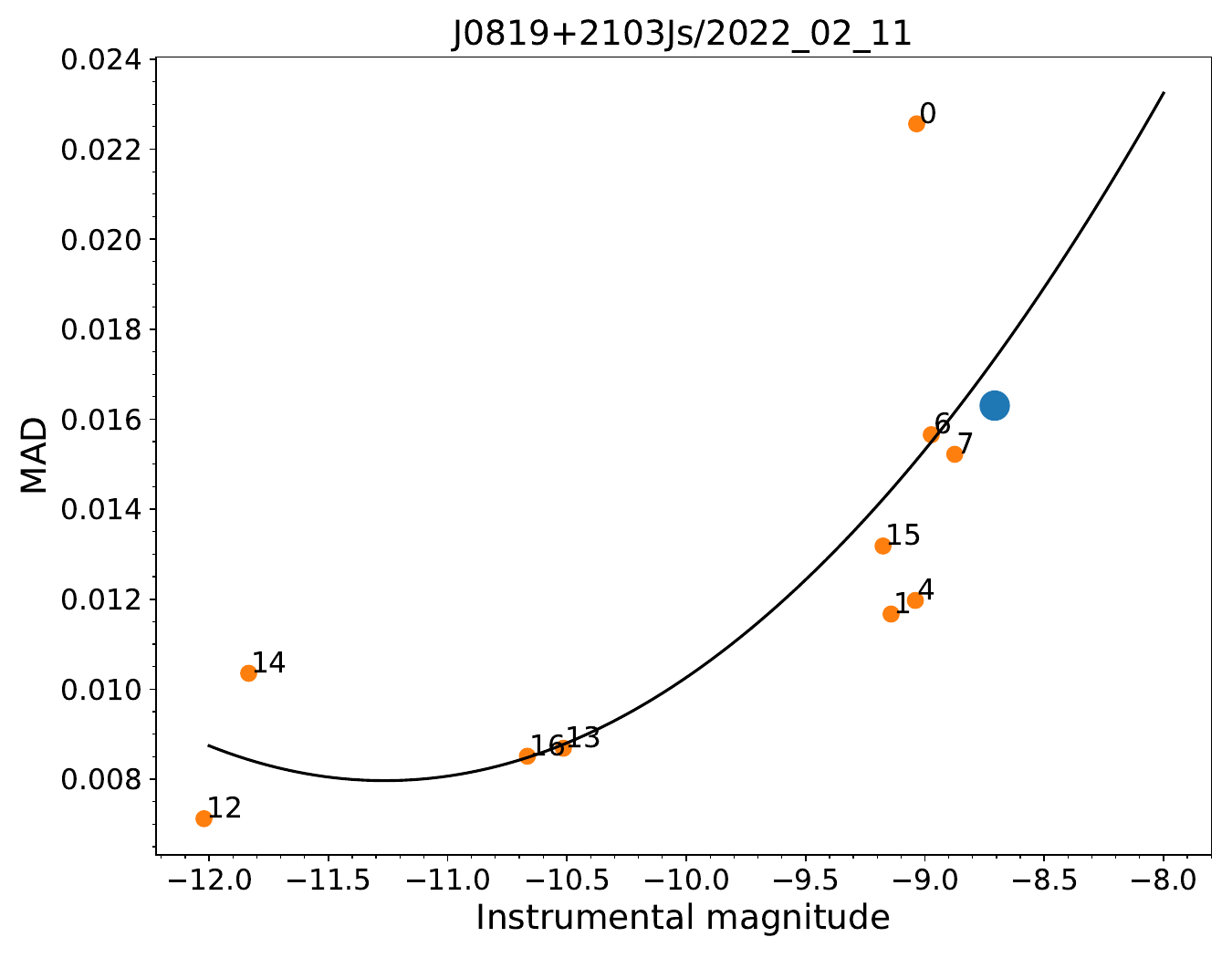}
    %J1316+0312
    \includegraphics[width=0.5\columnwidth]{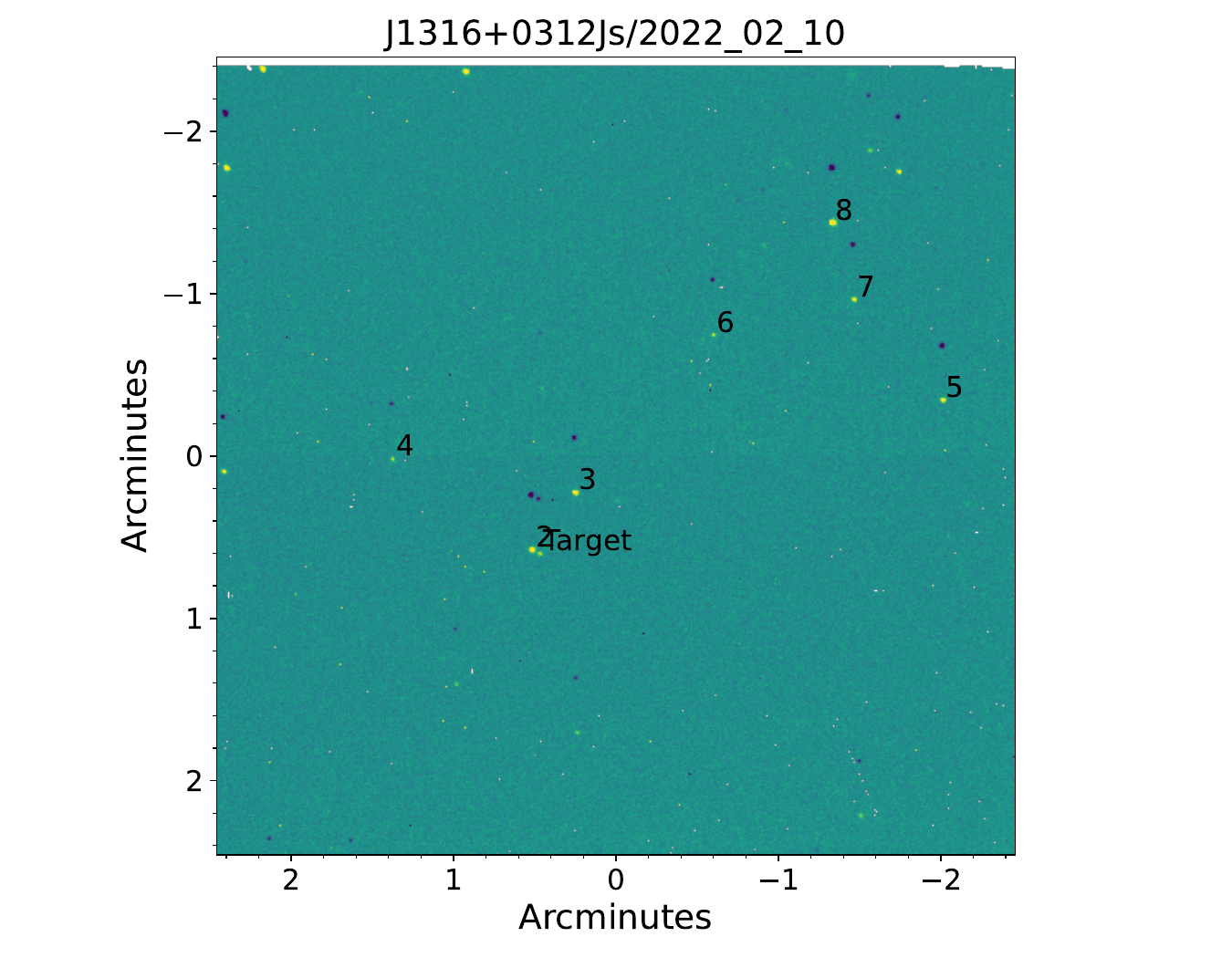}
    \includegraphics[width=0.5\columnwidth]{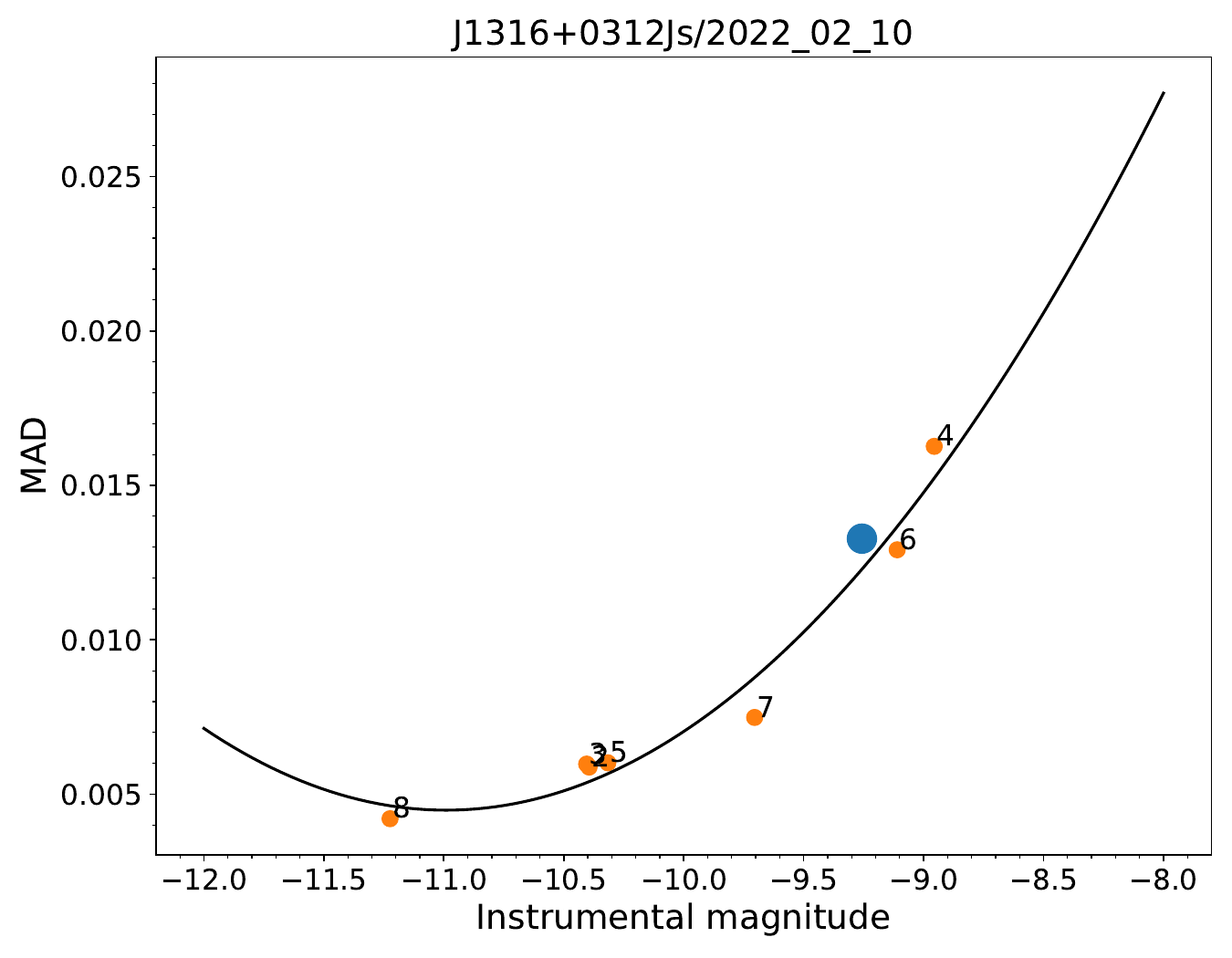}
    %J2255-3118
    \includegraphics[width=0.5\columnwidth]{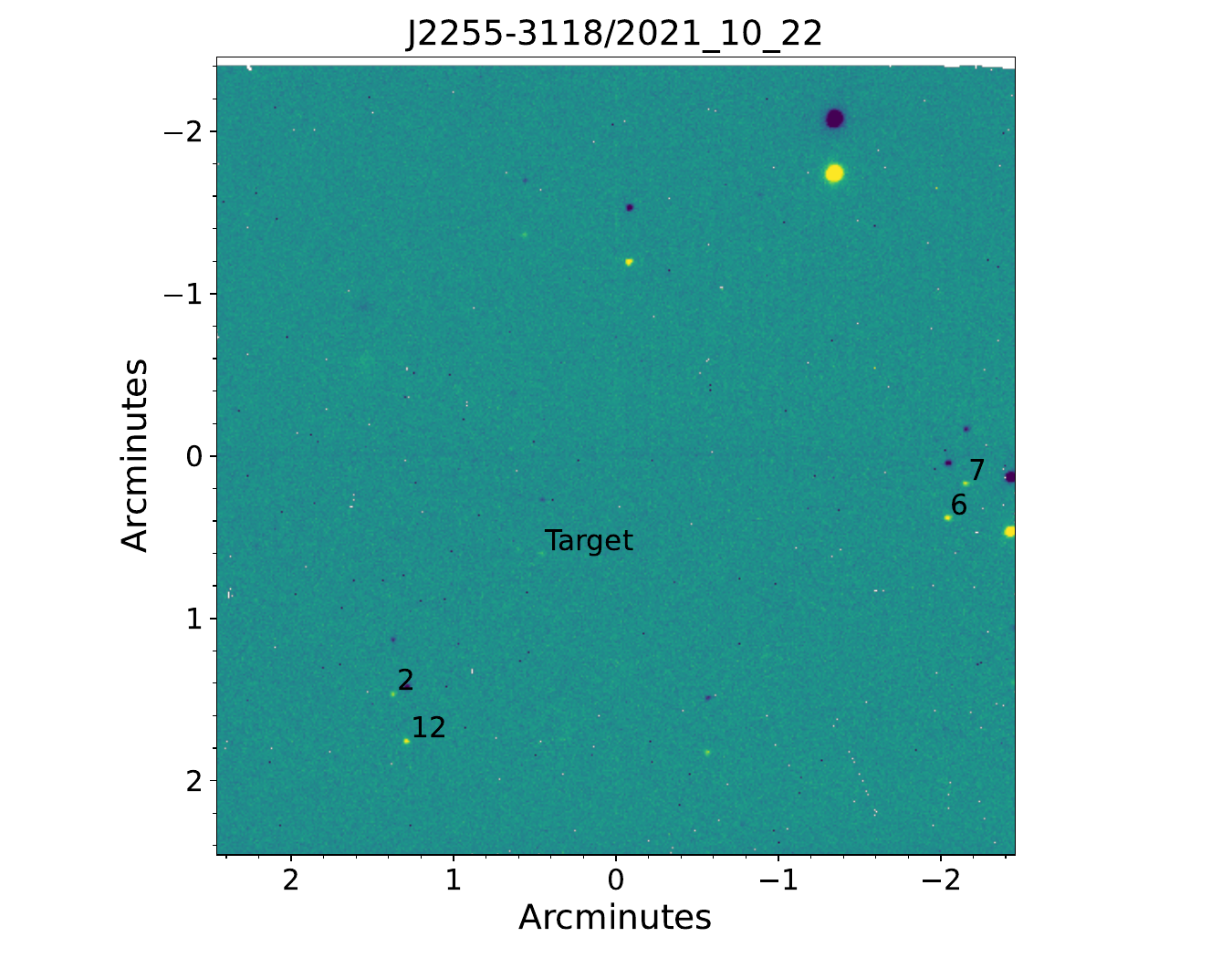}
    \includegraphics[width=0.5\columnwidth]{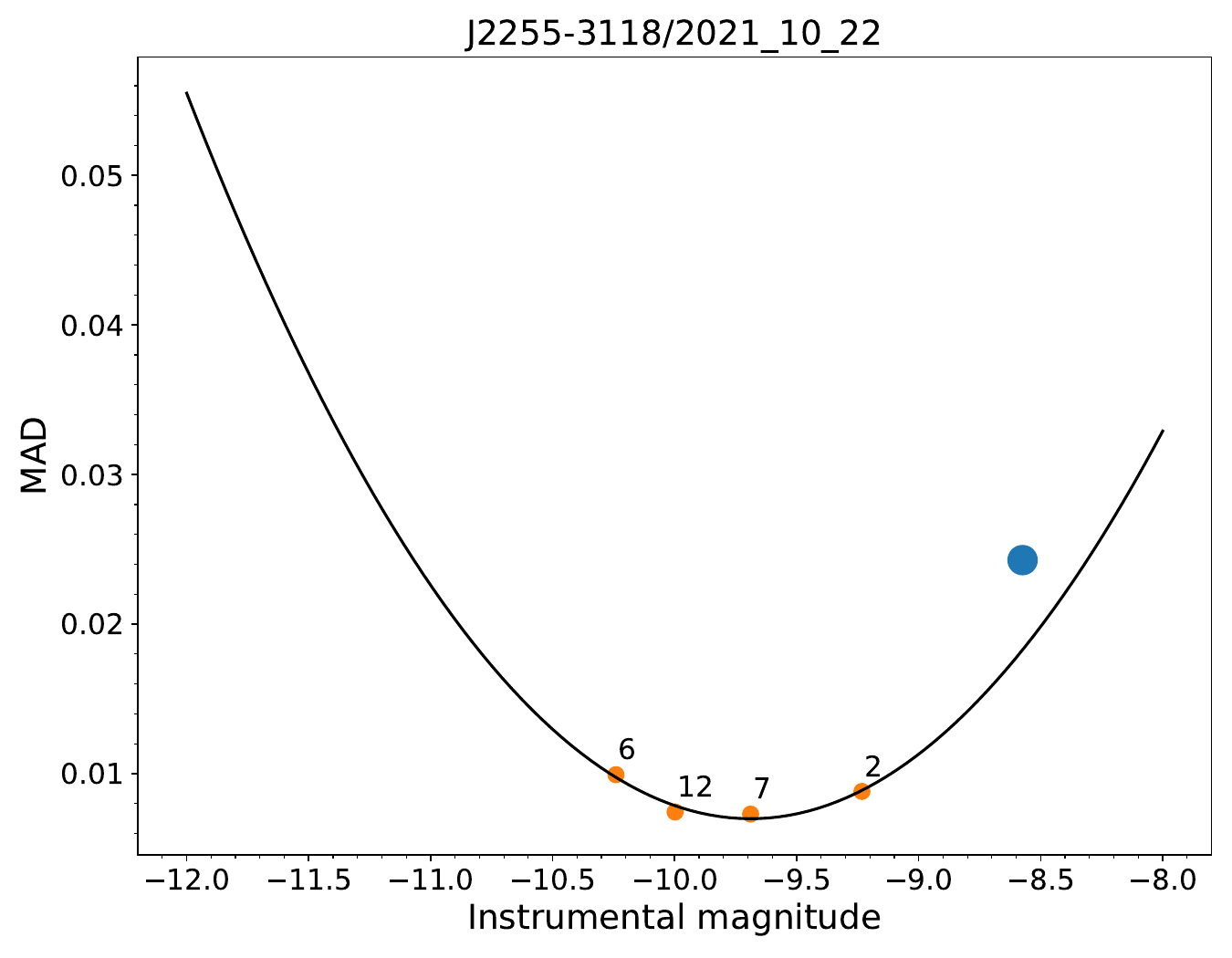}
    %J2323-0152
    \includegraphics[width=0.5\columnwidth]{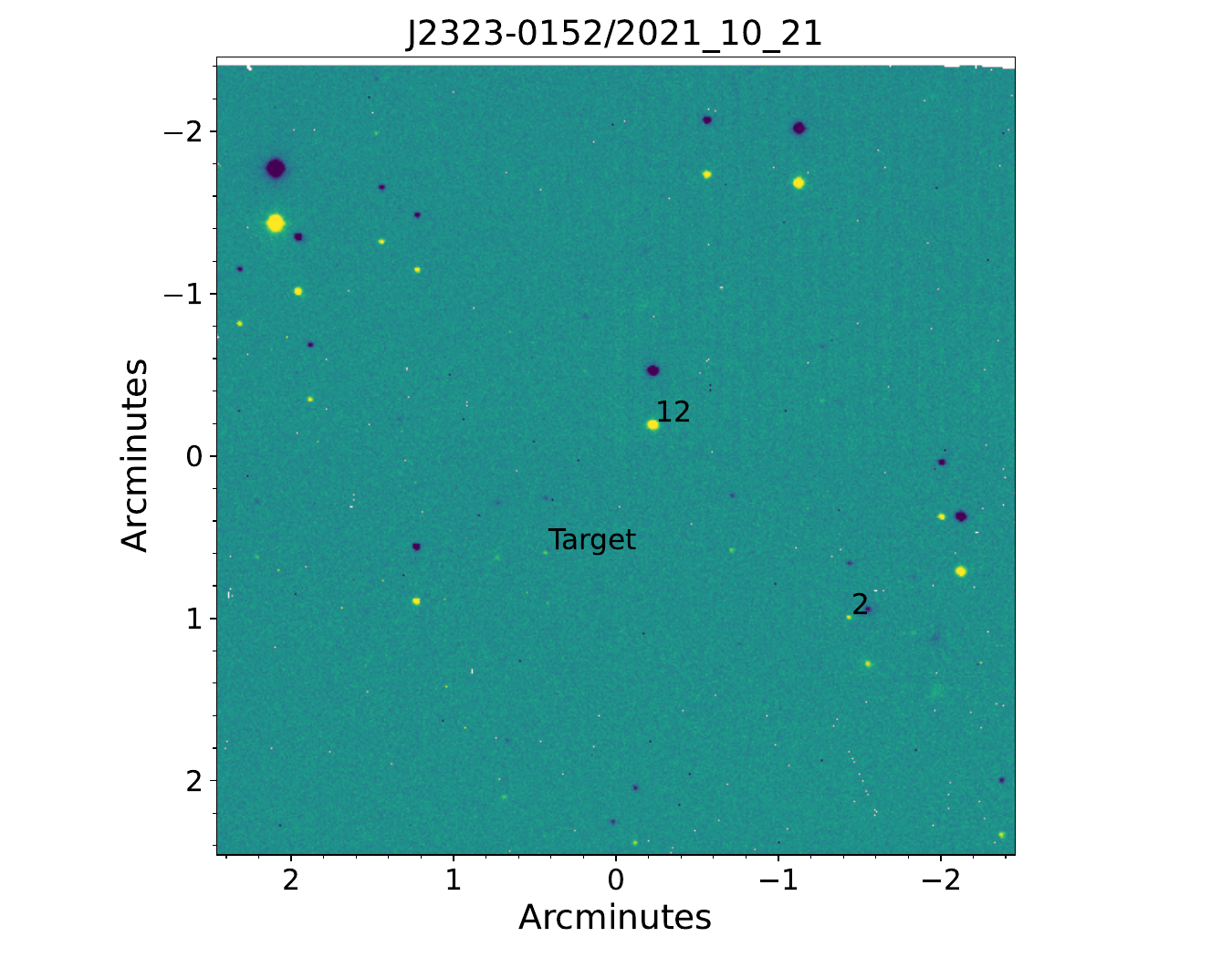}
    \includegraphics[width=0.5\columnwidth]{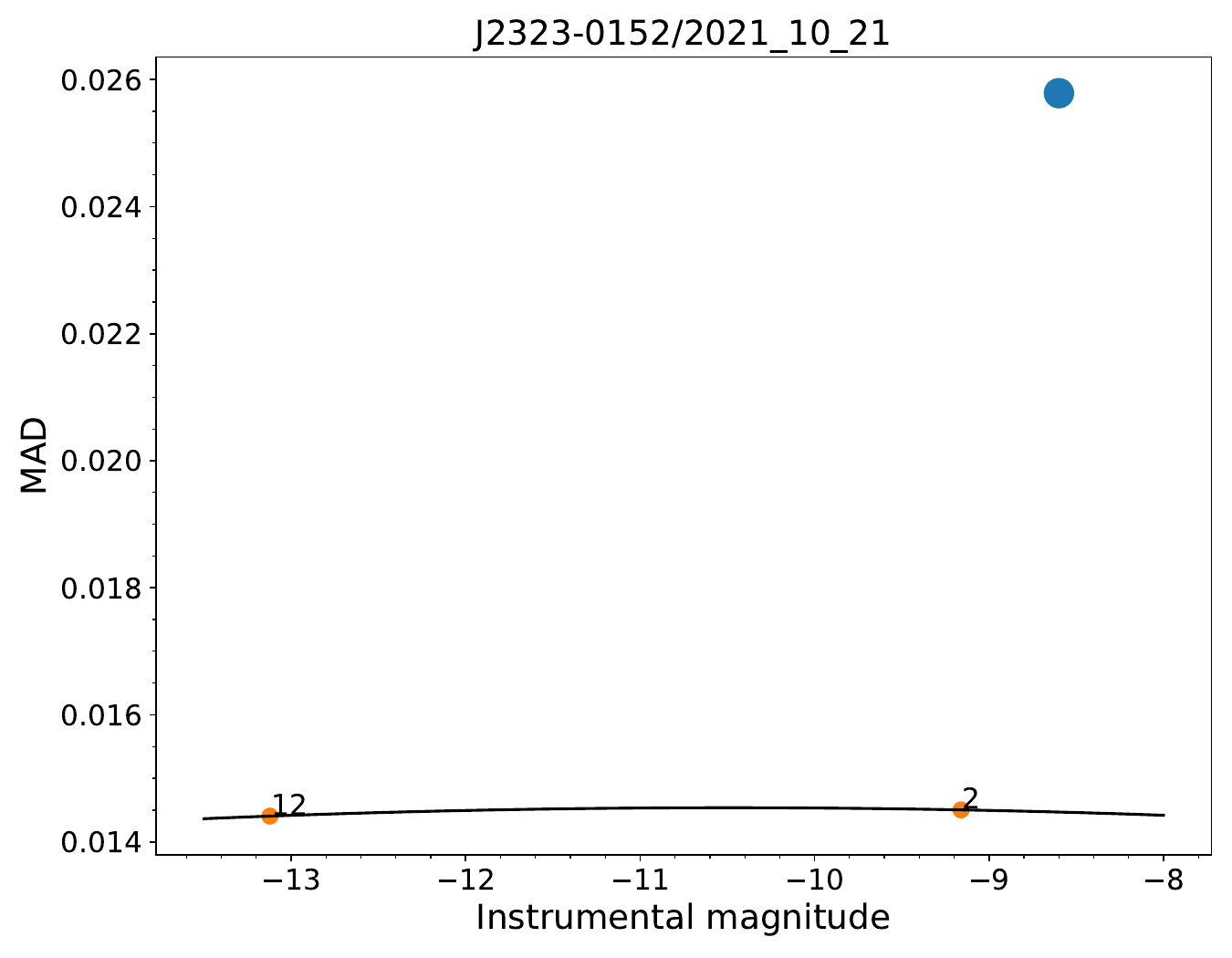}
    \includegraphics[width=0.5\columnwidth]{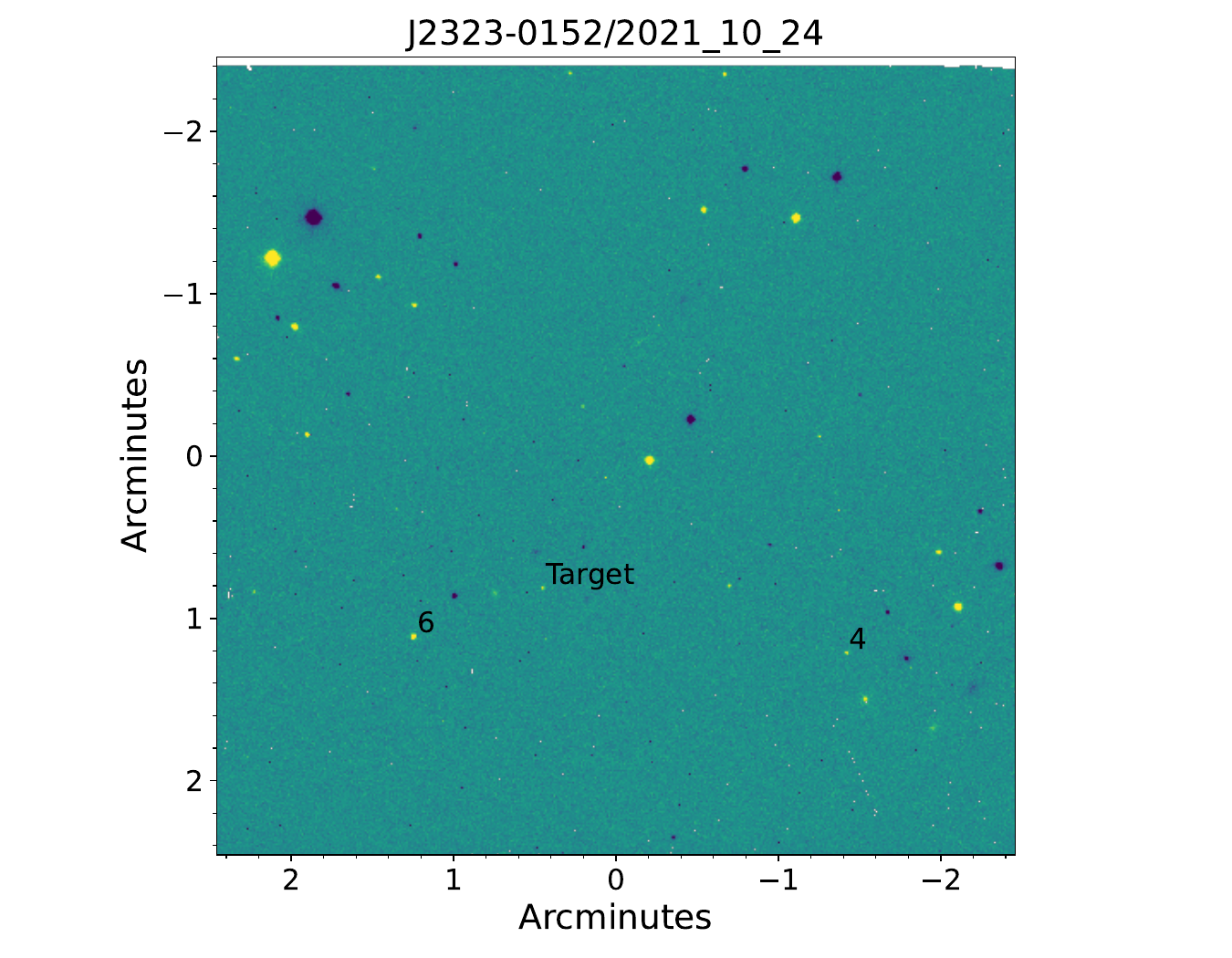}
    \includegraphics[width=0.5\columnwidth]{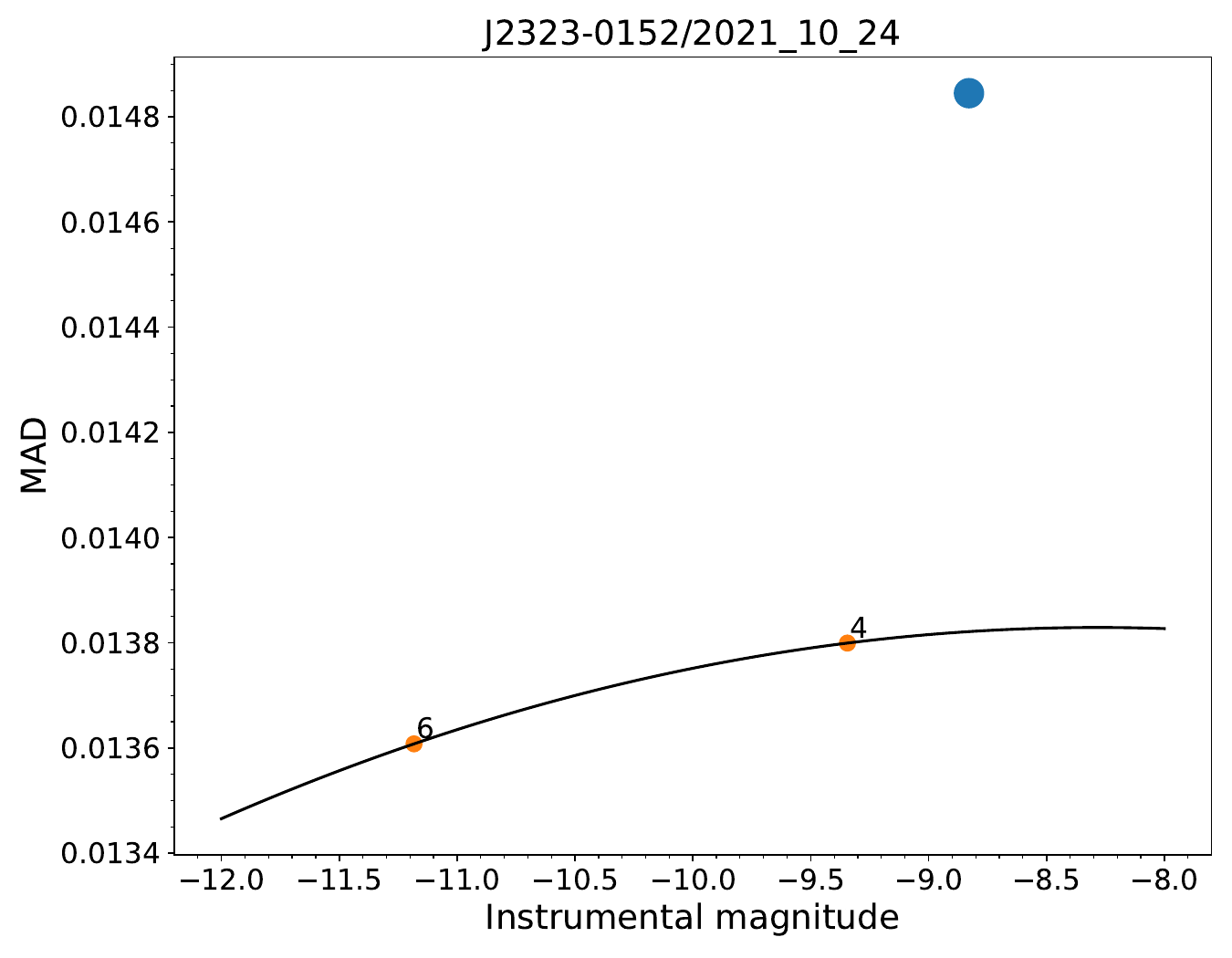}
    \includegraphics[width=0.5\columnwidth]{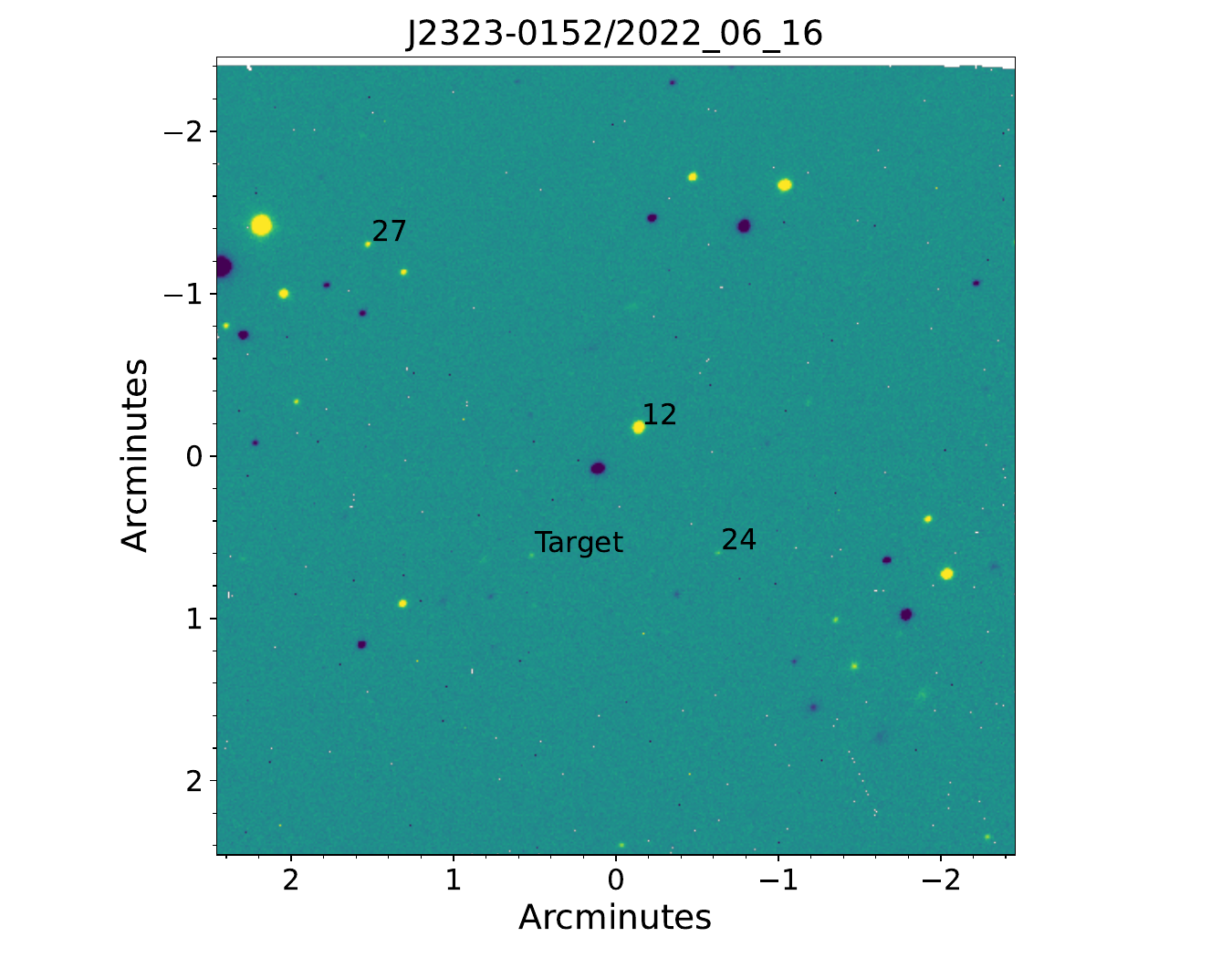}
    \includegraphics[width=0.5\columnwidth]{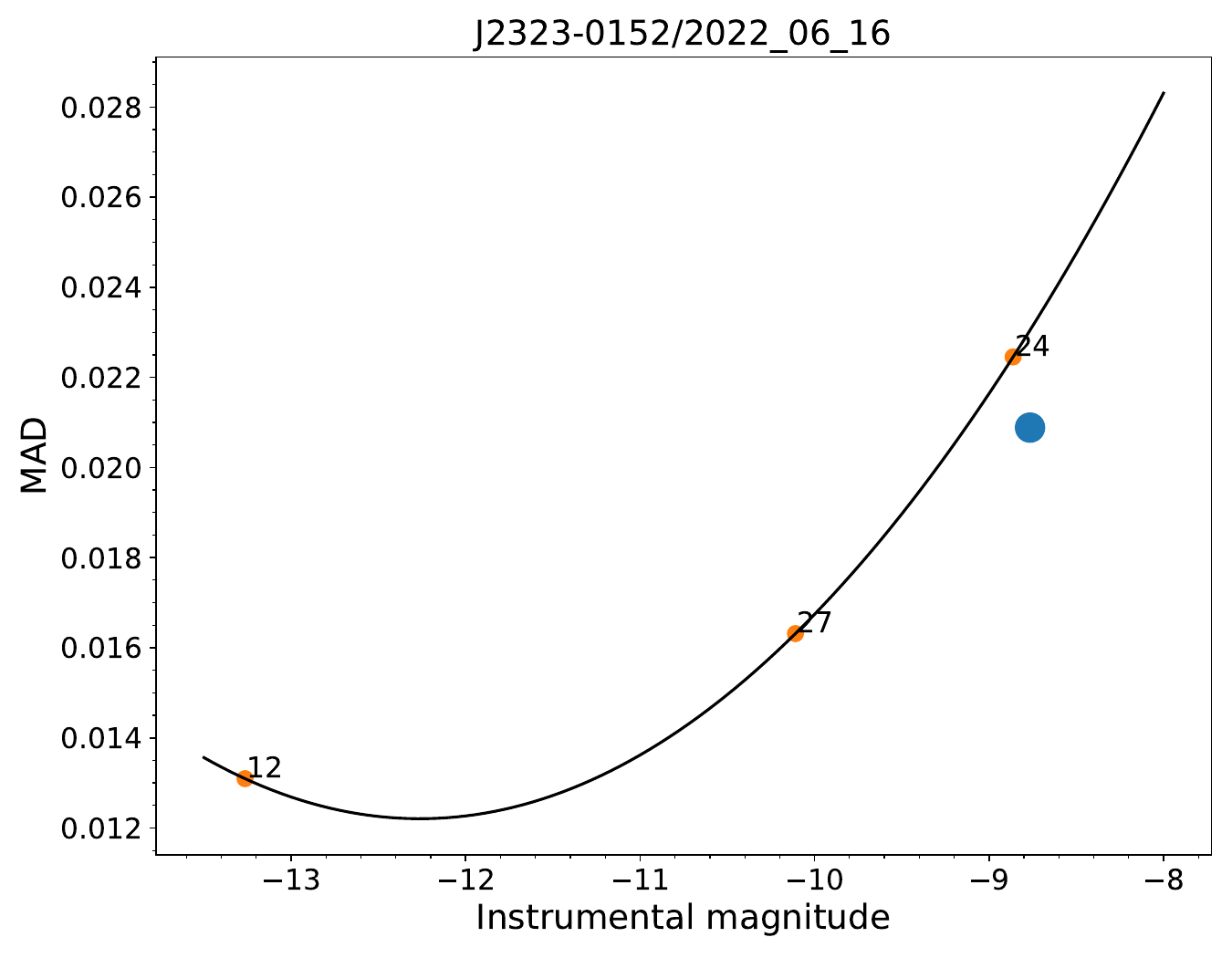}
    \includegraphics[width=0.5\columnwidth]{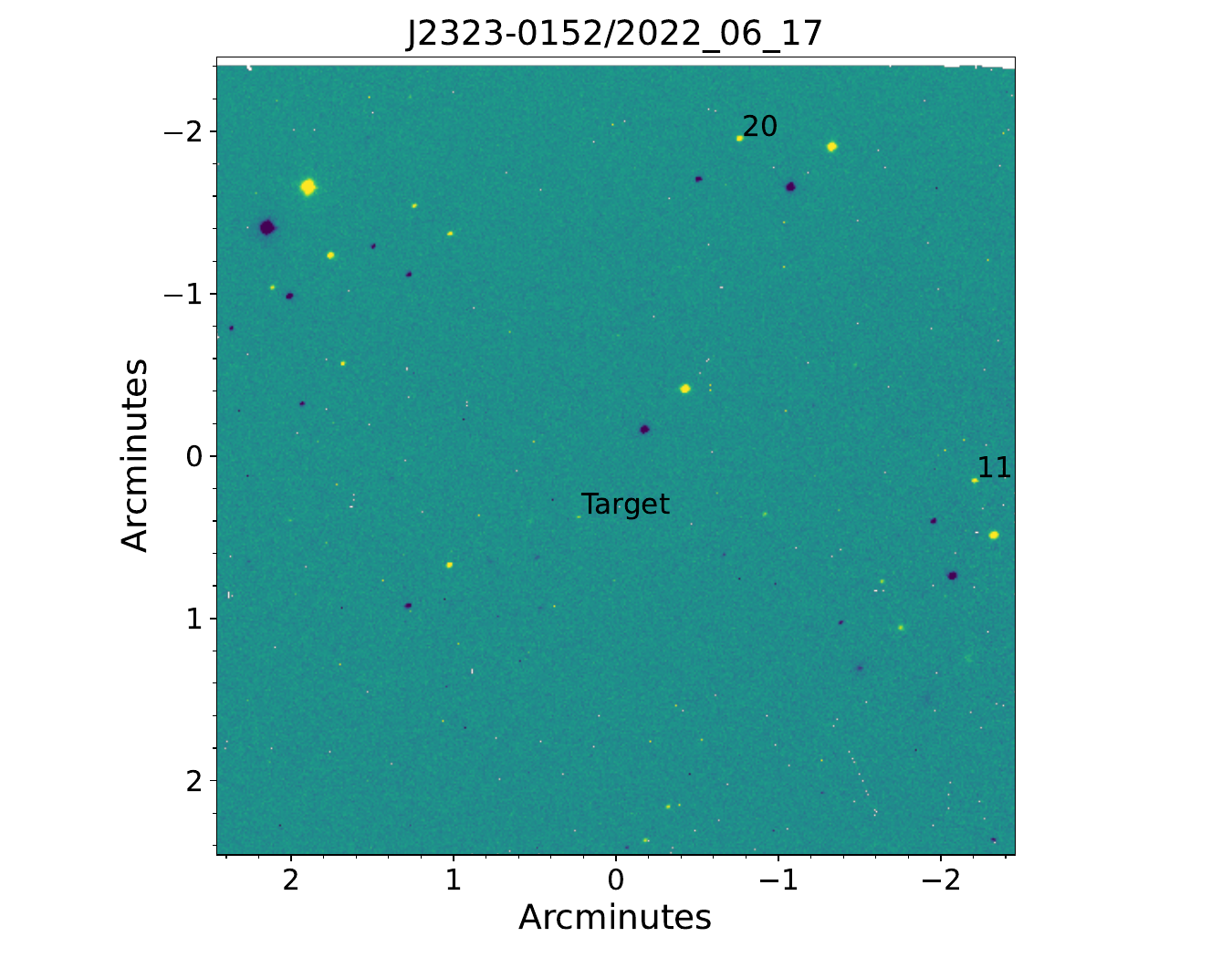}
    \includegraphics[width=0.5\columnwidth]{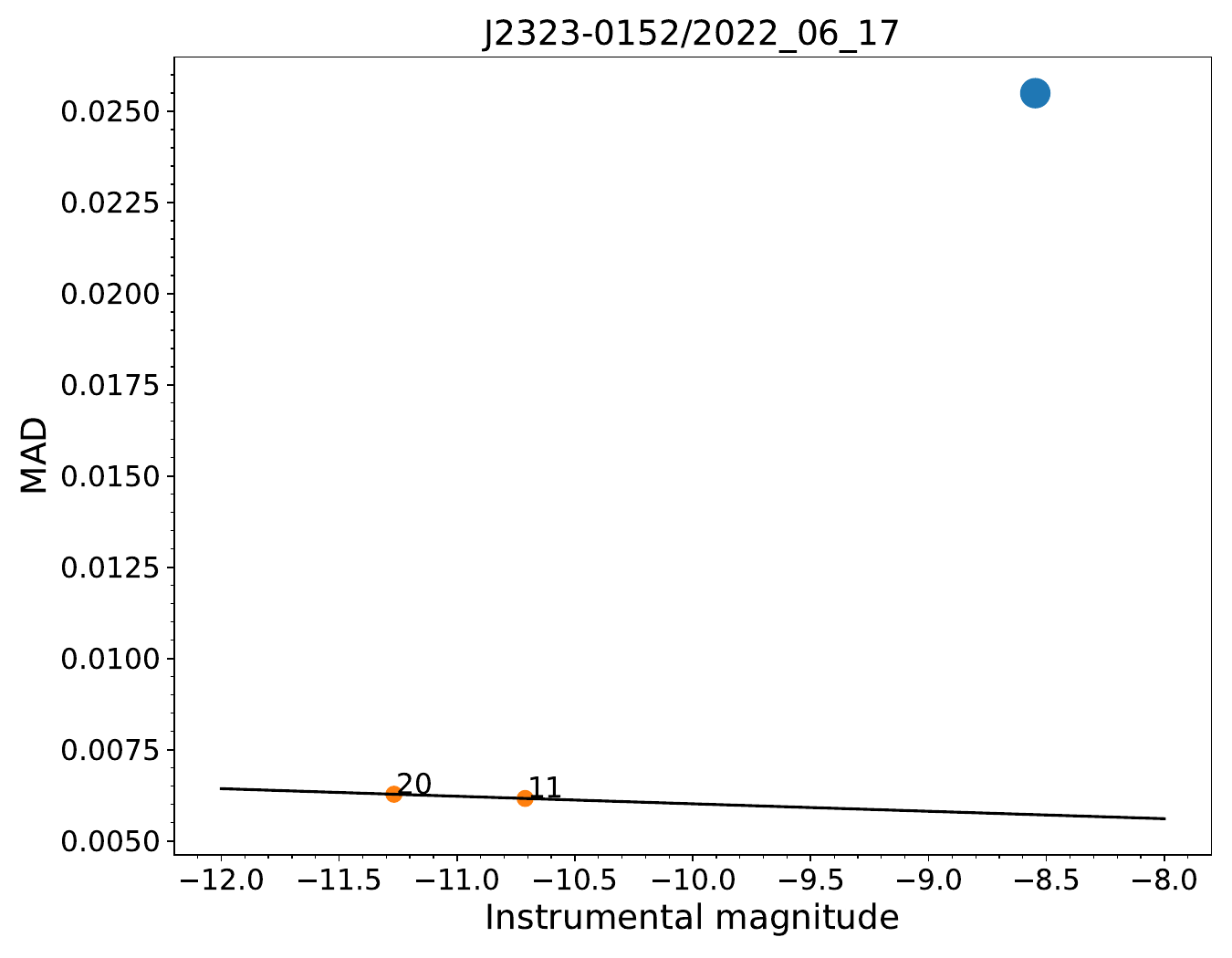}
    \includegraphics[width=0.5\columnwidth]{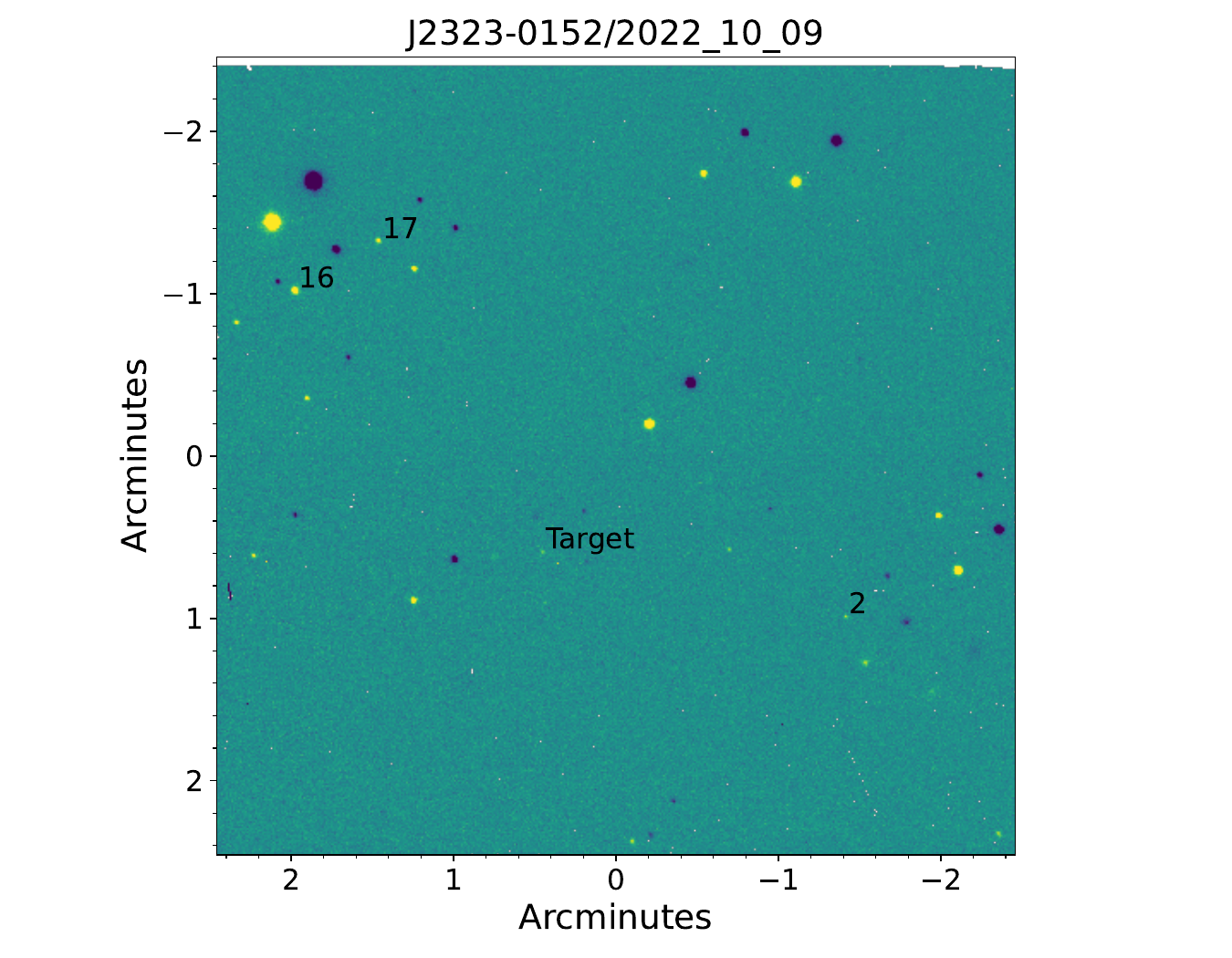}
    \includegraphics[width=0.5\columnwidth]{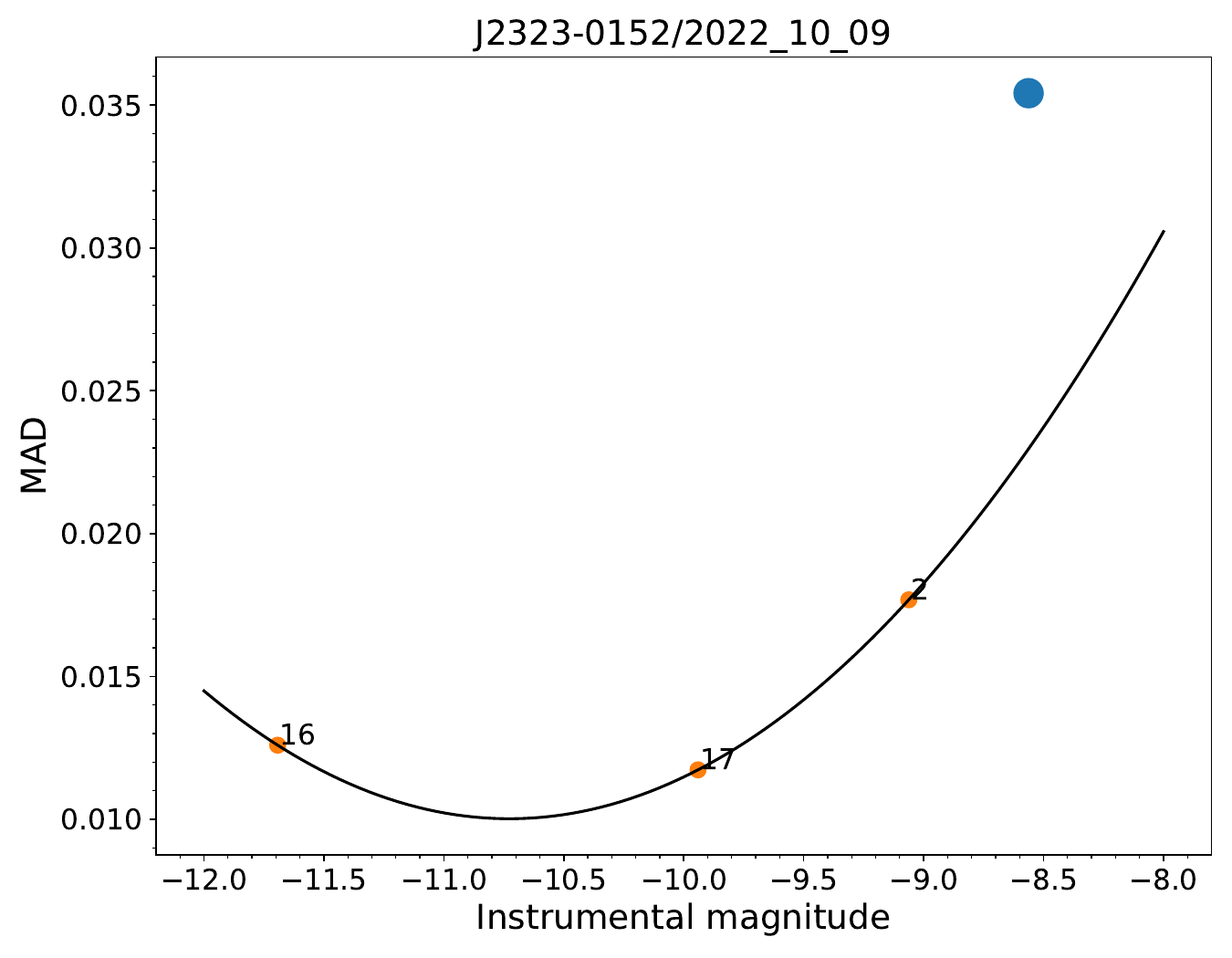}
    \contcaption{Targets and their selected reference stars.}
    \label{fig:stars_mag}
\end{figure*}

\begin{figure*}
    \includegraphics[width=0.5\columnwidth]{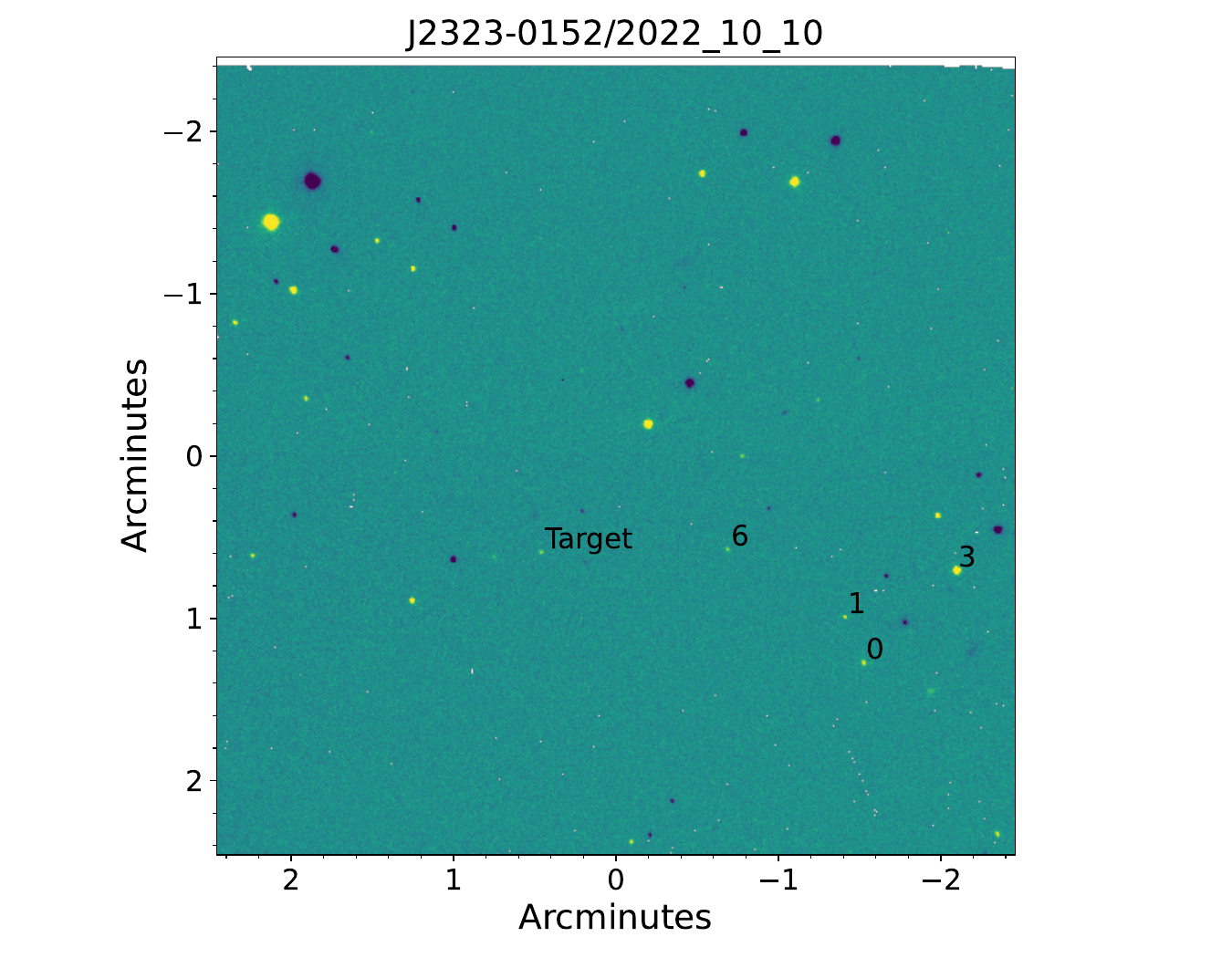}
    \includegraphics[width=0.5\columnwidth]{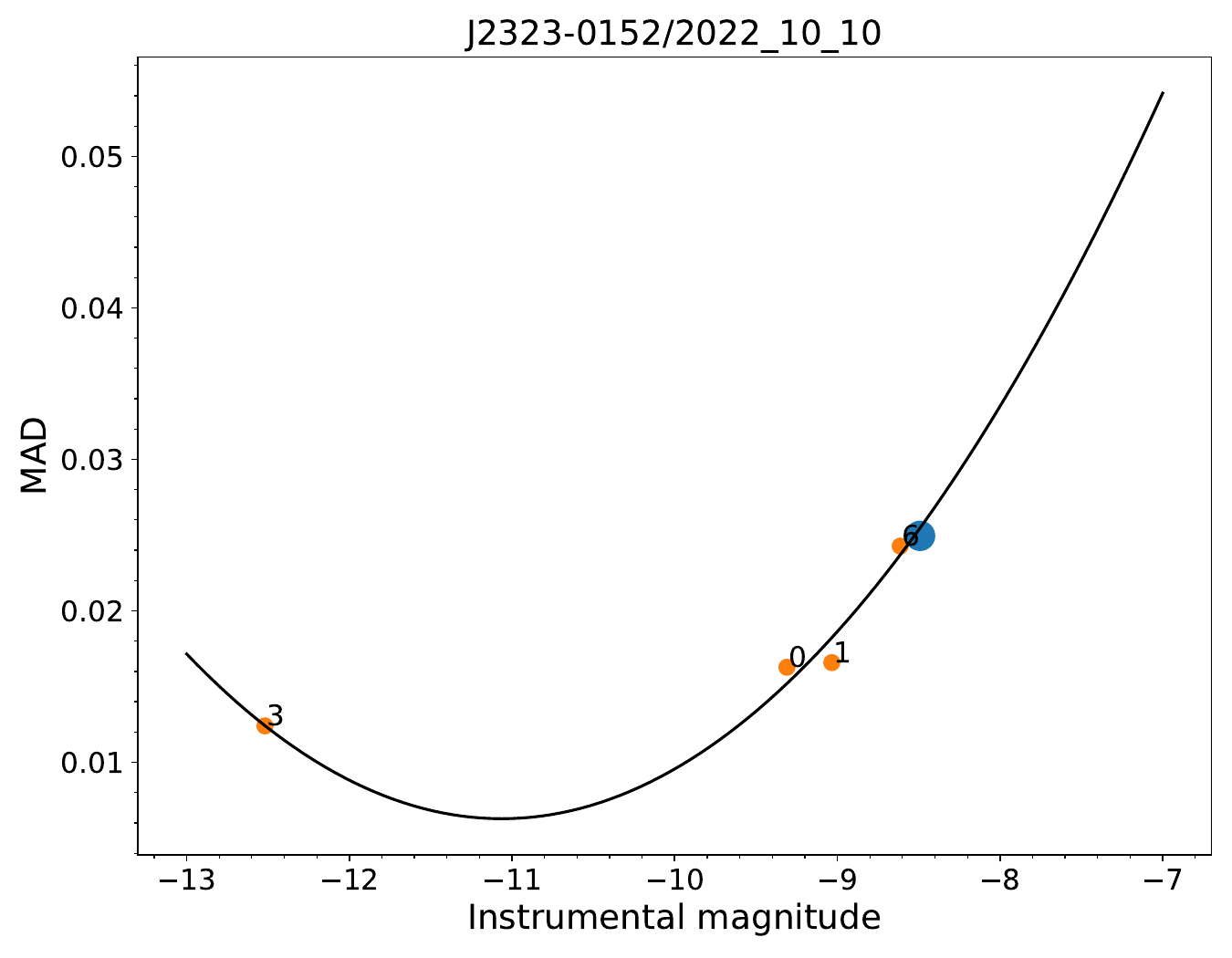}
    %2M1119-1137ABKsJs
    \includegraphics[width=0.5\columnwidth]{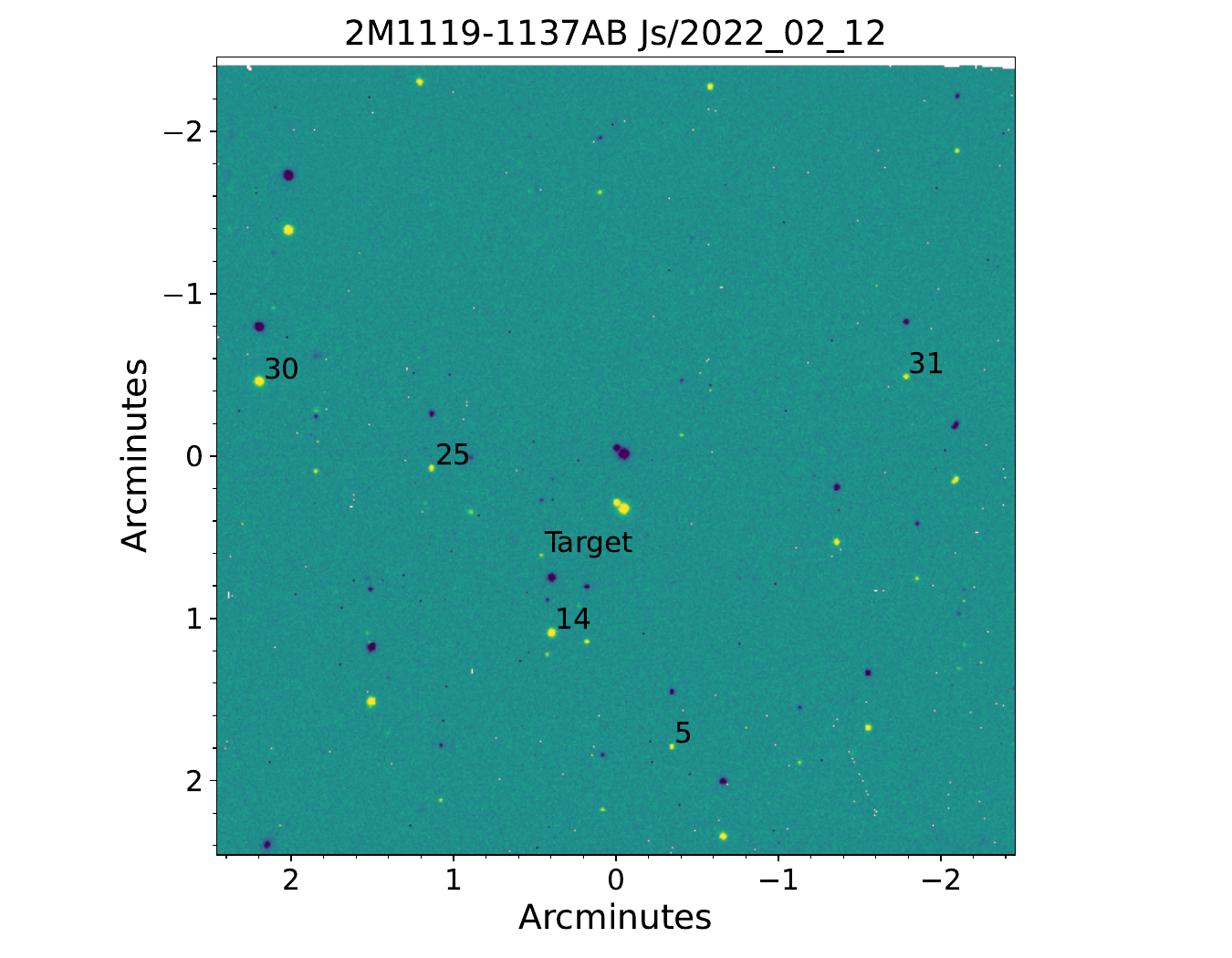}
    \includegraphics[width=0.5\columnwidth]{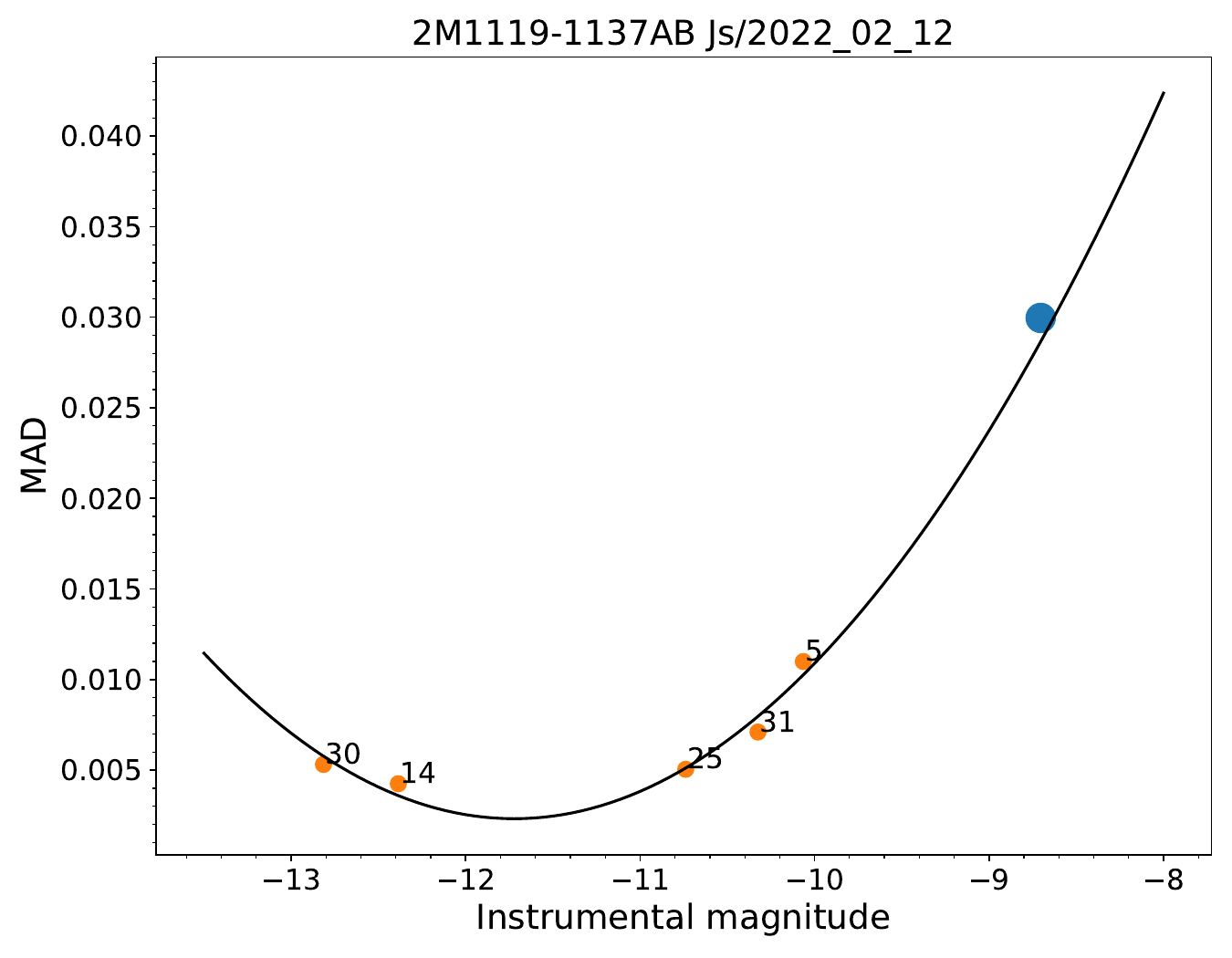}
    \includegraphics[width=0.5\columnwidth]{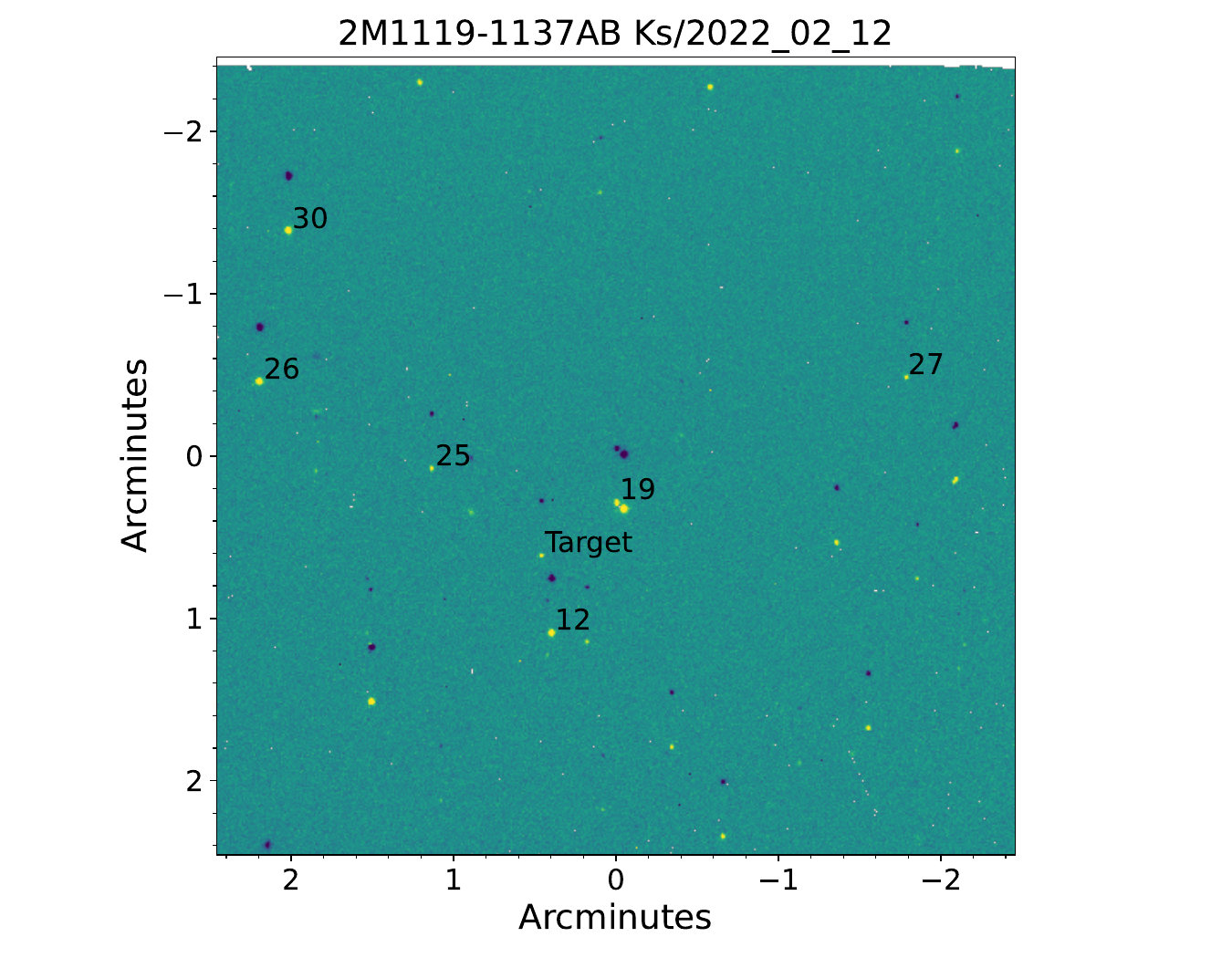}
    \includegraphics[width=0.5\columnwidth]{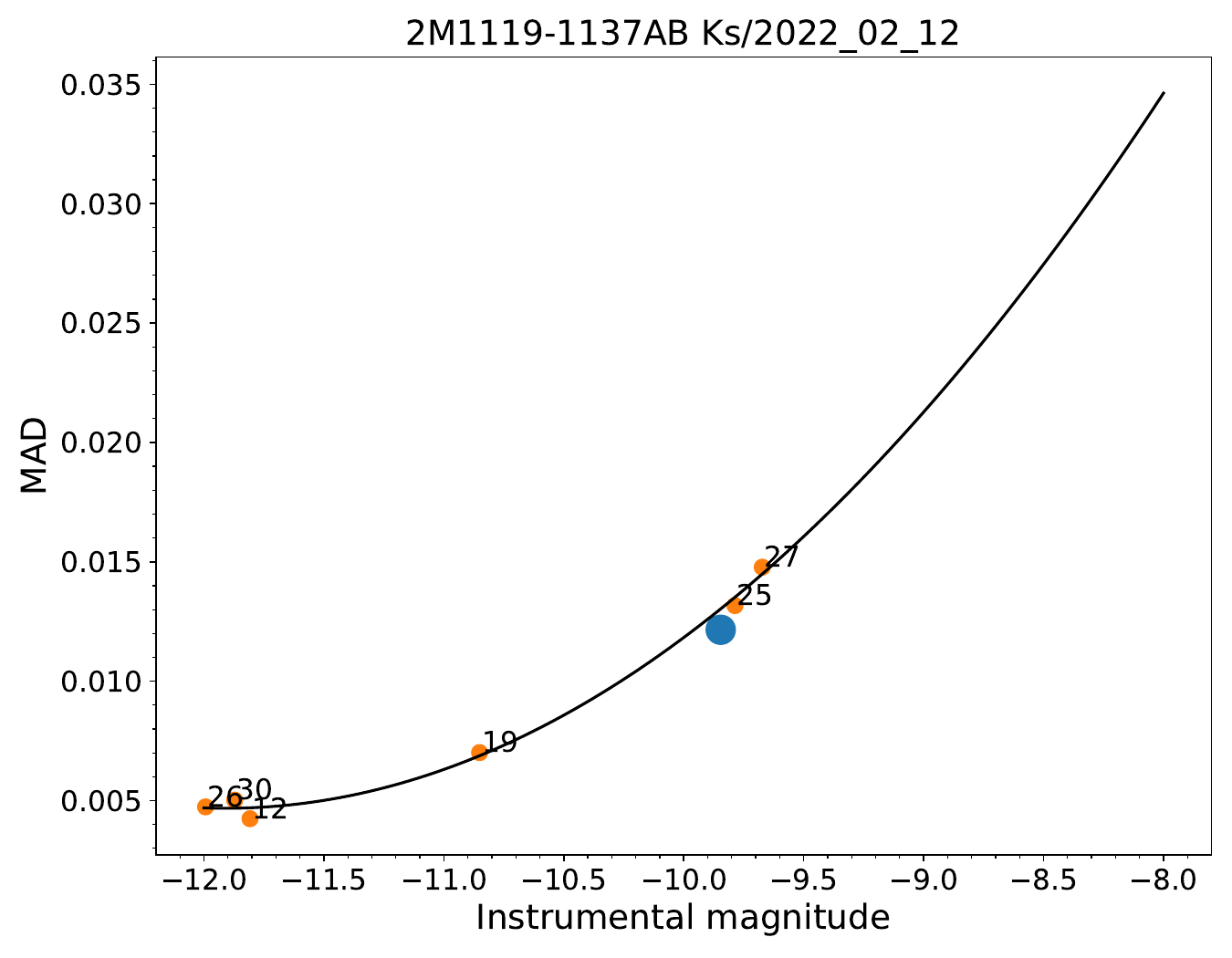}
    \includegraphics[width=0.5\columnwidth]{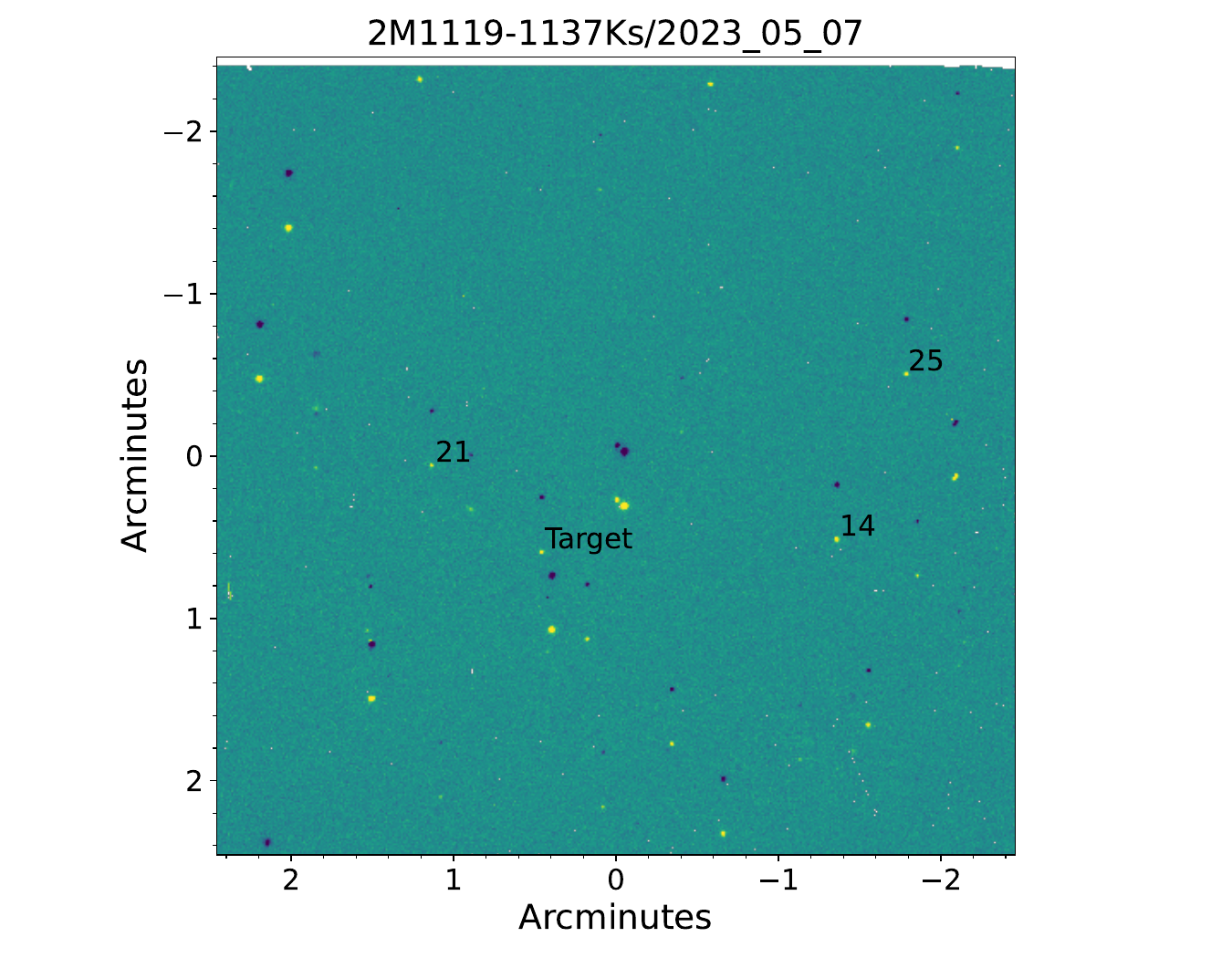}
    \includegraphics[width=0.5\columnwidth]{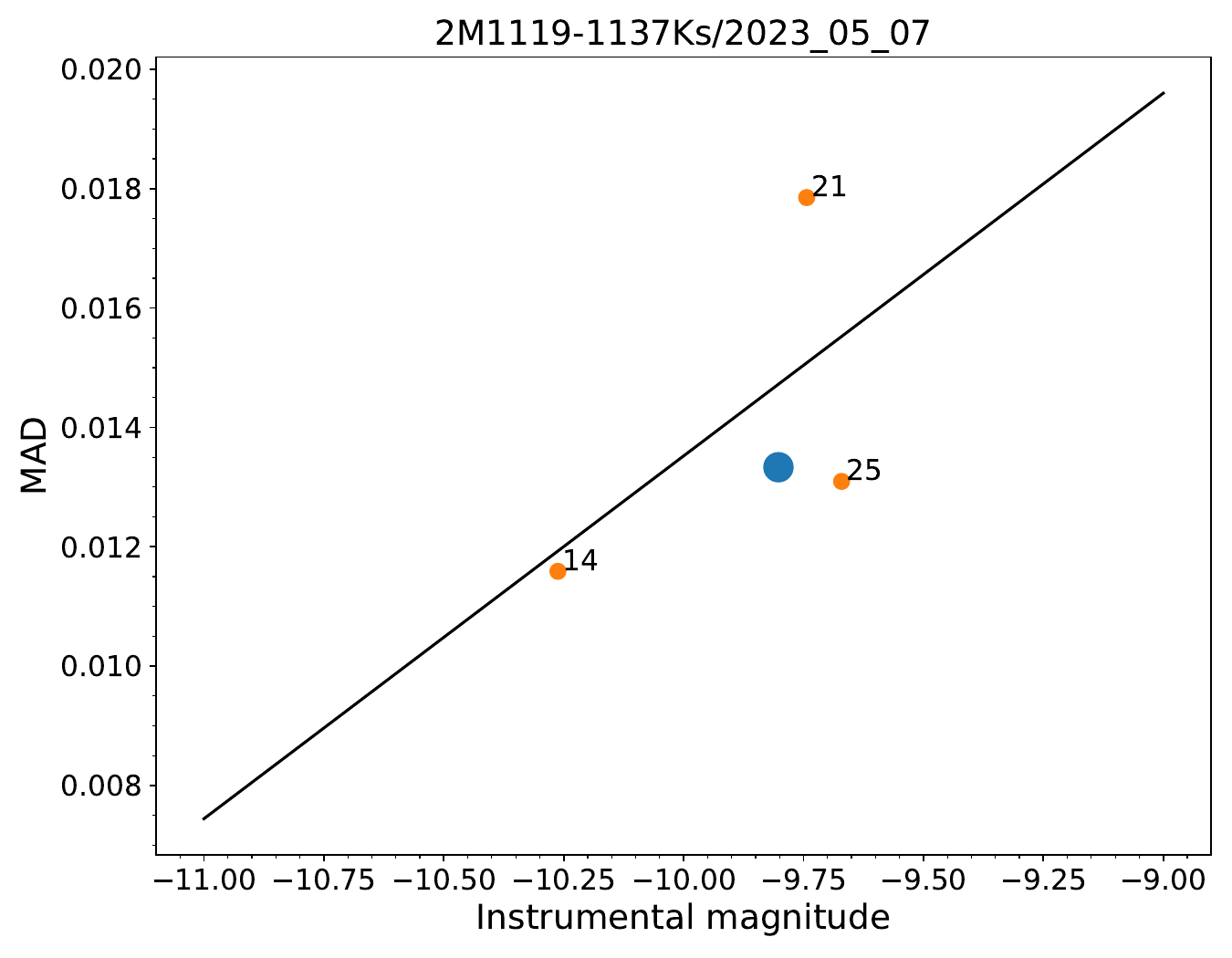}
    \includegraphics[width=0.5\columnwidth]{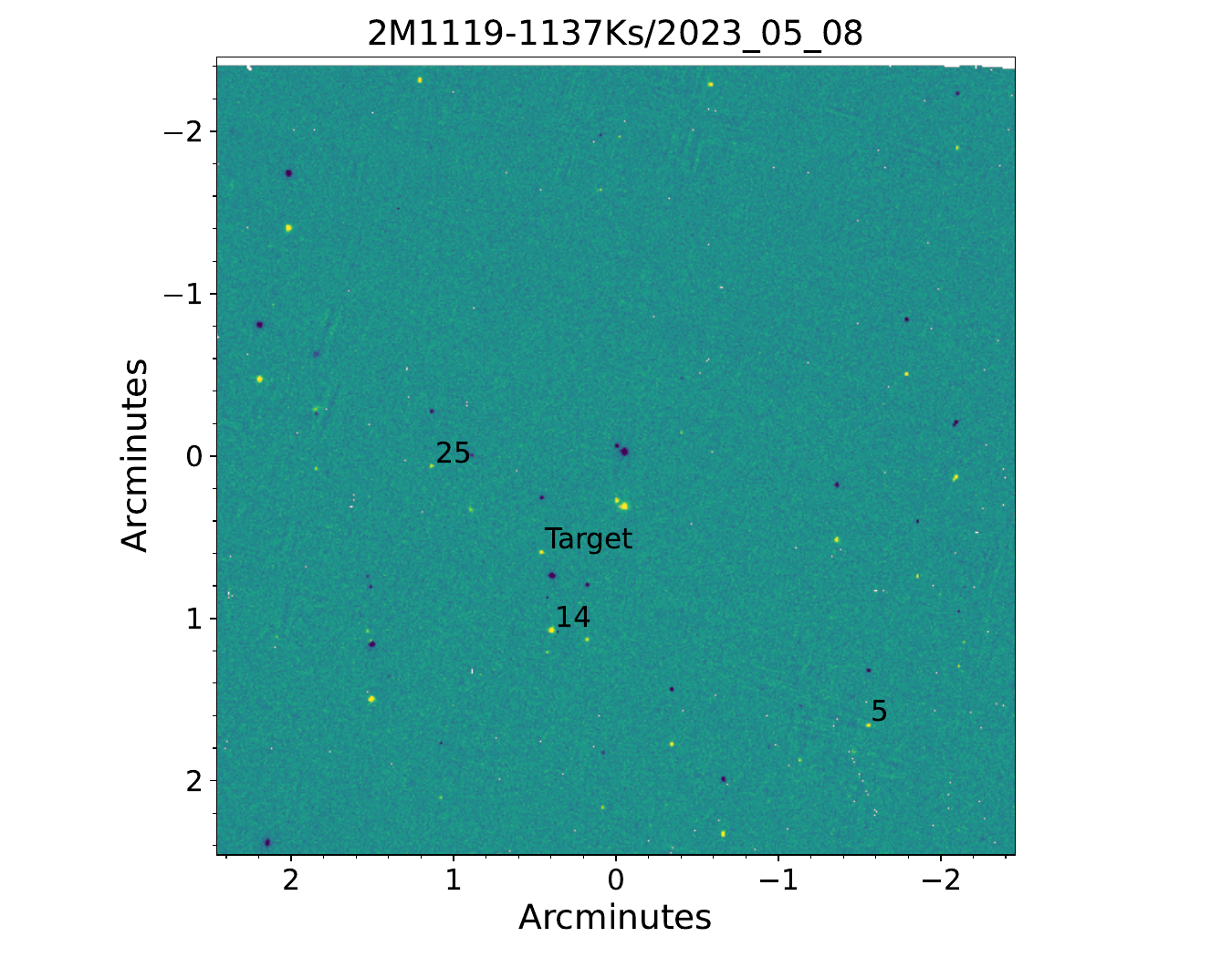}
    \includegraphics[width=0.5\columnwidth]{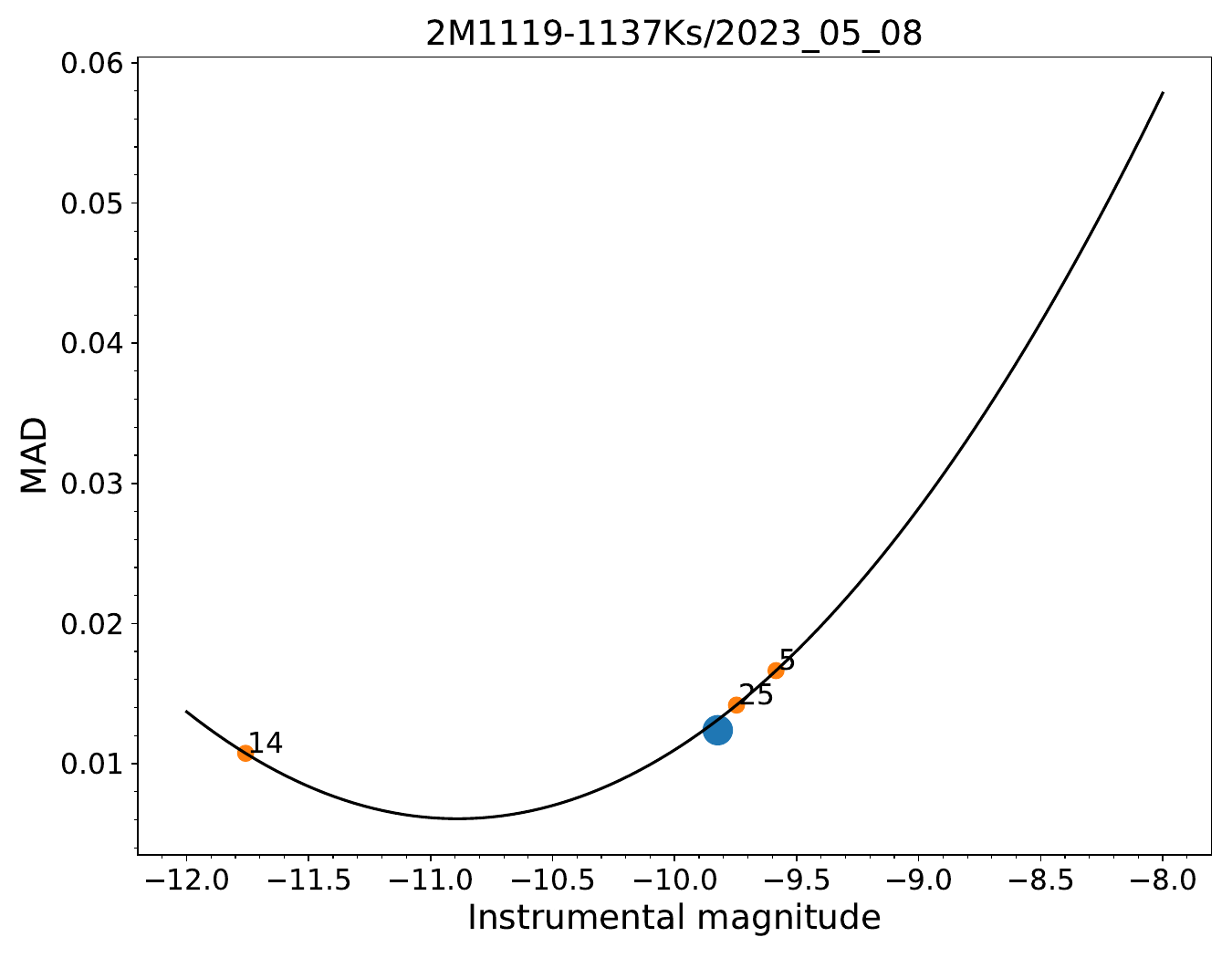}
    \includegraphics[width=0.5\columnwidth]{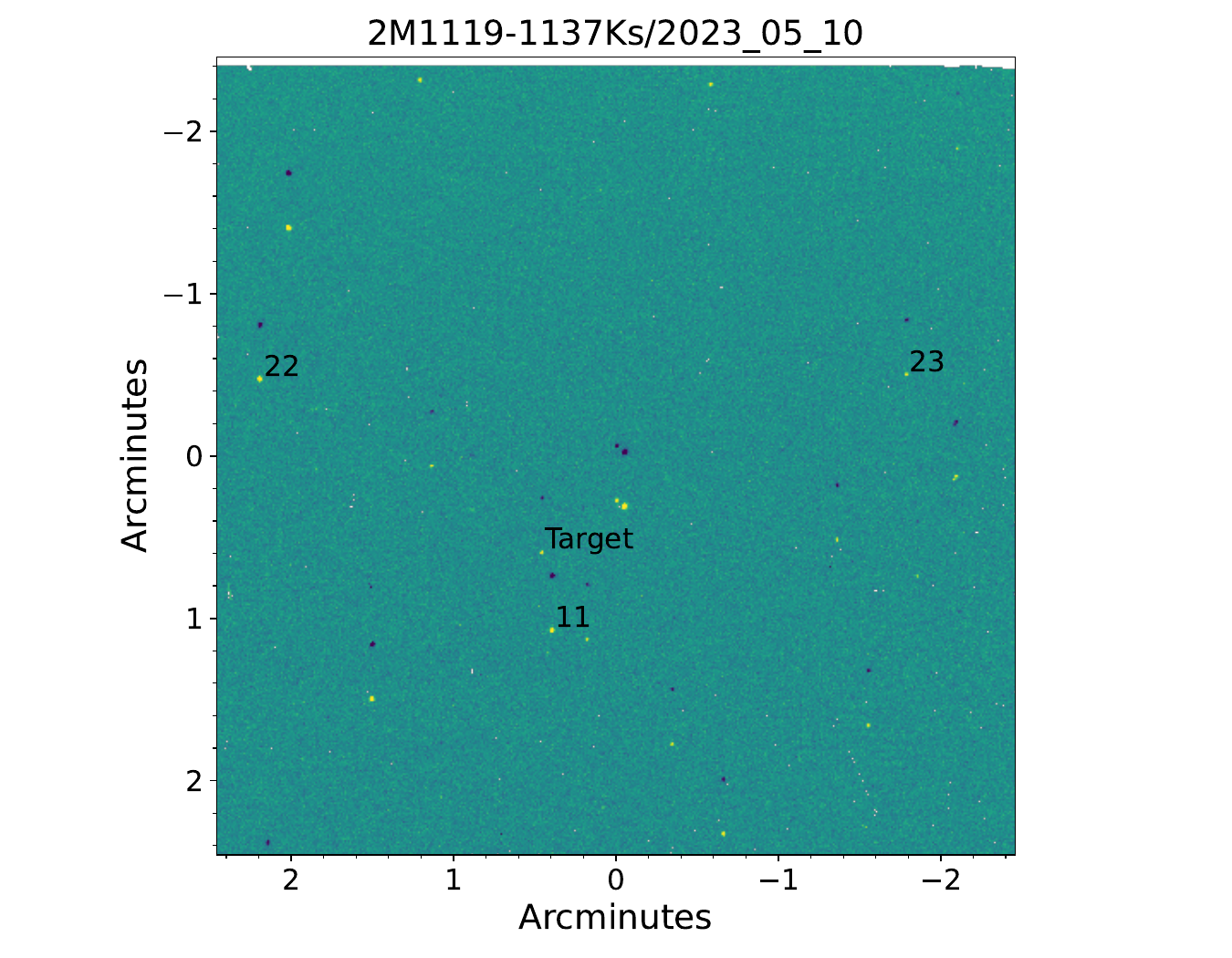}
    \includegraphics[width=0.5\columnwidth]{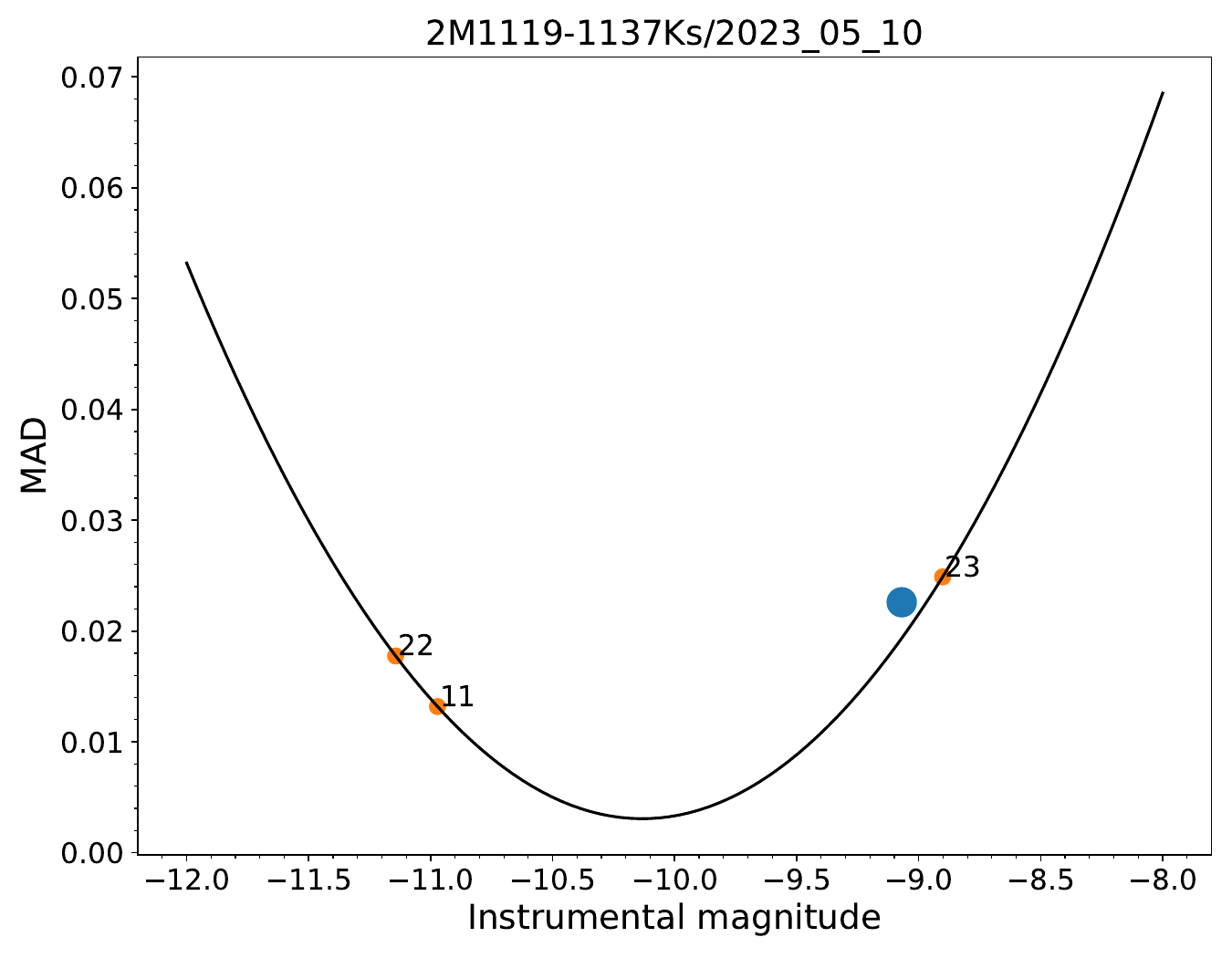}
    %W1147-2040
    \includegraphics[width=0.5\columnwidth]{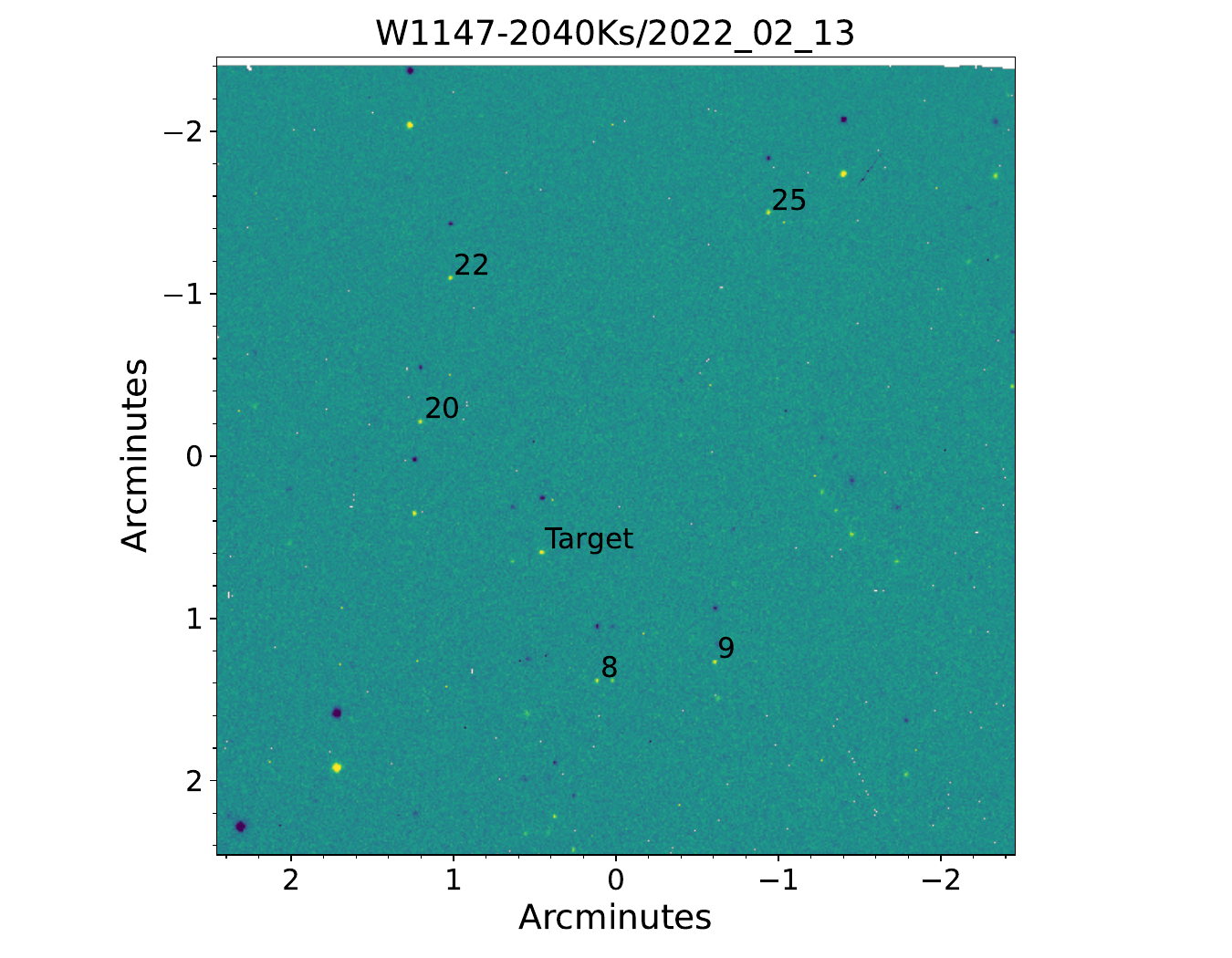}
    \includegraphics[width=0.5\columnwidth]{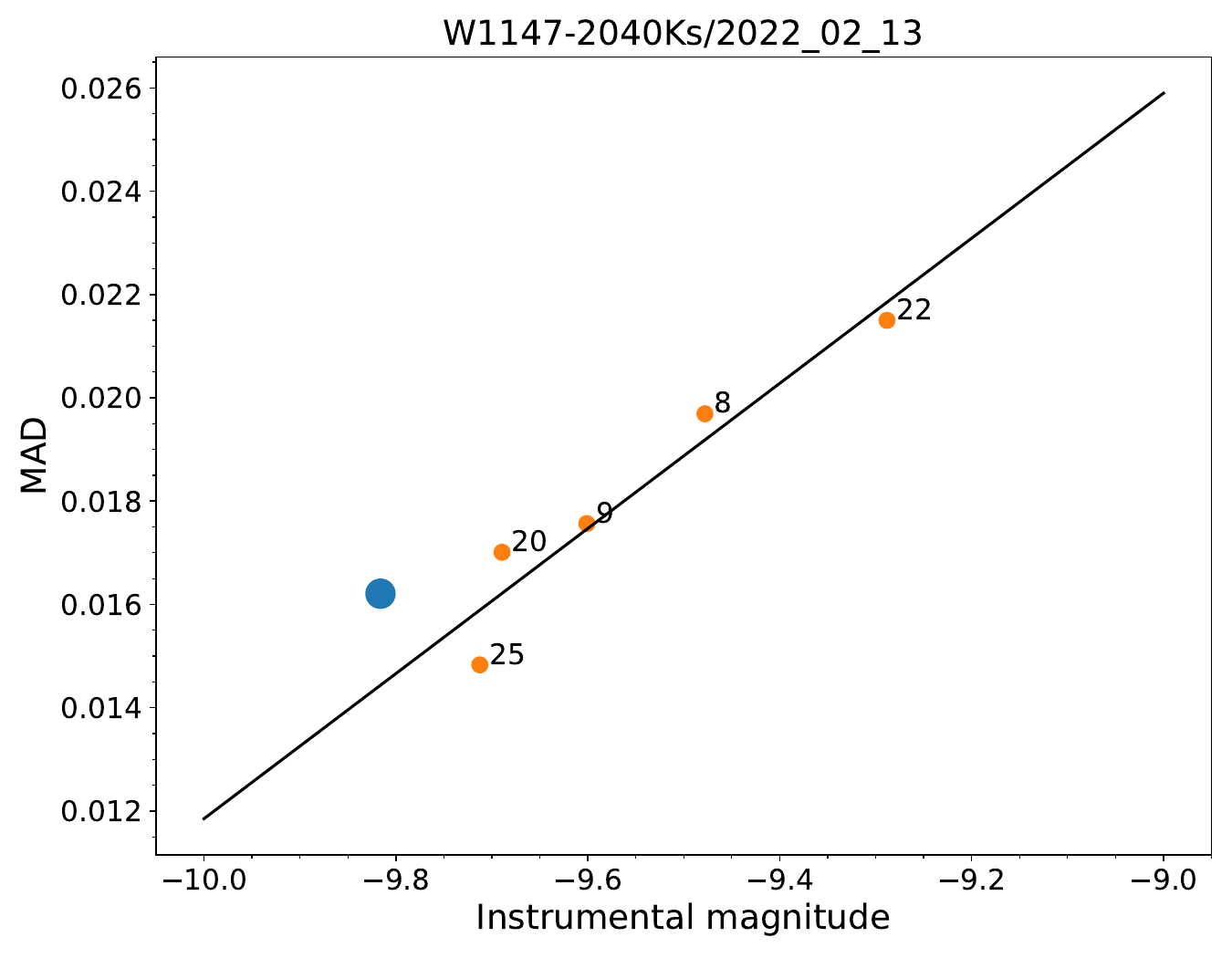}
    \includegraphics[width=0.5\columnwidth]{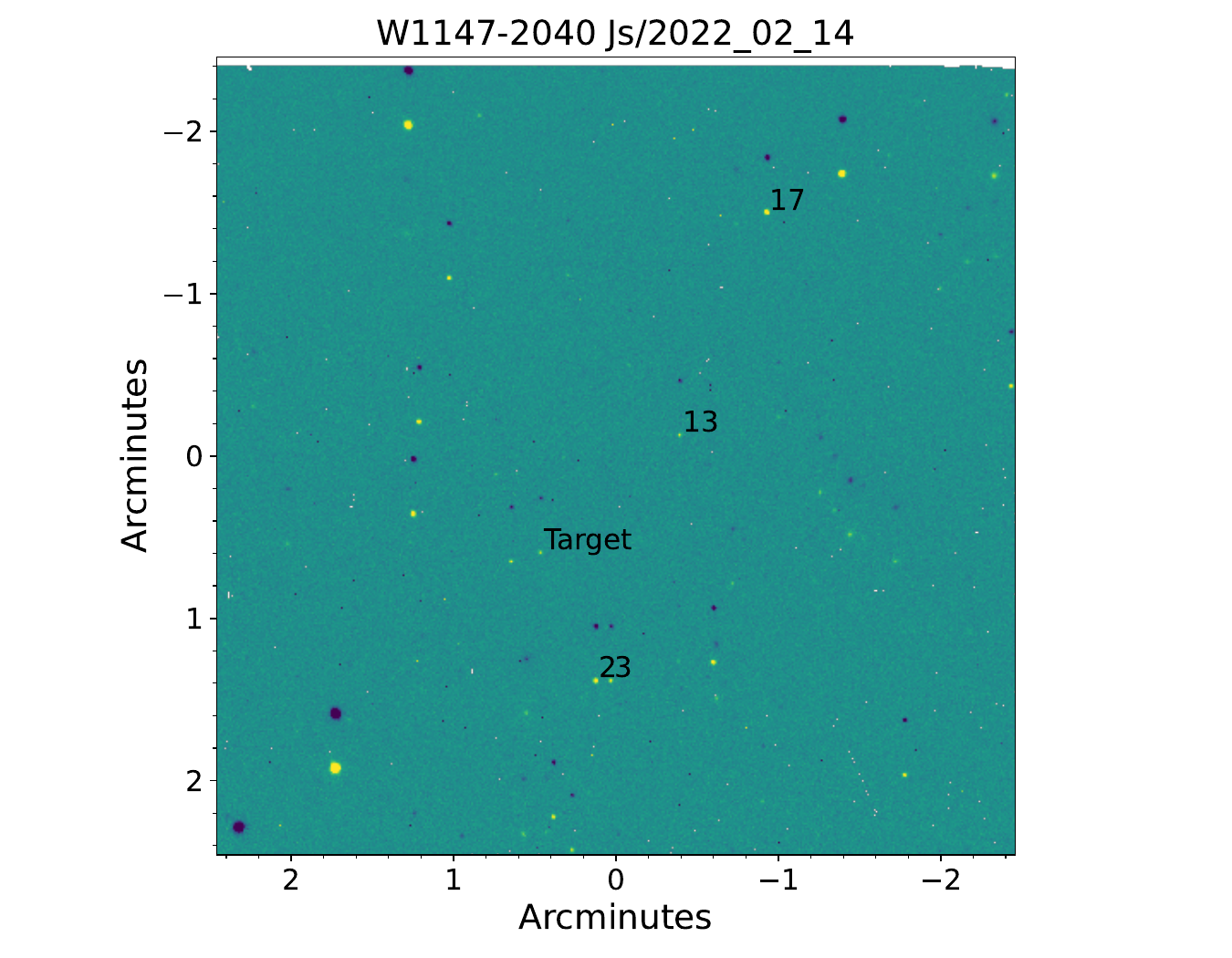}
    \includegraphics[width=0.5\columnwidth]{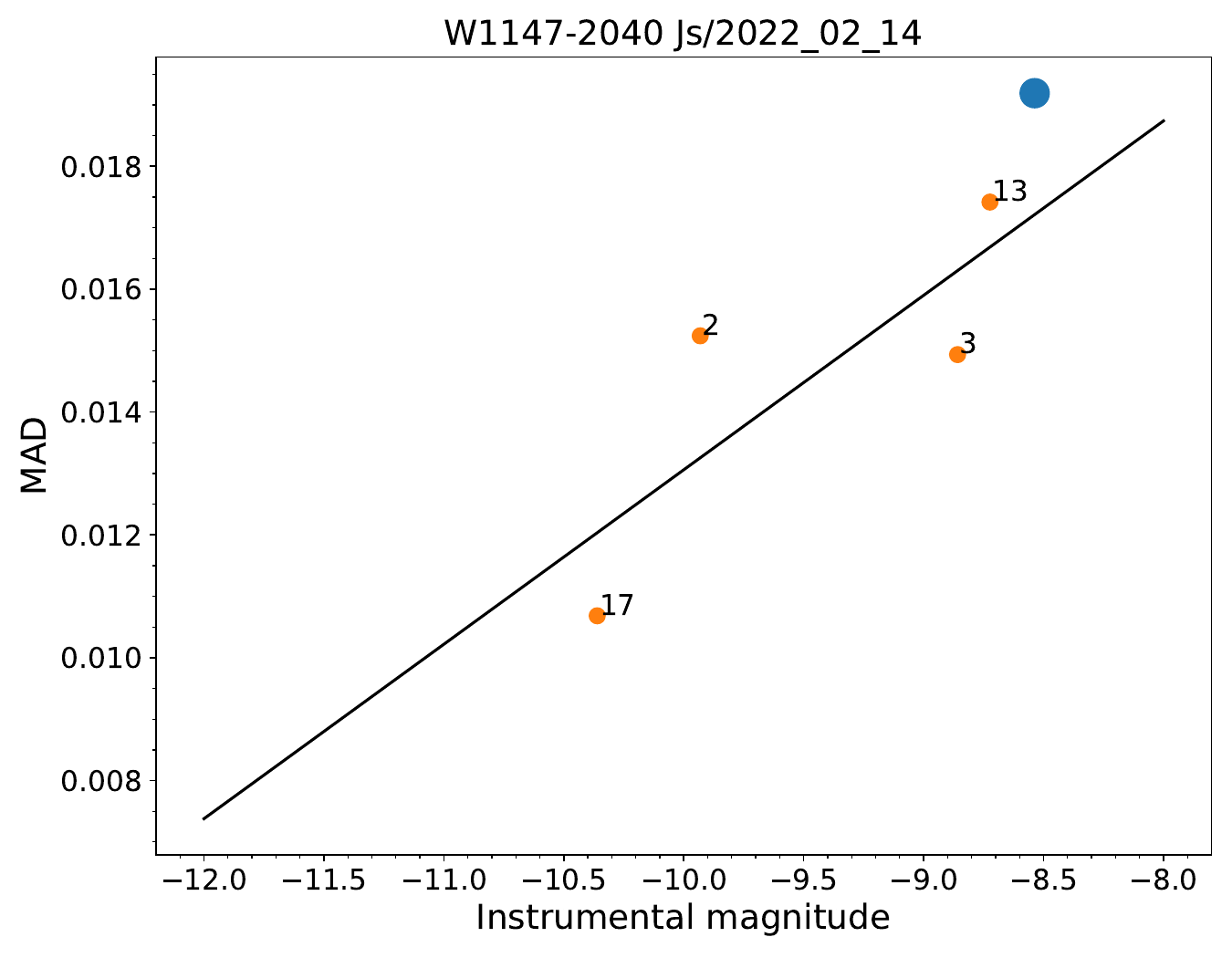}
    \includegraphics[width=0.5\columnwidth]{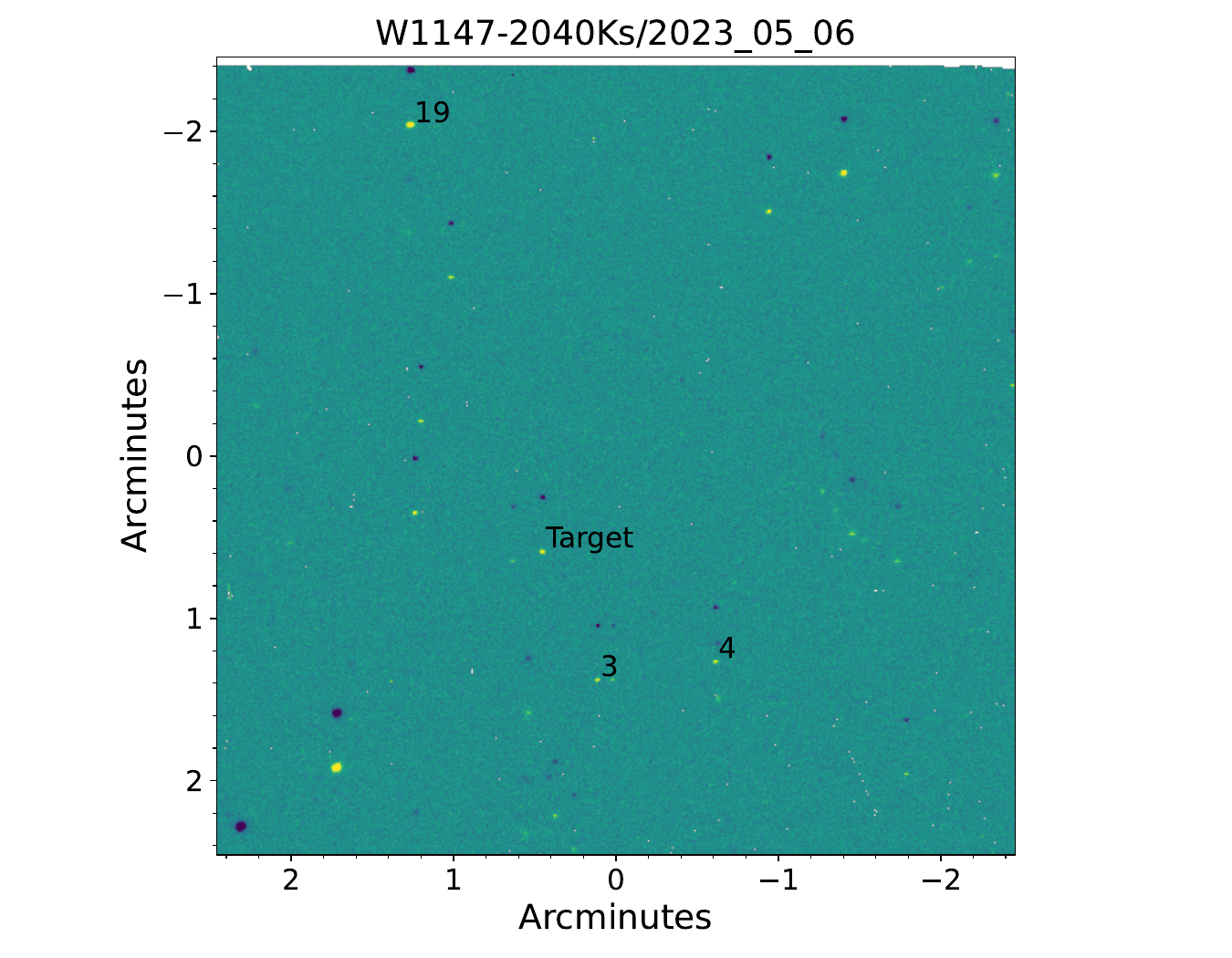}
    \includegraphics[width=0.5\columnwidth]{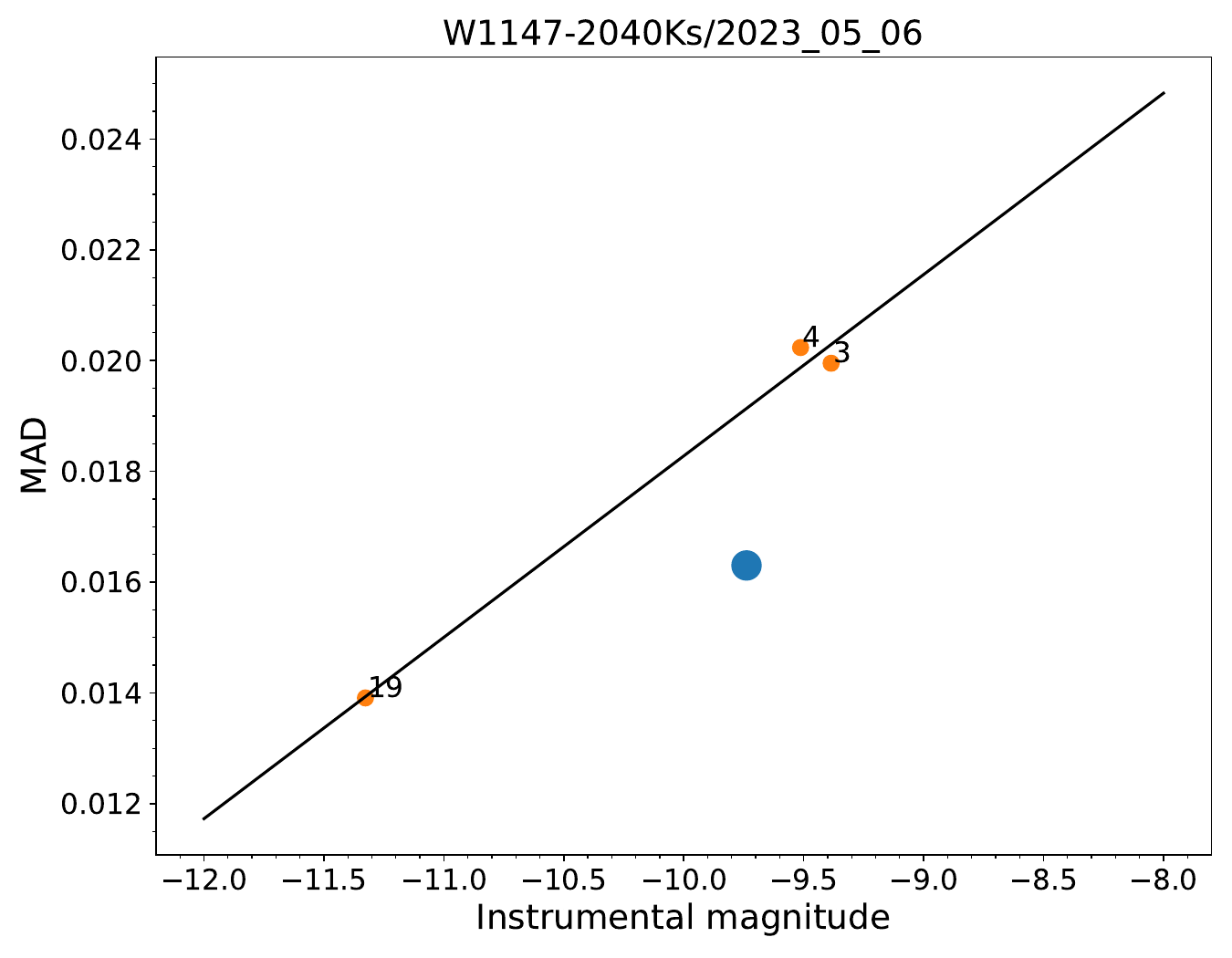}
    \includegraphics[width=0.5\columnwidth]{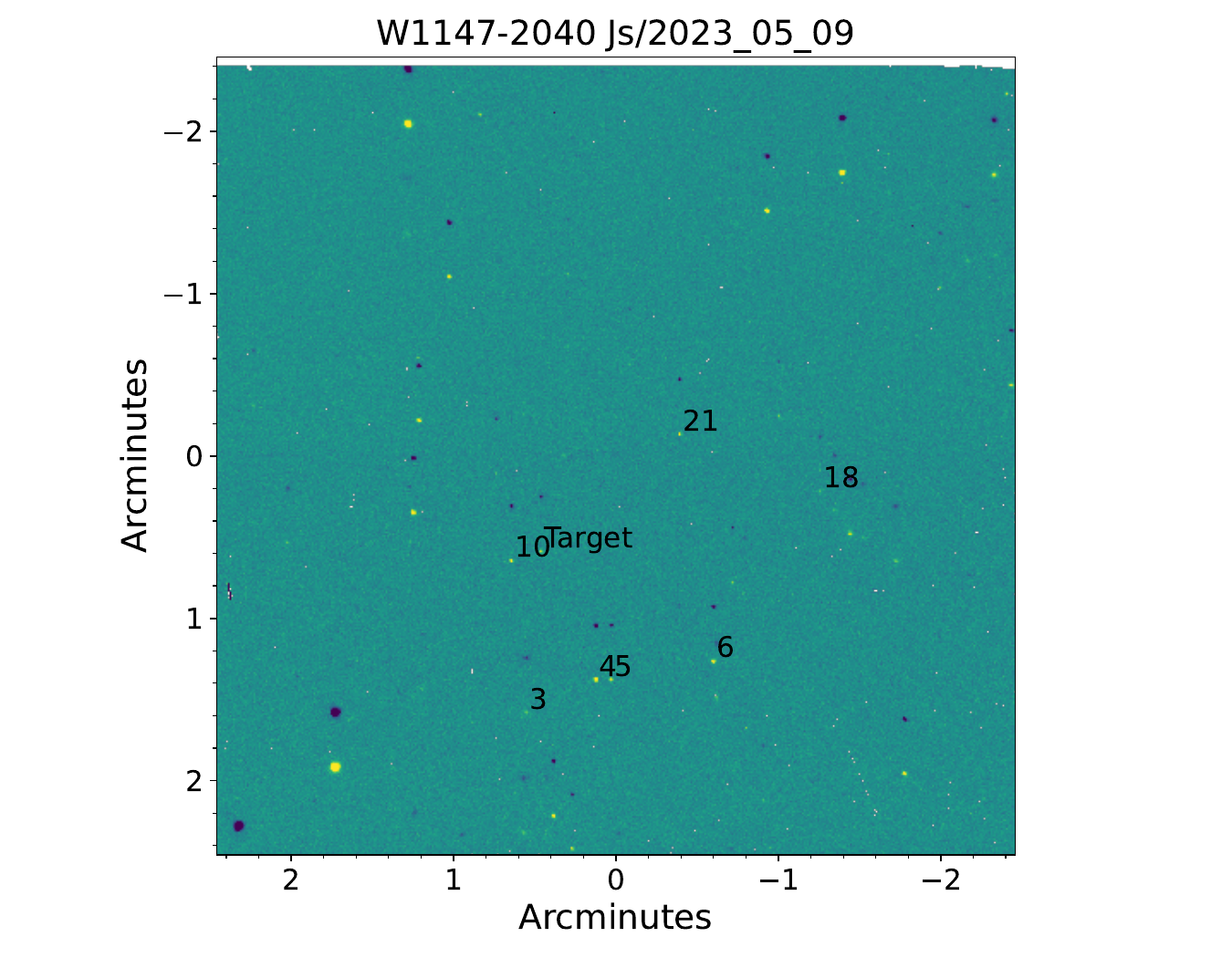}
    \includegraphics[width=0.5\columnwidth]{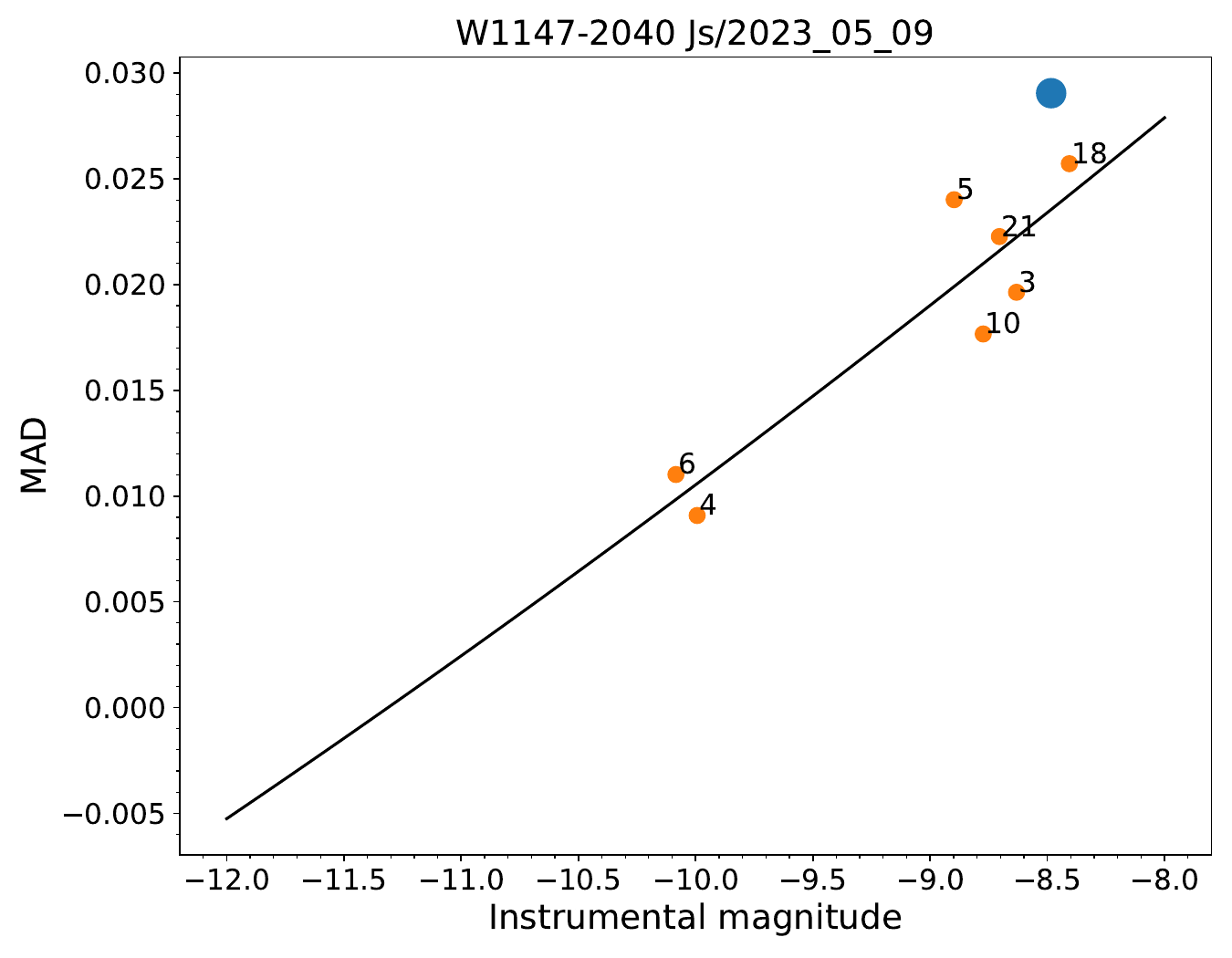}
    \includegraphics[width=0.5\columnwidth]{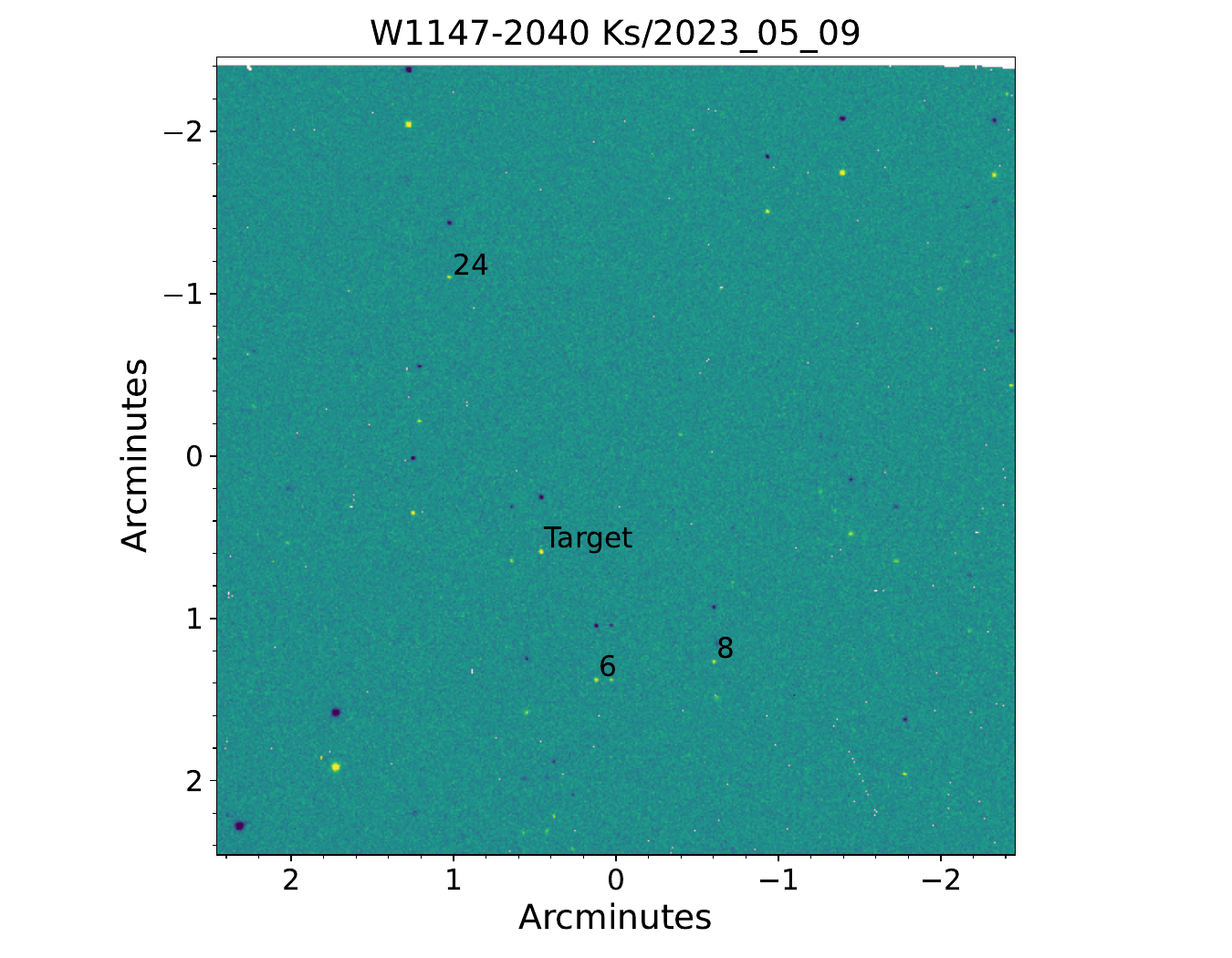}
    \includegraphics[width=0.5\columnwidth]{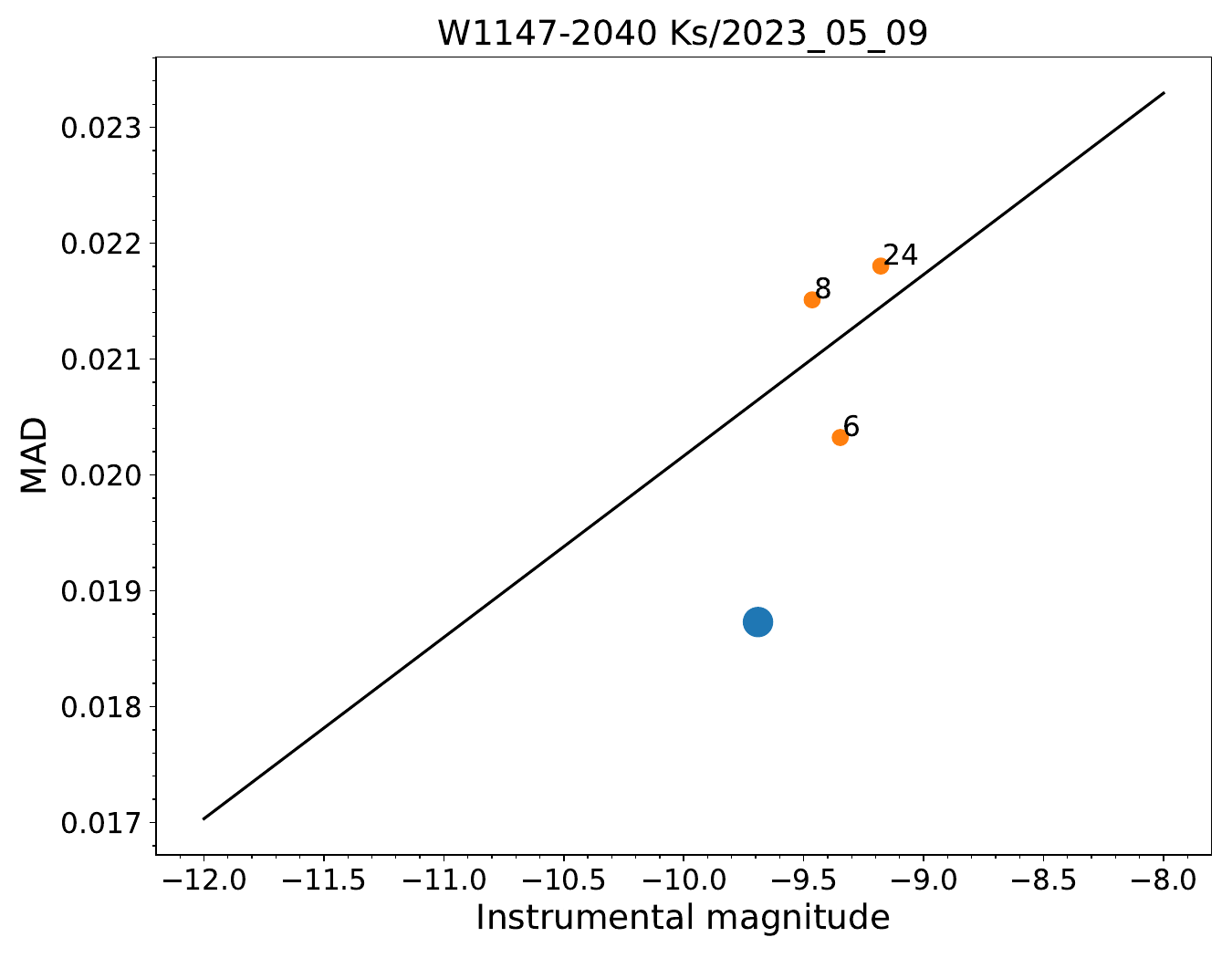}
    %PSO168
    \includegraphics[width=0.5\columnwidth]{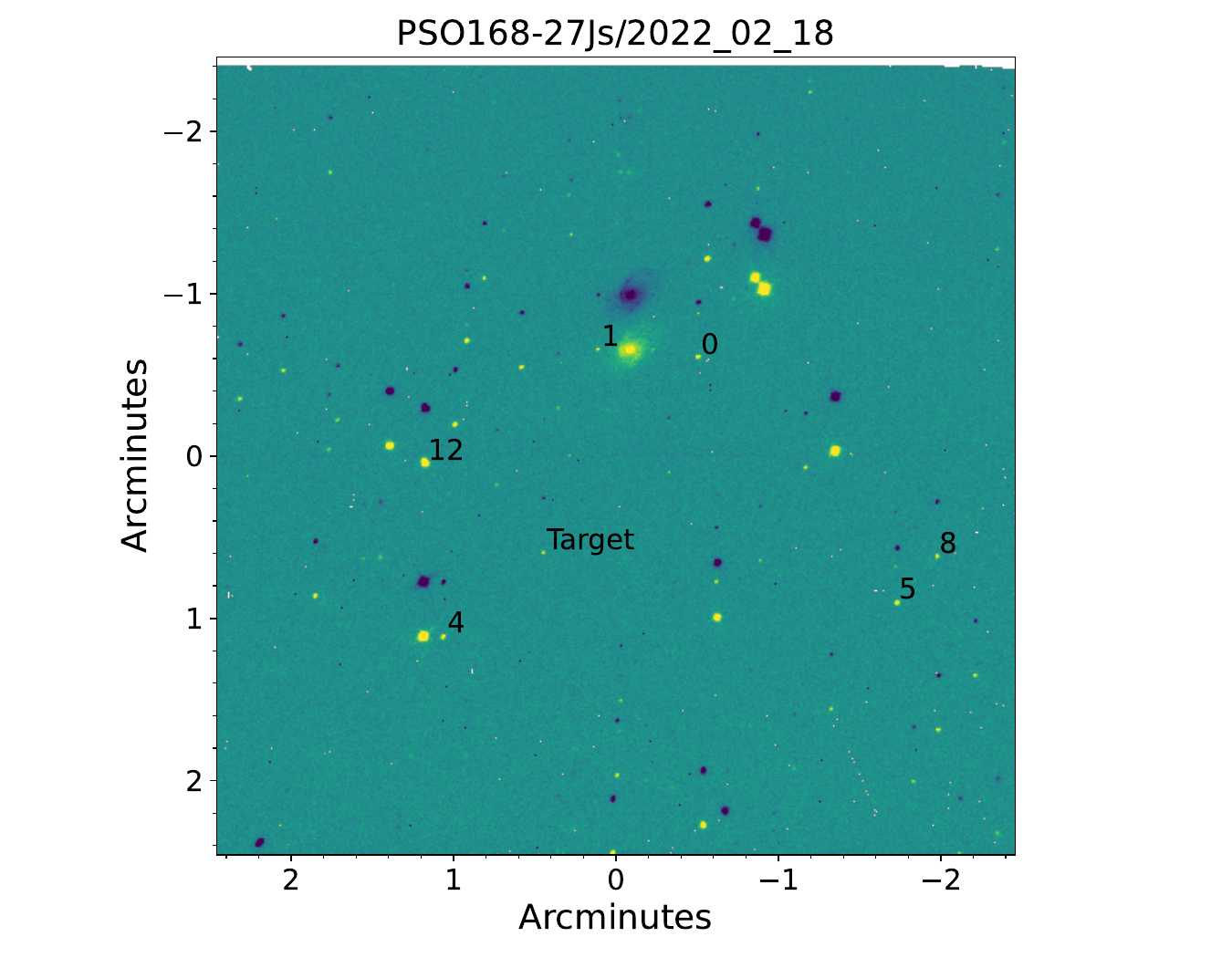}
    \includegraphics[width=0.5\columnwidth]{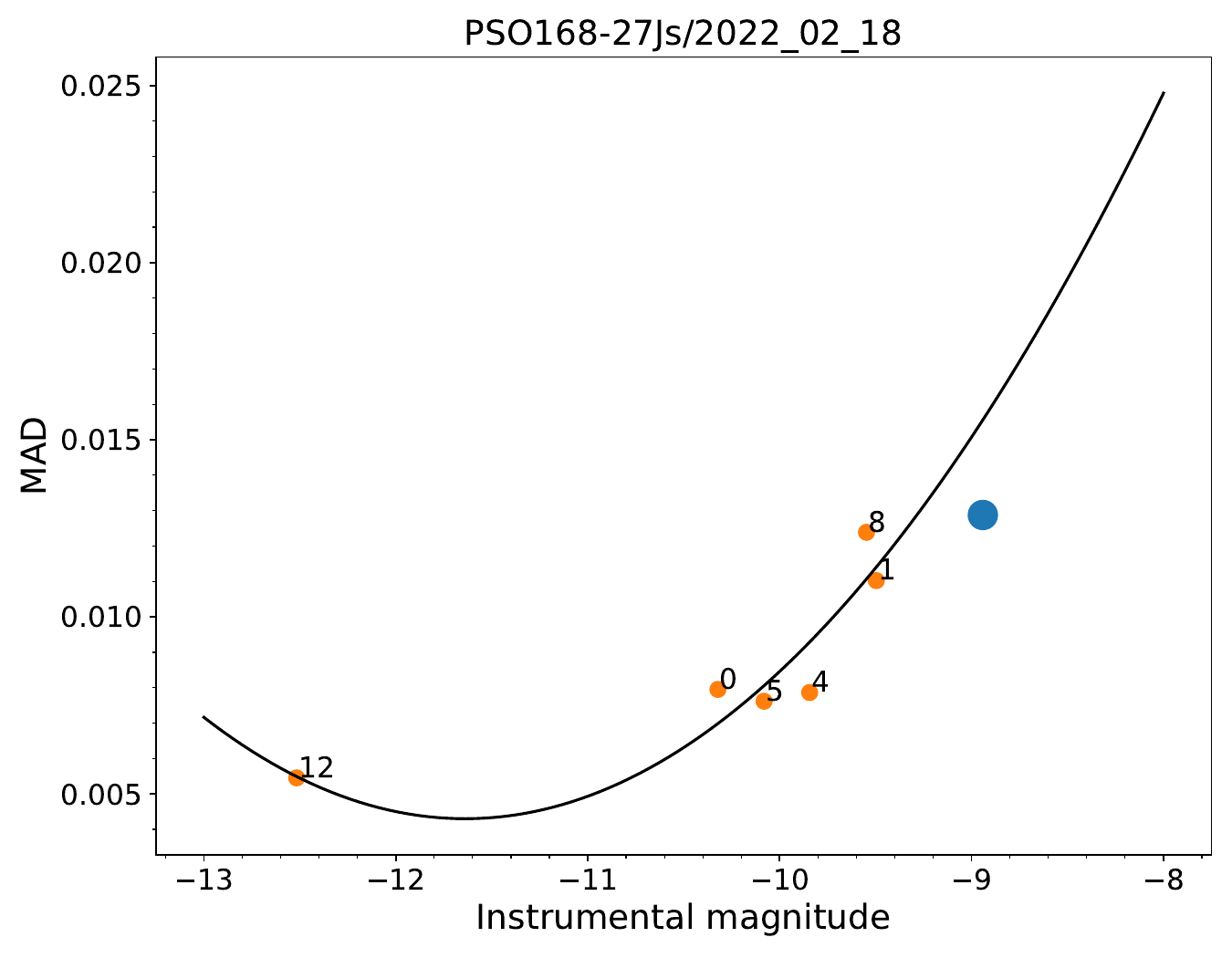}
    \includegraphics[width=0.5\columnwidth]{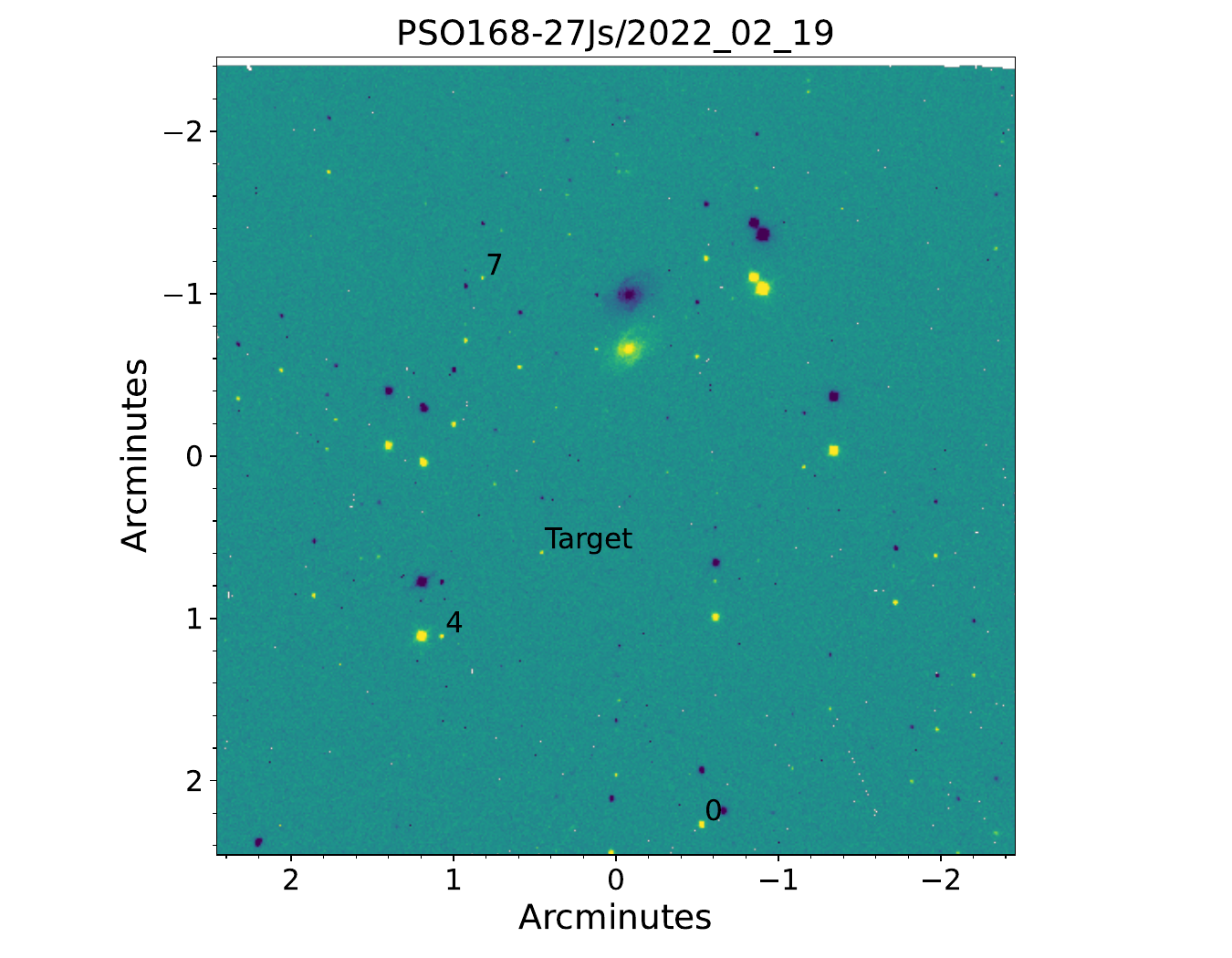}
    \includegraphics[width=0.5\columnwidth]{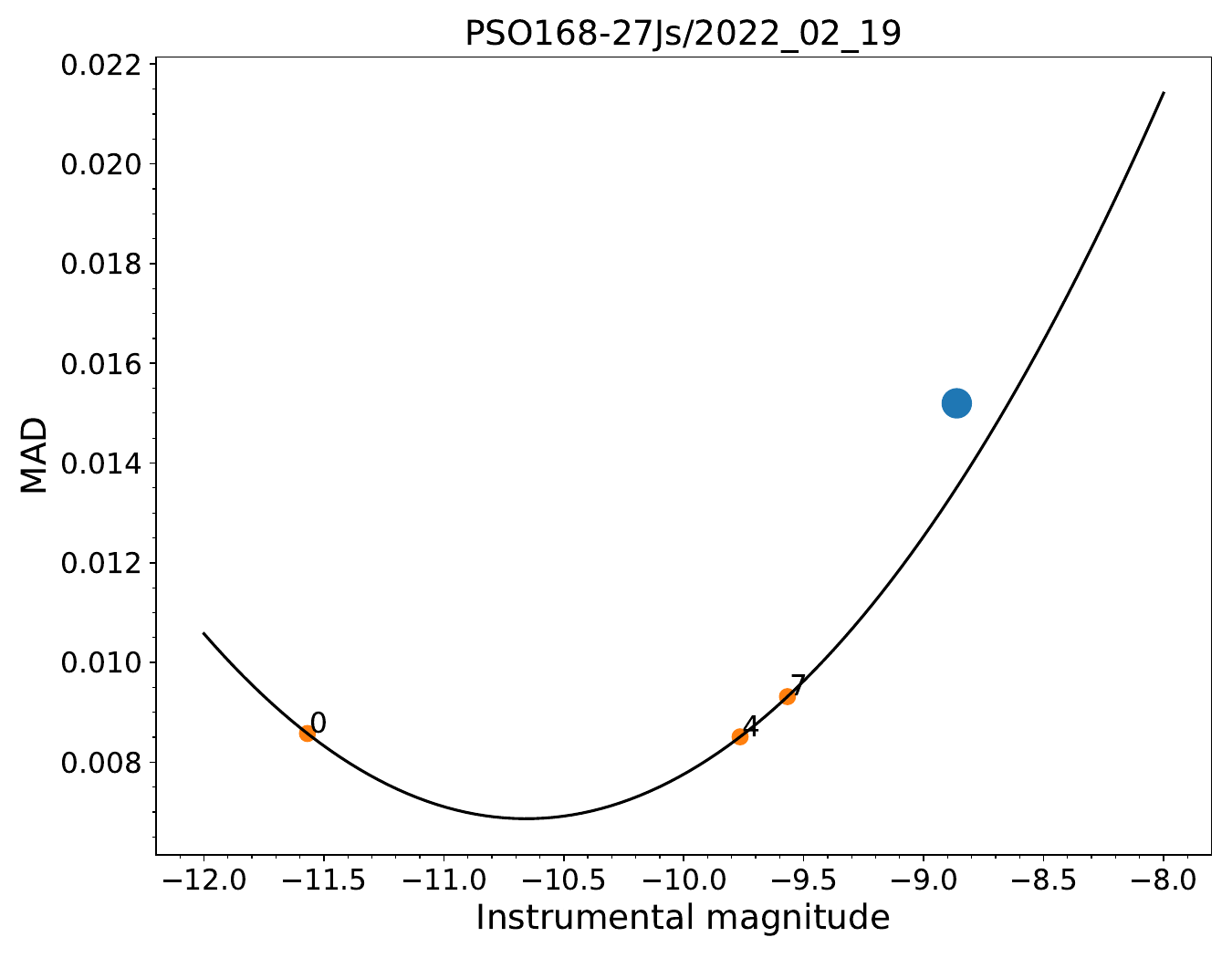}
    %2M1555+1532
    \includegraphics[width=0.5\columnwidth]{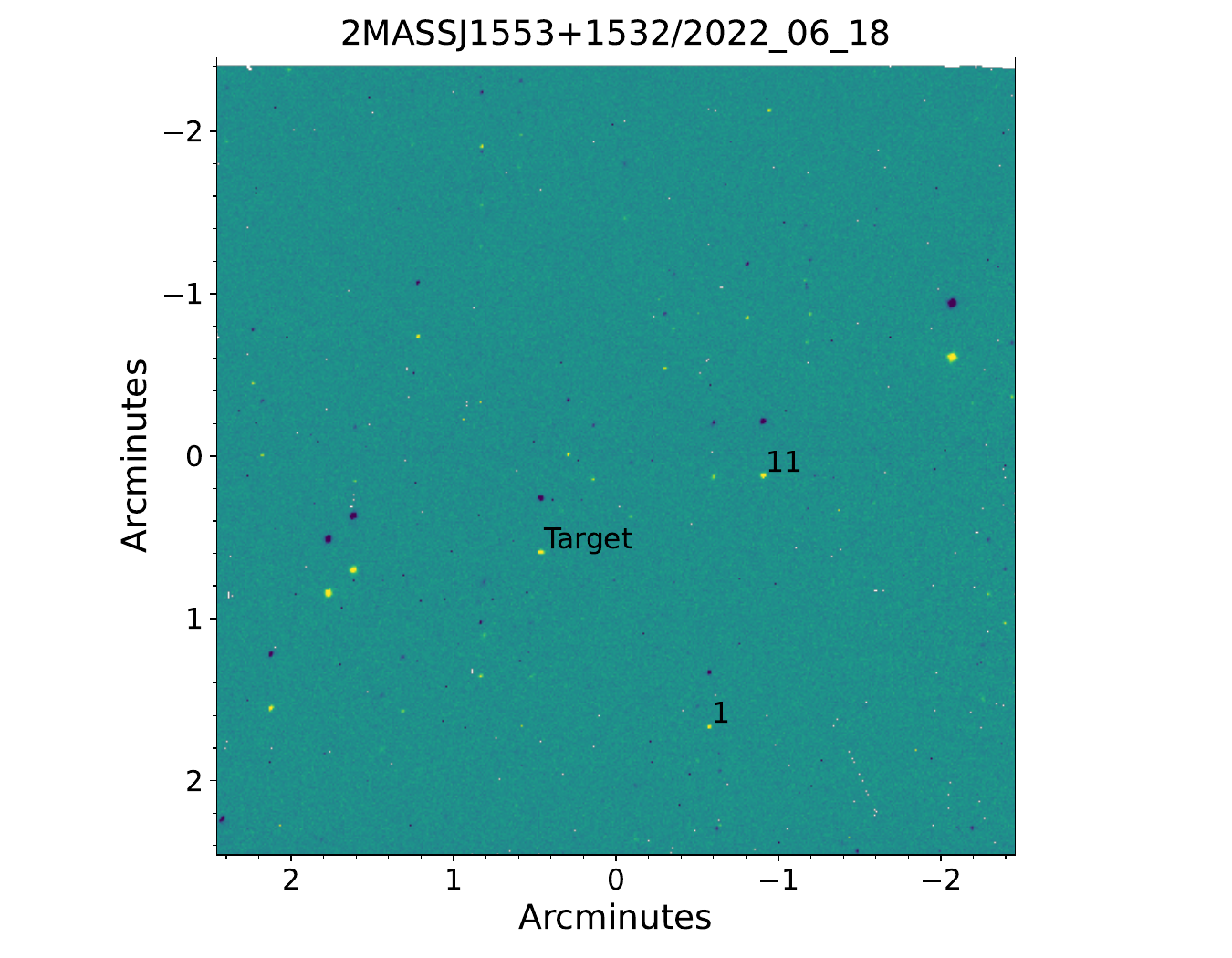}
    \includegraphics[width=0.5\columnwidth]{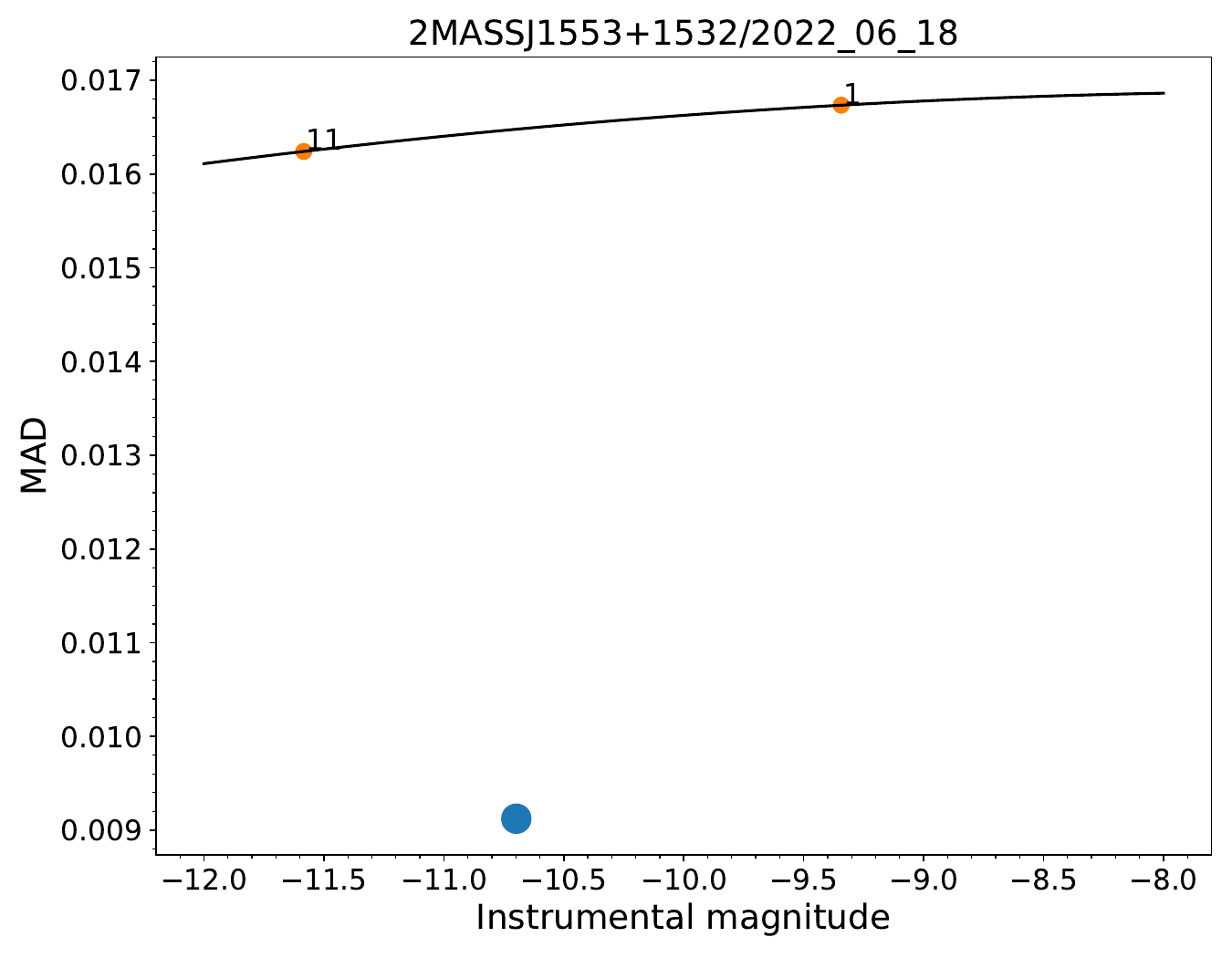} 
    \contcaption{Targets and their selected reference stars.}
    \label{fig:stars_mag}
\end{figure*}

\begin{figure*}
    \includegraphics[width=0.5\columnwidth]{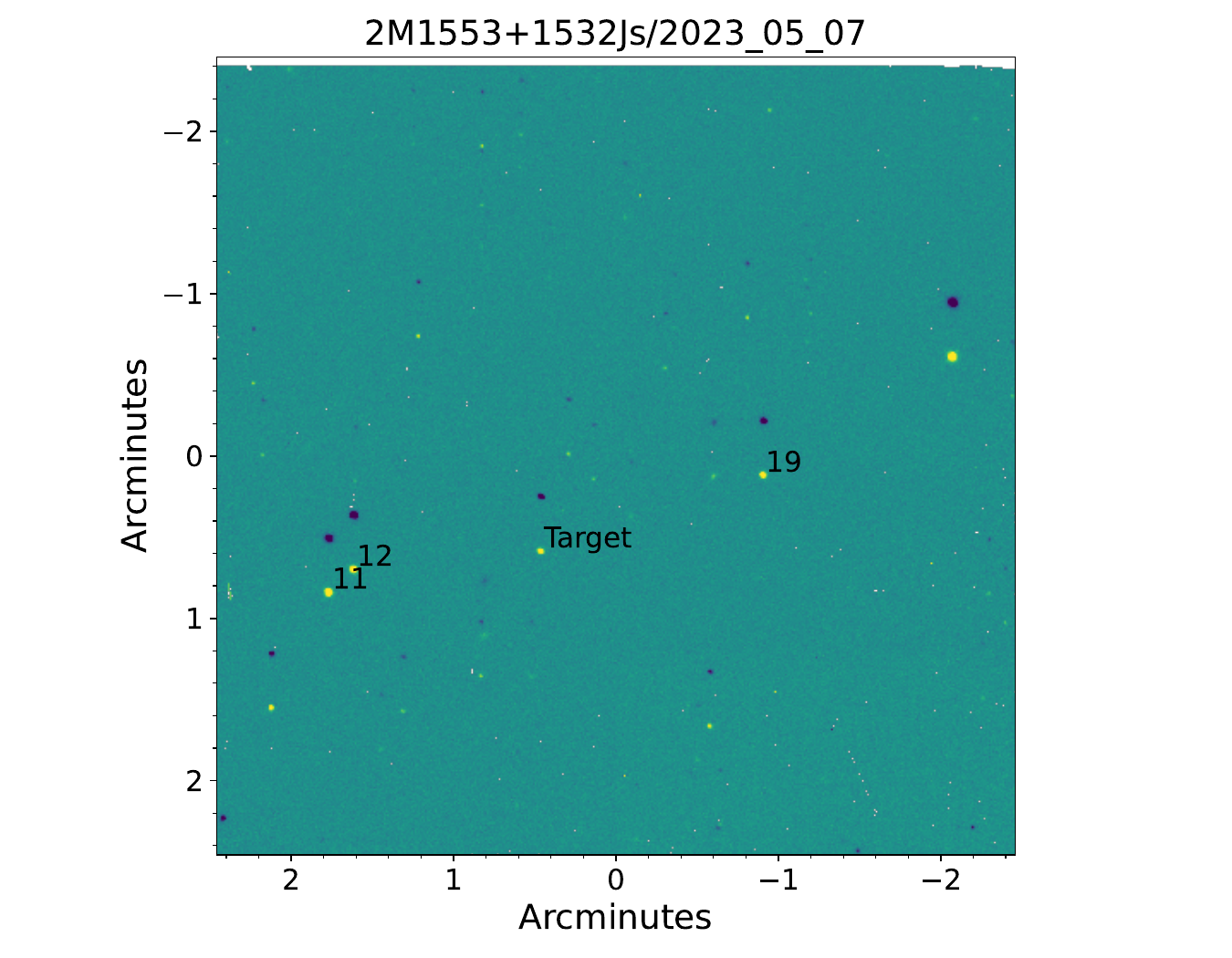}
    \includegraphics[width=0.5\columnwidth]{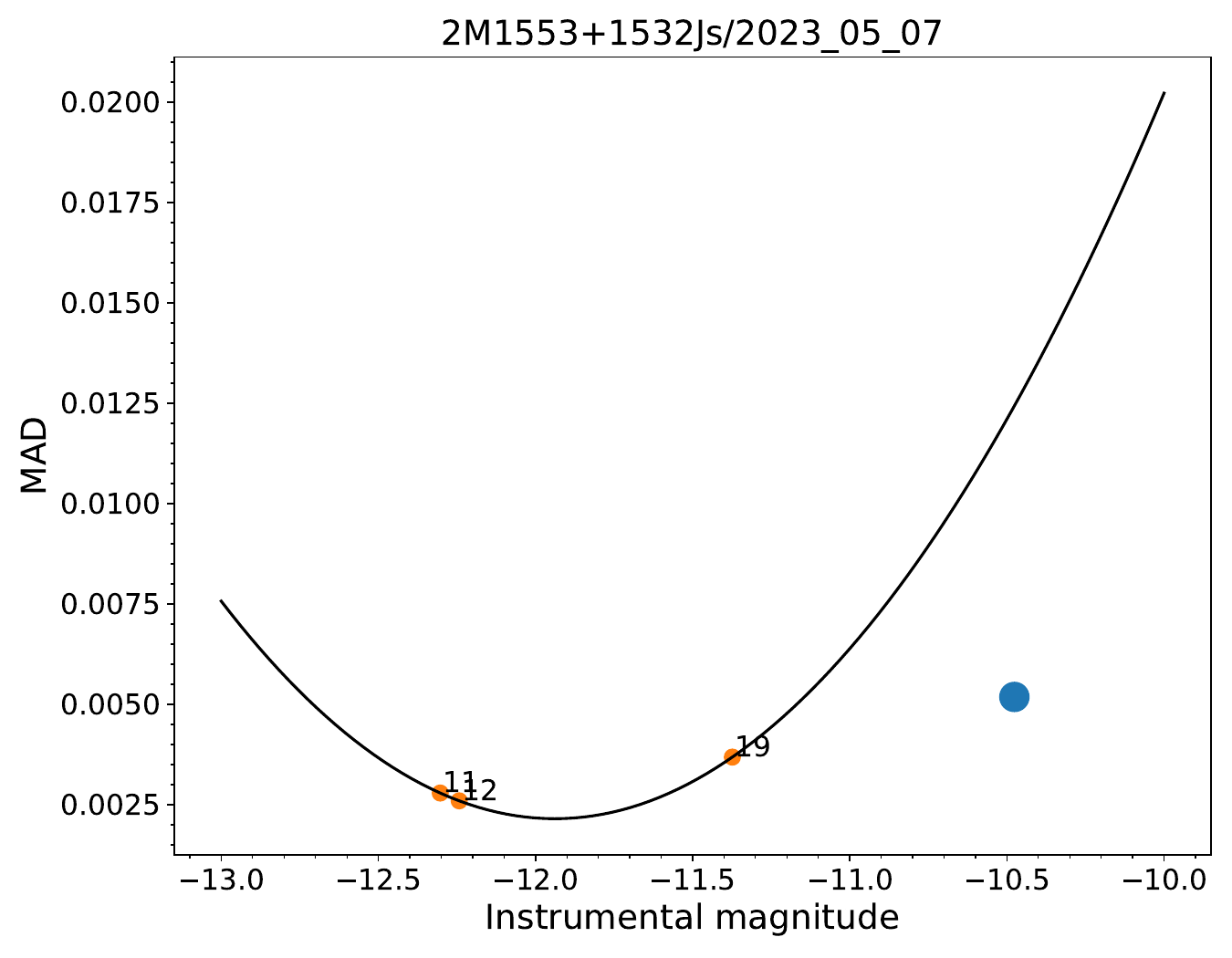} 
    \includegraphics[width=0.5\columnwidth]{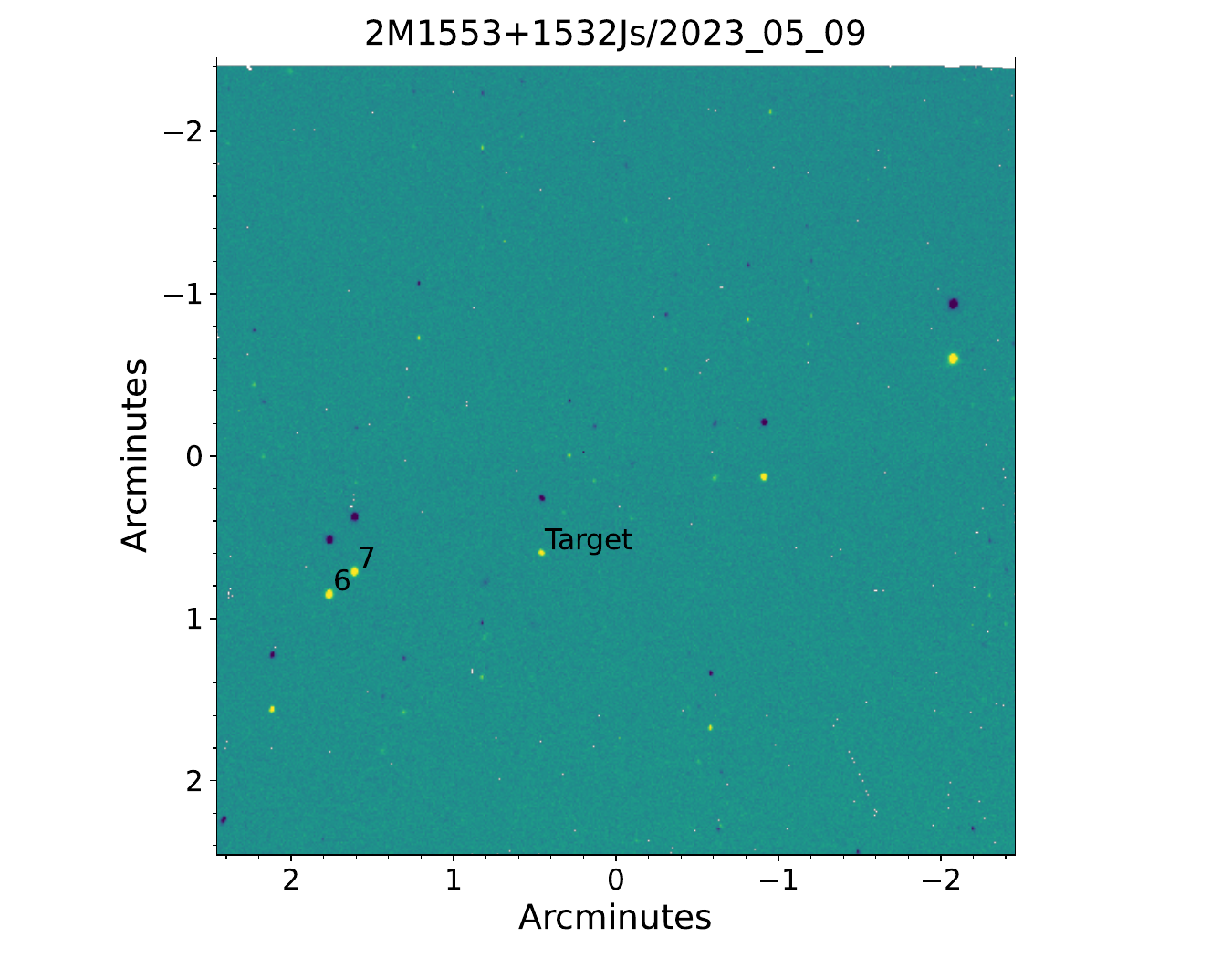}
    \includegraphics[width=0.5\columnwidth]{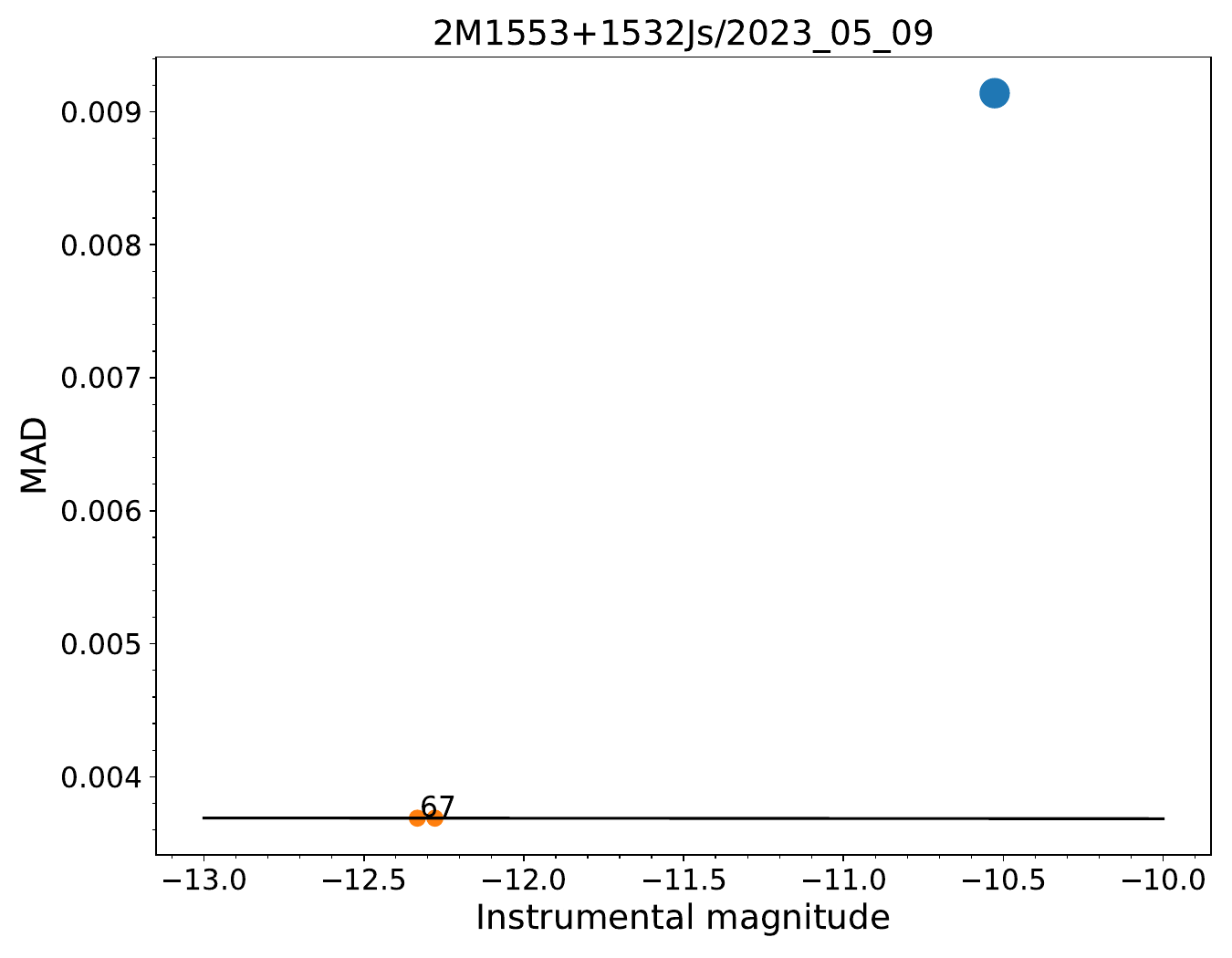} 
    \includegraphics[width=0.5\columnwidth]{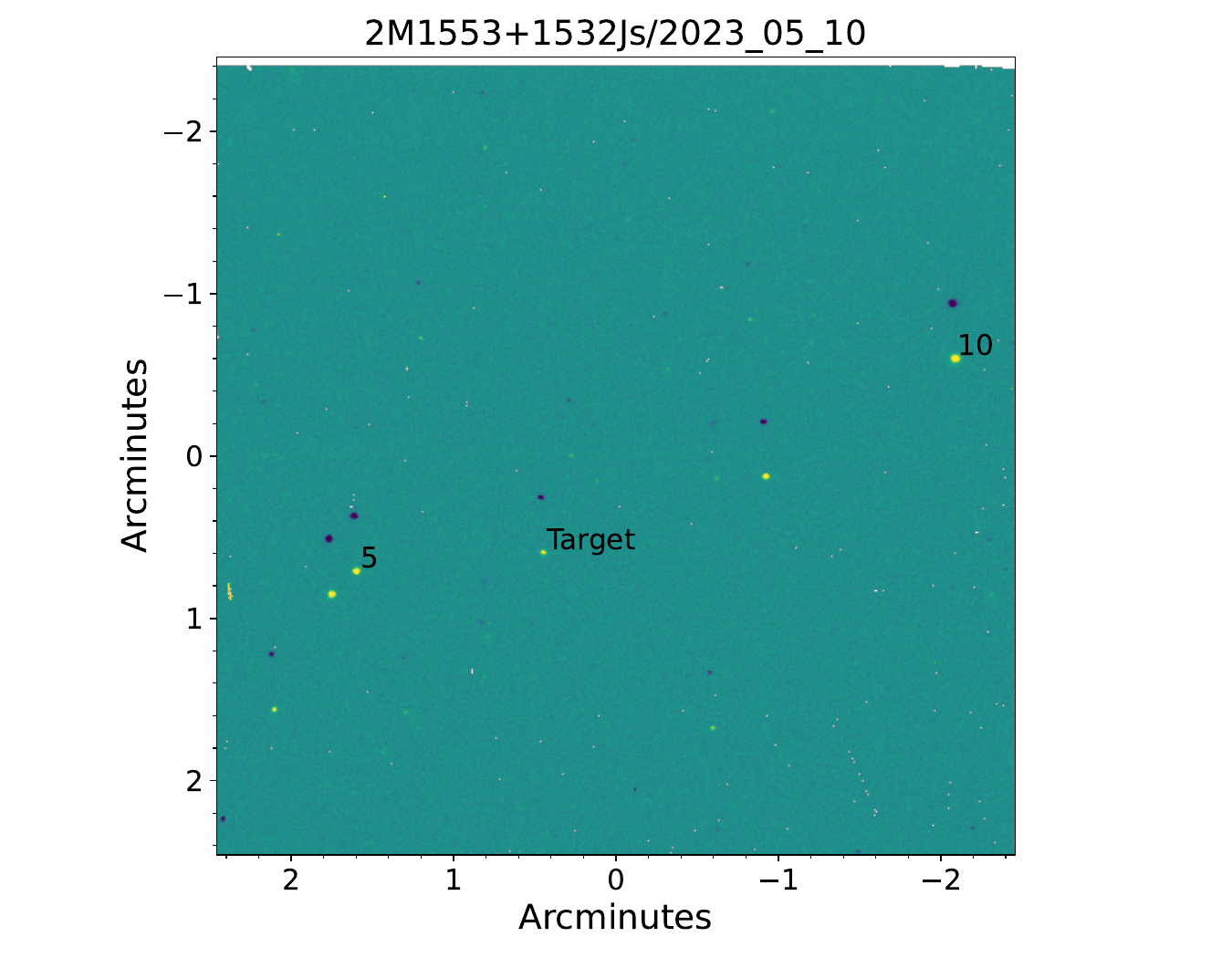}
    \includegraphics[width=0.5\columnwidth]{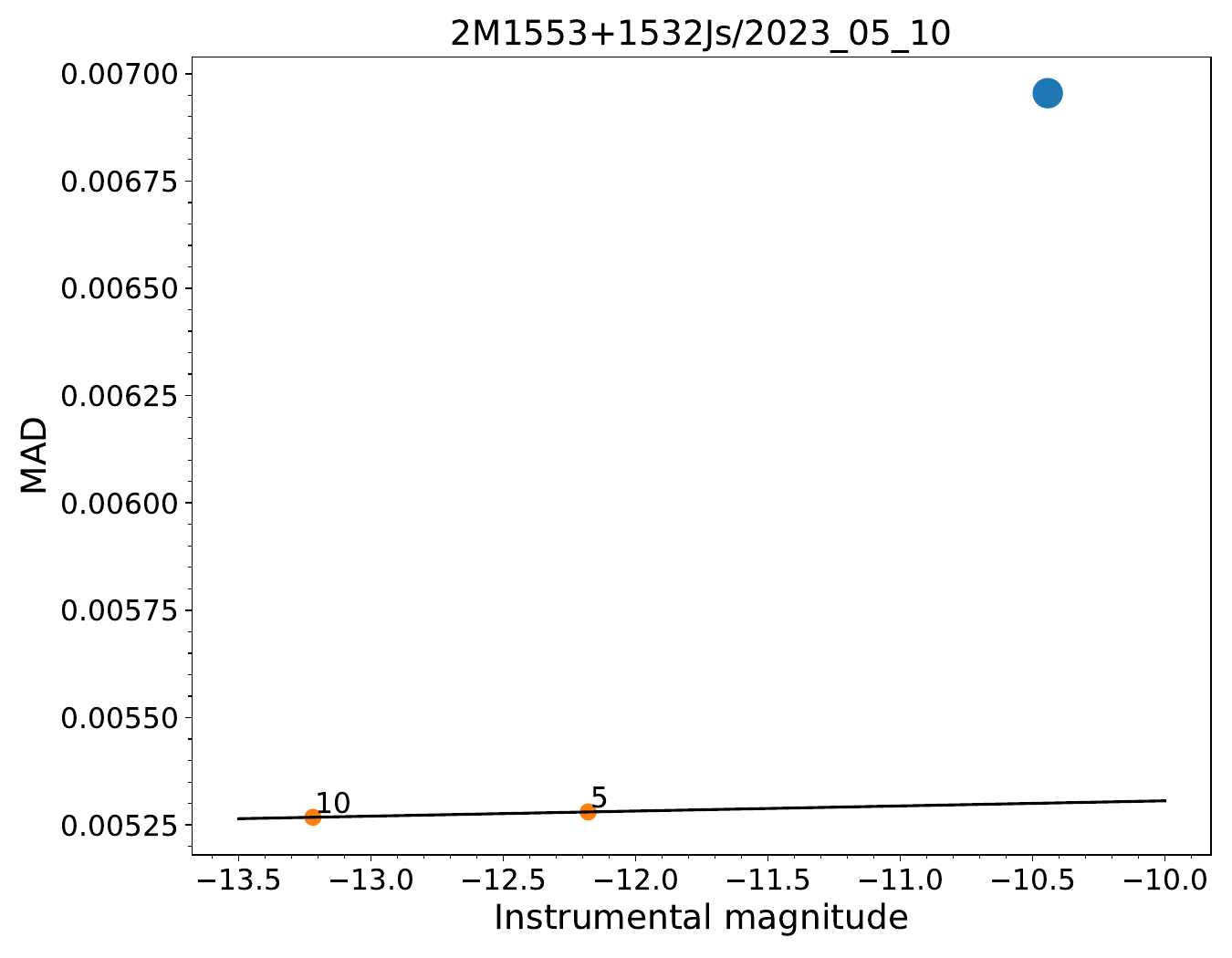} 
    %W1636-0743
    \includegraphics[width=0.5\columnwidth]{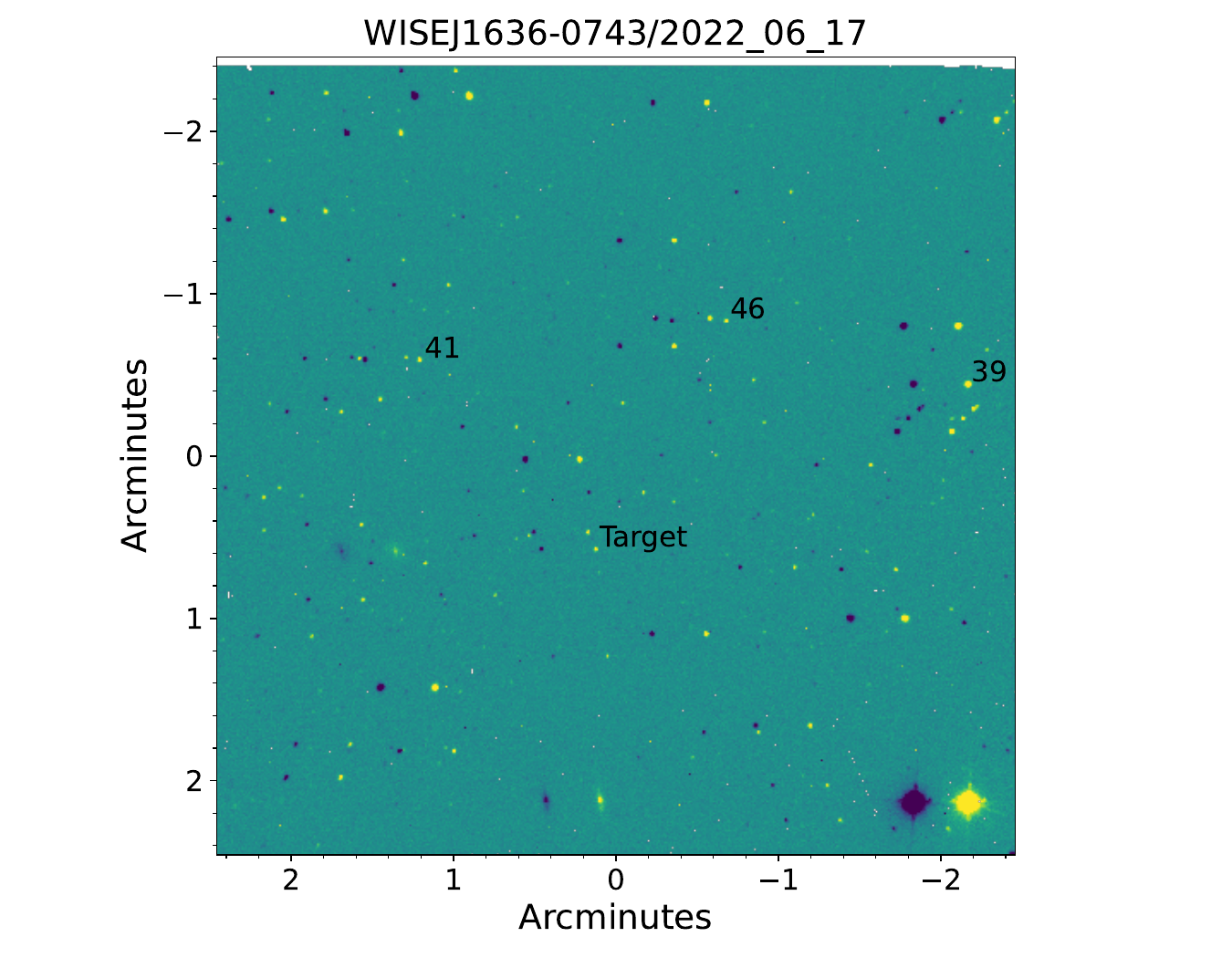}
    \includegraphics[width=0.5\columnwidth]{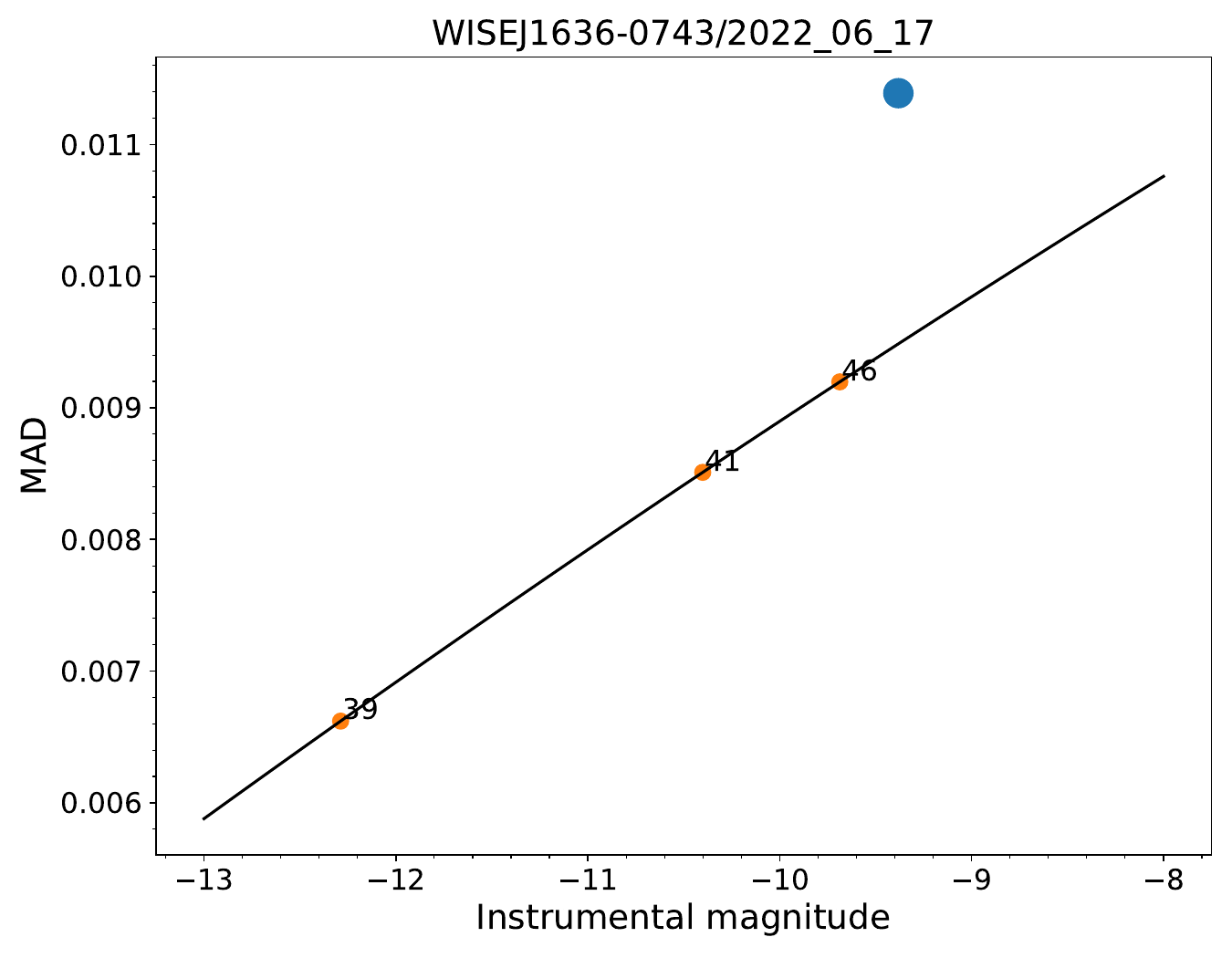} 
    \includegraphics[width=0.5\columnwidth]{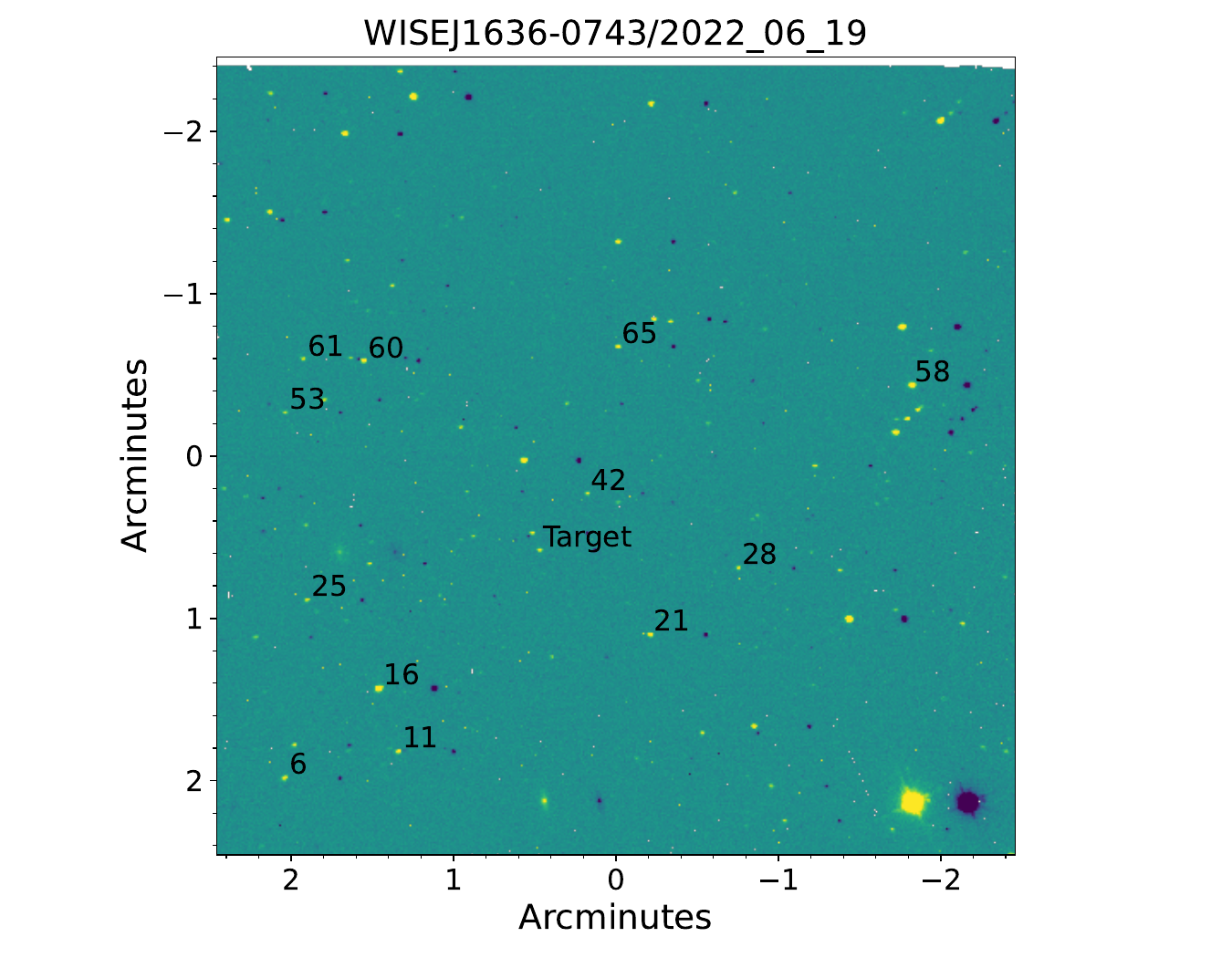}
    \includegraphics[width=0.5\columnwidth]{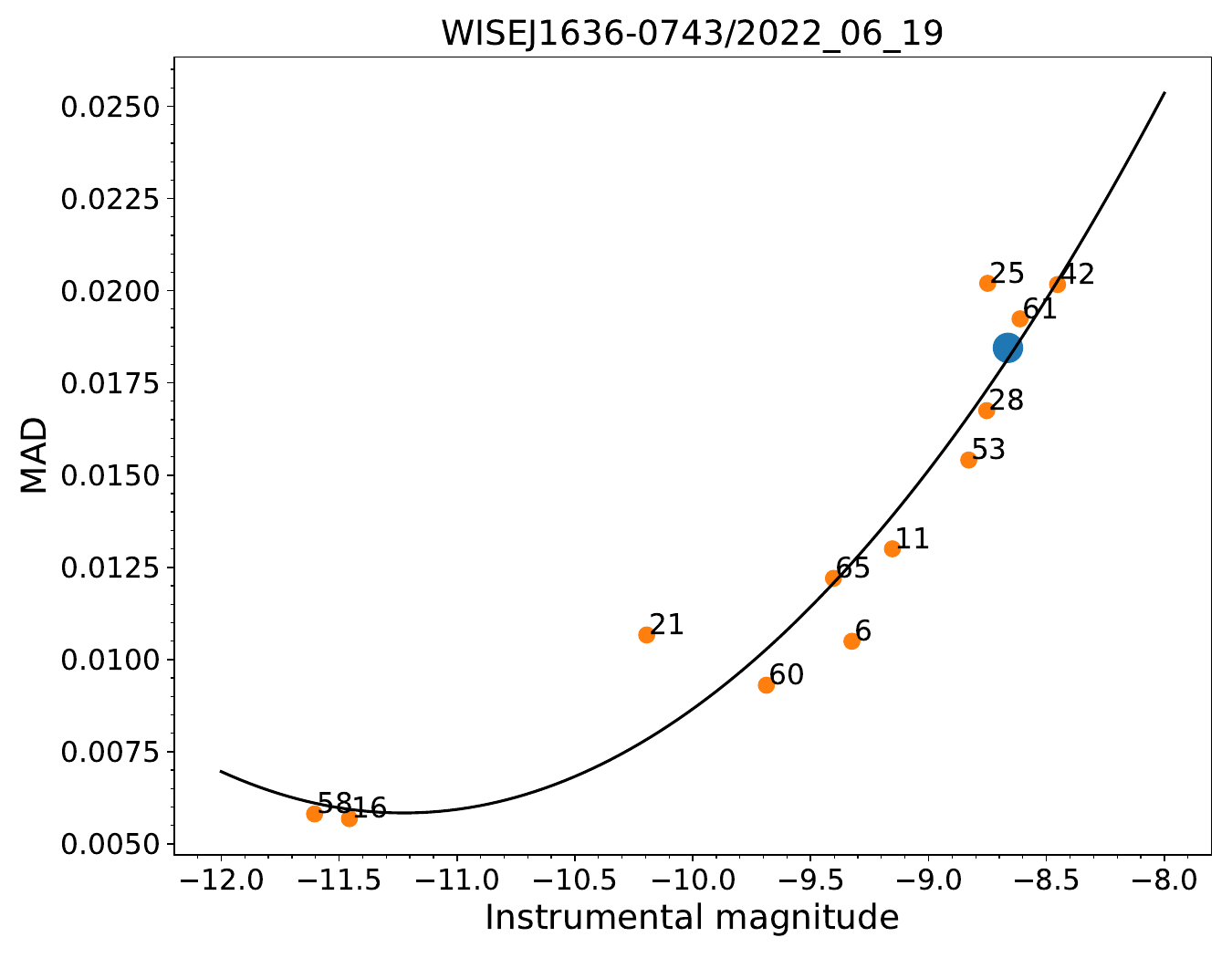} 
    \includegraphics[width=0.5\columnwidth]{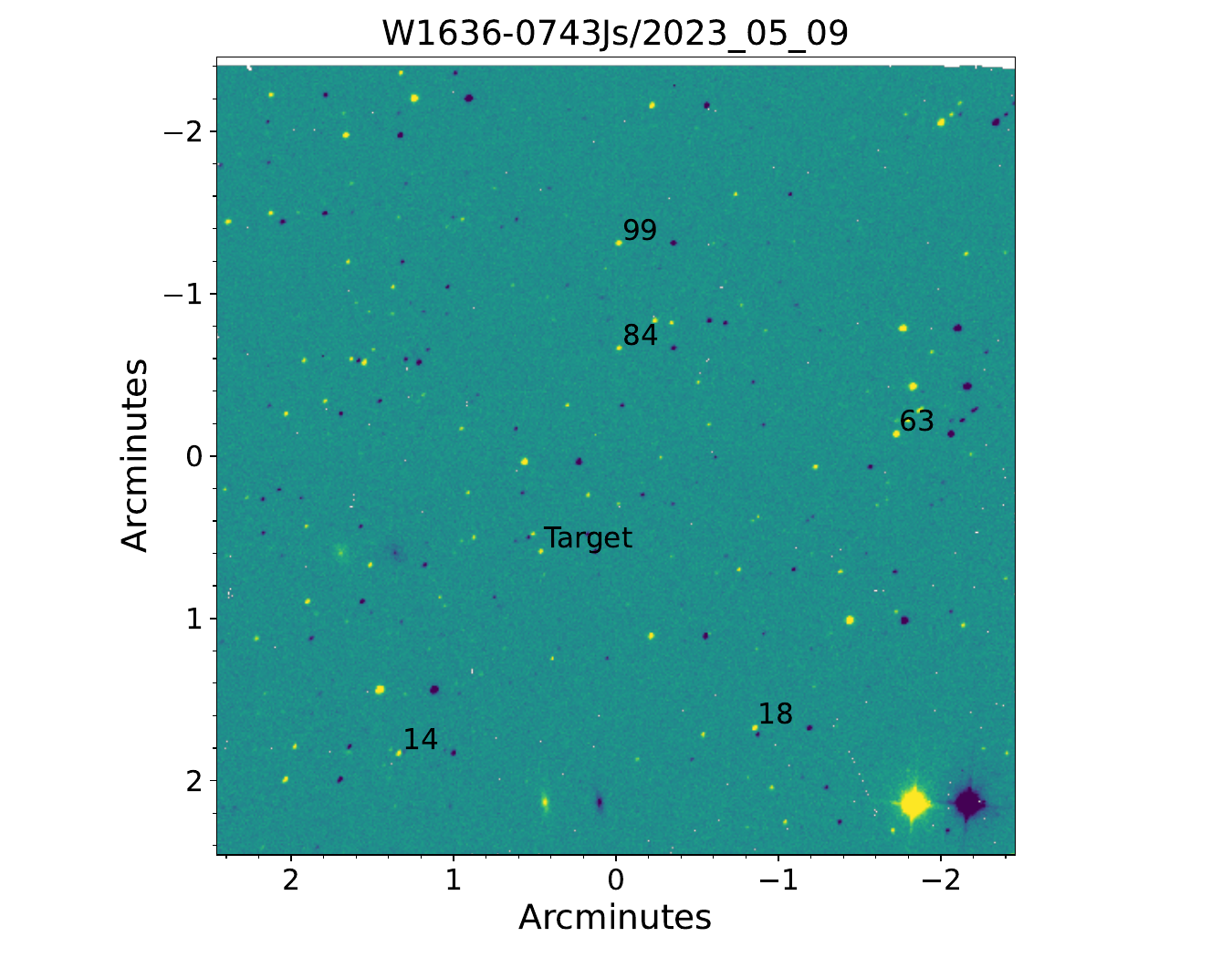}
    \includegraphics[width=0.5\columnwidth]{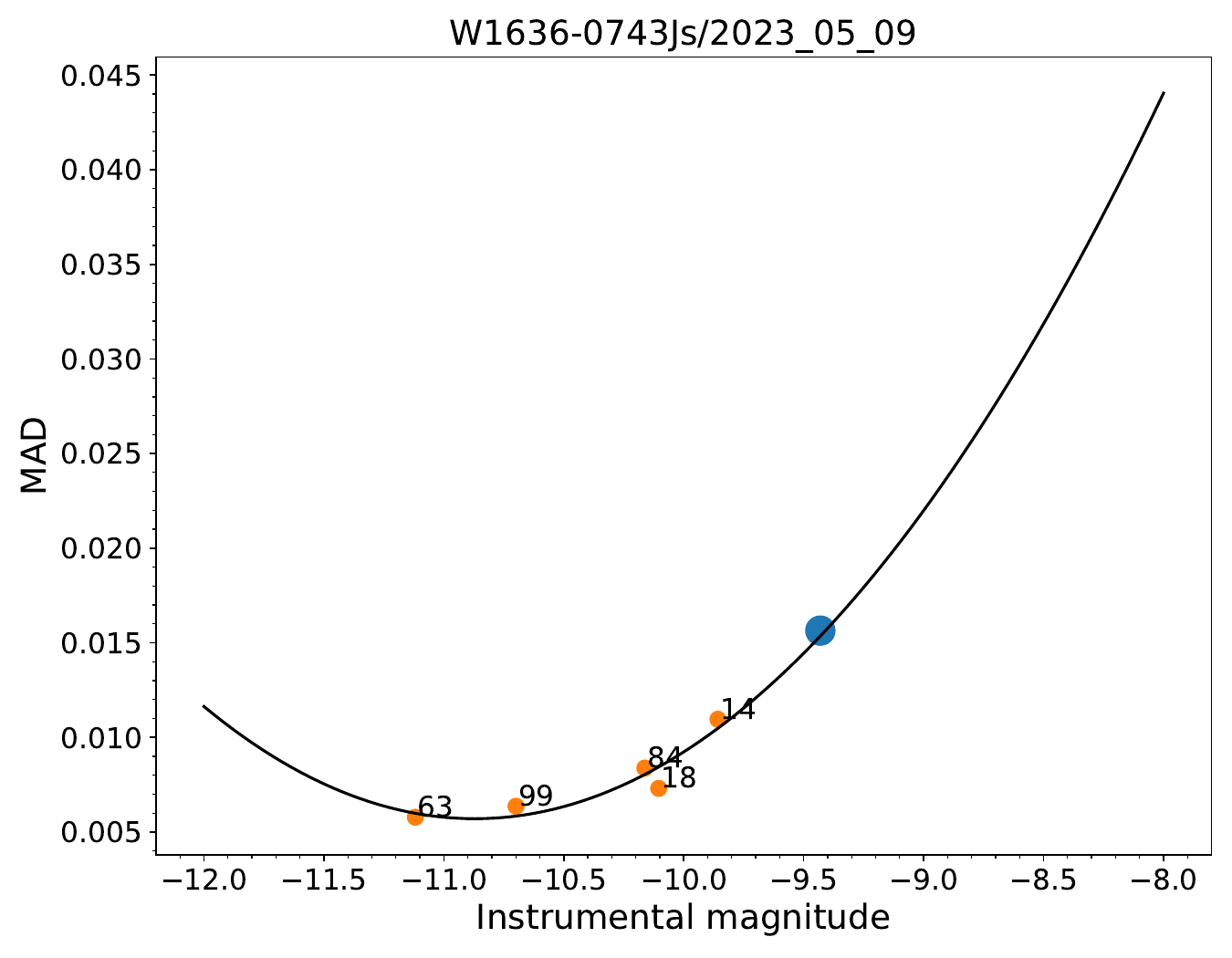} 
    \includegraphics[width=0.5\columnwidth]{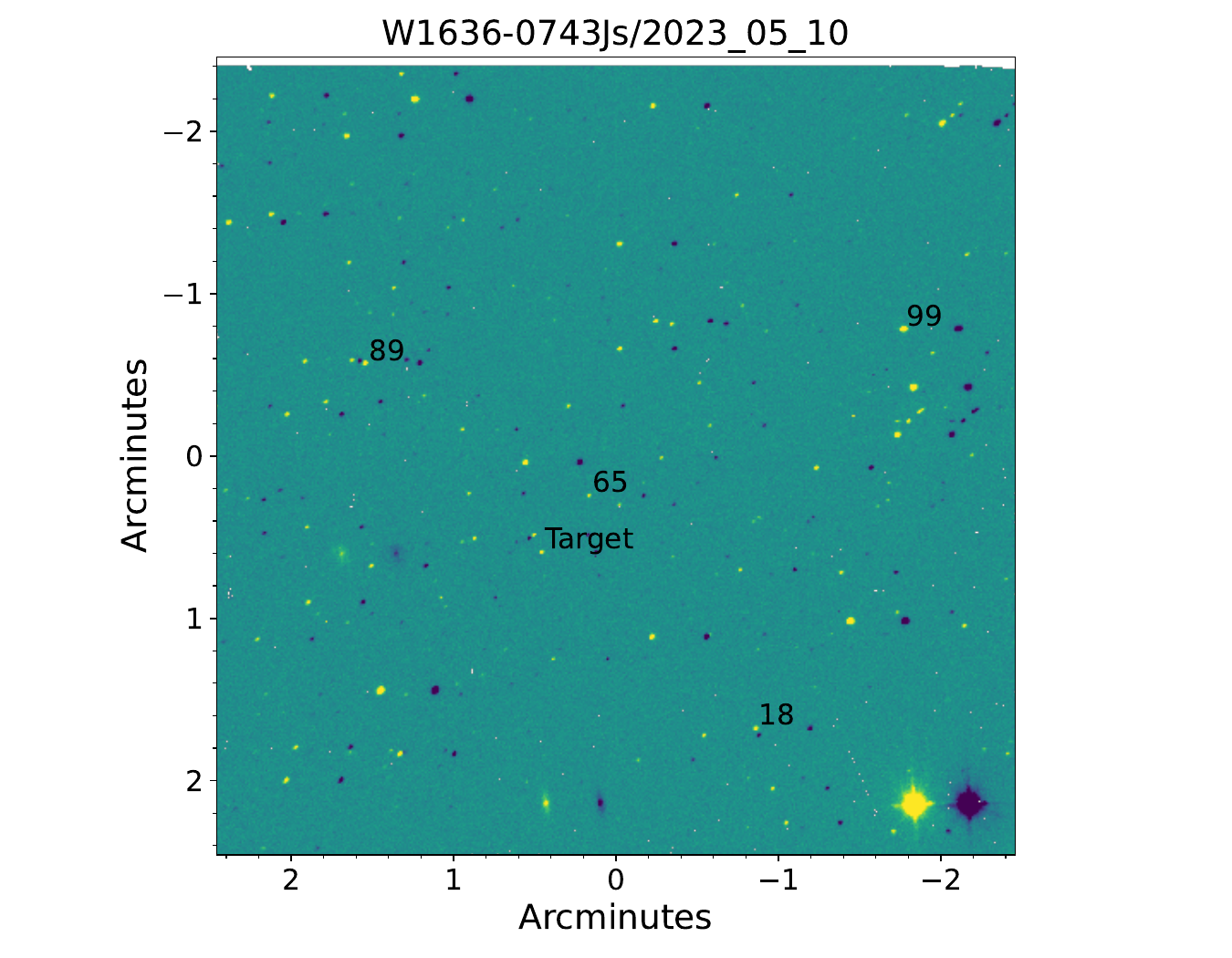}
    \includegraphics[width=0.5\columnwidth]{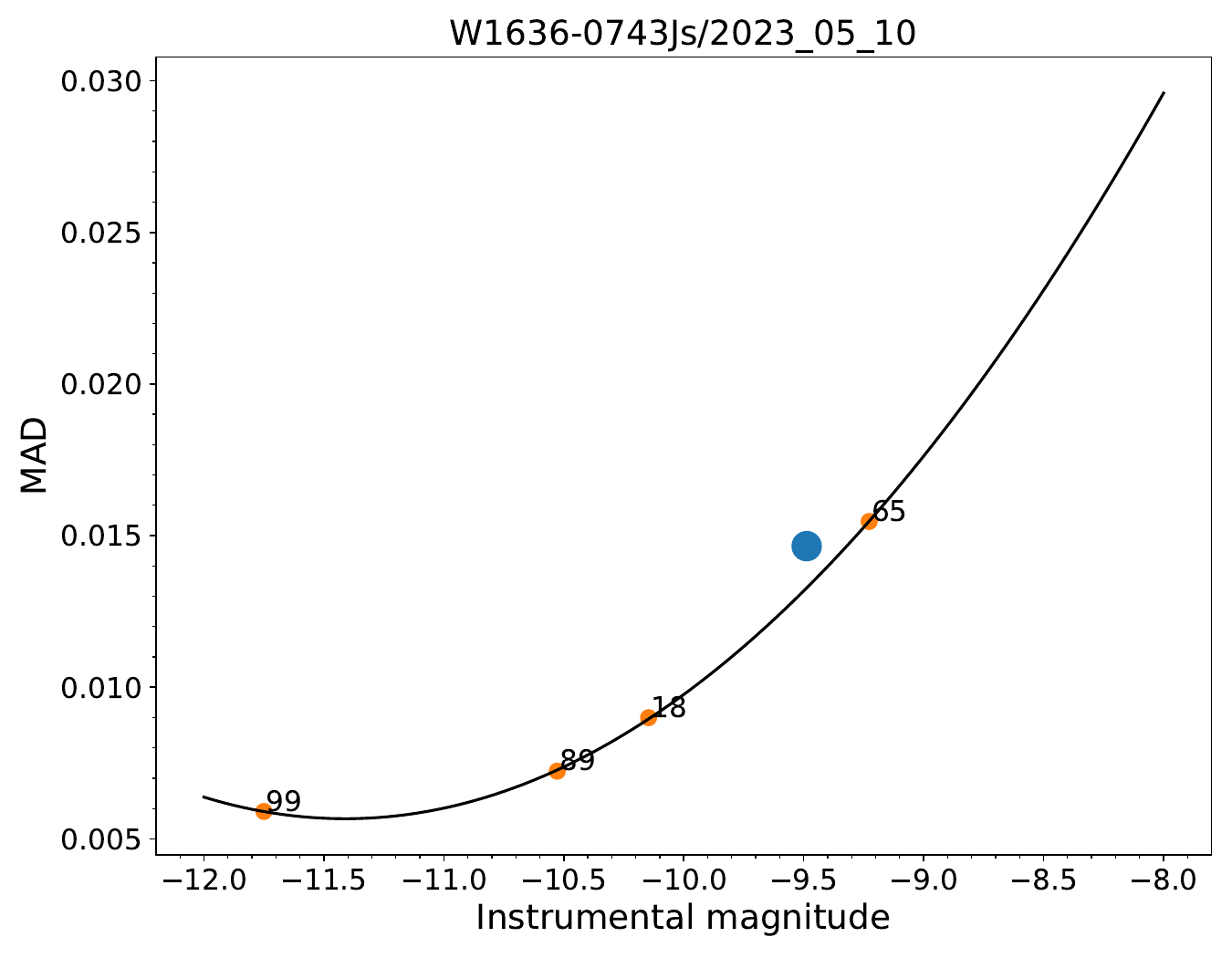} 
    %SDSS1521+0131
    \includegraphics[width=0.5\columnwidth]{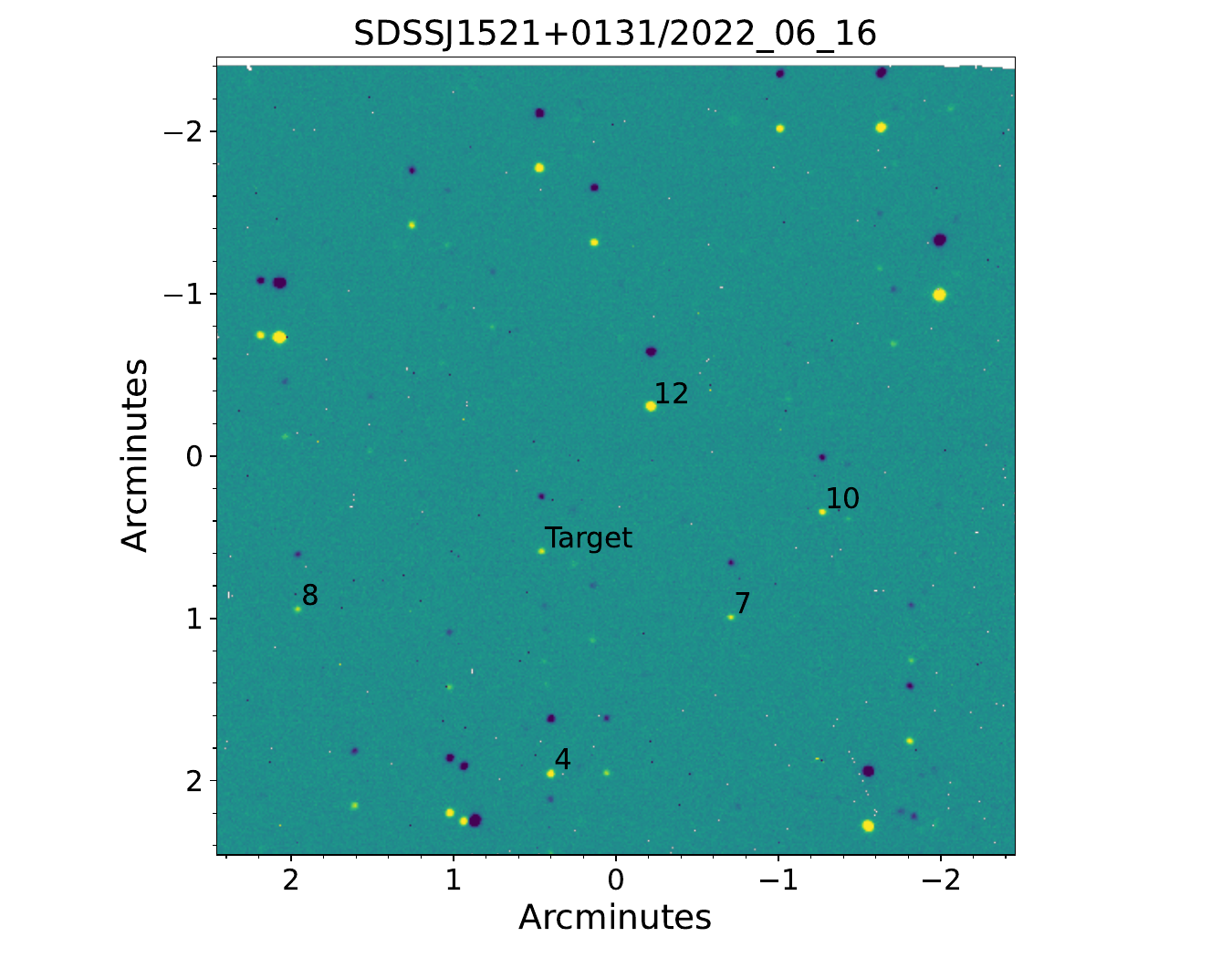}
    \includegraphics[width=0.5\columnwidth]{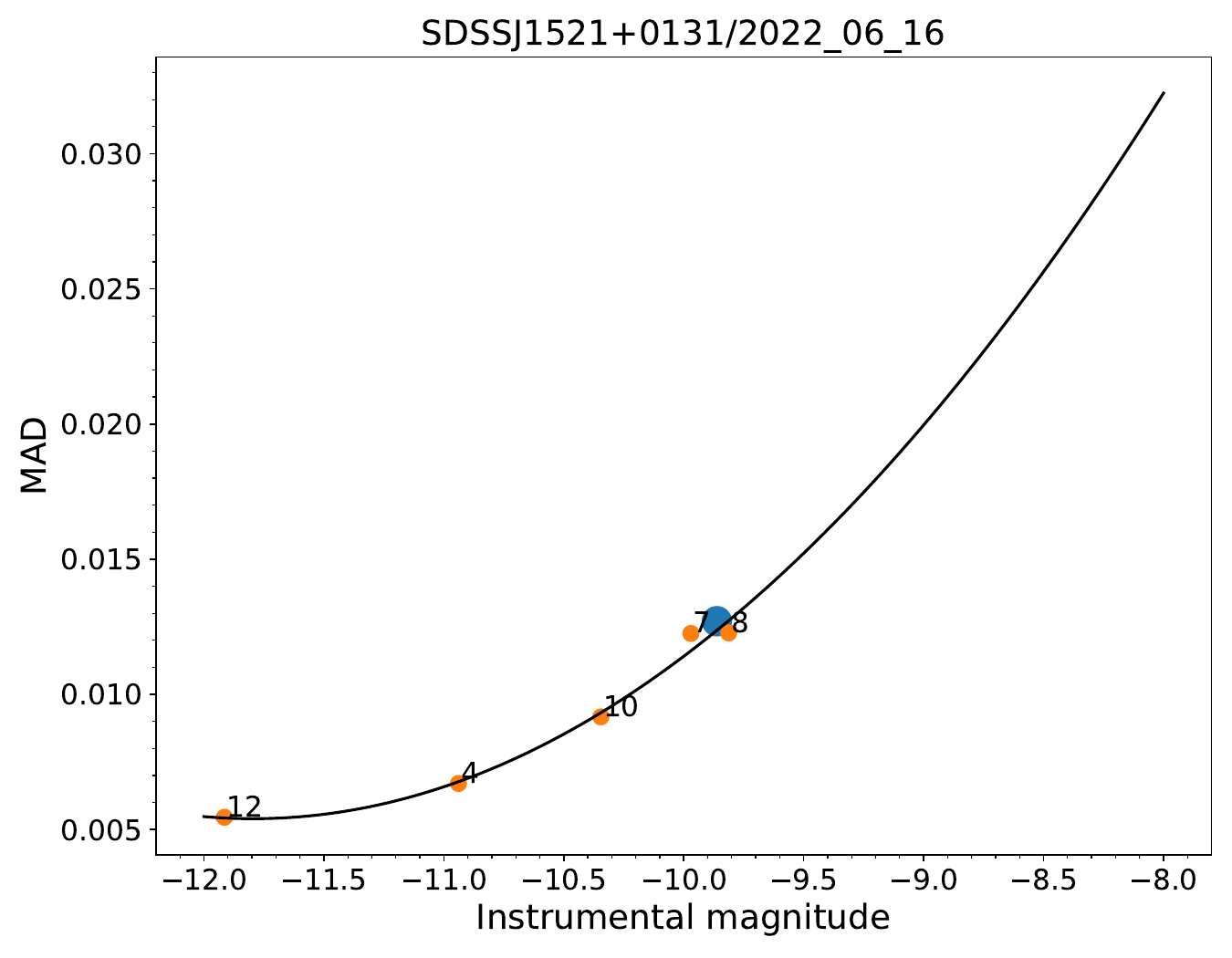} 
    \includegraphics[width=0.5\columnwidth]{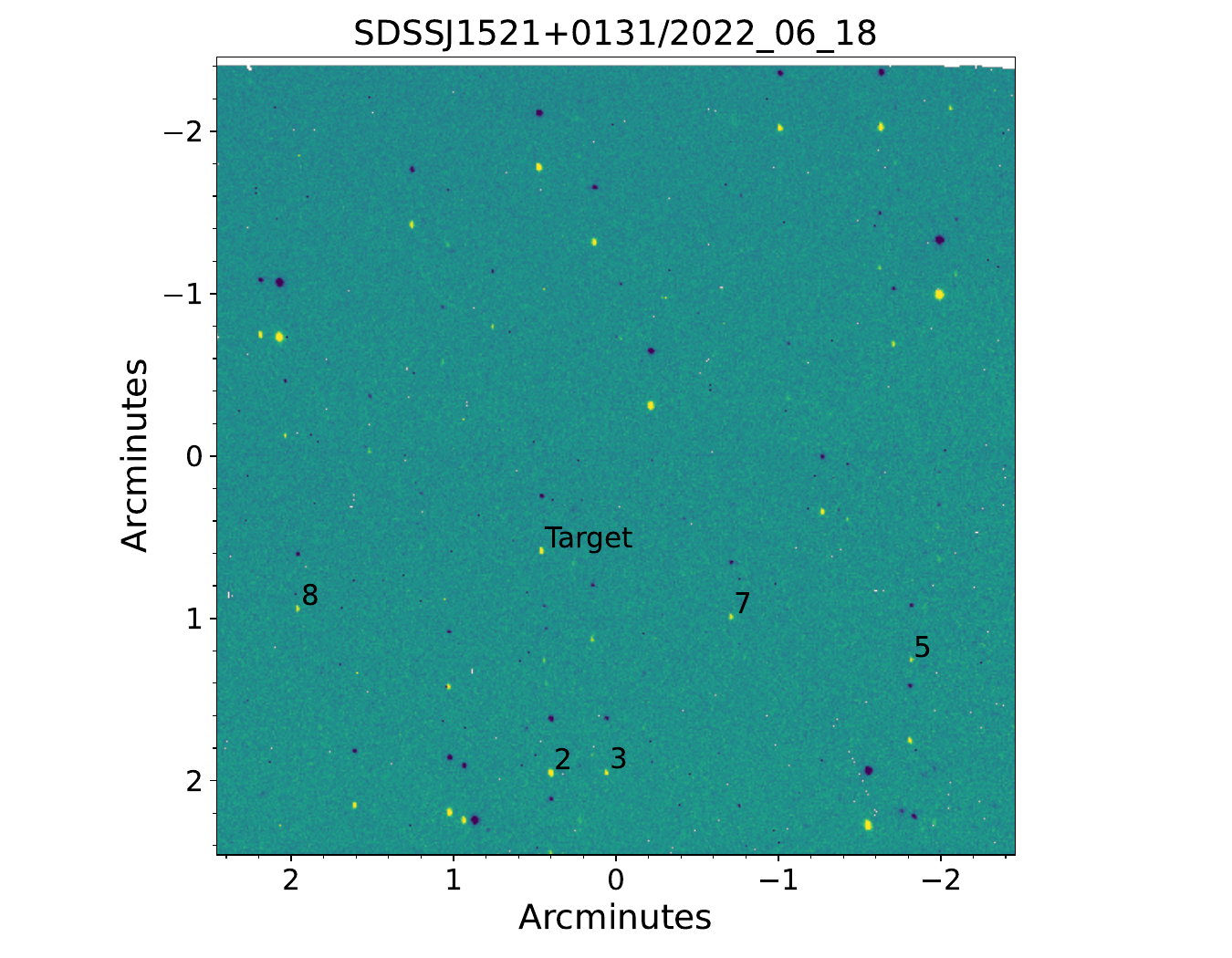}
    \includegraphics[width=0.5\columnwidth]{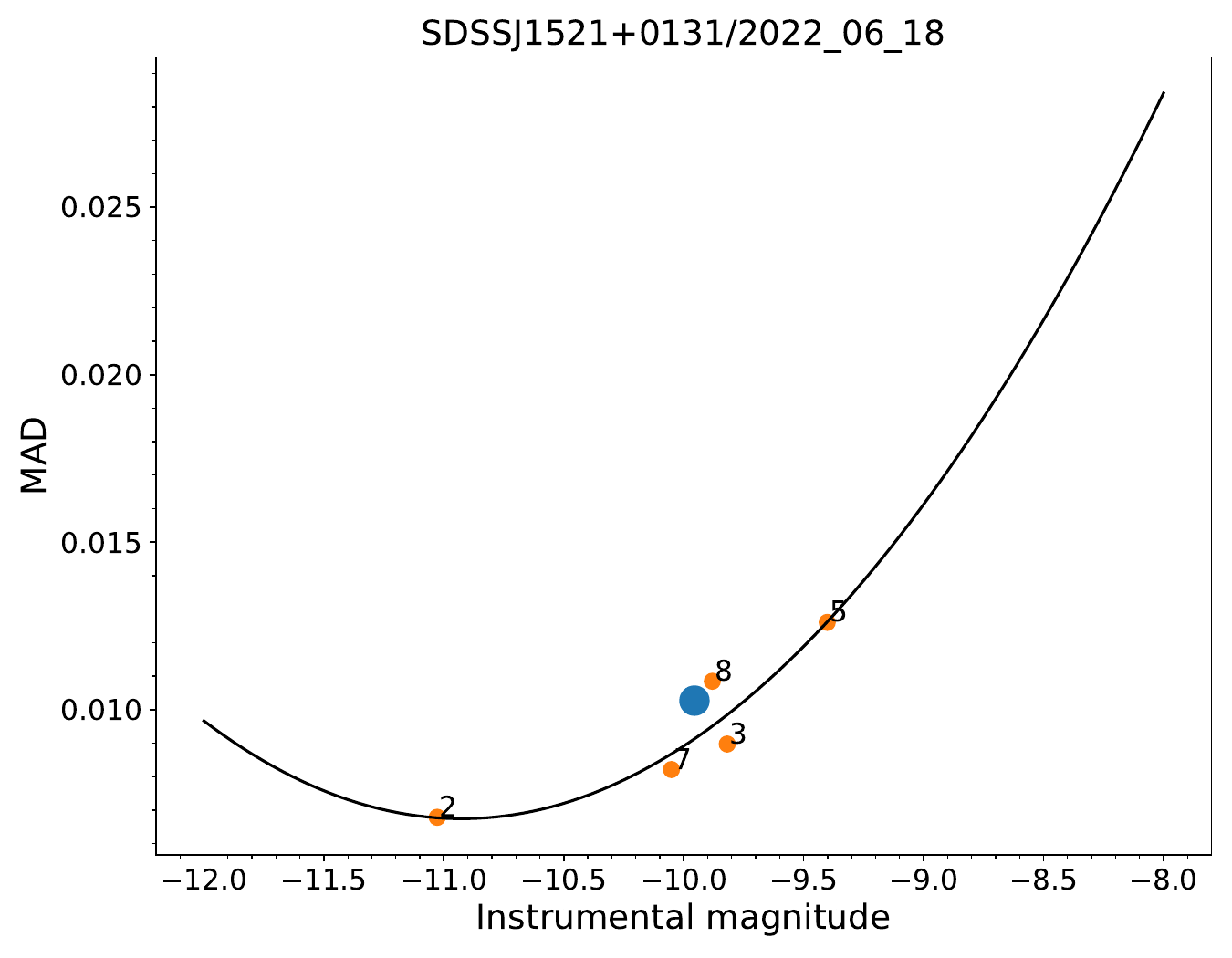} 
    \includegraphics[width=0.5\columnwidth]{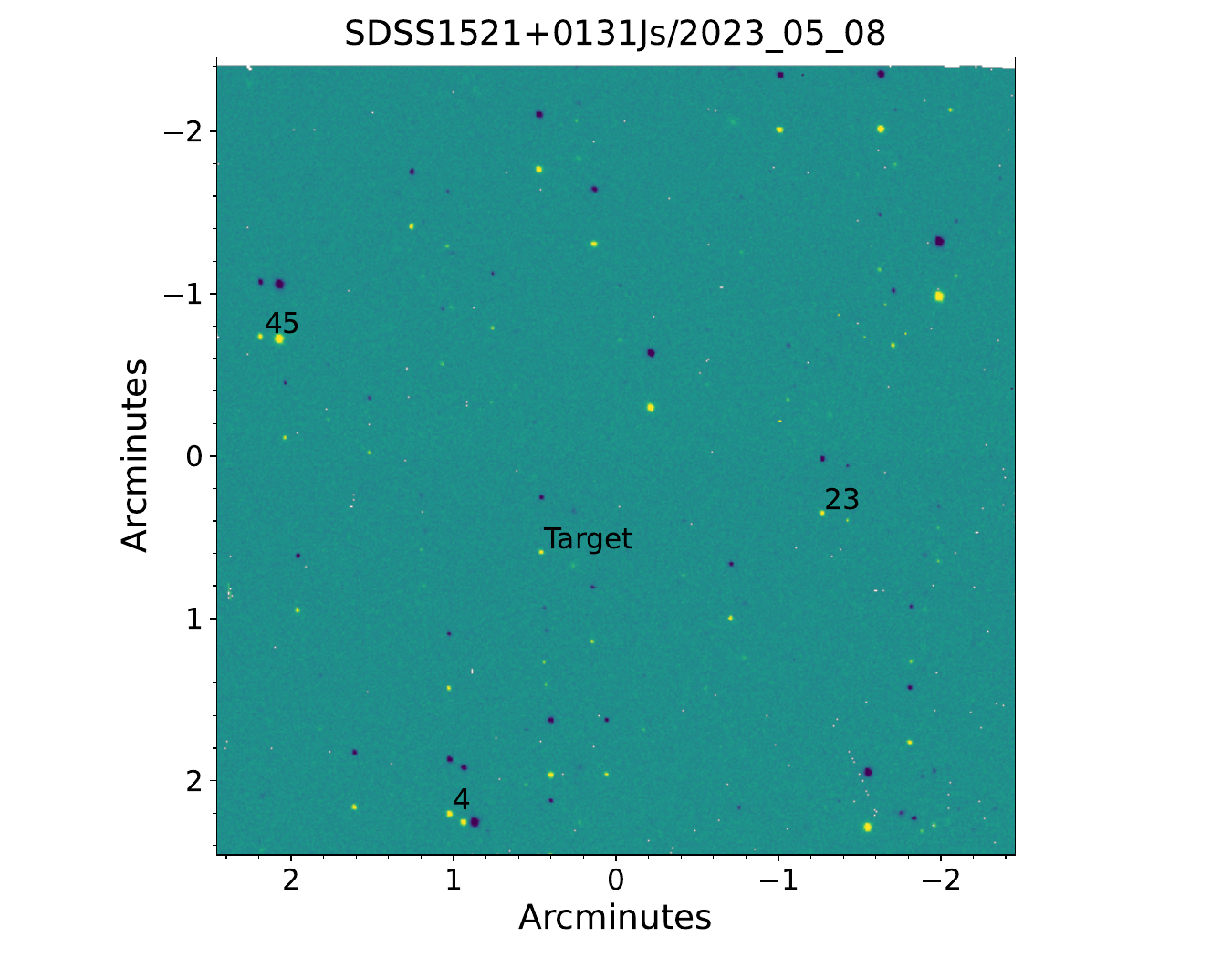}
    \includegraphics[width=0.5\columnwidth]{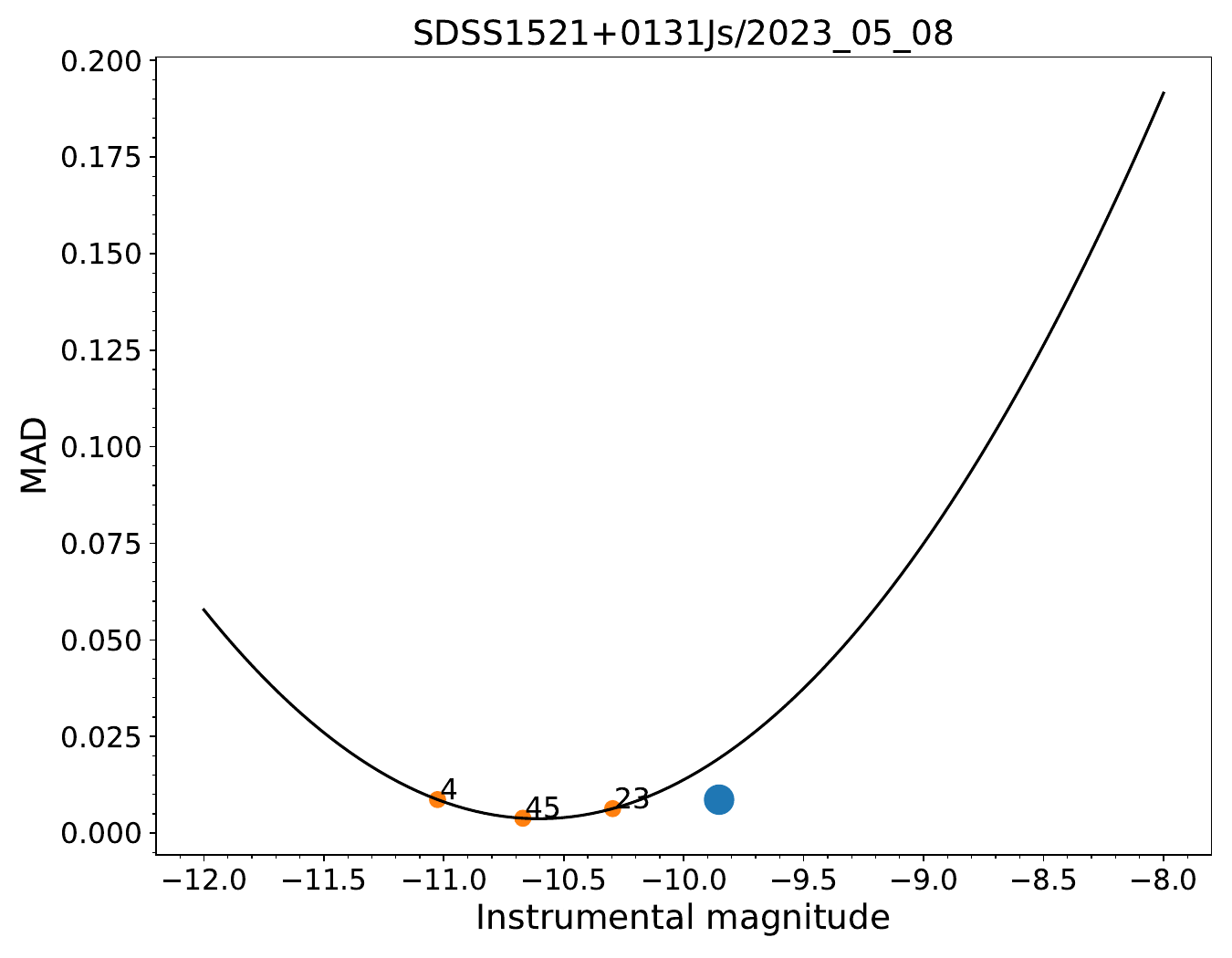} 
    \contcaption{Targets and their selected reference stars.}
    \label{fig:stars_mag}
\end{figure*}

\section{Potentially variable light curves}
Potentially variable light curves, periodograms, and sensitivity plots.
\label{apd:potential}
\begin{figure*}
    \includegraphics[width=1.8\columnwidth]{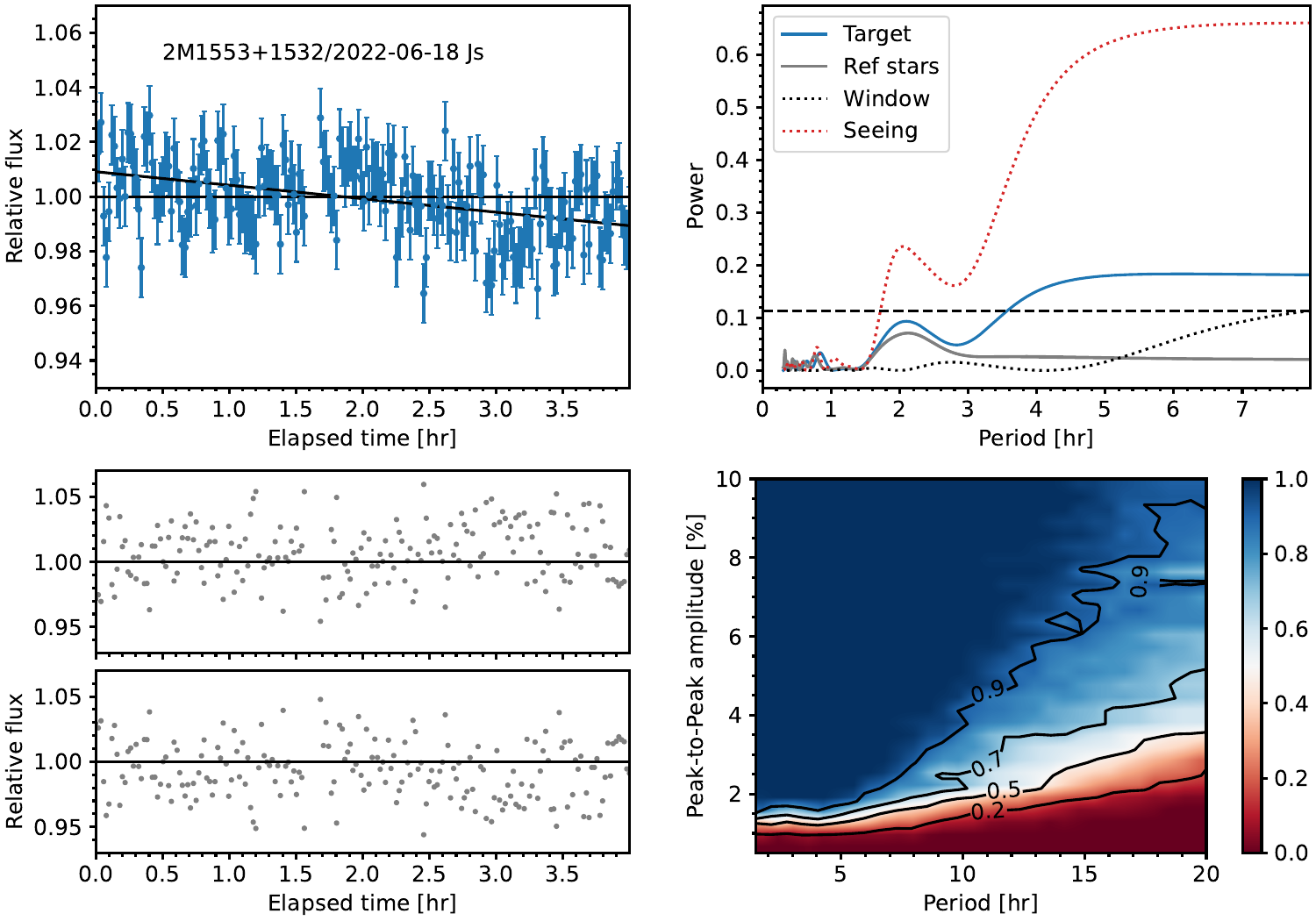}
    \includegraphics[width=1.8\columnwidth]{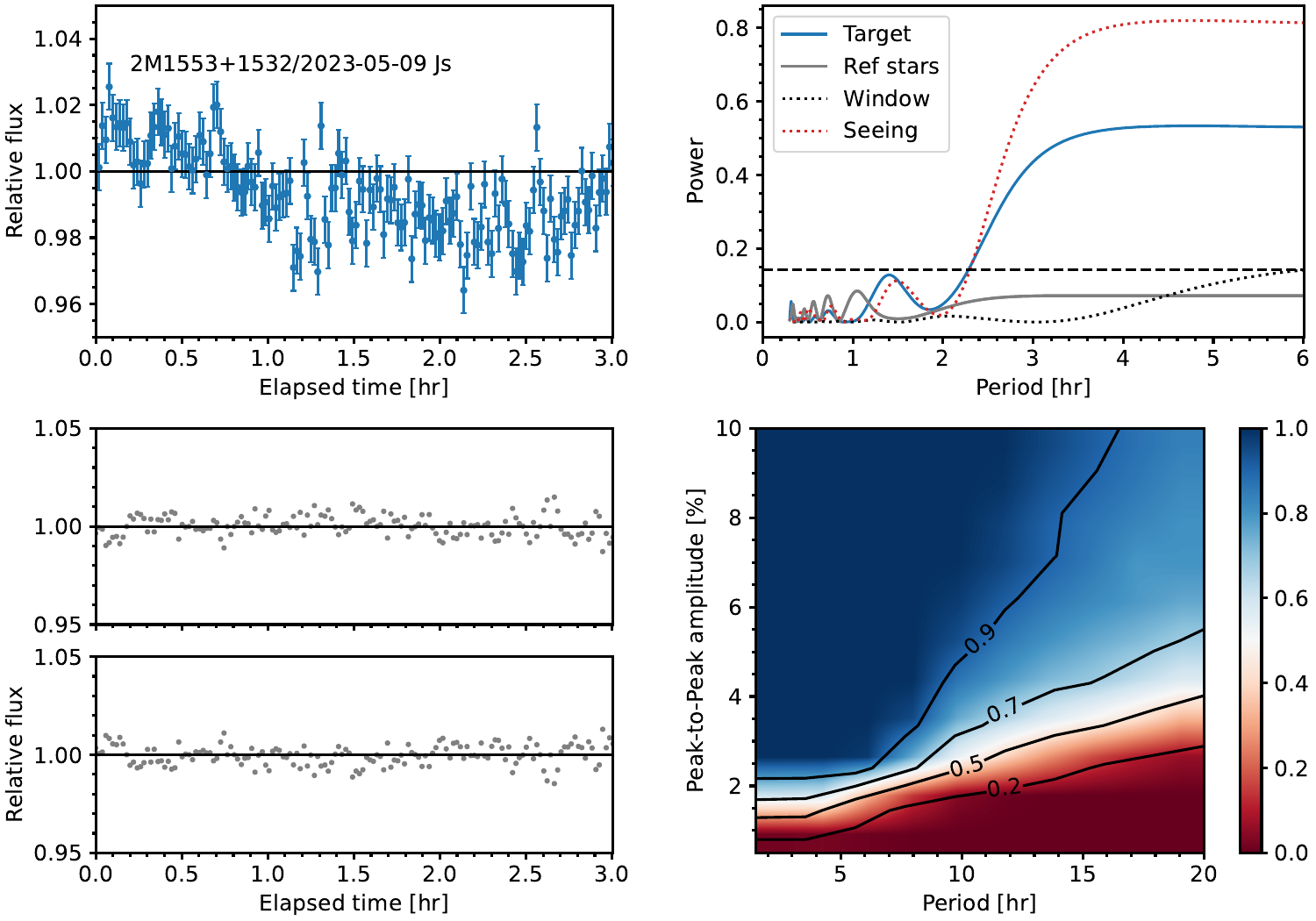}
    \caption{Potentially variable light curves, periodograms, and sensitivity plots of 2M1553+1532, including the detrended light curves and periodograms of the reference stars. While 2M1553+1532 clearly shows a variability above the 1\% FAP level in the periodogram of both nights, it overlaps with the periodogram of the seeing curve, making it a suspicious variable.}
    \label{fig:potential_curves}
\end{figure*}

\begin{figure*}
    \includegraphics[width=1.8\columnwidth]{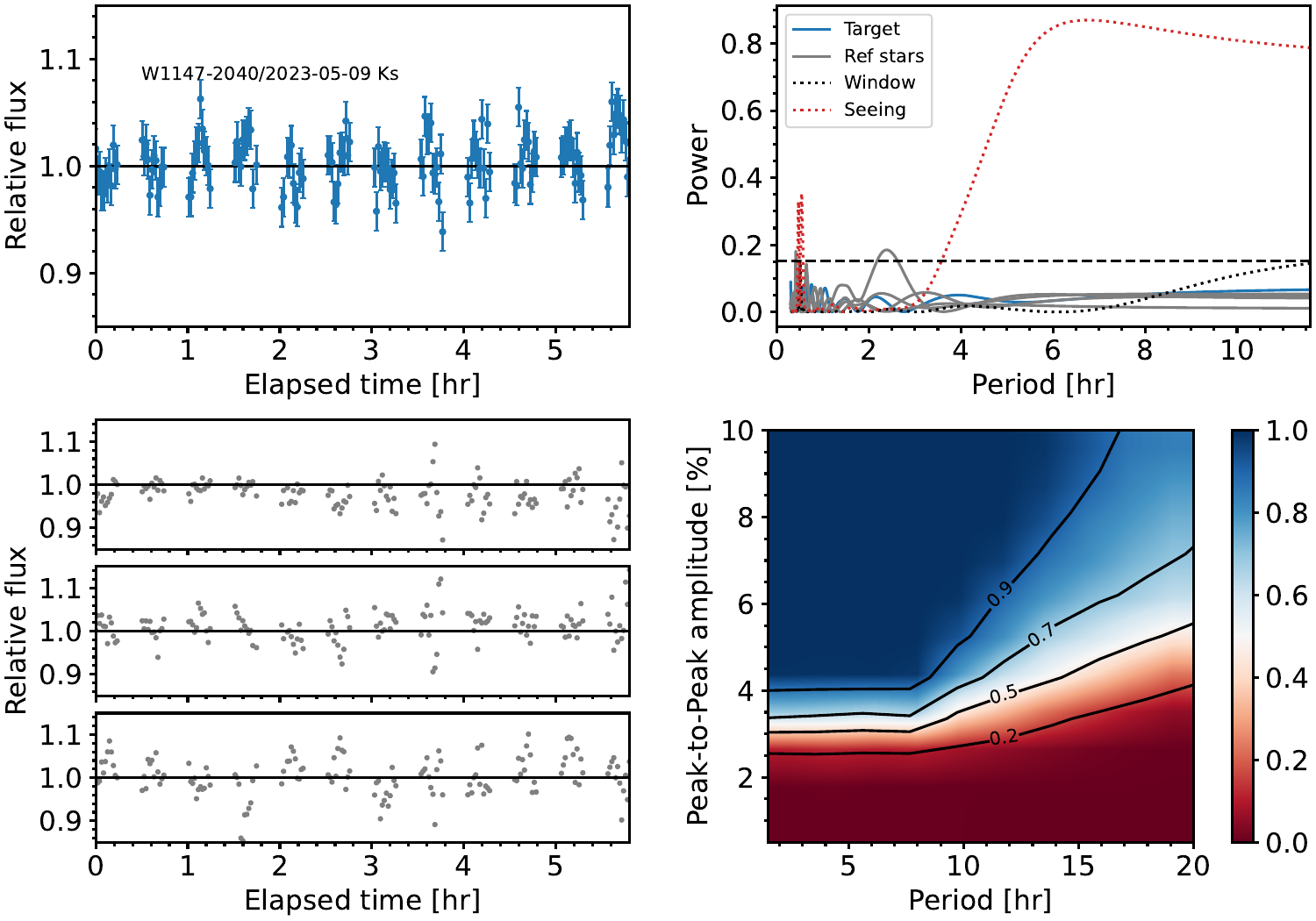}
    \includegraphics[width=1.8\columnwidth]{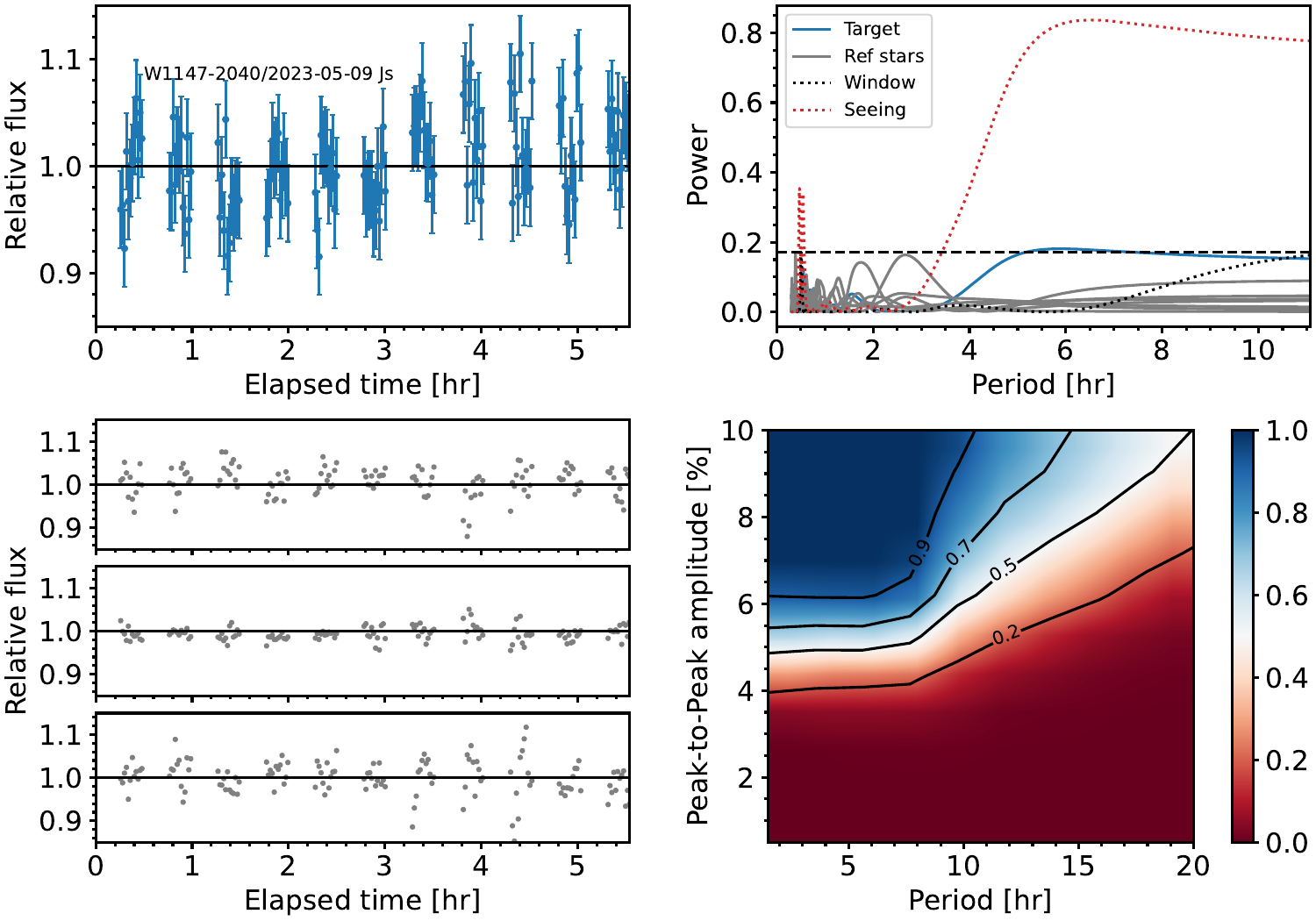}
    \caption{Interleaved $K_S$- and $J_S$-band light curves, periodograms, and sensitivity plots of W1147-2040 on 2023-05-09, including detrended light curves and periodograms of their reference stars. W1147-2040 shows potential variation in the $J_S$ band which is probably correlated with the seeing.}
    \label{fig:potential_curves}
\end{figure*}

\begin{figure*}
    \includegraphics[width=2\columnwidth]{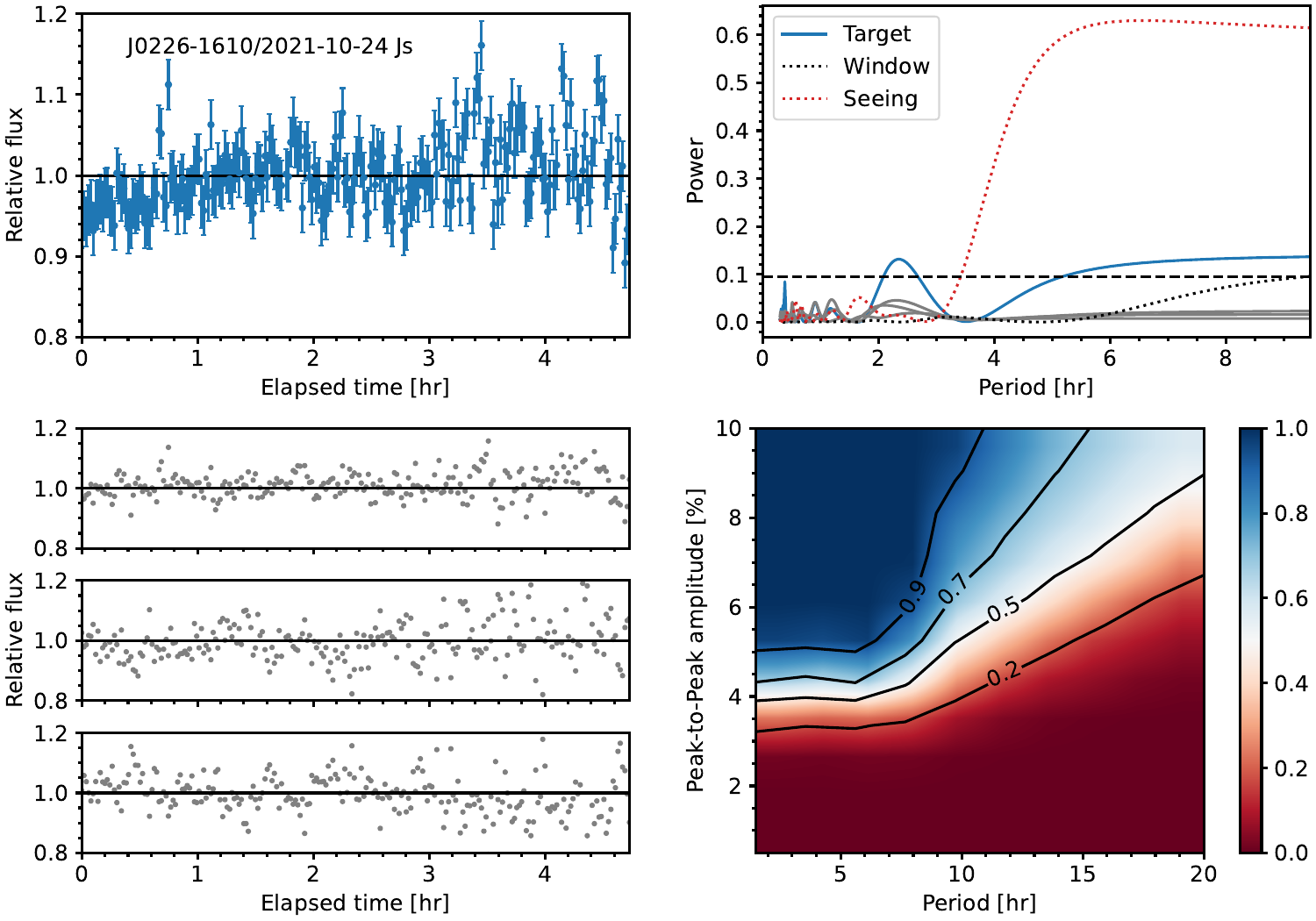}
    \caption{Potentially variable light curve, periodogram, and sensitivity plot of J0226-1610, including detrended light curves and periodograms of its reference stars. While it shows variability slightly above the 1\% FAP level in the periodogram, we are cautious about the variability due to large scatter in the detrended light curves of the reference stars}
    \label{fig:potential_curves}
\end{figure*}

\clearpage
\section{Non-variable light curves}
\label{apd:non-variable}
Non-variable light curves, periodograms and sensitivity plots. 

\begin{figure*}
    \centering
    \includegraphics[width=1.8\columnwidth]{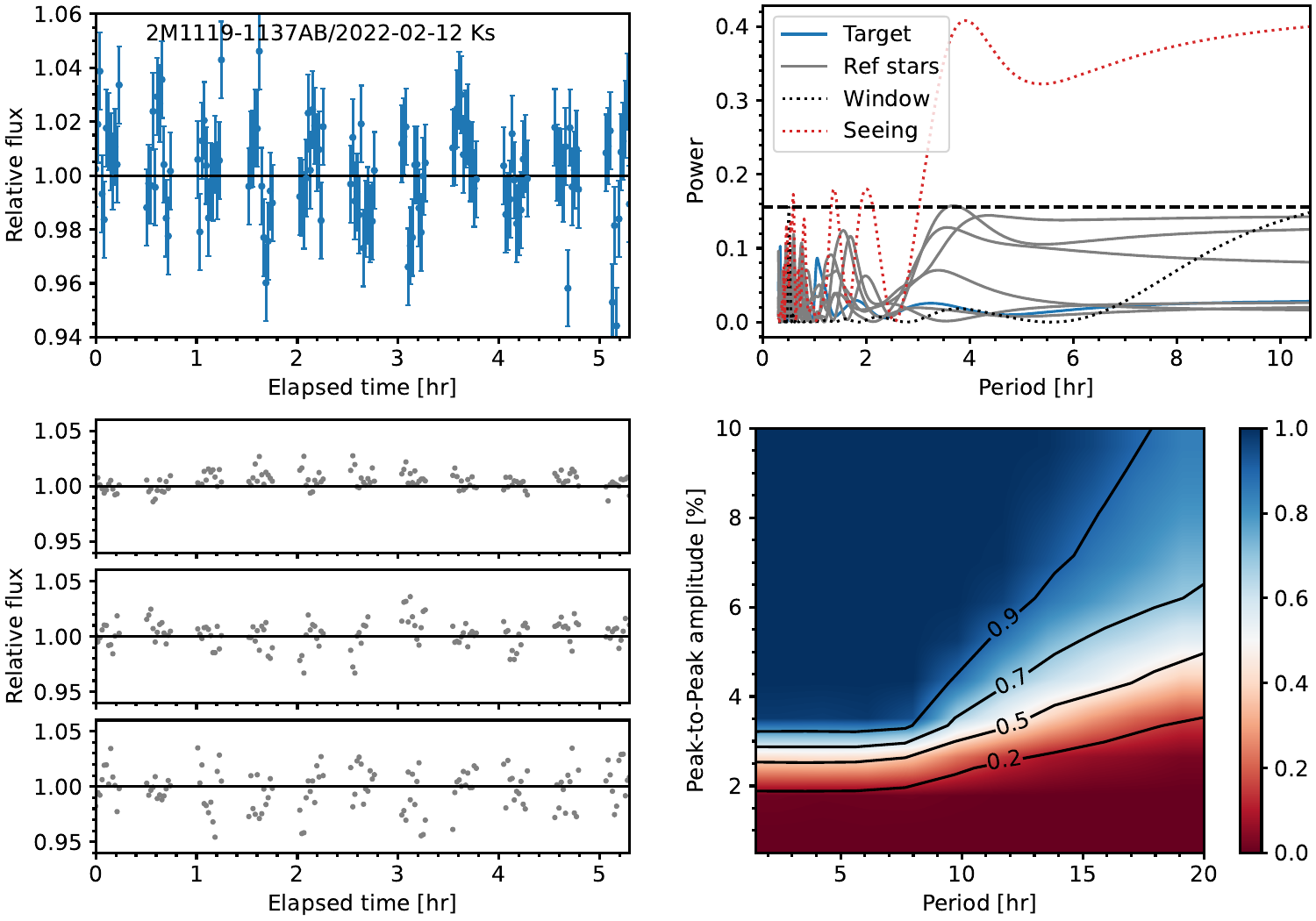}
    \includegraphics[width=1.8\columnwidth]{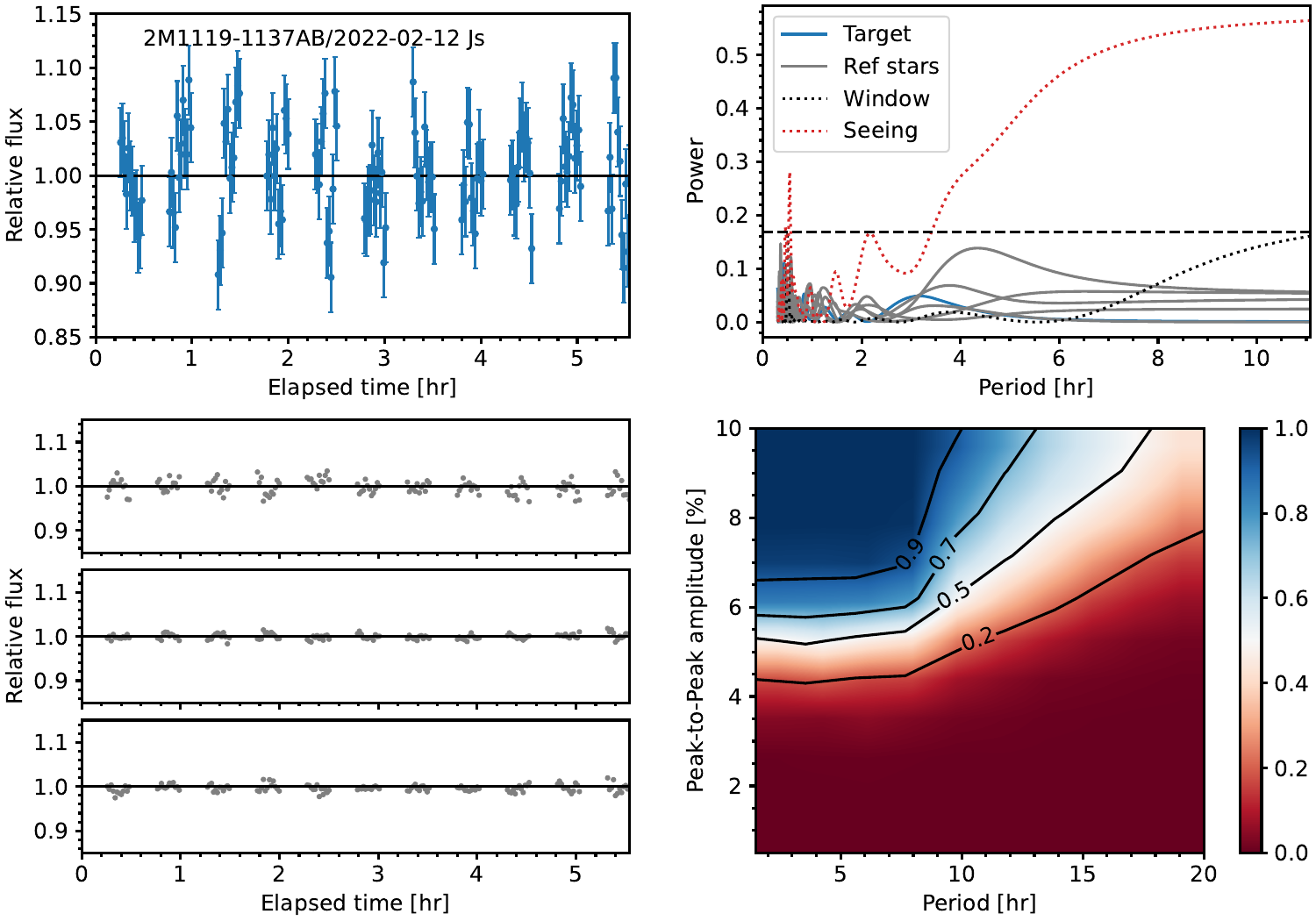}
    \caption{Interleaved $J_S$- and $K_S$-band non-variable light curves, periodograms, and sensitivity plots of a variable object 2M1119-1137AB, including detrended light curves and periodograms of its reference stars.}
    \label{fig:apd_2M1119-1137AB}
\end{figure*}

\begin{figure*}
    \centering
    \includegraphics[width=1.8\columnwidth]{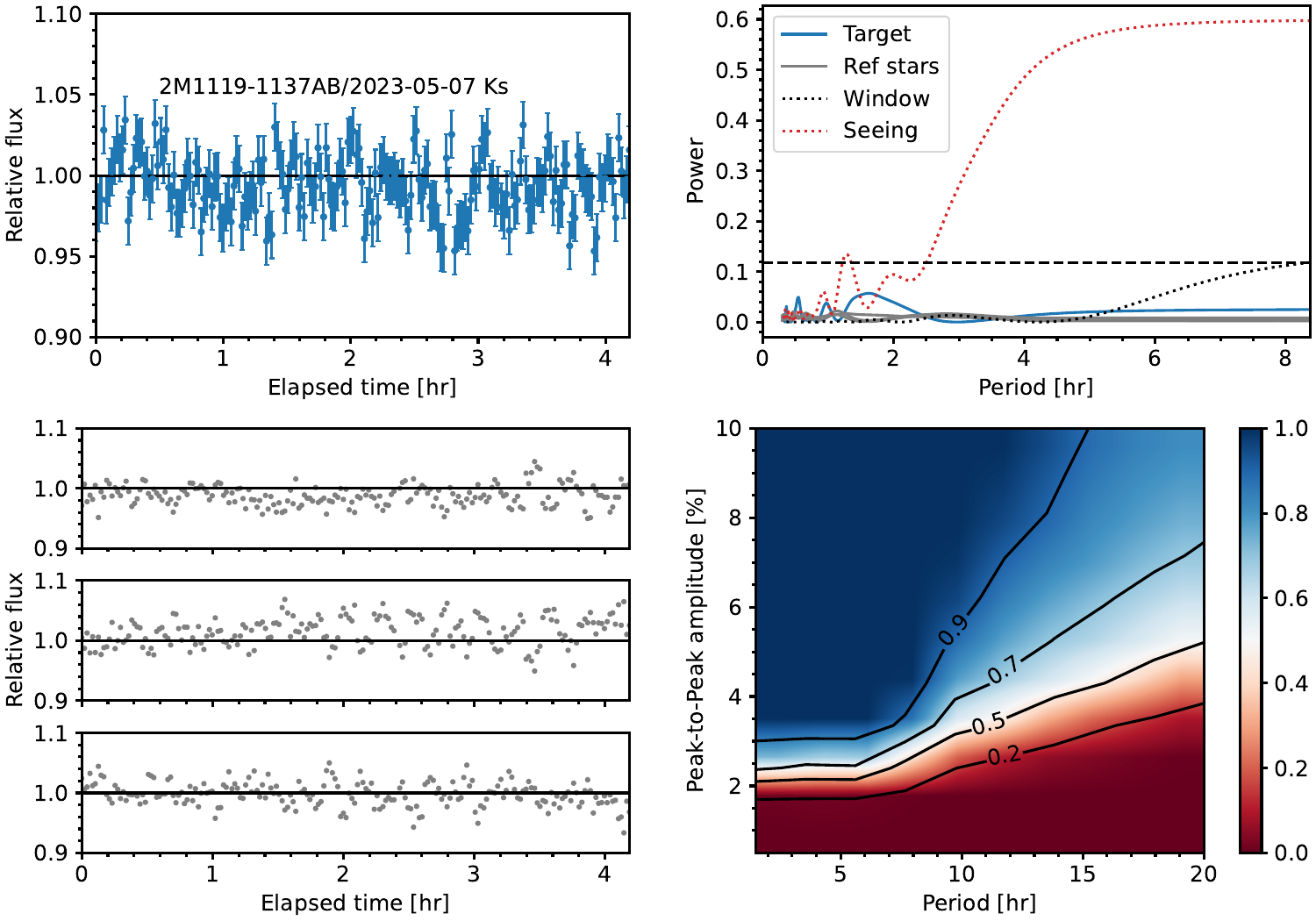}
    \includegraphics[width=1.8\columnwidth]{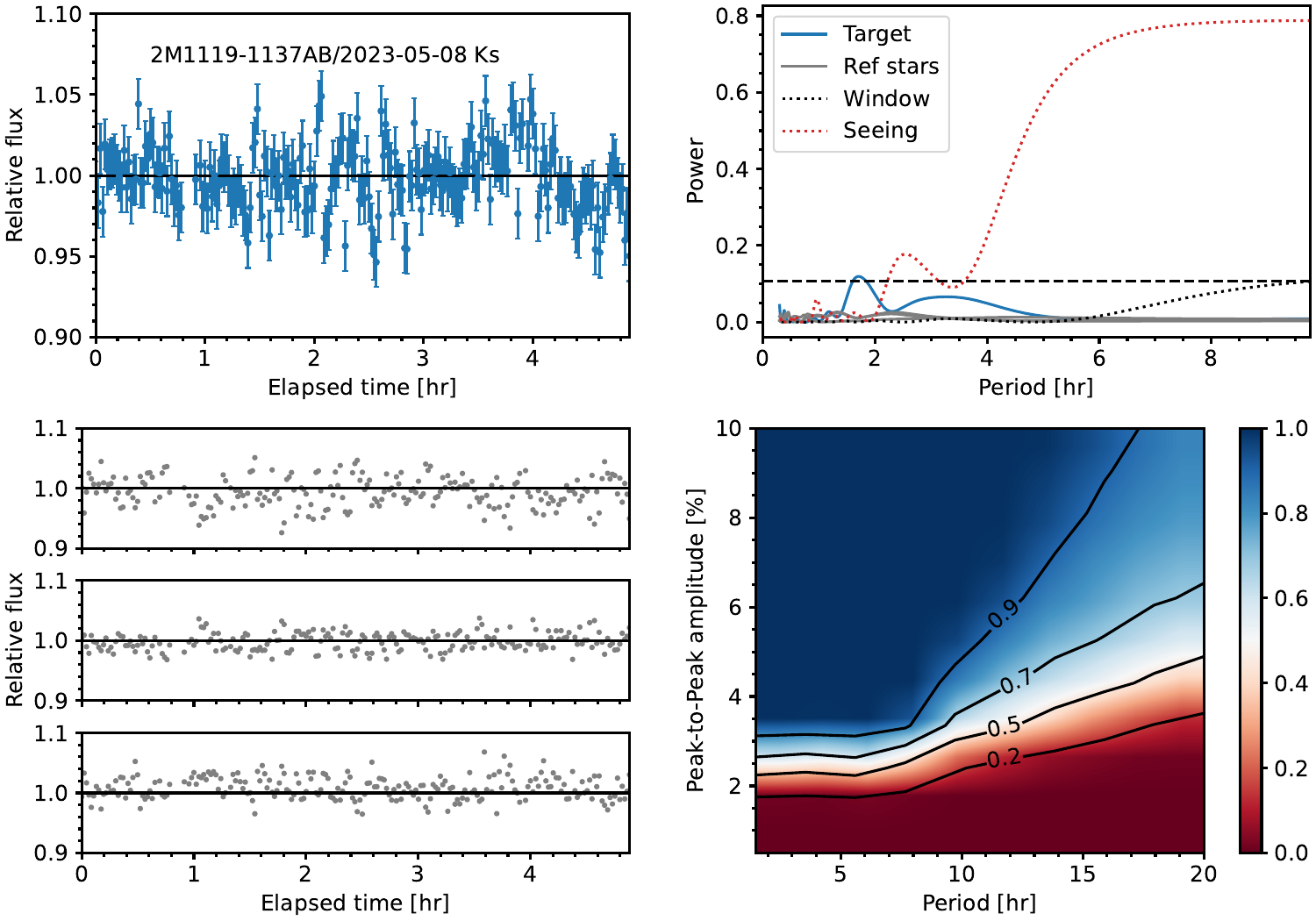}
    \caption{$K_S$-band non-variable light curves, periodograms, and sensitivity plots of a variable object 2M1119-1137AB, including detrended light curves and periodograms of its reference stars.}
    \label{fig:apd_2M1119-1137AB2}
\end{figure*}

\begin{figure*}
	\includegraphics[width=1.8\columnwidth]{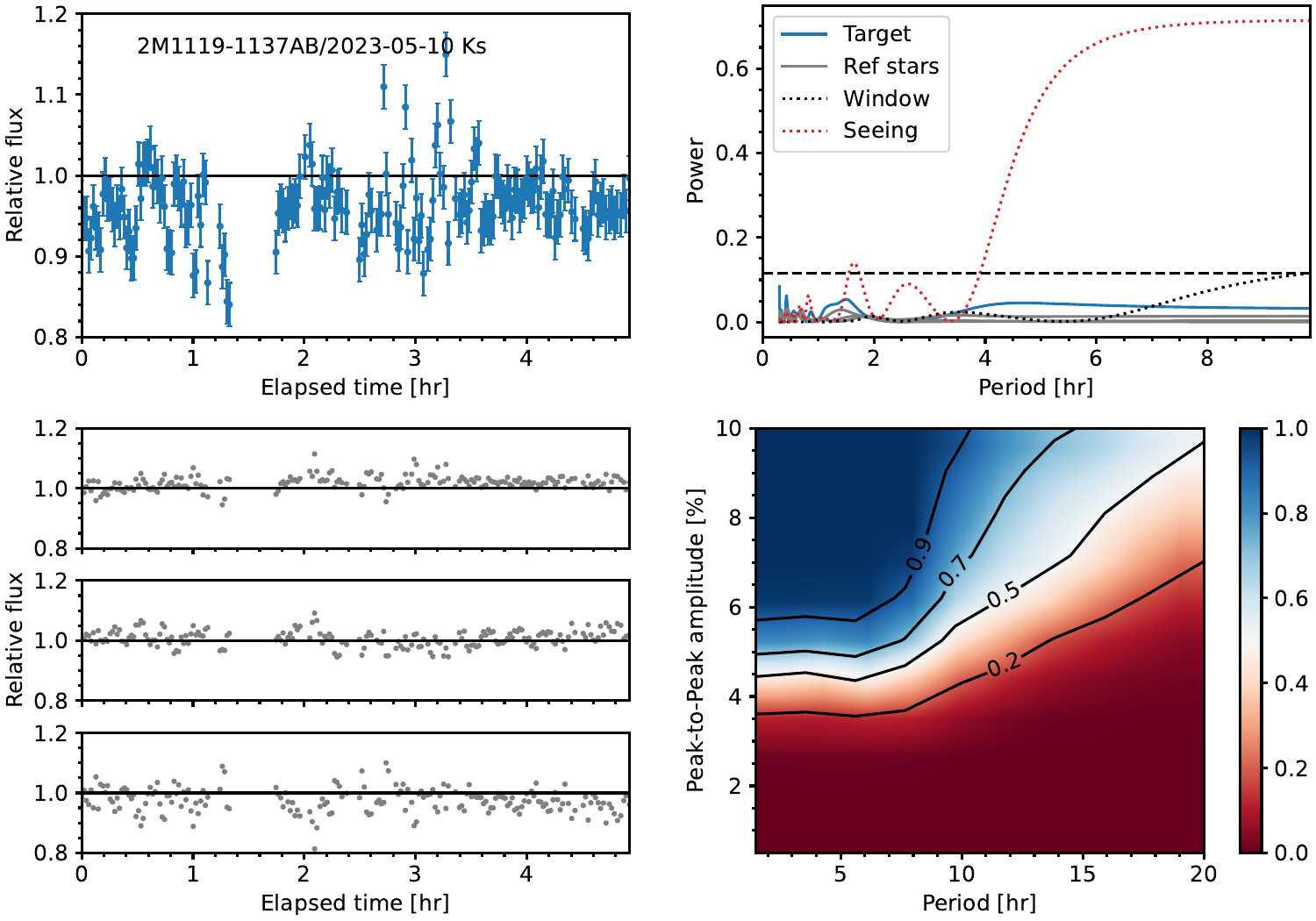}
    \includegraphics[width=1.8\columnwidth]{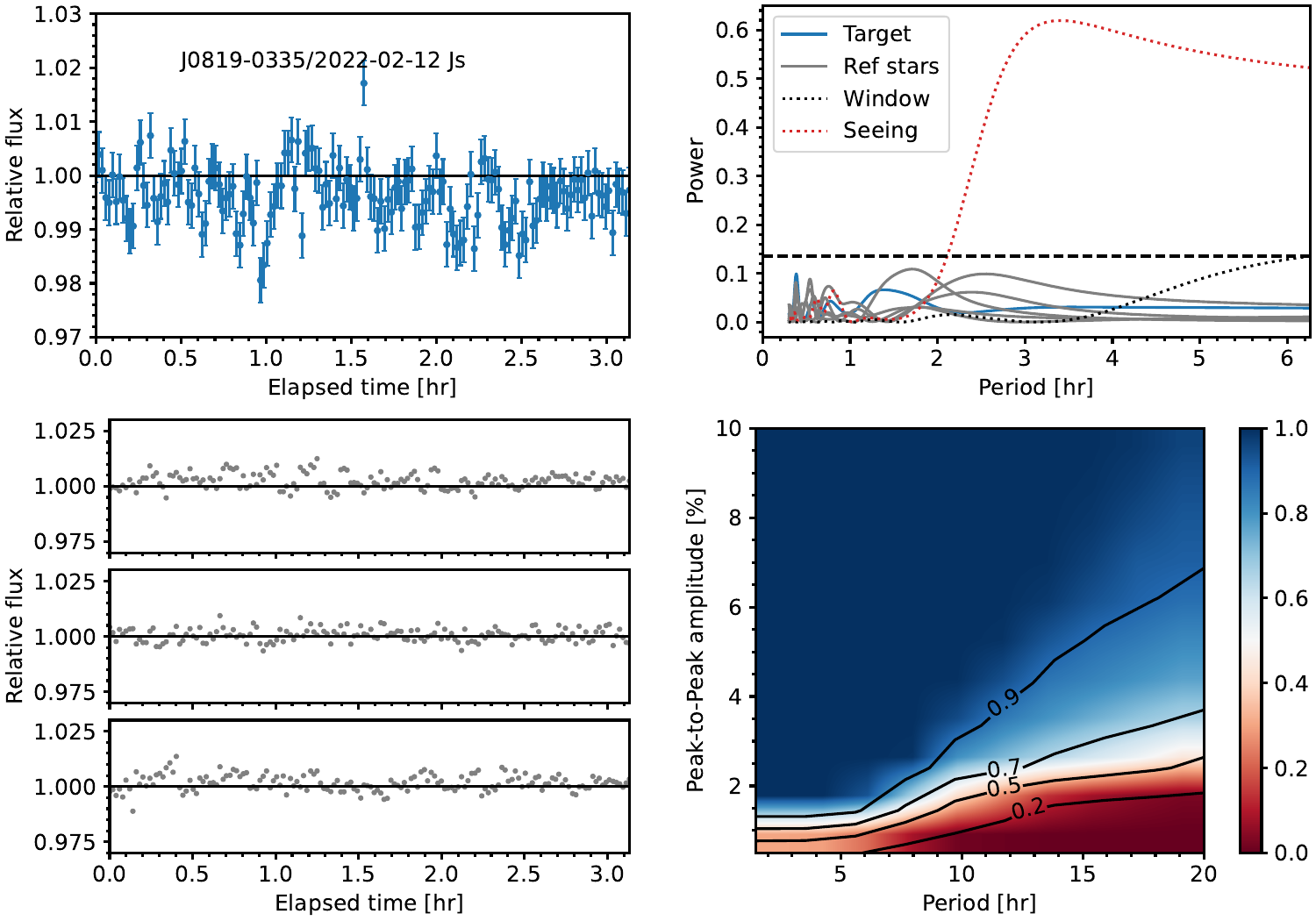}
    \caption{Non-variable light curves, periodograms, and sensitivity plots of two variable objects 2M1119-1137AB and J0819-0335, including detrended light curves and periodograms of their reference stars}
    \label{fig:apd_2M1119_J0819}
\end{figure*}

\begin{figure*}
    \centering
    \includegraphics[width=1.8\columnwidth]{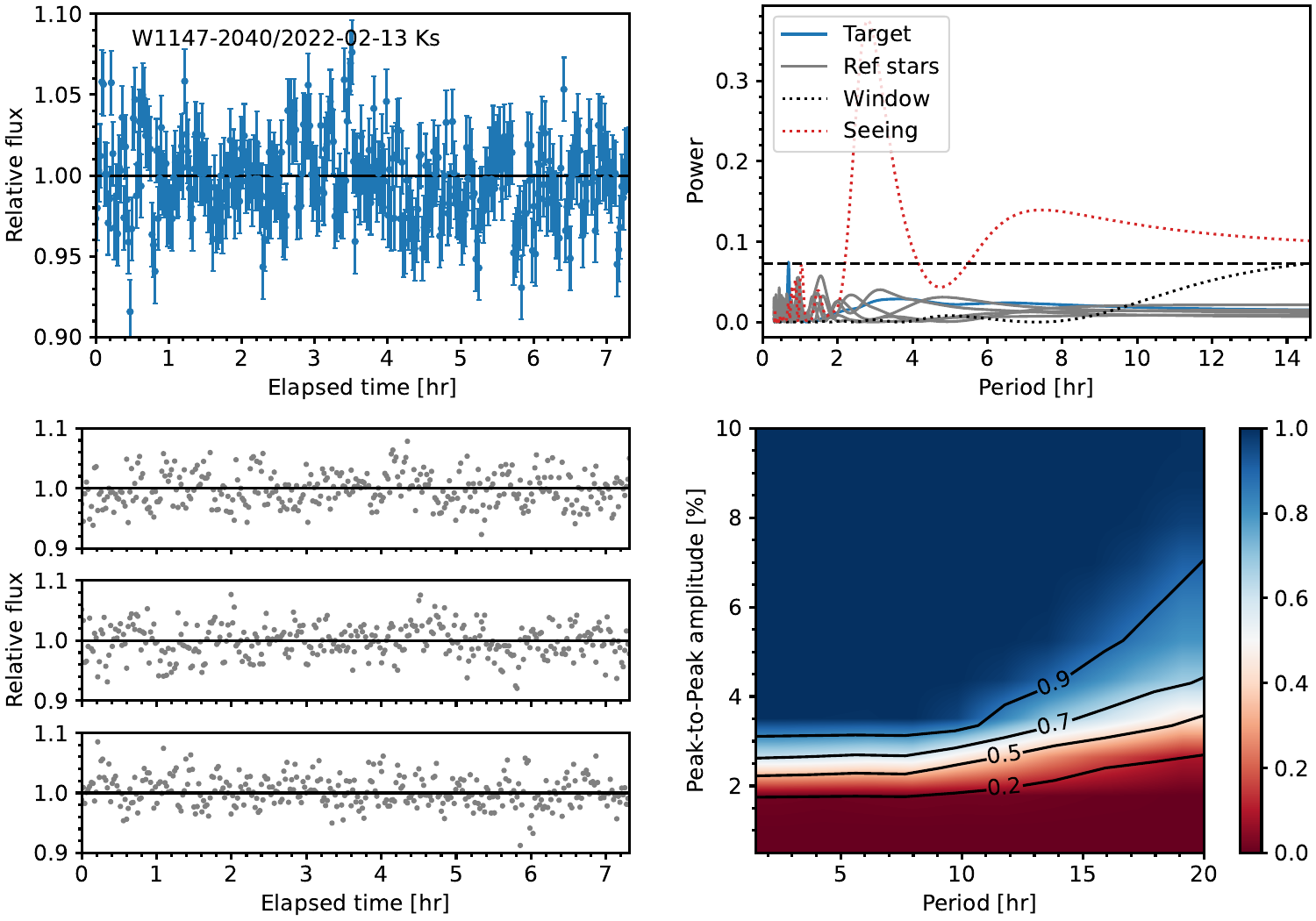}
    \includegraphics[width=1.8\columnwidth]{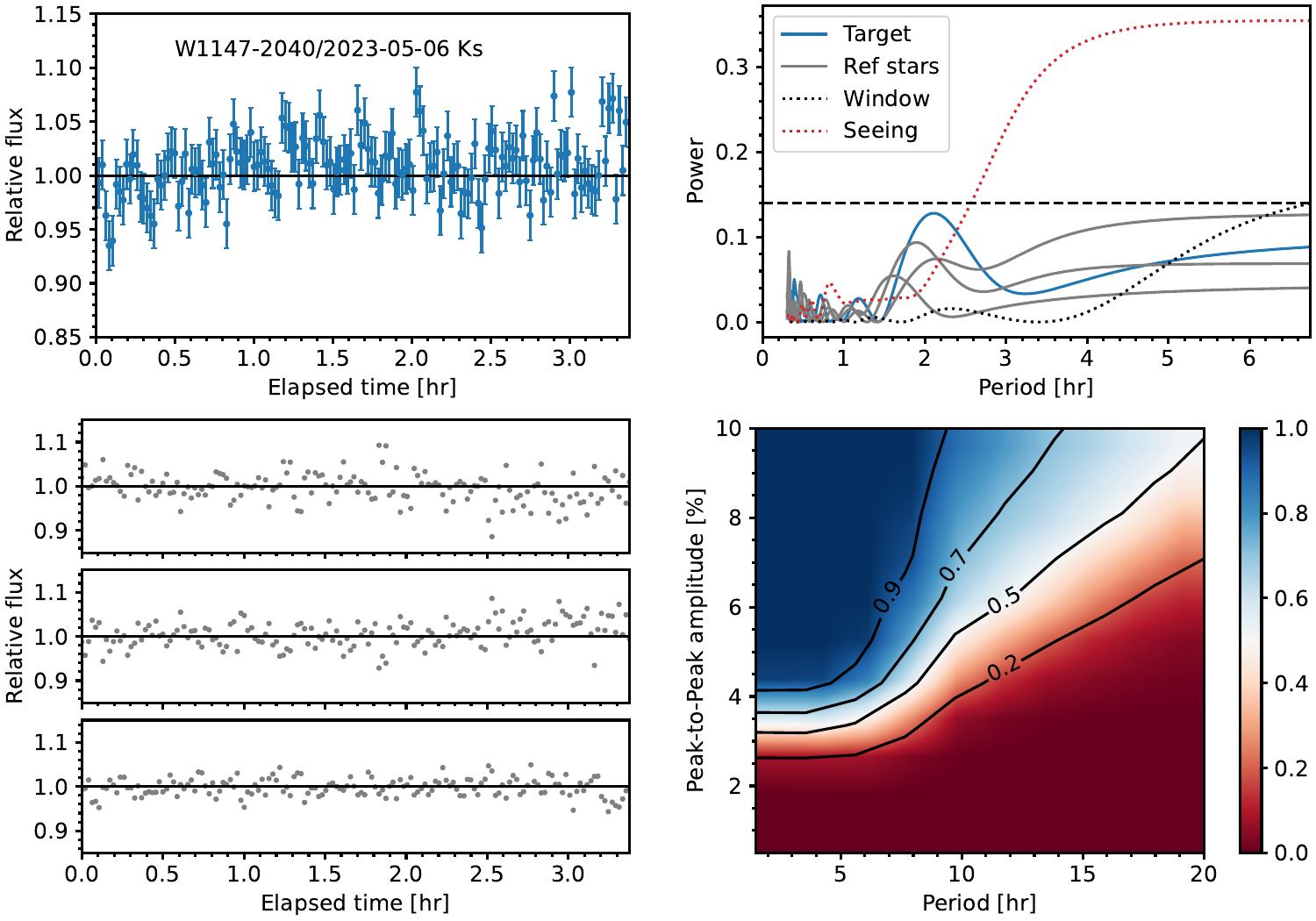}
    \caption{Non-variable light curves, periodograms, and sensitivity plots of a variable object W1147-2040, including detrended light curves and periodograms of its reference stars}
    \label{fig:apd_W1147-2040}
\end{figure*}

\begin{figure*}
    \centering
    \includegraphics[width=1.8\columnwidth]{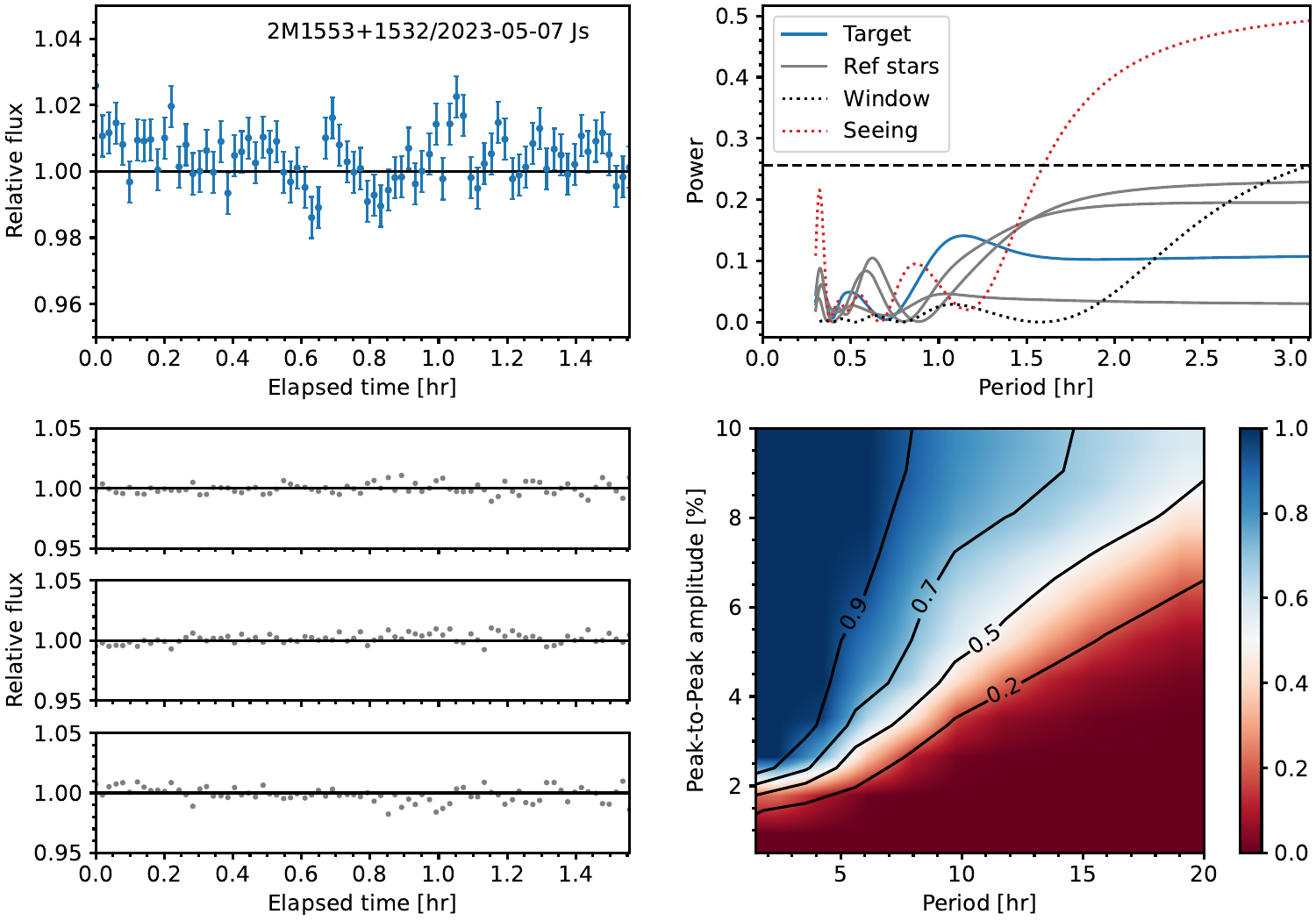}
    \includegraphics[width=1.8\columnwidth]{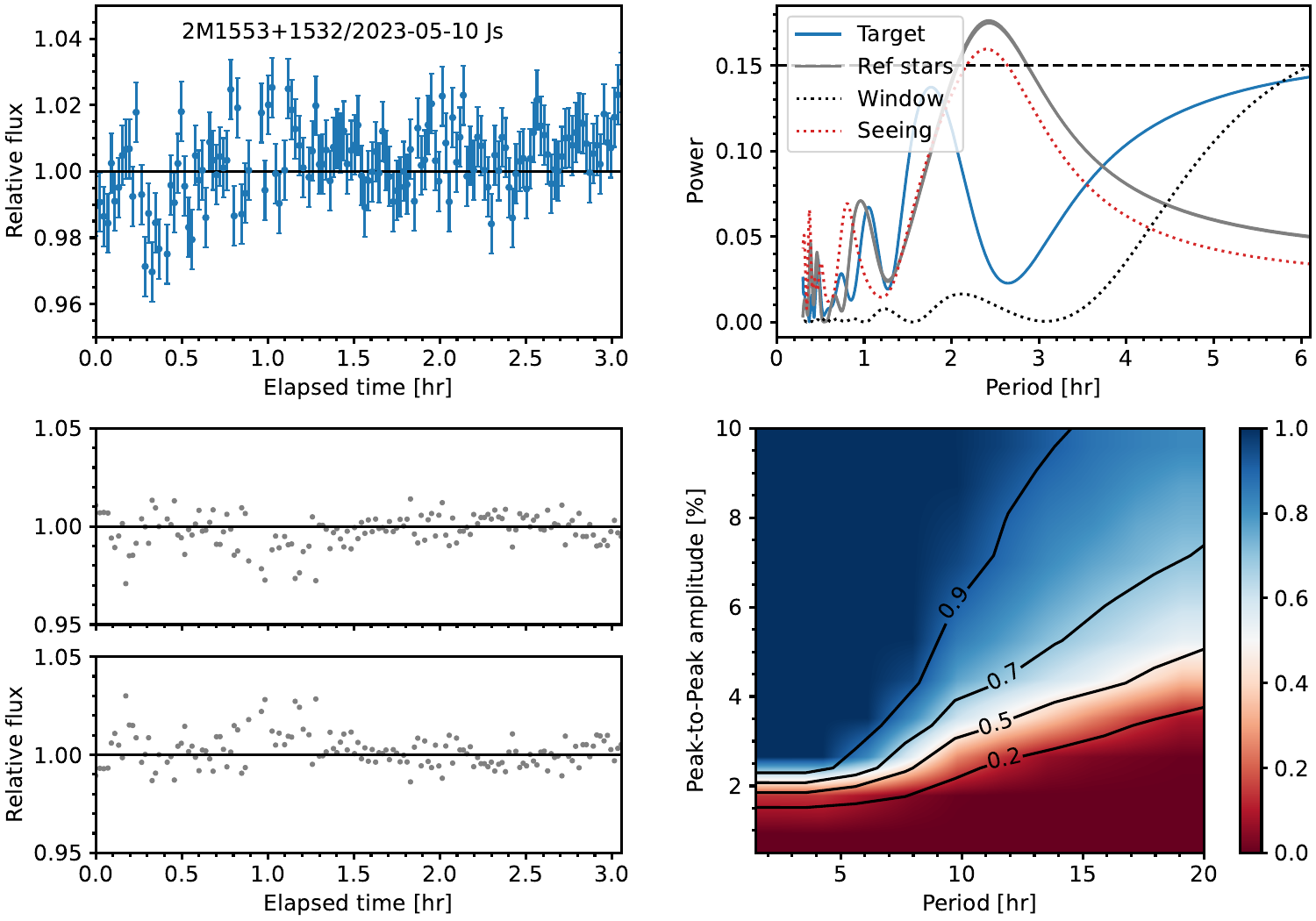}
    \caption{Non-variable light curves, periodograms, and sensitivity plots of a variable object 2M1552+1532, including detrended light curves and periodograms of its reference stars}
    \label{fig:apd_2M1553_non}
\end{figure*}

\begin{figure*}
    \centering
    \includegraphics[width=1.8\columnwidth]
    {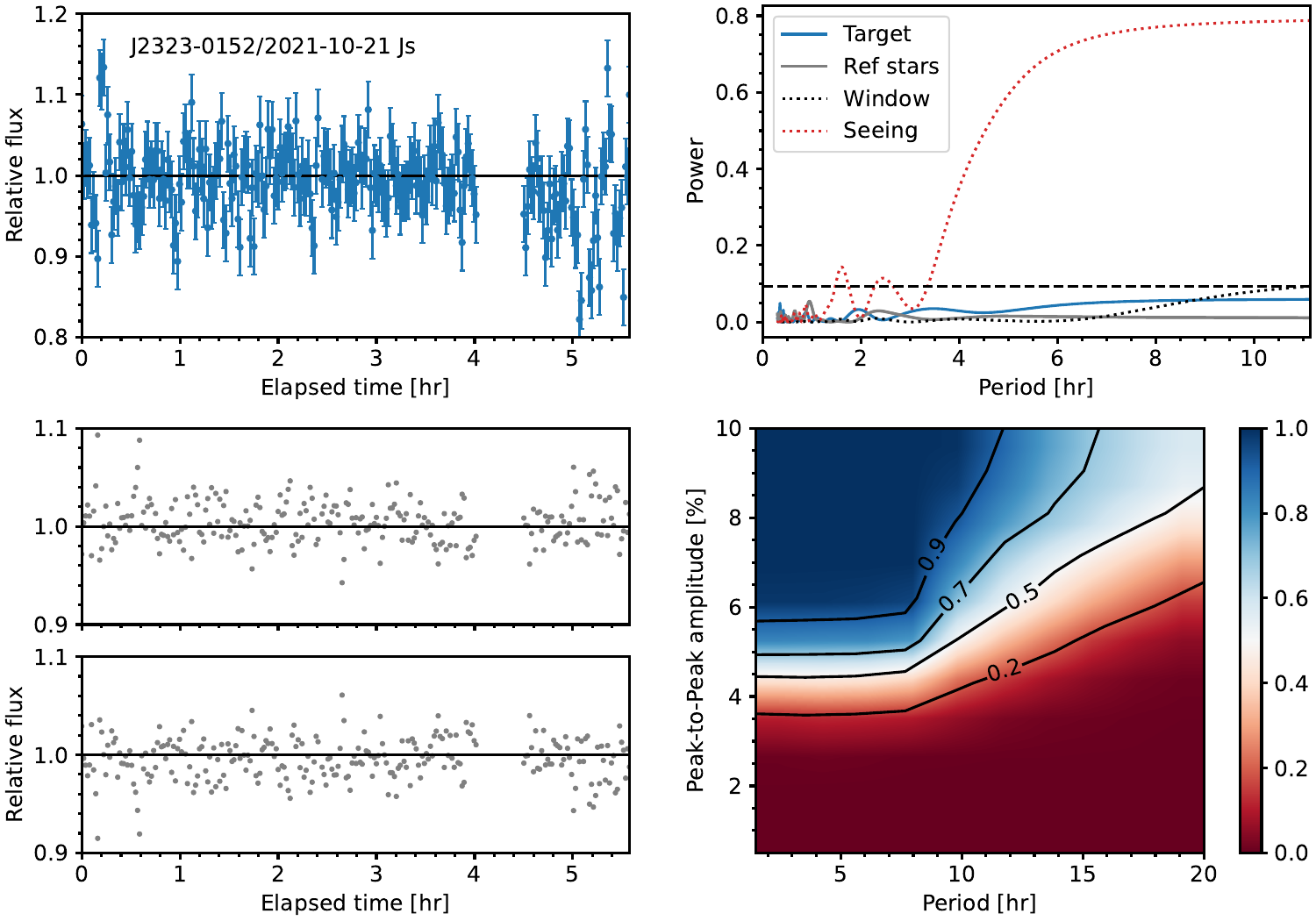}
    \includegraphics[width=1.8\columnwidth]{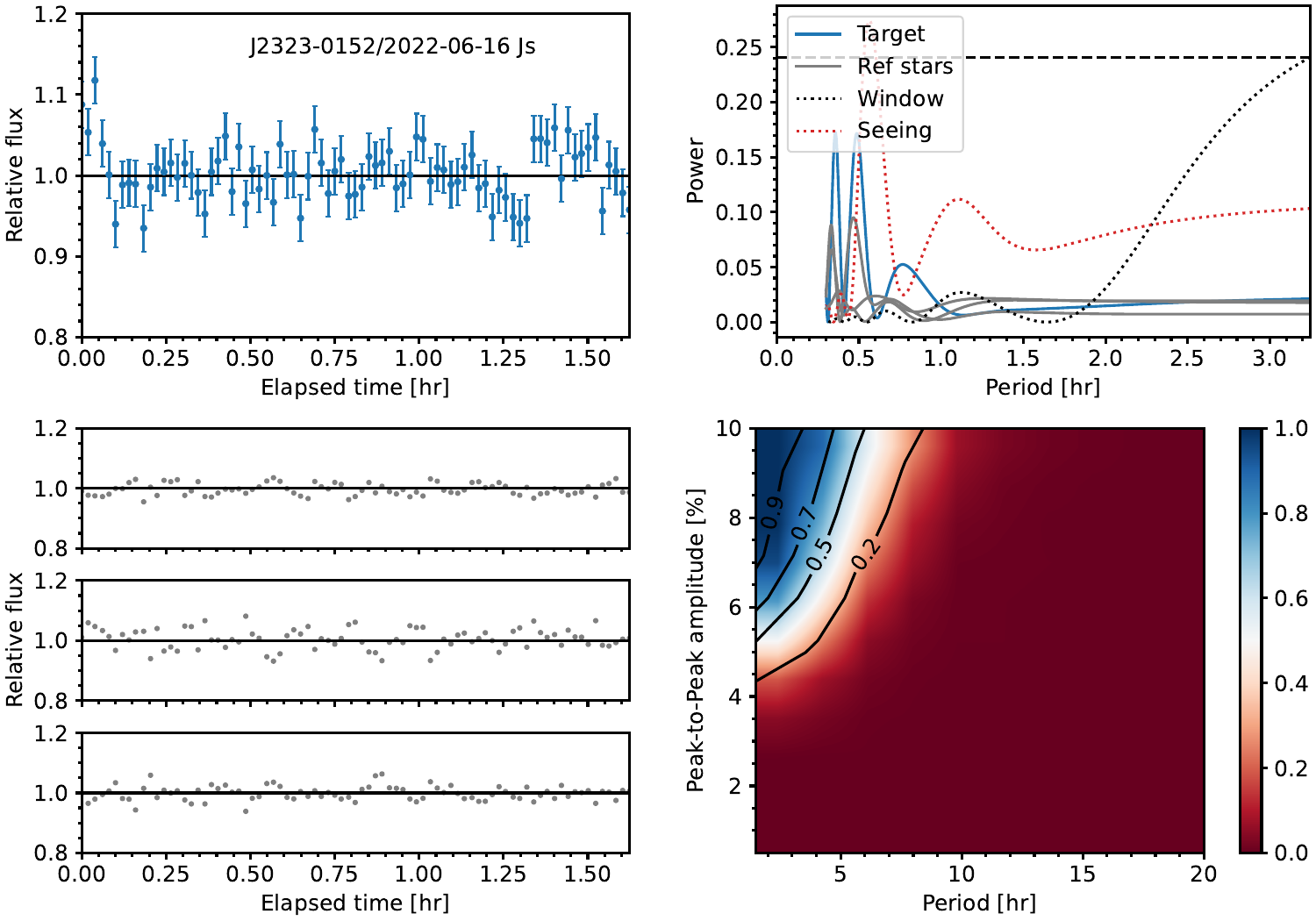}
    \caption{Non-variable light curves, periodograms, and sensitivity plots of a variable object J2323-0152, including detrended light curves and periodograms of its reference stars.}
    \label{fig:apd_J2323-0152}
\end{figure*}

\begin{figure*}
    \centering
    \includegraphics[width=1.8\columnwidth]{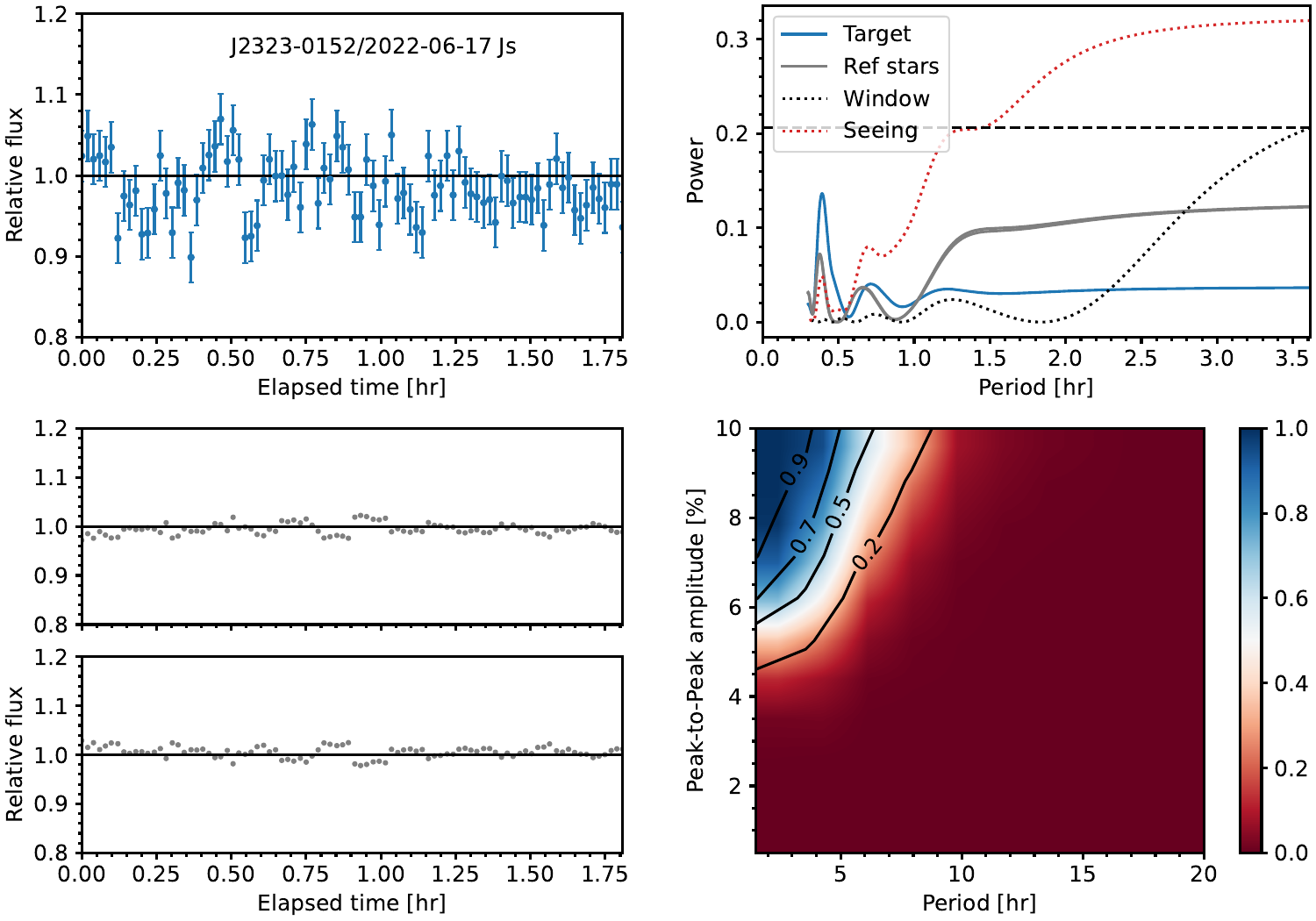}
    \includegraphics[width=1.8\columnwidth]{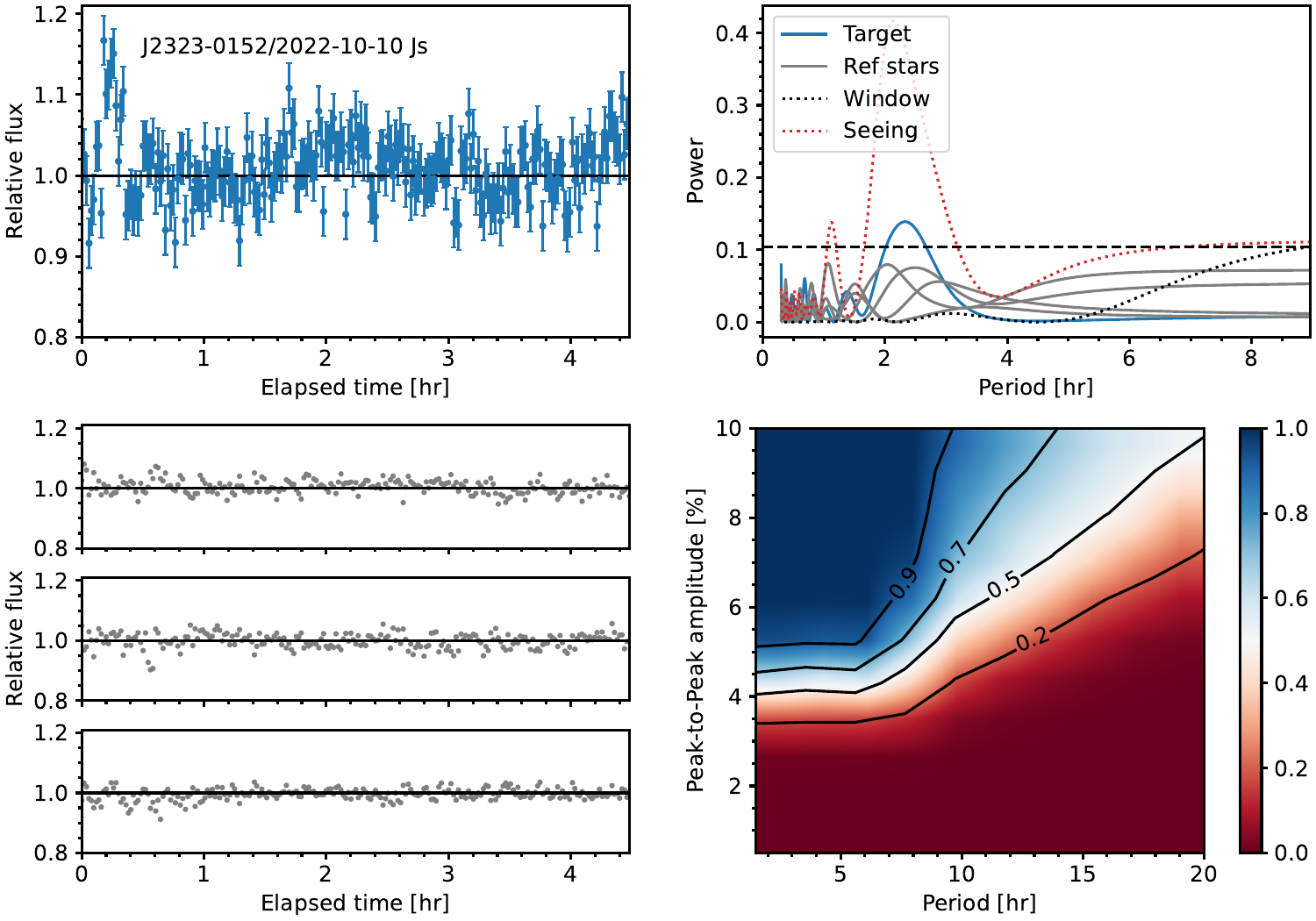}
    \contcaption{Non-variable light curves, periodograms, and sensitivity plots of a variable object J2323-0152, including detrended light curves and periodograms of its reference stars.}
    \label{fig:apd_J2323-0152}
\end{figure*}

\begin{figure*}
    \centering
    \includegraphics[width=1.8\columnwidth]{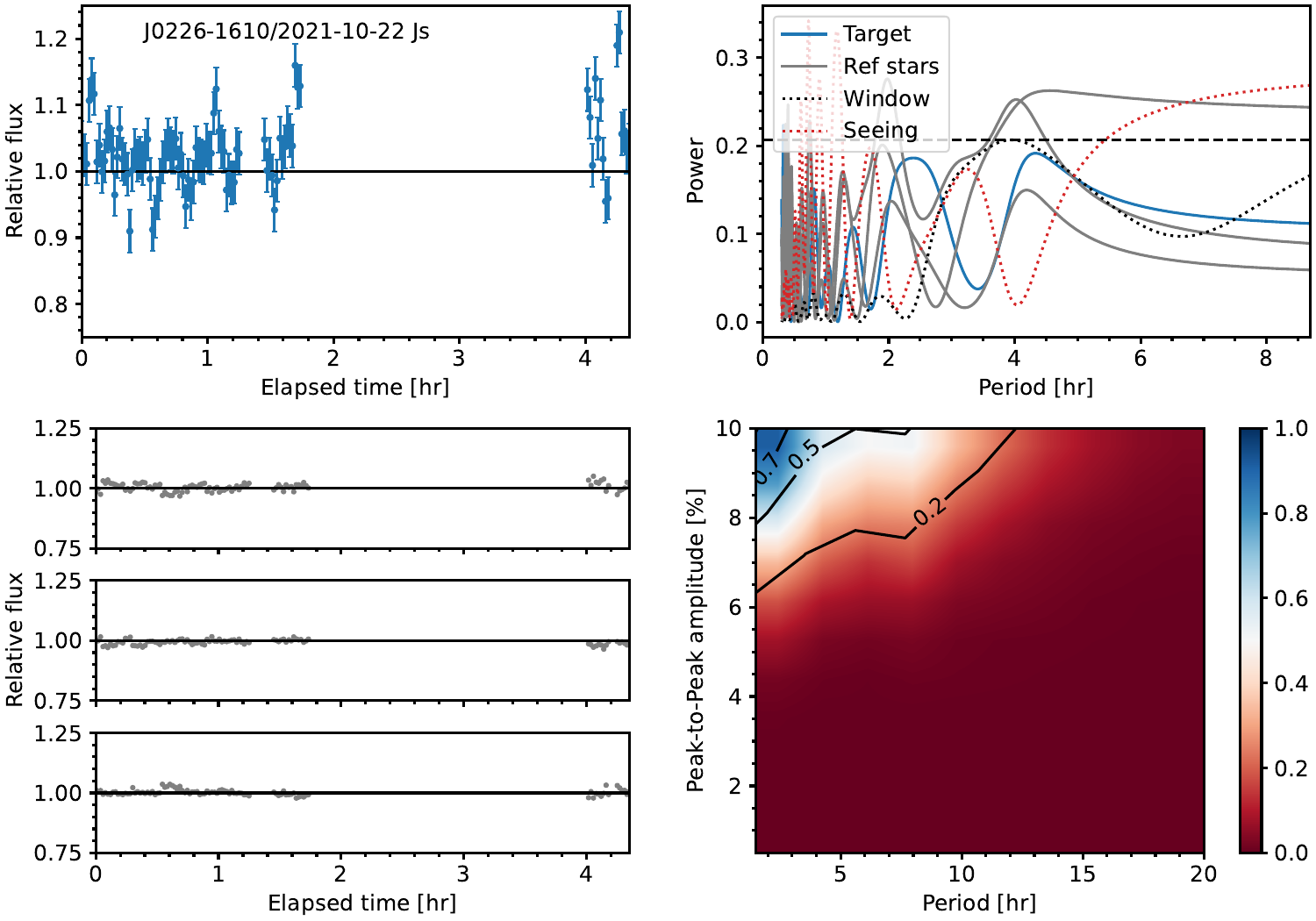}
    \includegraphics[width=1.8\columnwidth]{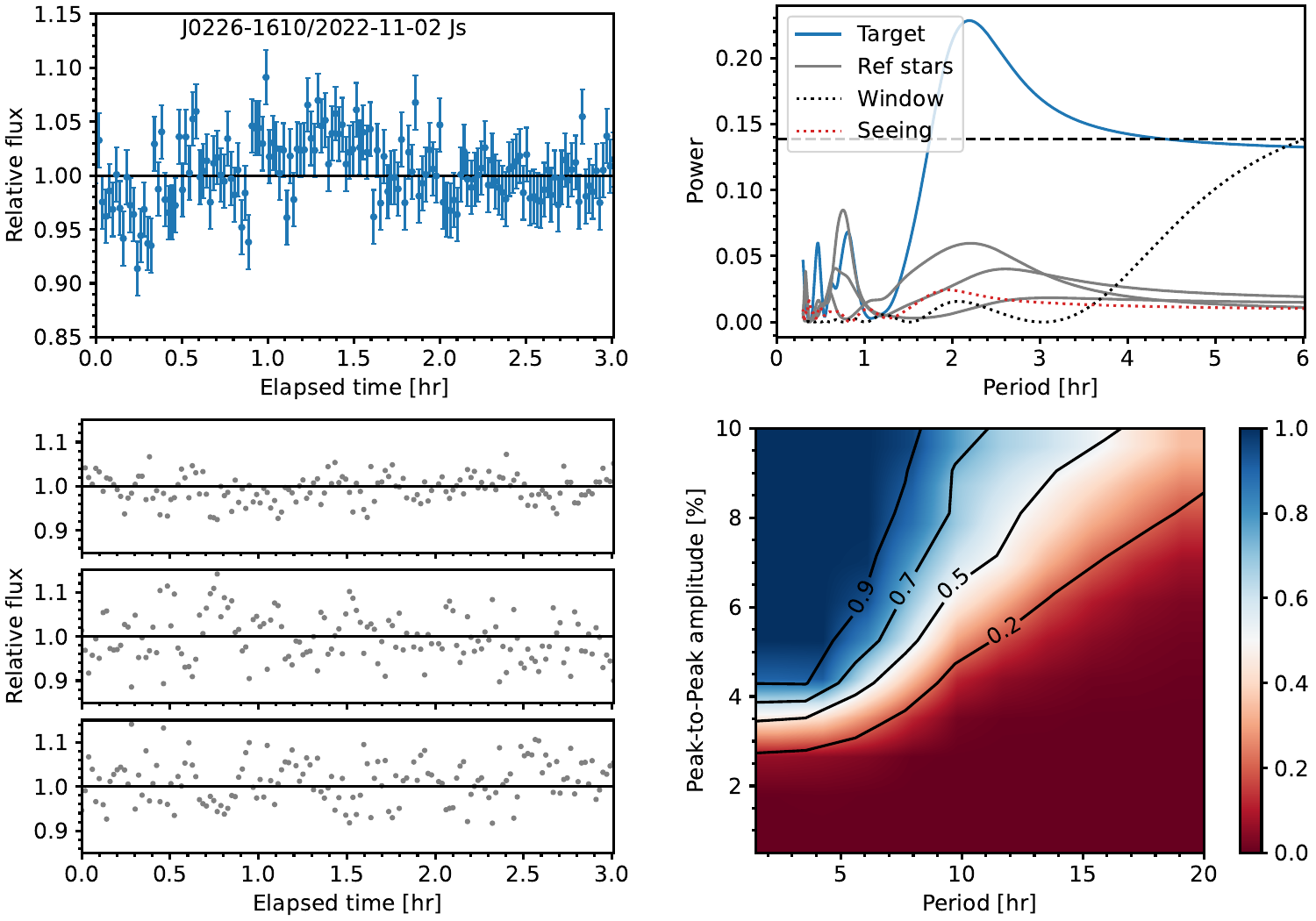}
    \caption{Non-variable light curves, periodograms, and sensitivity plots of a variable candidate J0200-1610, including detrended light curves and periodograms of its reference stars.}
    \label{fig:apd_J0200-1610}
\end{figure*}

\begin{figure*}
    \centering
    \includegraphics[width=1.8\columnwidth]{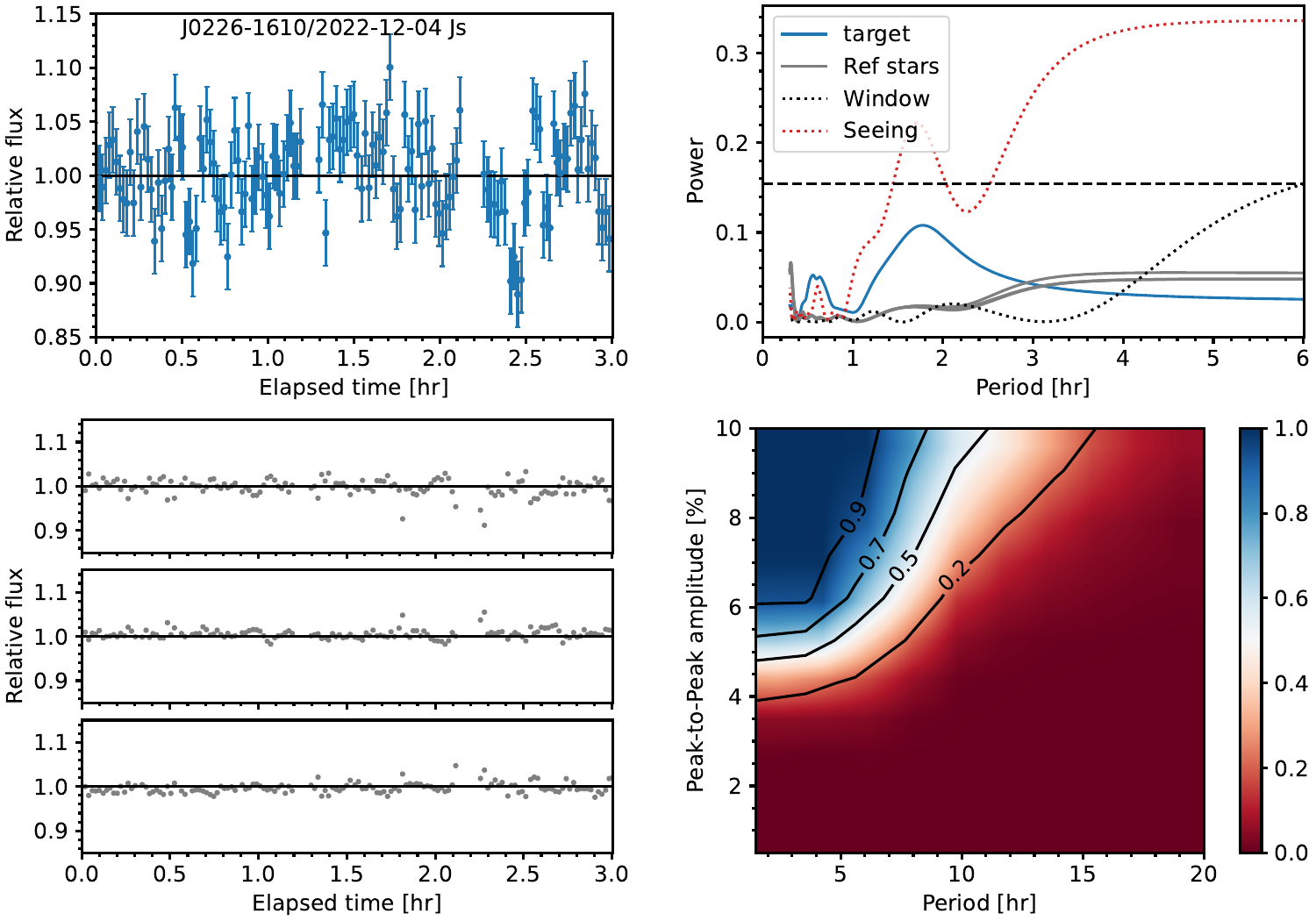}
    \contcaption{Non-variable light curves, periodograms, and sensitivity plots of a potentially variable candidate J0200-1610, including detrended light curves and periodograms of its reference stars.}
    \label{fig:apd_J0200-1610}
\end{figure*}

%non-variable objects
\begin{figure*}
    \centering
    \includegraphics[width=1.8\columnwidth]{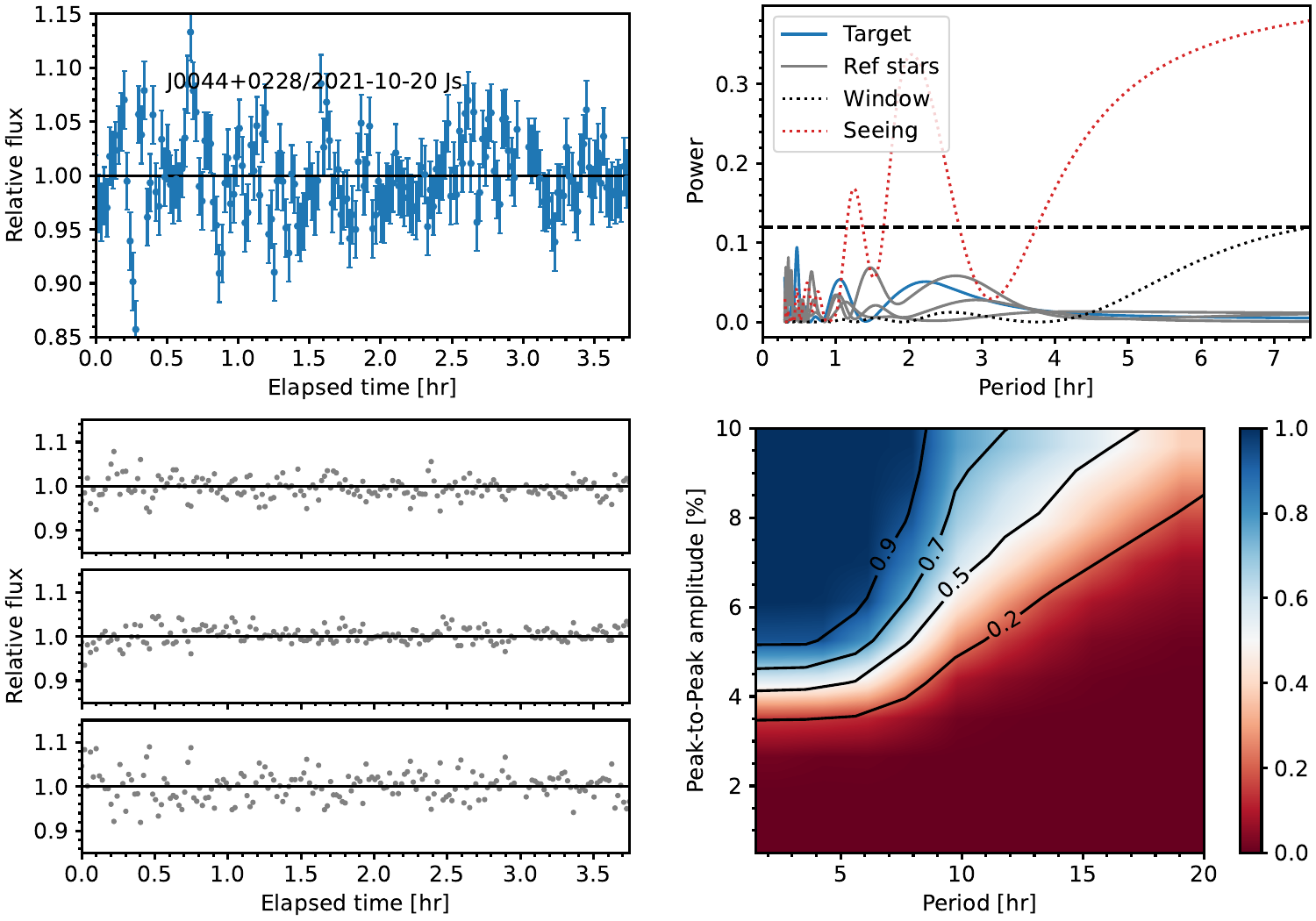}
    \includegraphics[width=1.8\columnwidth]{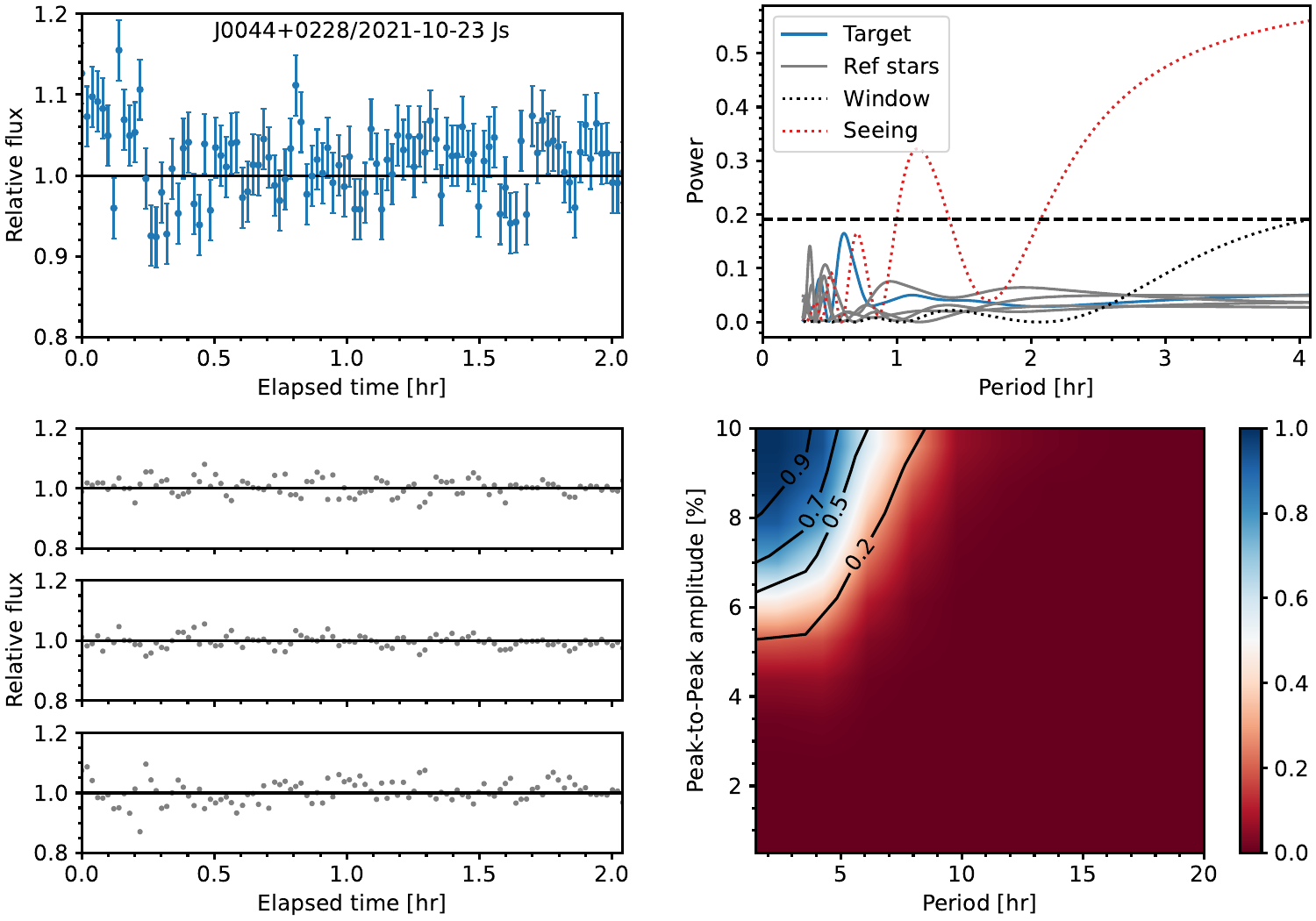}
    \caption{Light curves, periodograms, and sensitivity plots of a non-variable object J0044+0228, including detrended light curves and periodograms of its reference stars.}
    \label{fig:apd_J0044}
\end{figure*}

\begin{figure*}
    \centering
    \includegraphics[width=1.8\columnwidth]{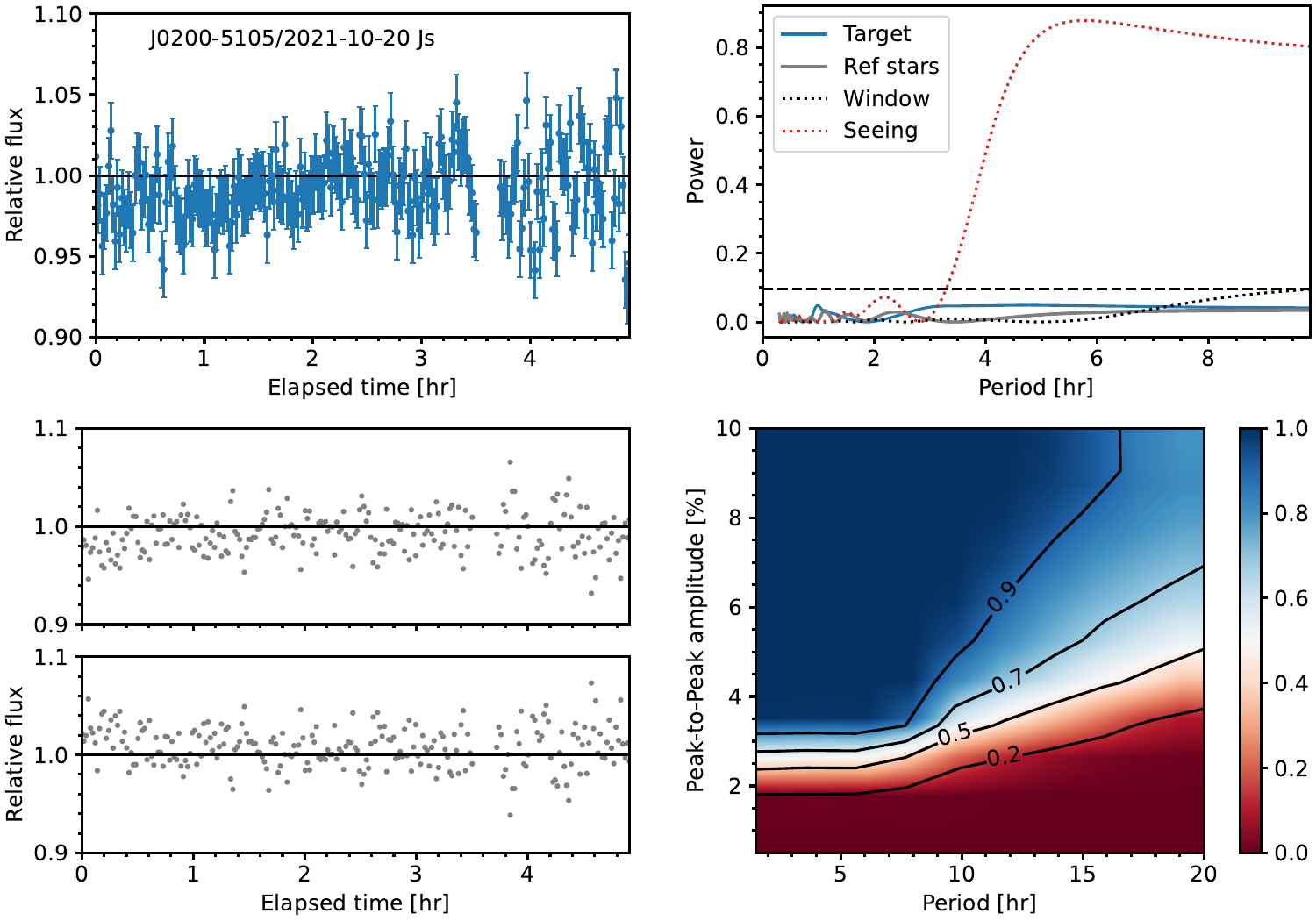}
    \includegraphics[width=1.8\columnwidth]{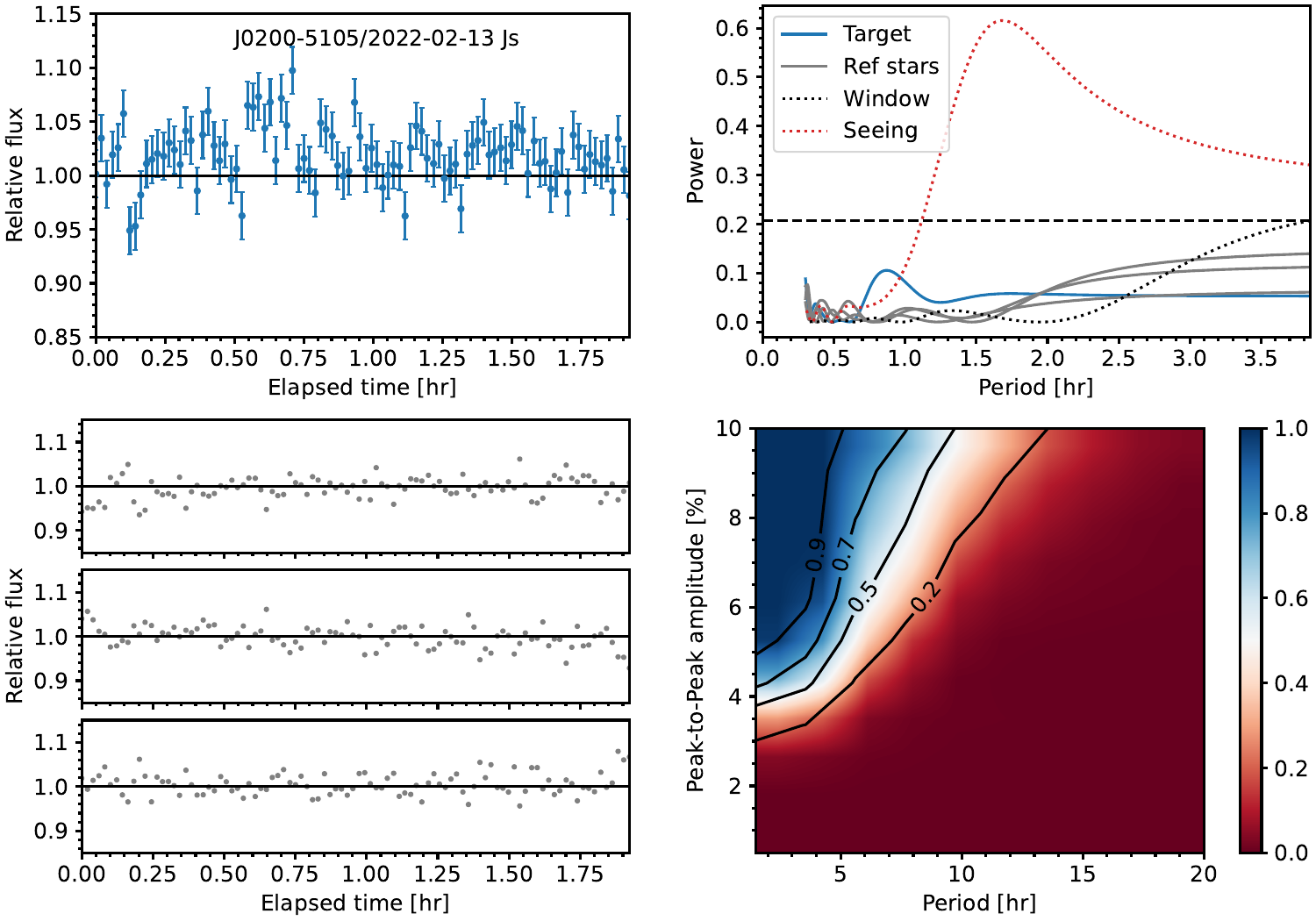}
    \caption{Light curves, periodograms, and sensitivity plots of a non-variable object J0200-5105, including detrended light curves and periodograms of its reference stars.}
    \label{fig:apd_J0200-5105}
\end{figure*}

\begin{figure*}
    \centering
    \includegraphics[width=1.8\columnwidth]{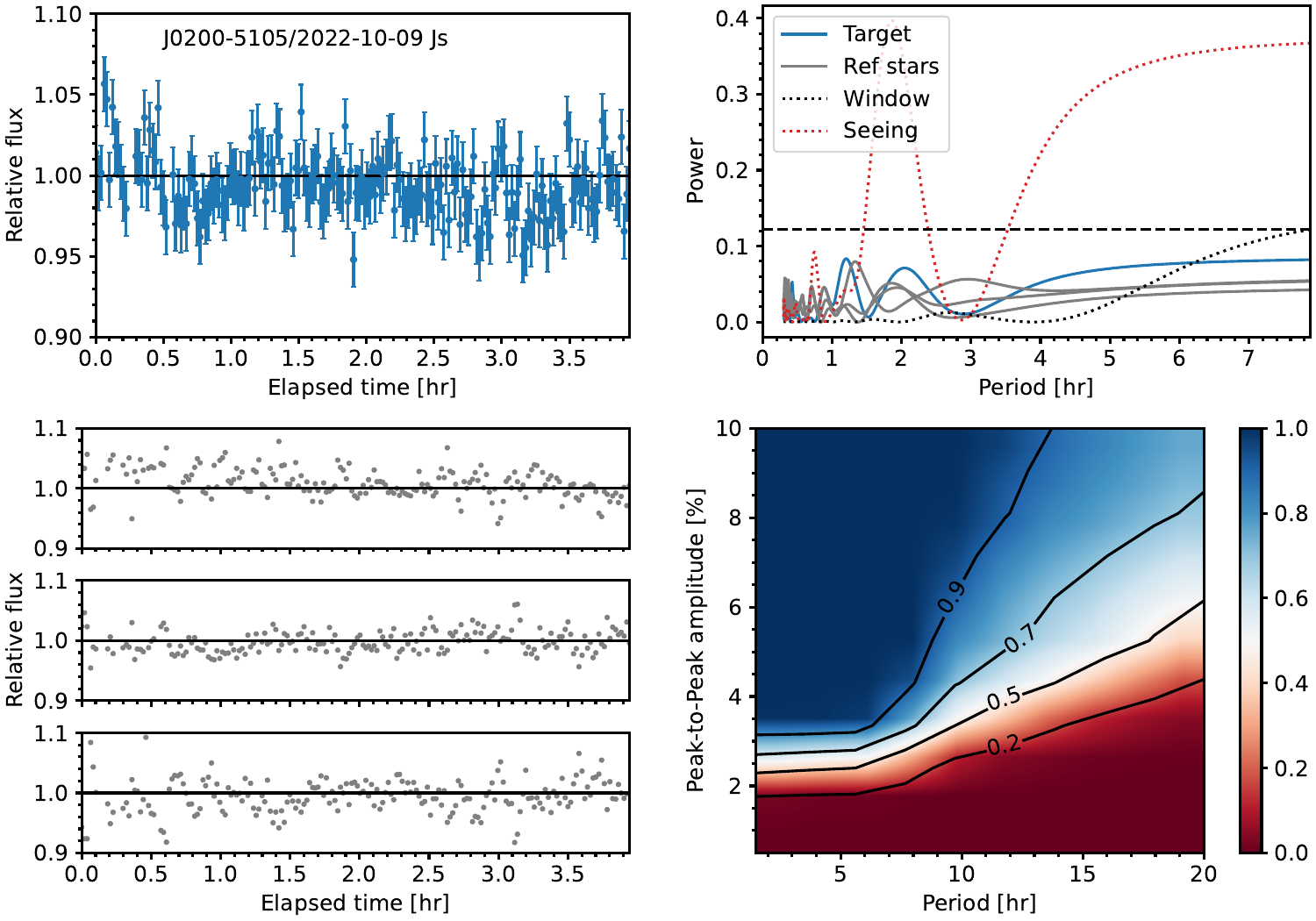}
    \includegraphics[width=1.8\columnwidth]{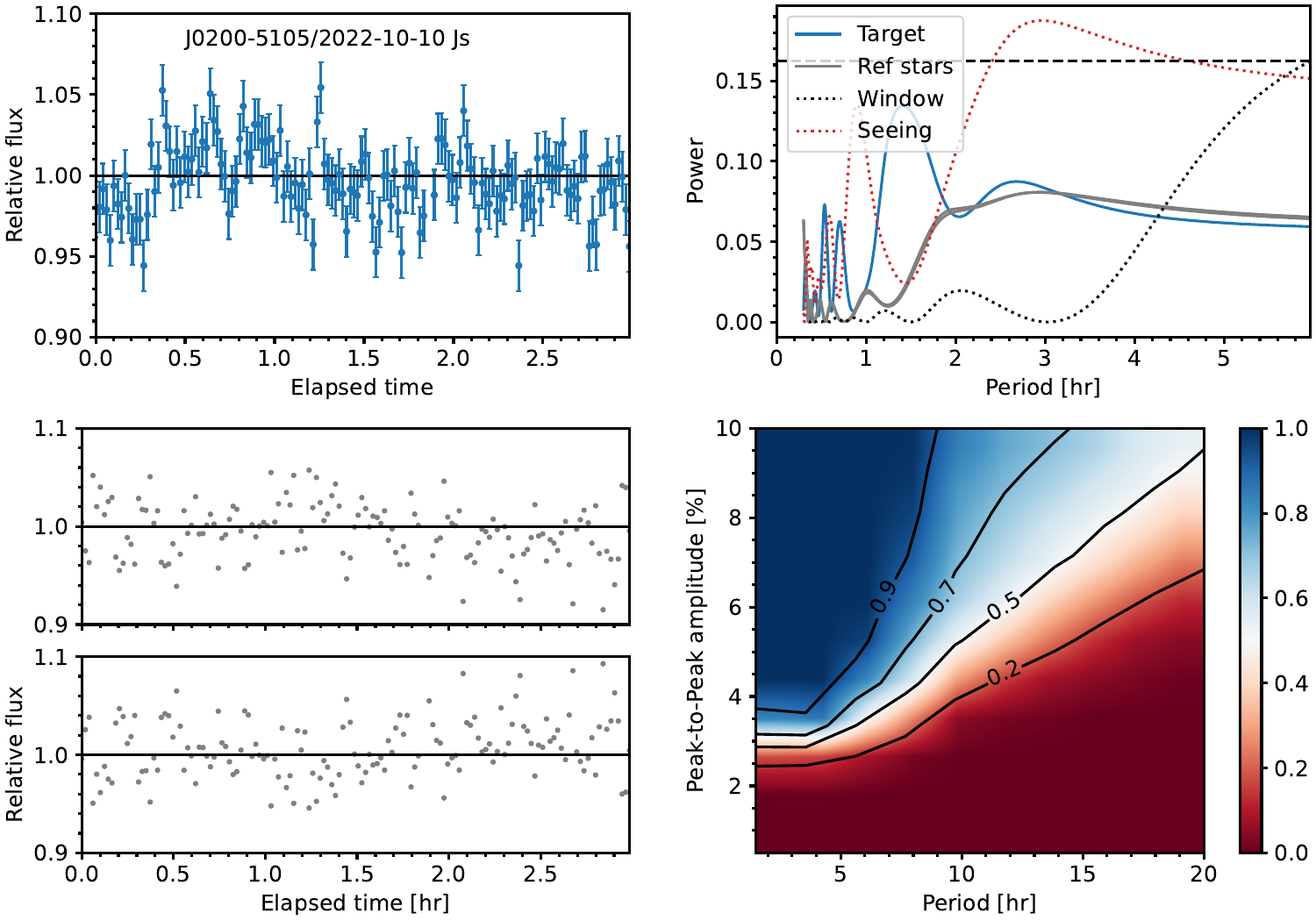}
    \contcaption{Light curves, periodograms, and sensitivity plots of a non-variable object J0200-5105, including detrended light curves and periodograms of its reference stars.}
    \label{fig:apd_J0200-5105}
\end{figure*}

\begin{figure*}
    \centering
    \includegraphics[width=1.8\columnwidth]{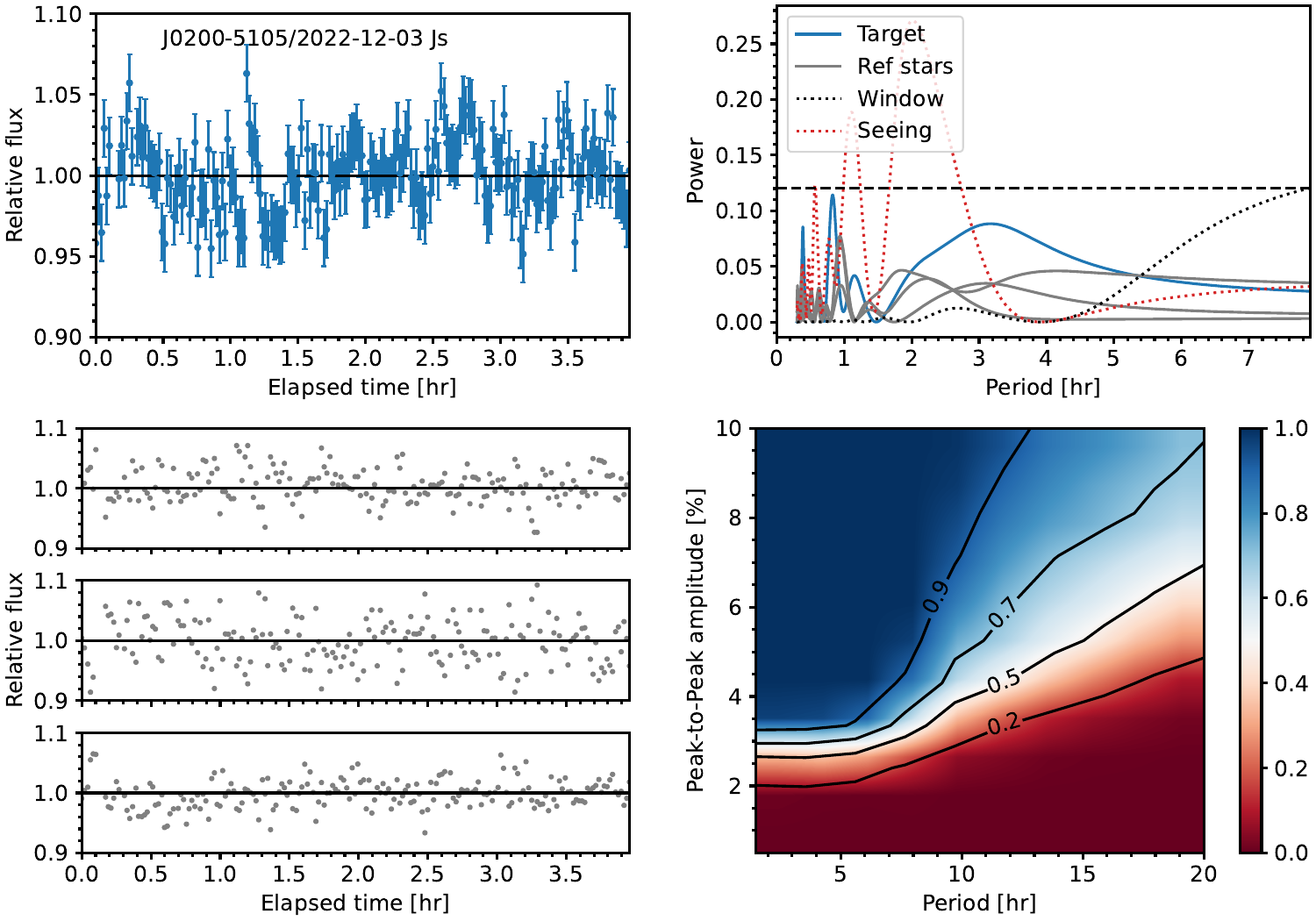}
    \contcaption{Light curves, periodograms, and sensitivity plots of a non-variable object J0200-5105, including detrended light curves and periodograms of its reference stars.}
    \label{fig:apd_J0200-5105}
\end{figure*}

\begin{figure*}
    \centering
    \includegraphics[width=1.8\columnwidth]{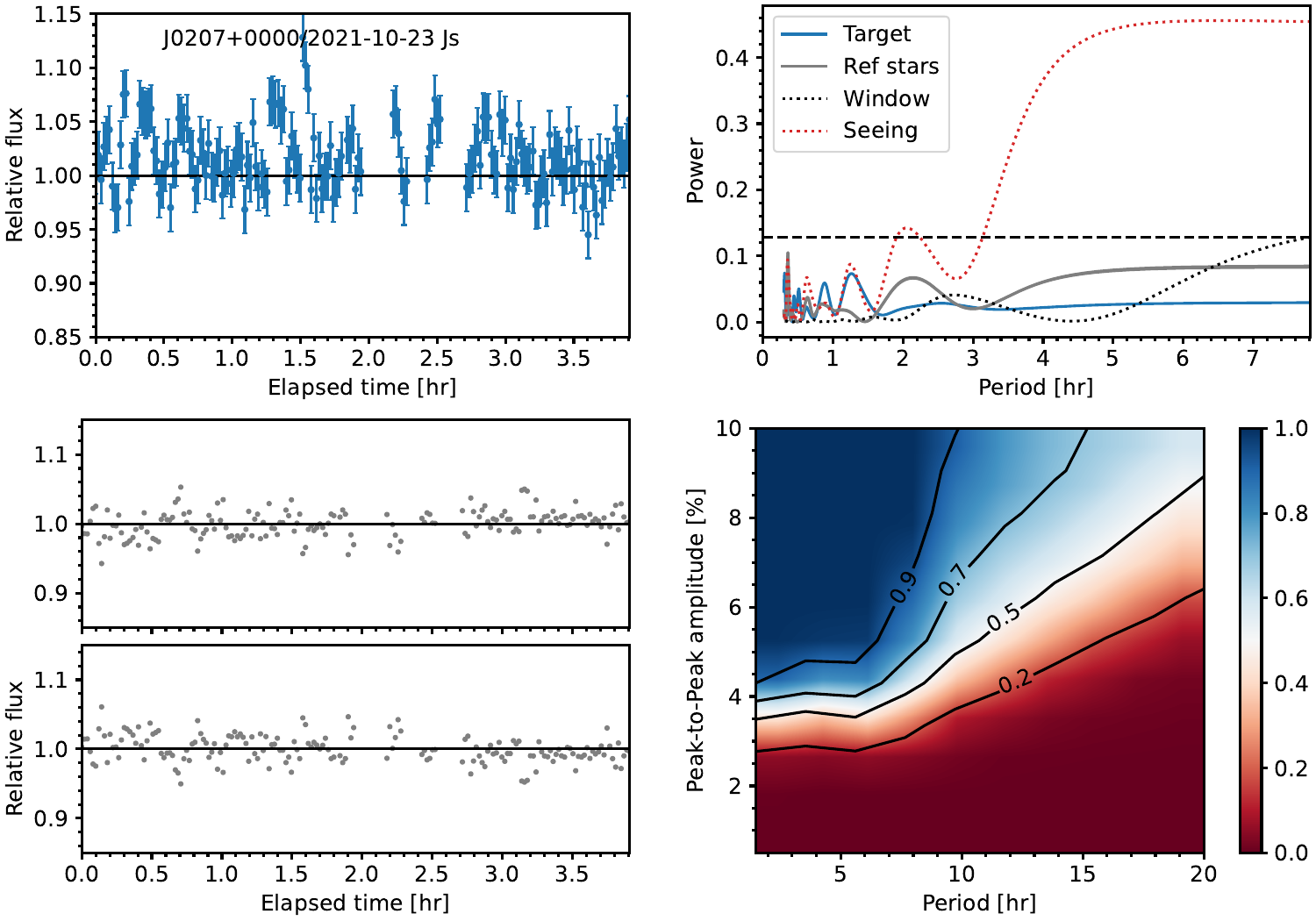}
    \includegraphics[width=1.8\columnwidth]{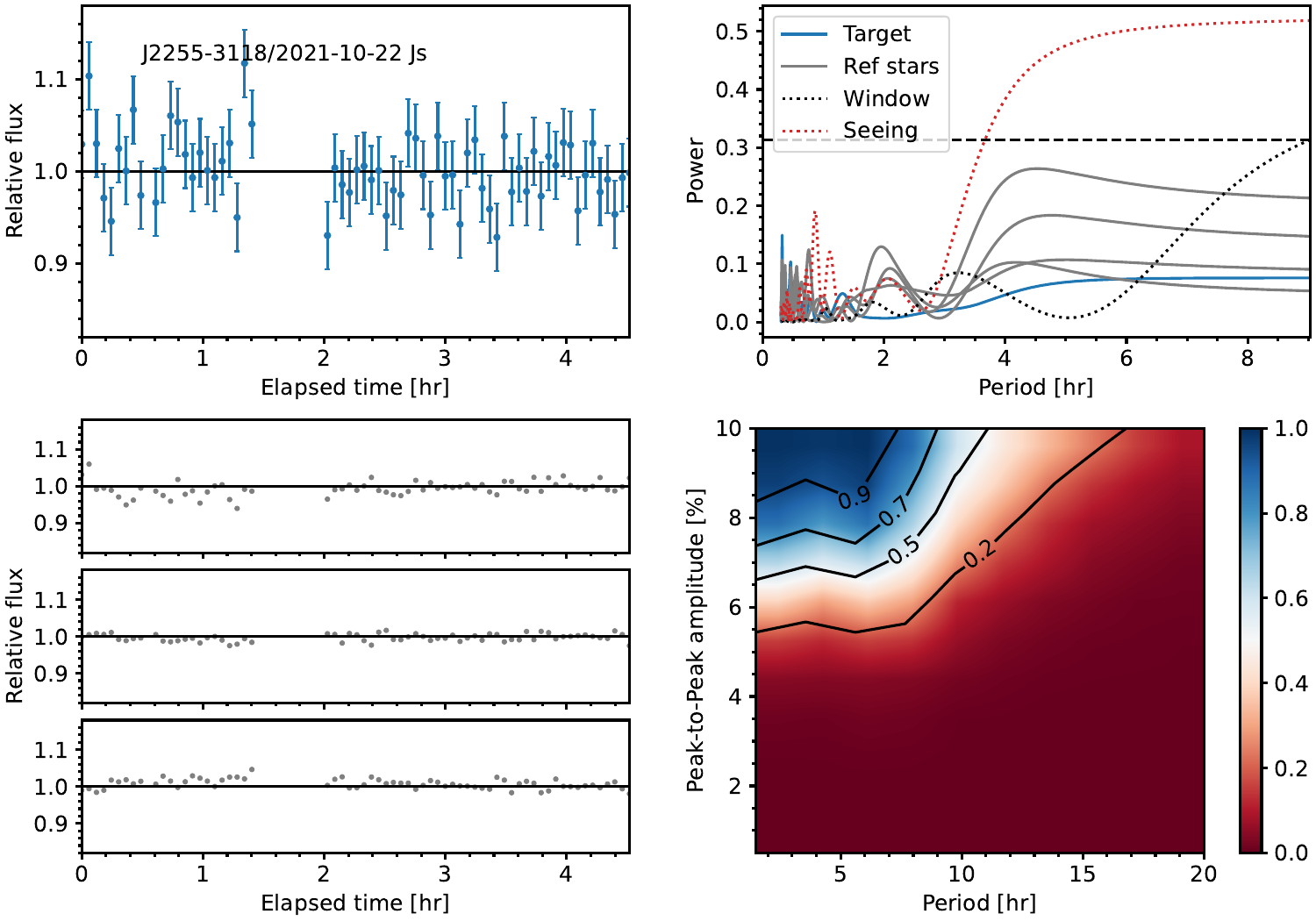}
    \caption{Light curves, periodograms, and sensitivity plots of non-variable objects J0207+0000 and J2255-3118, including detrended light curves and periodograms of their reference stars.}
    \label{fig:apd_J0207_J2255}
\end{figure*}

\begin{figure*}
    \centering
    \includegraphics[width=1.8\columnwidth]{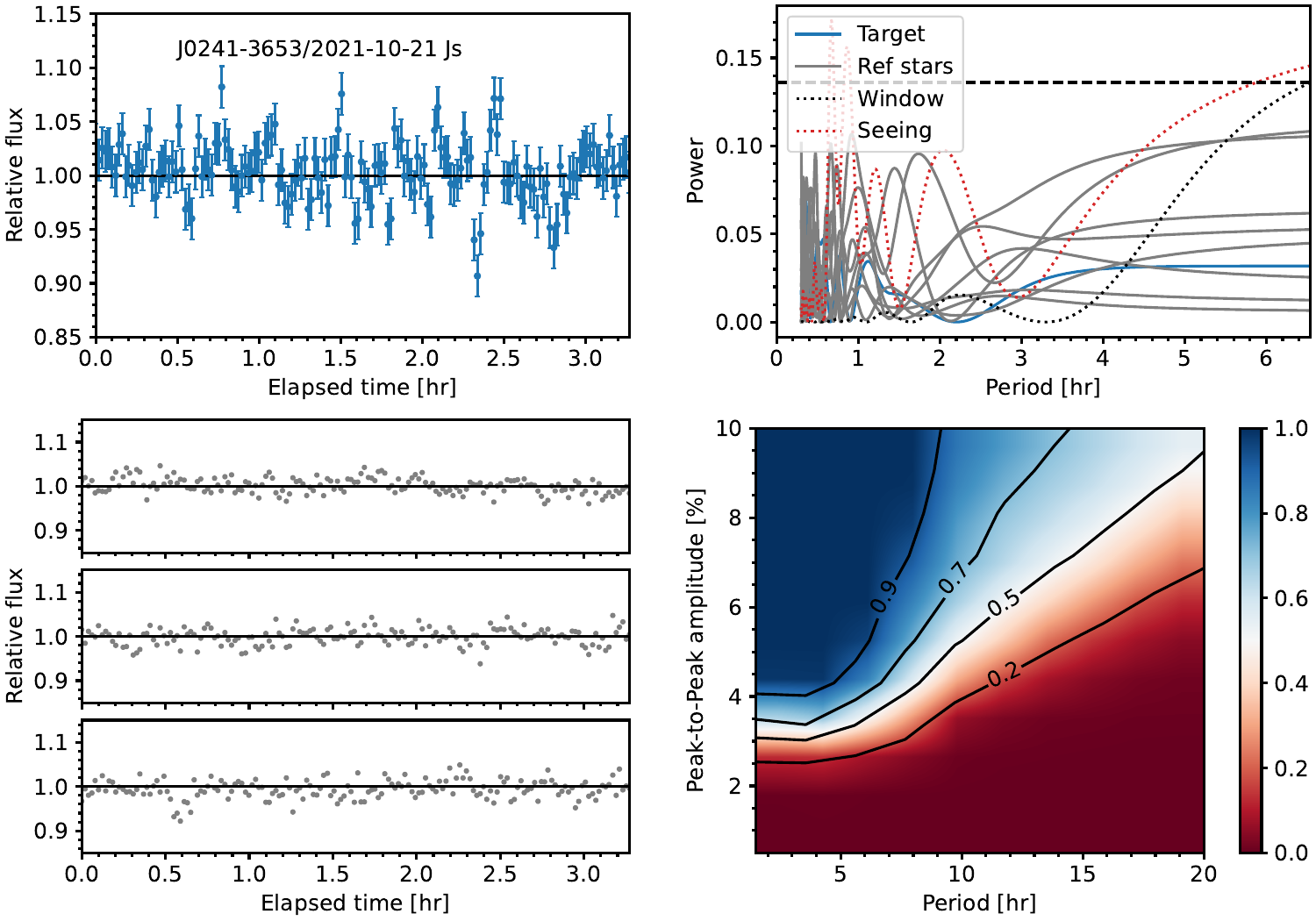}
    \includegraphics[width=1.8\columnwidth]{figures/appendix/J0241-3653_2021_10_21curves.pdf}
    \caption{Light curves, periodograms, and sensitivity plots of a non-variable object J0241-3653, including detrended light curves and periodograms of its reference stars.}
    \label{fig:apd_J0241}
\end{figure*}

\begin{figure*}
    \centering
    \includegraphics[width=1.8\columnwidth]{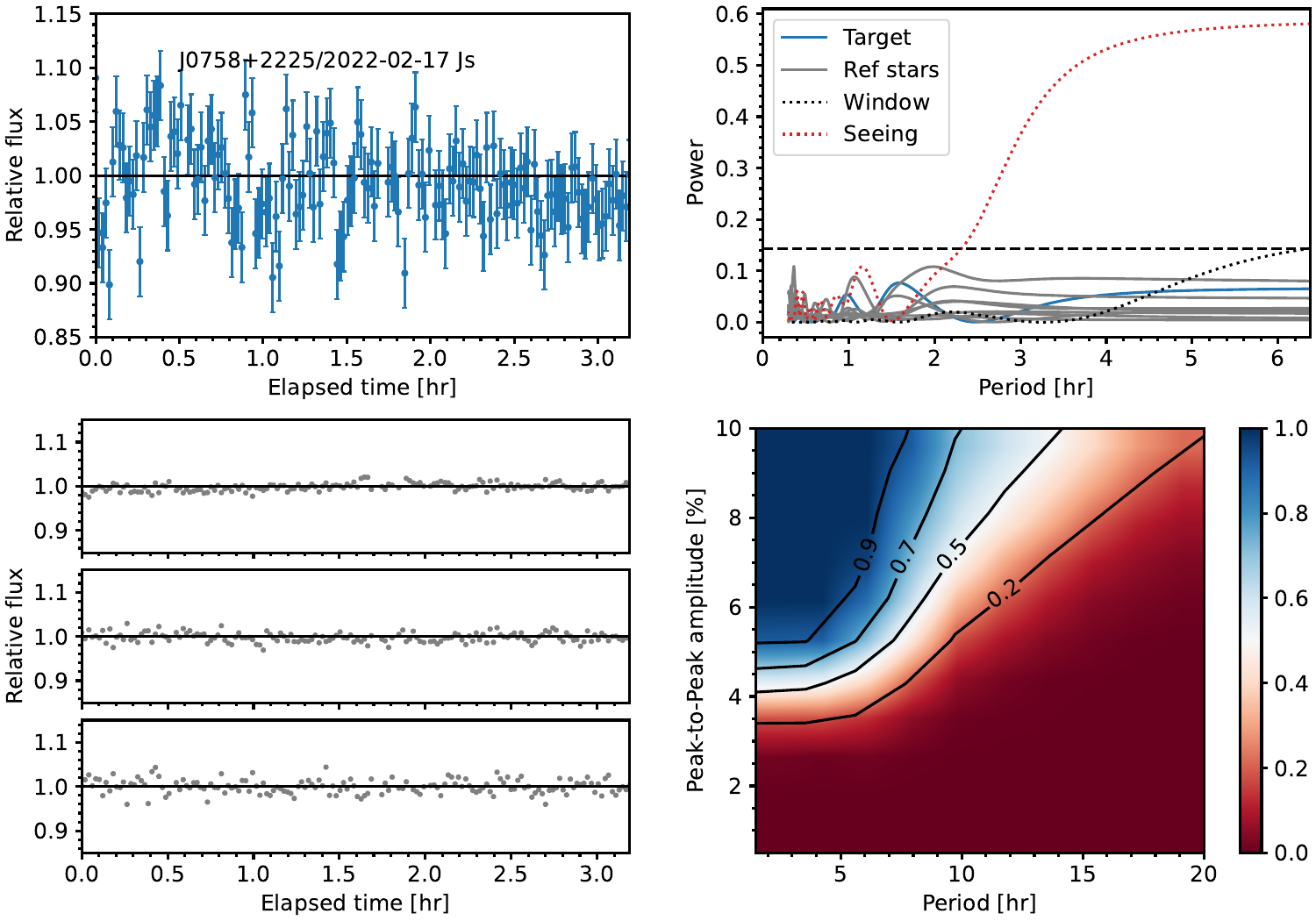}
    \includegraphics[width=1.8\columnwidth]{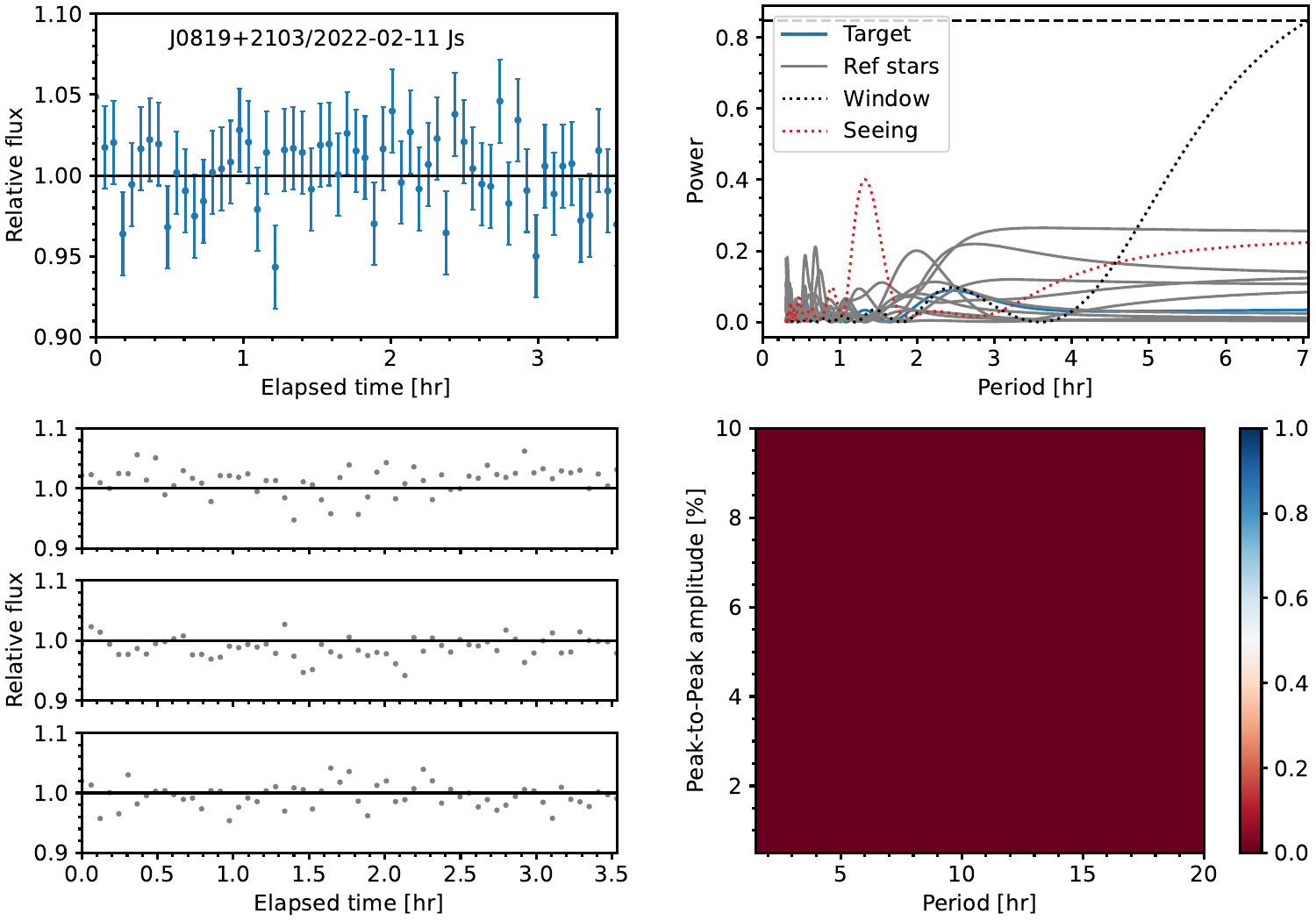}
    \caption{Light curves, periodograms, and sensitivity plots of non-variable objects J0758+2225 and J0819+2103, including detrended light curves and periodograms of their reference stars.}
    \label{fig:apd_J0758_J0819}
\end{figure*}

\begin{figure*}
    \centering
    \includegraphics[width=1.8\columnwidth]{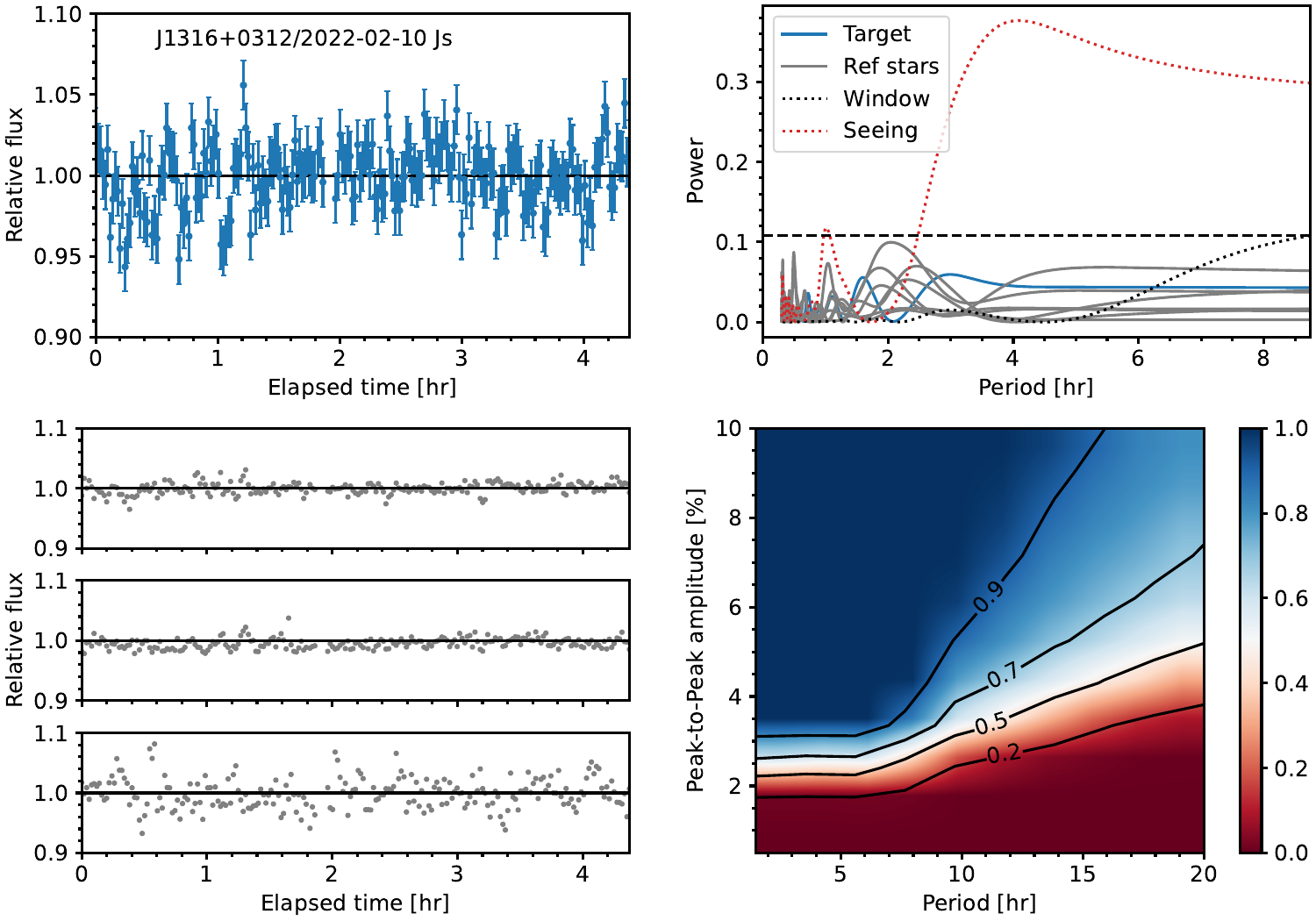}
    \includegraphics[width=1.8\columnwidth]{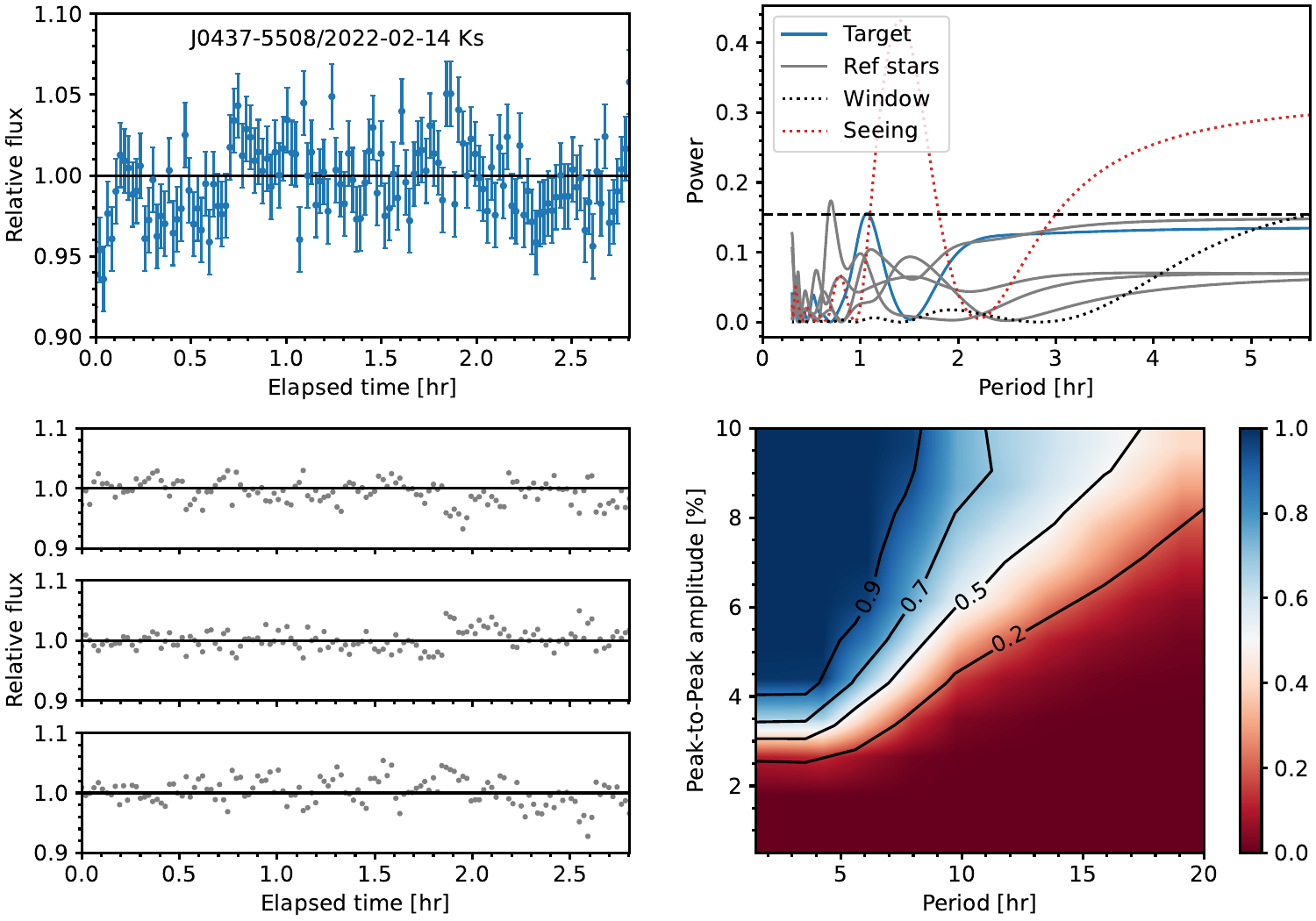}
    \caption{Light curves, periodograms, and sensitivity plots of non-variable objects J1316+0312 and J0437-5509, including detrended light curves and periodograms of their reference stars.}
    \label{fig:apd_J1316_J0437}
\end{figure*}

\begin{figure*}
    \centering
    \includegraphics[width=1.8\columnwidth]{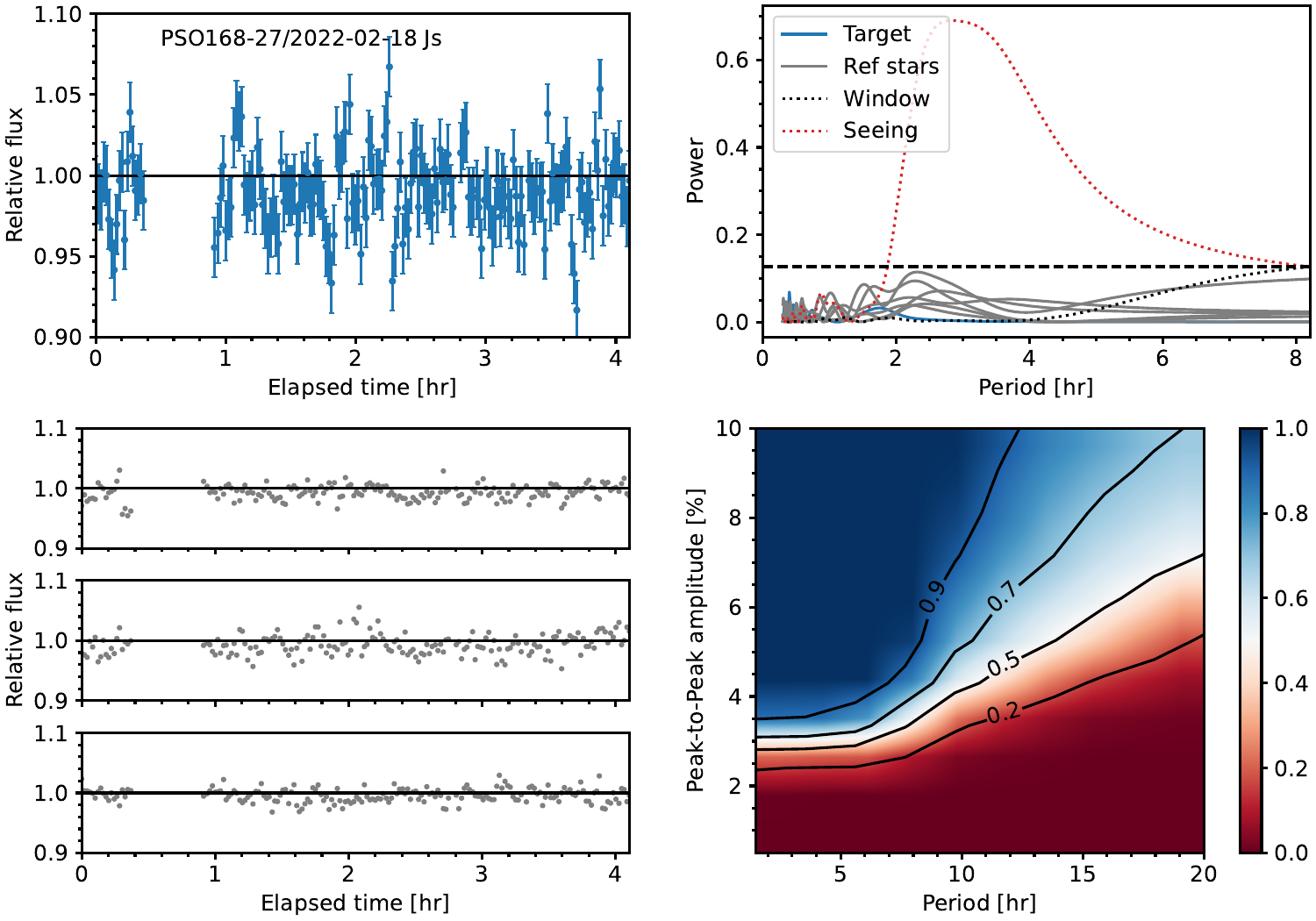}
    \includegraphics[width=1.8\columnwidth]{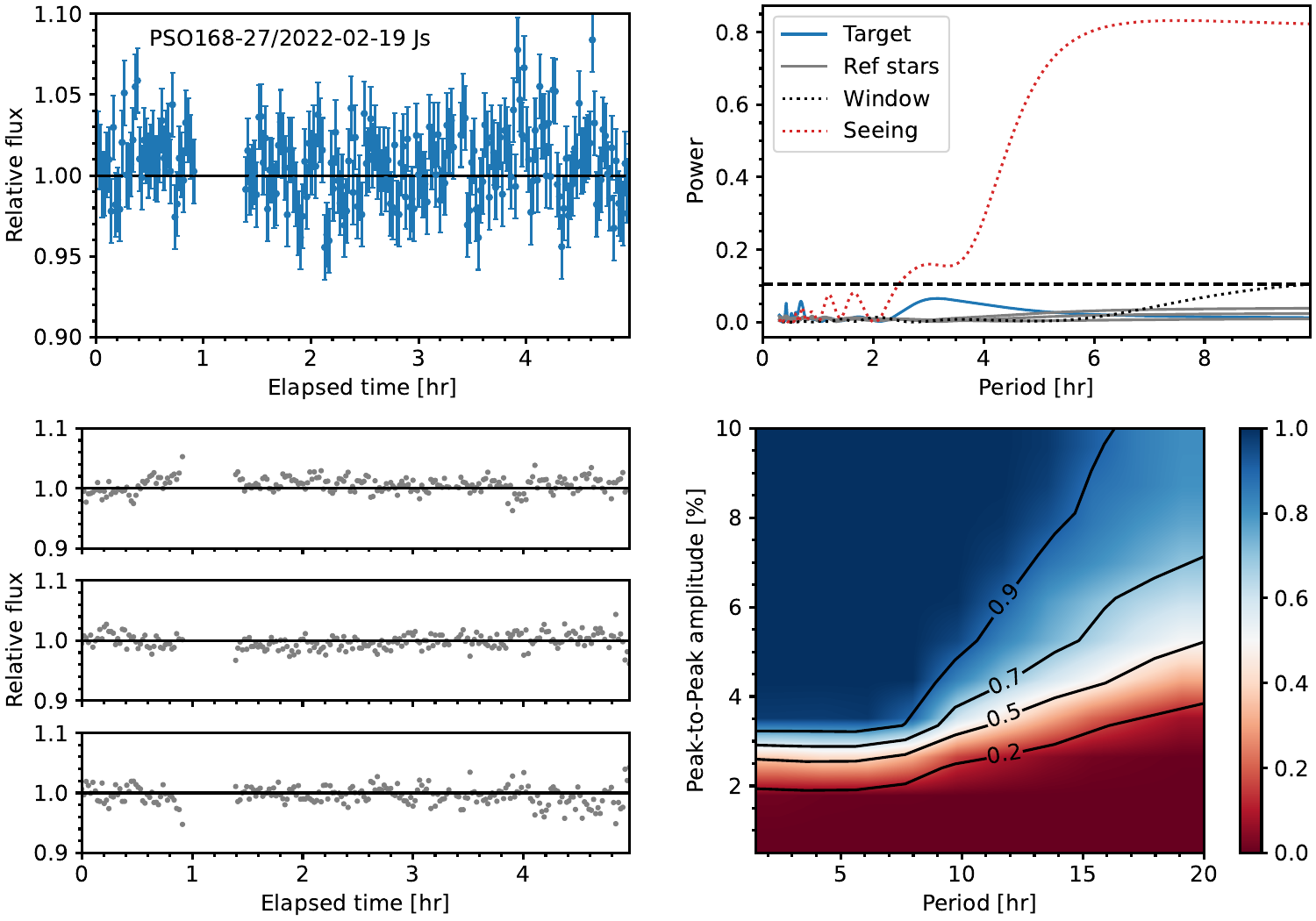}
    \caption{Light curves, periodograms, and sensitivity plots of a non-variable object PSO168-27, including detrended light curves and periodograms of its reference stars.}
    \label{fig:apd_PSO168}
\end{figure*}

\begin{figure*}
    \centering
    \includegraphics[width=1.8\columnwidth]{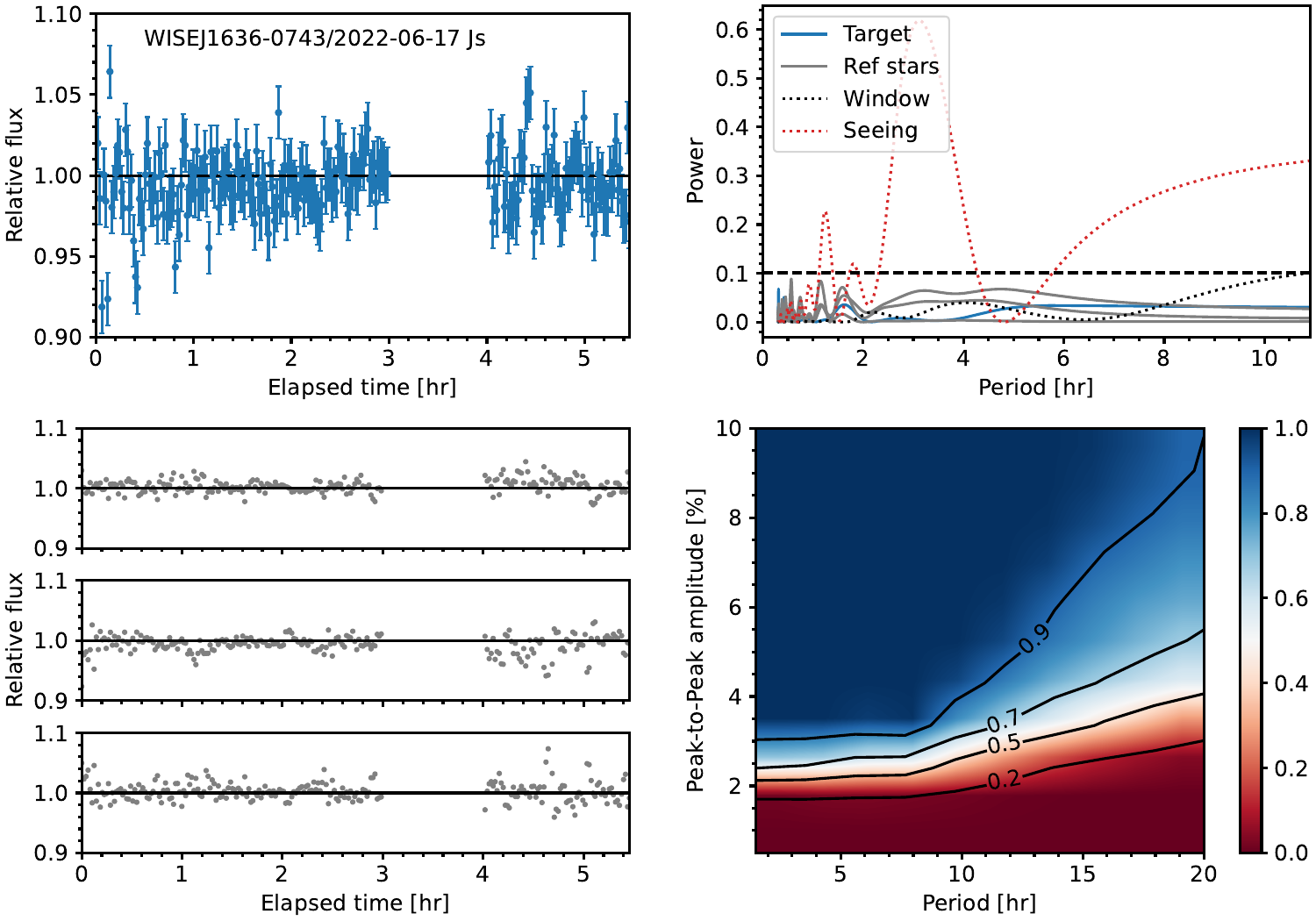}
    \includegraphics[width=1.8\columnwidth]{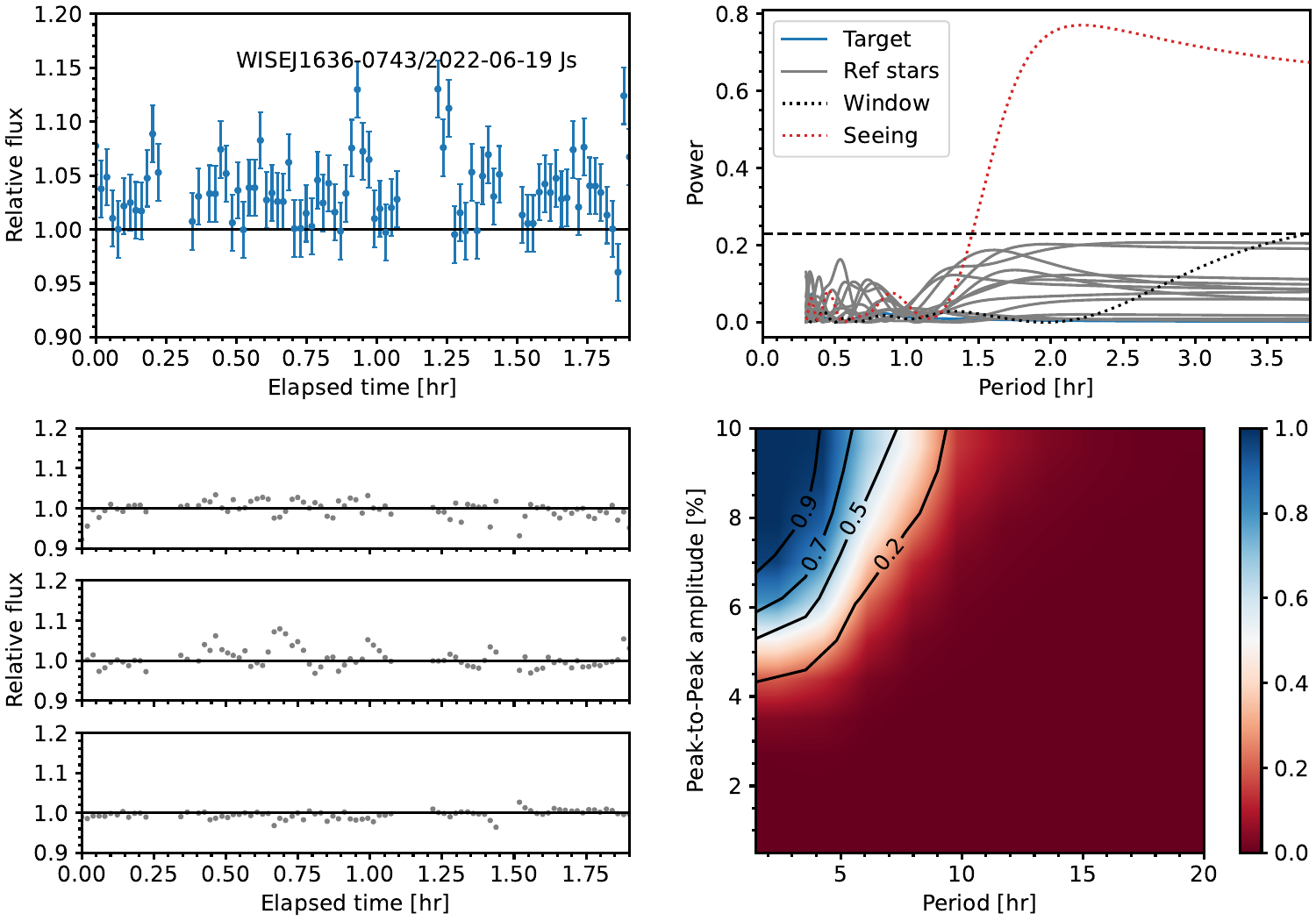}
    \caption{Light curves, periodograms, and sensitivity plots of a non-variable object WISEJ1636-0743, including detrended light curves and periodograms of its reference stars.}
    \label{fig:apd_J1636}
\end{figure*}

\begin{figure*}
    \centering
    \includegraphics[width=1.8\columnwidth]{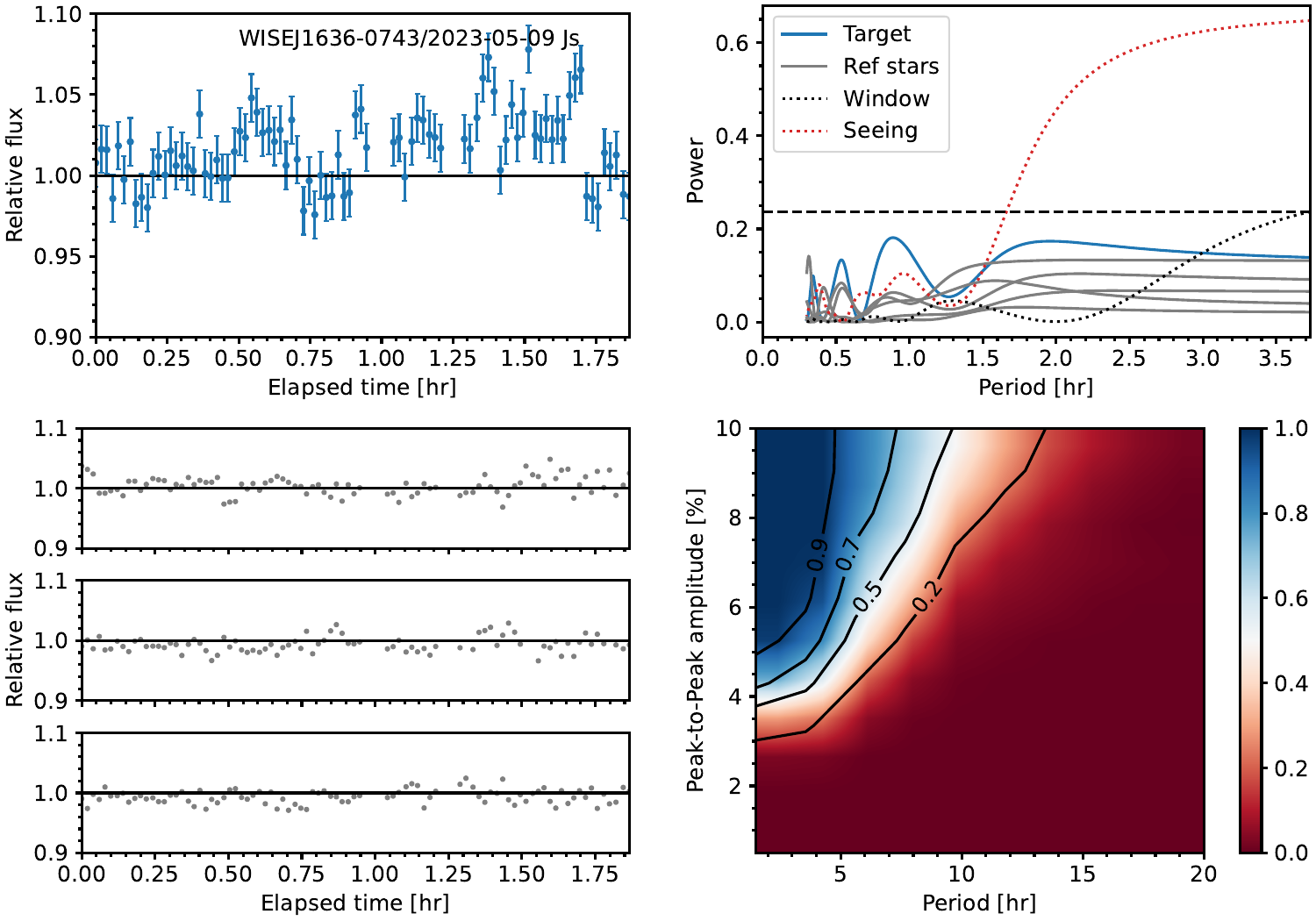}
    \includegraphics[width=1.8\columnwidth]{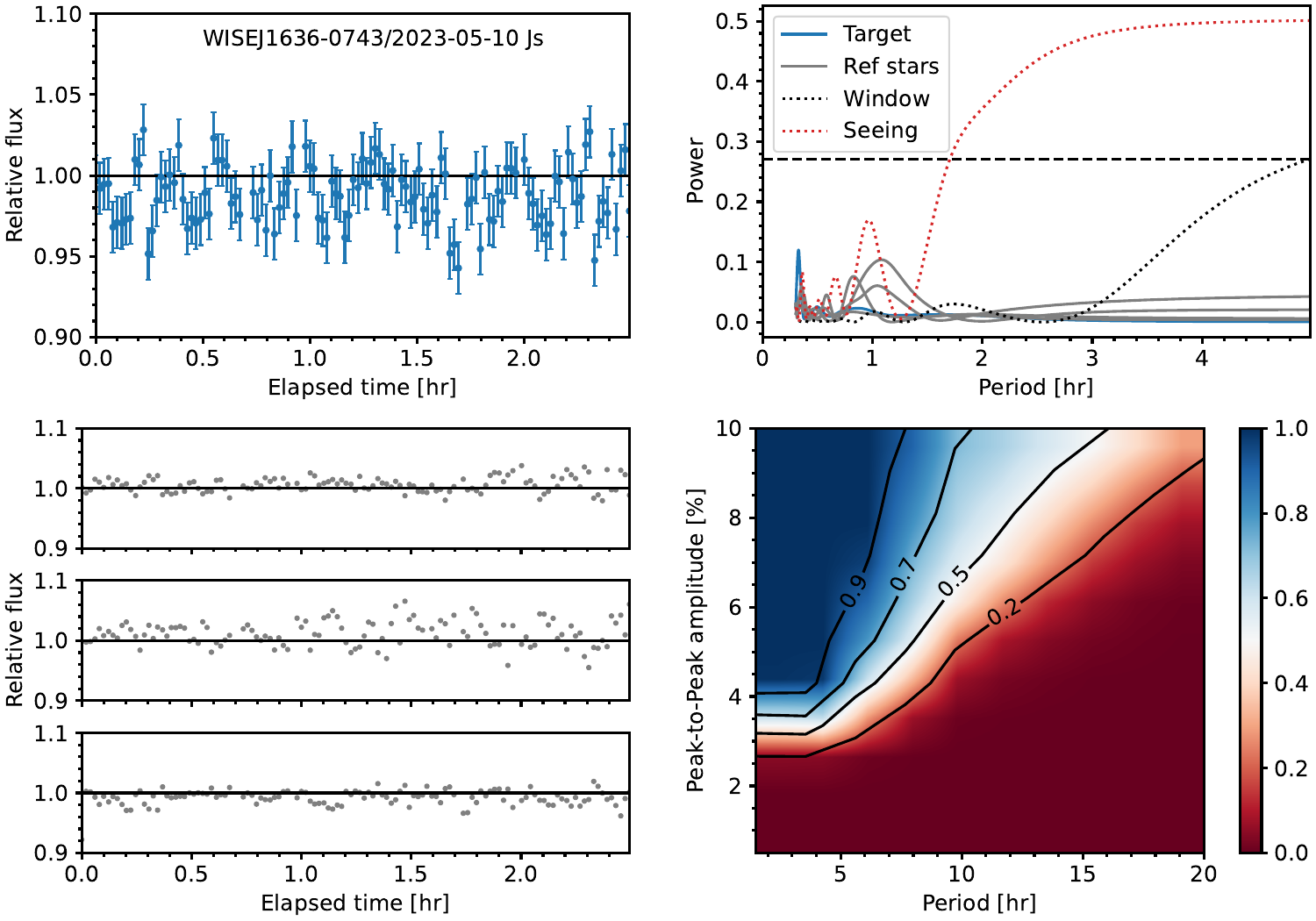}
    \contcaption{Light curves, periodograms, and sensitivity plots of a non-variable object WISEJ1636-0743, including detrended light curves and periodograms of its reference stars.}
    \label{fig:apd_J1636_2}
\end{figure*}

\begin{figure*}
    \centering
    \includegraphics[width=1.8\columnwidth]{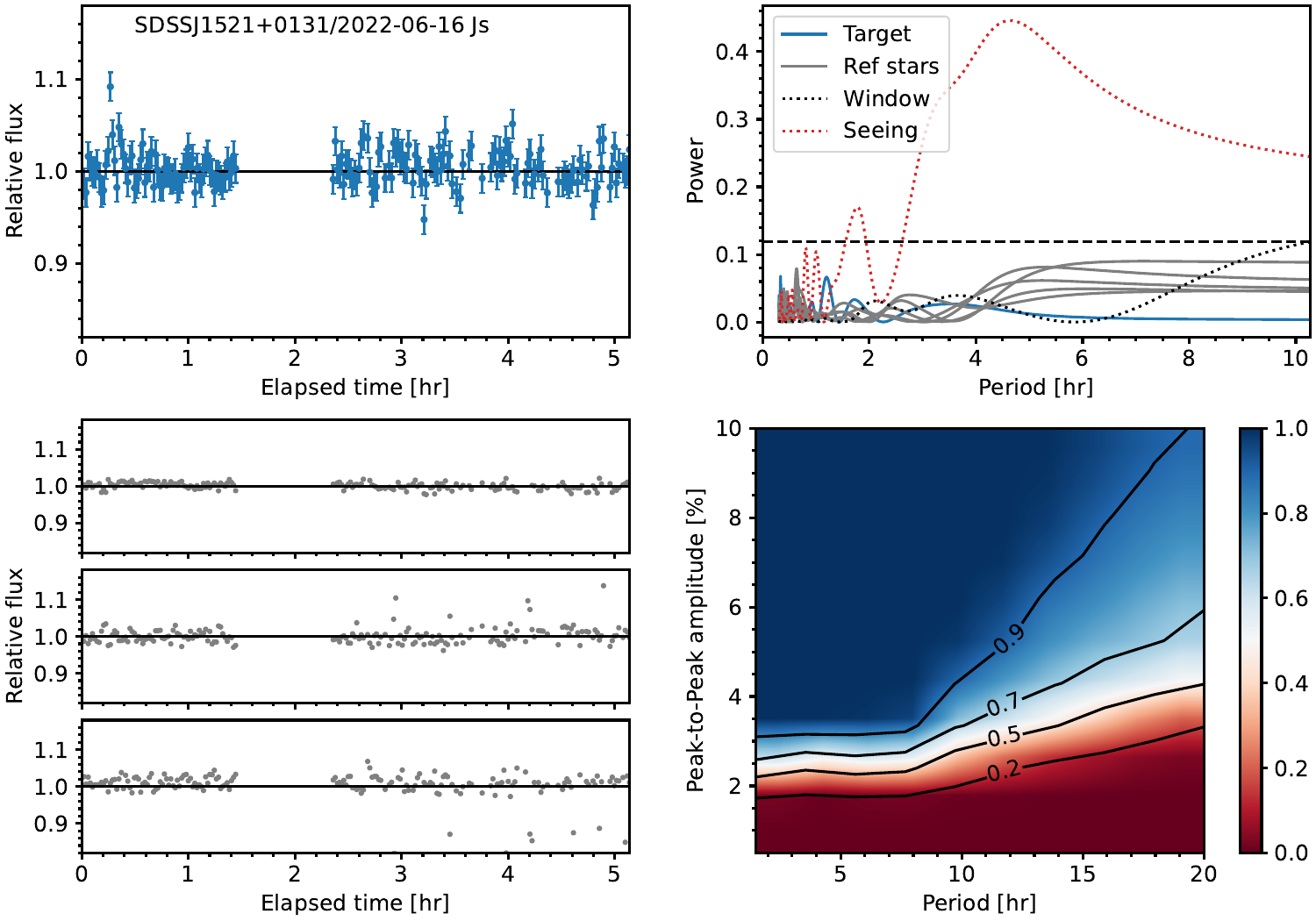}
    \includegraphics[width=1.8\columnwidth]{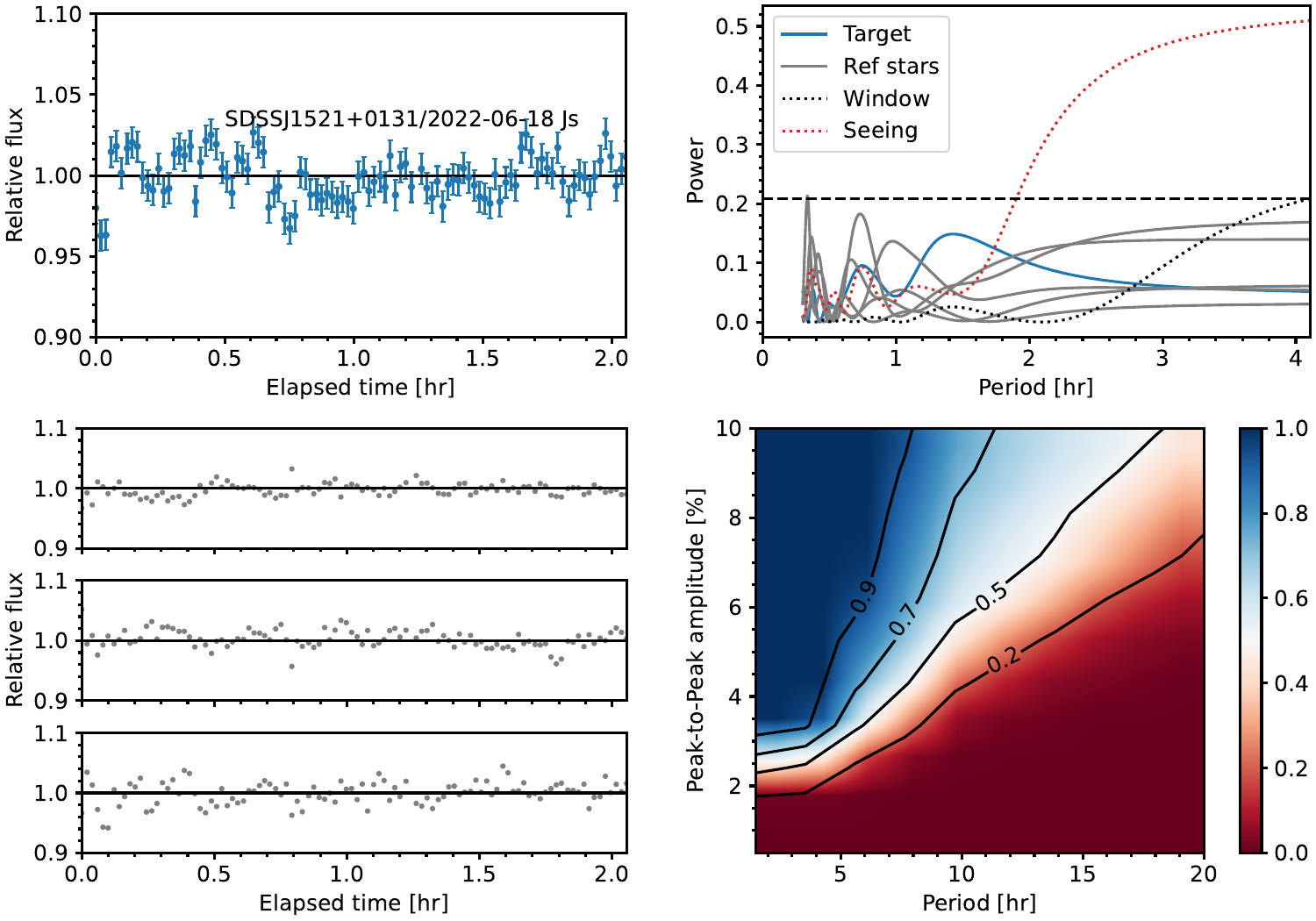}
    \caption{Light curves, periodograms, and sensitivity plots of a non-variable object SDSS1521+0131, including detrended light curves and periodograms of its reference stars.}
    \label{fig:apd_J1521}
\end{figure*}

\begin{figure*}
    \centering
    \includegraphics[width=1.8\columnwidth]{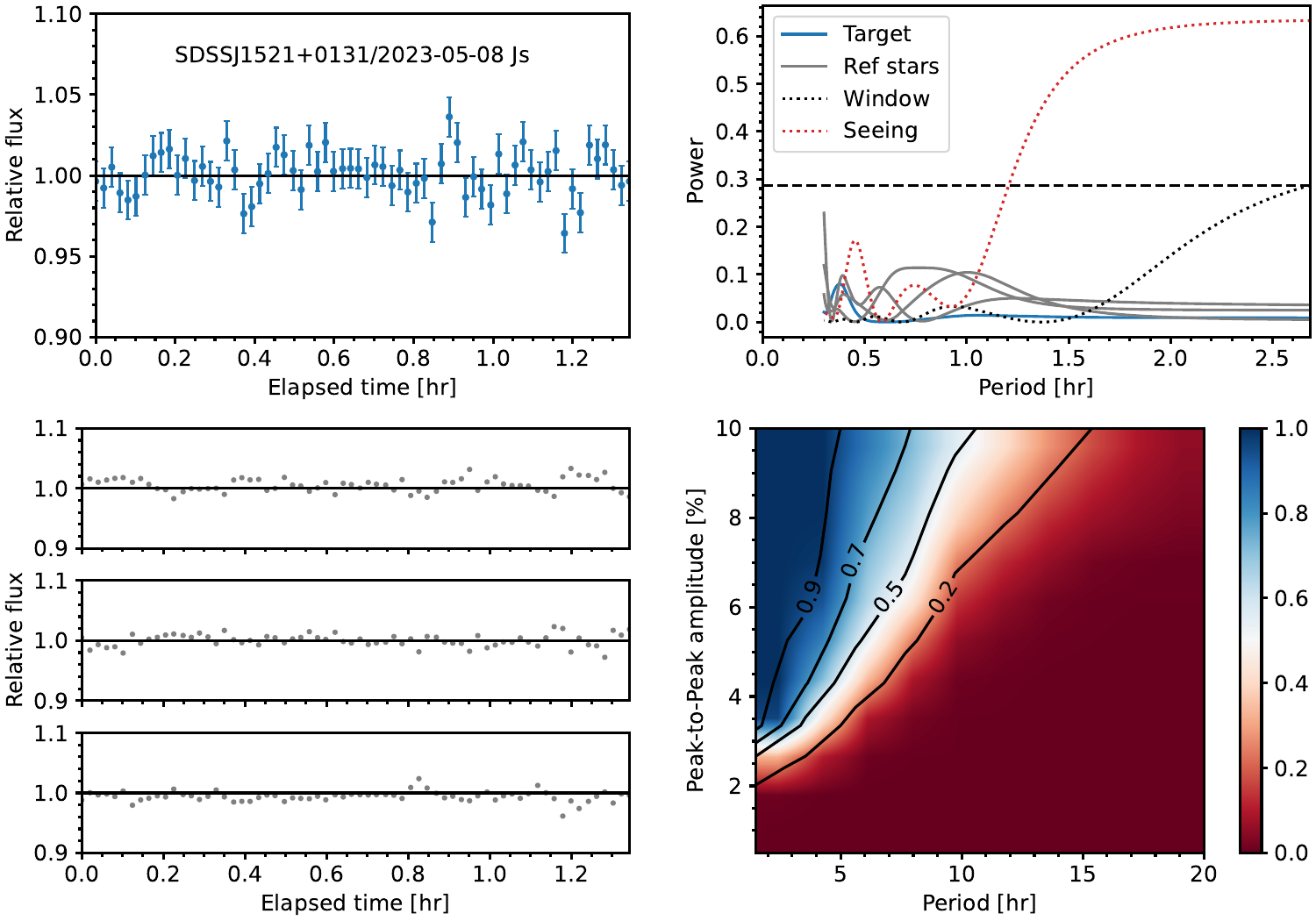}
    \contcaption{Light curves, periodograms, and sensitivity plots of a non-variable object SDSS1521+0131, including detrended light curves and periodograms of its reference stars.}
    \label{fig:apd_J15212}
\end{figure*}

%%%%%%%%%%%%%%%%%%%%%%%%%%%%%%%%%%%%%%%%%%%%%%%%%%

% Don't change these lines
\bsp	% typesetting comment
\label{lastpage}
\end{document}